\documentclass[12pt]{article}
\usepackage{amsmath}
\usepackage{mathtools} 
\usepackage[T1]{fontenc}
\usepackage{amssymb}
\usepackage{amsfonts}
\usepackage{amsthm}
\usepackage{mathrsfs}
\usepackage[all]{xy}
\usepackage{tensor}
\usepackage{comment}
\usepackage{bm}
\usepackage[normalem]{ulem}
\usepackage[dvipsnames]{xcolor}
\usepackage[colorlinks=true, citecolor=Blue, linkcolor=Blue]{hyperref}
\usepackage{ragged2e}
\usepackage{tikz}
\usepackage{xcolor}
\usetikzlibrary{calc}
\usetikzlibrary{patterns}
\usetikzlibrary{decorations.pathmorphing}
\usepackage{physics}
\usepackage{upgreek}
\usetikzlibrary{intersections,patterns}

\newcommand{\EternalBHBase}{%
	\coordinate (SfL) at (-2, 2); 
	\coordinate (SfR) at ( 2, 2); 
	\coordinate (SpL) at (-2,-2); 
	\coordinate (SpR) at ( 2,-2); 
	\coordinate (Bif) at ( 0, 0); 
	\coordinate (i0L) at (-4,0);  
	\coordinate (i0R) at ( 4,0);  
	
	\draw[decorate] (SfL) -- (SfR);
	\draw[decorate] (SpL) -- (SpR);
	
	\draw[dashed] (SpL) -- (Bif);
	\draw[dashed] (SpR) -- (Bif);
	
	\draw (i0L) -- (SfL);
	\draw (i0L) -- (SpL);
	\draw (SfR) -- (i0R);
	\draw (SpR) -- (i0R);
	
	\node[right] at ( 3.2, 1.4) {$\mathscr{I}^{+}$};
	\node[right] at ( 3.2,-1.4) {$\mathscr{I}^{-}$};
	
	\node[below] at (i0R) {$i^{0}$};
	
	\fill (Bif) circle (1.3pt);
	\node[below] at (0,-0.18) {$\mathscr{B}$};
}

\hypersetup{
	breaklinks=true,   
	colorlinks=true,   
}

\usepackage{diagbox}  
\usepackage{makecell} 
\usepackage{hhline} 
\usepackage{stackengine}

\usepackage{enumerate}

\usepackage[labelfont={small, bf, sf}, textfont={small,sf}]{caption}
\numberwithin{equation}{section} 
\usepackage{authblk}
\usepackage[in]{fullpage}
\usepackage{cite}

\newcommand{\beq}{\begin{equation}}
	\newcommand{\eeq}{\end{equation}}
\newcommand{\bes}{\begin{subequations}}
	\newcommand{\ees}{\end{subequations}}
\newcommand{\bea}{\begin{eqnarray}}
	\newcommand{\eea}{\end{eqnarray}}
\newcommand{\Eqref}[1]{Eq.~\eqref{#1}}

\newcommand{\be}{\begin{equation}}
	\newcommand{\ee}{\end{equation}}

\newcommand{\lie}{\pounds}

\newcommand{\M}{\widehat{\mathcal{M}}}

\newcommand{\Pnull}{\hat{\mathscr{P}}}
\newcommand{\Sgen}{\bar{S}_{\mathrm{gen}}}

\newcommand{\Srel}{S_{\mathrm{rel}}}
\newcommand{\HH}{\hat{\Omega}} 
\newcommand{\cc}{u}     
\newcommand{\D}{\Delta} 

\newcommand{\psihat}{\hat{\psi}}

\newtheorem*{theorem-non}{Theorem}

\newcommand{\hateq}{\mathrel{\mathop {\widehat=} }} 

\newcommand{\scri}{\mathscr{I}}
\newcommand{\red}[1]{\textcolor{red}{#1}}

\def\be{\begin{equation}}
	\def\ee{\end{equation}}

\newcount\hh
\newcount\mm
\mm=\time
\hh=\time
\divide\hh by 60
\divide\mm by 60
\multiply\mm by 60
\mm=-\mm
\advance\mm by \time
\def\hhmm{\number\hh:\ifnum\mm<10{}0\fi\number\mm}

\def\be{\begin{equation}}
	\def\ee{\end{equation}}

\newcommand{\EF}[1]{\textcolor{magenta}{\textsf{[EF: #1]}}}

\numberwithin{equation}{section}

\usepackage{cleveref}
\creflabelformat{equation}{(#2#1#3)}
\crefrangelabelformat{equation}{(#3#1#4)–(#5#2#6)}

\crefname{equation}{Eq.}{Eqs.}
\creflabelformat{equation}{#2#1#3}

\crefname{section}{Section}{Sections}
\crefname{appendix}{Appendix}{Appendices}
\crefname{figure}{Fig.}{Figs.}

\crefname{definition}{Def.}{Defs.}
\crefname{prop}{Prop.}{Props.}
\crefname{lemma}{Lemma}{Lemmas}
\crefname{corollary}{Cor.}{Cors.}
\crefname{thm}{Theorem}{Theorems}
\crefname{remark}{Remark}{Remarks}

\begin{document}
	
	\title{Subregion algebras in classical and quantum gravity}
	\date{\today}
	
	\author[1]{Venkatesa Chandrasekaran\thanks{venc@stanford.edu}}
	\author[2]{\'Eanna \'E. Flanagan\thanks{eef3@cornell.edu}}
	
	\affil[1]{\small \it Leinweber Institute for Theoretical Physics, Stanford University, Stanford, CA 94305, USA}
	\affil[2]{\small \it Department of Physics, Cornell University, Ithaca, NY, 14853, USA}

	\maketitle
	
	\begin{abstract}
		We study the kinematics and dynamics of subregion algebras in classical and perturbative quantum gravity associated with portions of null surfaces such as event horizons and finite causal diamonds. We construct half-sided supertranslation generators by extending subregion phase spaces of the event horizon to include doubled pairs of corner edge modes obtained from splitting the horizon, namely relative boosts and null translations of the respective corners. These edge modes carry a corner symplectic form and give rise to canonical charges generating half-sided boosts and translations. We show that the null translation generator is necessarily two-sided in the complementary translation edge modes. The charges act nontrivially on gravitationally dressed local observables on the horizon, such that the horizon subalgebra naturally takes the form of a crossed product by the associated automorphism group. 
        
        Quantizing the extended phase space after linearizing around a black hole background, we obtain for each horizon cut a Type II$_{\infty}$ von Neumann algebra equipped with a trace, whose von Neumann entropy coincides with the generalized entropy of that cut. The integrability of the half-sided null translation generator lifts to the existence of a self-adjoint operator that implements null time evolution on the Type~II$_\infty$ horizon subalgebras. The area operator is identified as the bulk implementation of the Connes cocycle flow for one-sided observables in excited states. The nesting property of the resulting one-parameter family of horizon subalgebras implies a generalized second law for non-stationary linearized perturbations of Killing horizons. Lastly, we use gravitational half-sided modular inclusion algebras to prove the quantum focusing conjecture in the perturbative quantum gravity regime.
	\end{abstract}
	
	\tableofcontents
	
	\section{Introduction}
	
	A central lesson of semiclassical gravity is that null surfaces behave similarly to ordinary thermodynamic systems \cite{Bardeen1973, Wald:1993nt}. They carry entropy and constrain the fundamental flow of information across spacetime through the generalized second law (GSL) and the quantum focusing conjecture (QFC) \cite{Hawking1975ParticleCreation, Bekenstein1973BlackHolesEntropy, Bekenstein1974GeneralizedSecondLaw, Wall:2011hj, BoussoEtAl2016QFC}. At the same time, from the viewpoint of local quantum field theory, null surfaces are natural places to localize subalgebras of observables and to study entanglement properties of QFT using modular theory \cite{Witten:2018zxz}. The goal of this paper is to unify these two perspectives in a single dynamical framework: we construct genuine gravitational subregion algebras in classical and quantum gravity associated to portions of null surfaces, and we show that their intrinsic algebraic properties encode generalized entropy, the GSL, and quantum focusing \cite{Bousso:2014sda, Bousso:2014uxa, Bousso:1999xy,BoussoEtAl2016QNECProof}.
	
	A further motivation for this work is conceptual: we would like to recast the abstract algebraic picture developed in previous work on large-$N$ algebras and generalized entropy into the more familiar language of canonical quantization and edge modes. In the large-$N$ story, the generalized entropy of a bifurcate Killing horizon was identified with the von Neumann entropy of a Type $\mathrm{II}_\infty$ factor obtained as a crossed product of a Type $\mathrm{III}_1$ algebra by its modular automorphism group \cite{Connes1973ClassificationTypeIII, Takesaki2003TOA2, Takesaki1973DualityCrossedProducts,Witten:2021unn}.\footnote{For a more recent comprehensive review of type classifications of von Neumann algebras see \cite{Sorce:2023fdx}.} While this gave a clean algebraic explanation of why generalized entropy behaves like an ordinary fine-grained von Neumann entropy, it was phrased largely in terms of the boundary theory and in abstract algebraic terms. One of the aims of the present paper is to build a bulk perturbative quantum gravity realization of that structure directly from the horizon phase space of gravity, wherein the central objects are canonical variables, symplectic forms, and edge modes at corners.
	
	In particular, on future event horizons $\mathscr{H}$ we can consider a series of cuts labeled by an affine parameter $u$.  For each horizon subregion of the form ${\mathscr{H}}_{>u}  = [u,\infty)\times \mathbb{S}^{d-2}$, we construct explicity the extended subregion phase space (\cref{sec:classicalalgebra} below), and from its canonical quantization we construct a crossed product algebra that we denote as $\widehat{\mathcal{A}}_{\mathscr{H}_{>u}}$ (\cref{sec:canonicalquant} below).  This algebra should be thought of as the horizon analogue of the large-$N$ algebra of \cite{Chandrasekaran:2022eqq}.
	
	The area operator, which previously entered only indirectly through the boundary ADM Hamiltonian in an abstract Type $\mathrm{II}_\infty$ factor, is here realized as a genuine corner charge $\hat{\mathscr{A}}$ conjugate to a boost edge mode (see \cref{sec:edgemodederiv} below), and its non-central action on ``bulk'' horizon observables arises from gravitational dressing  of subregions across a cut (see \cref{sec:classicalalgebra} below). In this way, the algebraic story of generalized entropy, Connes cocycle flow, and Type $\mathrm{II}_\infty$ traces is translated into the more geometric language of edge modes and canonical generators on null boundaries, making the connection between algebraic QFT and the horizon phase space of gravity manifest.
	
	From the semiclassical perspective, the Bekenstein--Hawking entropy,
	\begin{equation}
		S_{\rm BH} = \frac{\mathrm{Area}}{4G_N},
	\end{equation}
	is a purely coarse-grained quantity \cite{Hawking1975ParticleCreation, Bekenstein1973BlackHolesEntropy}. In principle, a full accounting of the corresponding microstates is provided by string theory, and in a few highly symmetric examples one can literally count them \cite{StromingerVafa1996MicroOrigin}. In general, however, such microscopic computations are not only extremely difficult but also outside the realm of describing the actual dynamics of microscopic degrees of freedom.
	
	The algebraic QFT approach allows us to paint a picture of the microscopic dynamics ``in between'' purely thermodynamic and fully microscopic, in the setting of perturbative quantum gravity. In previous work, the area term was reinterpreted as part of a fine-grained entropy: the von Neumann entropy of a Type $\mathrm{II}_\infty$ algebra describing a large-$N$ sector \cite{Chandrasekaran:2022eqq, Chandrasekaran:2022cip}. In that framework, the area contribution appears as the logarithm of the ``size'' of an infinite-dimensional algebra, made precise by the existence of a trace. The present paper builds on this perspective by constructing an explicit bulk realization of the relevant Type $\mathrm{II}_\infty$ horizon algebra in terms of canonical variables and edge modes, while also generalizing it to arbitrary subregions. 
	
	In a sense, the edge mode construction provides a hydrodynamic interpretation of black hole microstates. Specifically, we end up with an effective description of the UV theory controlled by charges/currents living at a horizon cut + bulk quantum fields, such that their crossed product algebra controls the generalized entropy. In this way, the area term is promoted from a coarse-grained quantity in a purely thermodynamic object to a genuine fine-grained entropy of a horizon subalgebra, with a concrete quantum mechanical interpretation in terms of gravitational edge modes.

	We next turn to a discussion of how to define subregion phase spaces in gravity. The basic problem is familiar, and two-fold. In ordinary QFT on a fixed background, a spacetime region $\mathcal{U}$ comes with a von Neumann algebra $\mathcal{A}(\mathcal{U})$ of local observables, and many properties of energy and entanglement can be phrased purely in terms of modular theory applied to this algebra. In gravity, however, diffeomorphism invariance makes ``the region'' itself dynamical: specifying a subregion requires gravitational dressing, and observables must be constructed so that they commute with constraints. This makes it nontrivial to even \emph{define} the algebra of observables associated to, say, the portion of an event horizon to the future of a cut.
	
	But ostensibly there's an even more non-trivial issue. Even if we could define such subregion algebras, can we construct the half-sided boost and translation generators needed to describe the dynamics of the subregion under relational time evolution? The reason this is non-trivial is that the subregion is an open subsystem, i.e. excitations can enter or leave the subregion under time evolution. Normally in such a setting one cannot integrate up Hamiltonian vector fields on phase space to get symmetry generators that act non-perturbatively\footnote{By ``non-perturbatively'' we just mean beyond linearized order in the flow parameter of the half-sided boost/translation. So for example if the perturbation is in powers of $G_N$, we want to be able to flow at least an $\mathcal{O}(1)$ amount in $G_N$ counting.} on all states, due to explicit time dependence \cite{WZ, CFP, Freidel:2021dxw, Ciambelli2019b, Ciambelli:2021nmv}.
    
	Relatedly, previous work \cite{Chandrasekaran:2022eqq} has shown that for large-$N$ theories, or in perturbative quantum gravity, one can recover a Type $\mathrm{II}_\infty$ ``large-$N$'' algebra whose von Neumann entropy agrees with the generalized entropy of a bifurcate Killing horizon. But these constructions are typically tied to special backgrounds (stationary black holes, de Sitter space) and to global horizons. What is missing is a quasi-local, dynamical picture of gravitational subregions that: (i) works directly on subregion phase spaces of the event horizon, including non-stationary configurations, (ii) identifies a canonical set of edge modes and symmetry generators that implement half-sided boosts and half-sided null translations on all states in the subregion phase space, even on excited states, and (iii) produces a family of Type $\mathrm{II}_\infty$ algebras associated with arbitrary cuts of the horizon upon quantization, with a natural trace and von Neumann entropy.
	
	This paper develops such a framework. At a high level, we show that if one treats gravity dynamically and keeps careful track of edge modes at the corners of a gravitational subregion, then:
	\begin{itemize}
		\item Classically, horizon subregions admit a crossed product phase space algebra generated by local, gravitationally dressed observables and a pair of canonical corner charges $(\hat{\mathscr{A}}, \hat{\mathscr{P}})$ that generate half-sided boosts and half-sided translations along the horizon.
		\item Upon quantization, the corresponding crossed product von Neumann algebra at each cut $u$ of the horizon is a Type $\mathrm{II}_\infty$ factor $\widehat{\mathcal{M}}_{\mathscr{H}_{>u}}$ equipped with a natural trace and a von Neumann entropy
		\begin{equation}
			S(\hat\psi; \widehat{\mathcal{M}}_{\mathscr{H}_{>u}}) = -\,\mathrm{tr}\!\left[\rho_{\hat\psi}(u)\log\rho_{\hat\psi}(u)\right].
		\end{equation}
		\item In perturbative quantum gravity this entropy coincides, up to a state-independent constant and a small smearing in the cut location, with the generalized entropy $S_{\rm gen}(u)$ of the horizon at that cut.
		\item The GSL follows from nesting properties of the subregion algebras, and quantum focusing follow from a gravitational analogue of the half-sided modular inclusion property.
	\end{itemize}
	
	At a high level, our framework builds off the fundamental question of how to carve out a gravitational subregion and assign it a phase space of its own.\footnote{Of course, this is a question that has been studied extensively. See for example \cite{Donnelly:2016auv, Speranza:2017gxd, Carrozza:2022xut, Pulakkat:2025eid, Ciambelli:2021nmv}. But our construction is different in several ways: (1) we use the edge modes to obtain integrable (surface deforming) symmetry generators that act non-trivially on non-stationary spacetimes; (2) we carry out canonical quantization of this construction; (3) we marry the covariant phase space formalism for gravitational subregions with the algebraic QFT formalism for subregion algebras and entropies.} We begin with a null surface, which we split into a subregion and its complement at a corner. In gravity, this split cannot be done trivially, because the gravitational constraints must still be satisfied across the corner. To implement the split consistently, we smear the corner into a thin “Cauchy splitting region” that thickens it into a short tube, with two nearby cuts that separate the subregion from its complement. On these cuts we then introduce gravitational edge modes, which keep track of the relative boost angles at the respective corners and shifts in the affine parameter locations of the corners, and thereby capture the way in which the constraints fail to factorize strictly at the corner.
	
	The phase space associated to the subregion therefore includes not just the naive ``bulk'' degrees of freedom but also the edge modes on the boundary of the subregion.  The symplectic form contains then both bulk and corner terms from the edge modes. In the standard covariant phase space construction these corner terms are effectively omitted, and this omission shows up as non-integrability of symmetry generators in generic, non-stationary configurations. Once the edge modes are included, the subregion can be isolated algebraically and a set of symmetry generators conjugate to the corner data can be found that are fully integrable while still acting non-trivially on phase space observables. In this description, half-sided flows move the subregion relative to its complement by transforming the edge modes on the future side of the corner, while keeping the bulk fields smooth across the split and holding the past corner data fixed.
	
	The rest of this introduction spells out this construction and its implications in more detail.

    \subsection{Classical gravity: edge modes and subregion phase spaces}

    \subsubsection*{Horizon phase space and half-sided supertranslations}

    We begin with a covariant phase space description of gravity restricted to a future event horizon $\mathscr{H}$.\footnote{See \cite{Harlow:2019yfa, CFP, WZ, Chandrasekaran:2021vyu, Speranza:2017gxd, Crnkovic:1987tz, LeeWald1990} for comprehensive expositions of the relevant formalism.} For the purposes of this introduction, the only structural input we will use is that the horizon degrees of freedom can be organized into canonical pairs of configuration space variables $\bm{\Psi}$ and conjugate momenta $\dot{\Psi}$ intrinsic to $\mathscr{H}$, so that the horizon symplectic form has the canonical form
    \begin{equation}
    	\Omega_{\mathscr{H}}
    	= \int_{\mathscr{H}} \delta \bm{\Psi}\wedge \delta \dot{\Psi}.
    	\label{eq:intro-symplectic}
    \end{equation}
    where $\wedge$ denotes the phase space wedge product.
    
    Fix a parameter $u$ along the null generator $\ell^a$ and a cut $S_0$ at $u=u_0$.  A supertranslation along the horizon is generated by a vector field of the form $\xi^a=f\ell^a$, which we decompose into an angle-dependent translation plus an angle-dependent boost:
    \[
    	f(u,x^A)=\alpha(x^A)+u\,\beta(x^A).
    \]
    To describe a subregion $\mathscr{H}_{>u_0}$, we consider the corresponding half-sided (or truncated) phase space action of the supertranslation which is turned on only to the future of the cut, denoted $\hat{\xi}_T$. Its action on the horizon data is
    \begin{equation}
    	\mathfrak{i}_{\hat{\xi}_T}\,\delta \bm{\Psi}
    	=
    	\bigl(\mathfrak{L}_{\hat{\xi}} \bm{\Psi}\bigr)\,H(u-u_0),
    	\label{eq:intro-half-sided-action}
    \end{equation}
    with $H(u-u_0)$ the Heaviside function, $\mathfrak{i}_{\hat{\xi}_T}$ the contraction map on phase space, and $\mathfrak{L}_{\hat{\xi}}$ the phase space Lie derivative along the full phase space vector field $\hat{\xi}$.
    
    A key point is that when one contracts the full horizon symplectic form with a half-sided flow on phase space, one generically produces (i) a would be generator supported on $\mathscr{H}_{>u_0}$ plus (ii) an additional corner term at $S_0$ that measures the failure of strict factorization at the corner.  Explicitly,
    \begin{equation}
    	-\mathfrak{i}_{\hat{\xi}_T}\Omega_{\mathscr{H}} =
    	\delta \mathcal{Q}_{\xi} +
    	\int_{S_0}
    	\left(i_{\xi}\bm{\mathcal{E}} -\delta \bm{\Psi}\,\star \mathfrak{L}_{\hat{\xi}}\Psi\right),
    	\label{eq:intro-half-sided-omega}
    \end{equation}
    where the first term should be thought of as the ``expected'' boundary variation of a symmetry generator associated to $\mathscr{H}_{>u_0}$, while the second term contains the usual flux obstruction term $i_{\xi}\bm{\mathcal{E}}$ localized at the cut. The obstruction term cancels off against the additional delta function term whenever the flux can be put in Dirichlet form.\footnote{See \cite{Chandrasekaran:2020wwn, Chandrasekaran:2021vyu, Chandrasekaran:2021hxc} for detailed discussions of the significance of the Dirichlet form of the flux term.}
    
    The overall lesson is simple: to obtain integrable half-sided boost/translation generators for non-stationary event horizons, one must properly account for the corner degrees of freedom that appear when the horizon is split into a subregion and its complement.\footnote{Standard covariant phase space approaches such as \cite{Iyer:1994ys, WZ, CFP} instead use a truncated symplectic form, which ends up missing the additional corner term in \Eqref{eq:intro-half-sided-omega} that cancels off the obstruction term. So they are not able to get actual symmetry generators on phase space, but rather just plain corner charges.}
    
    \subsubsection*{Corner symplectic form and horizon edge modes}
    
    The cancellation in \Eqref{eq:intro-half-sided-omega} is hinting at what the extended subregion phase space must contain.  Splitting $\mathscr{H}$ across a cut does not strictly factorize the gravitational data: the constraints couple the two sides, and the missing relational information is localized at the corner.
    
     We make this precise by introducing a thin Cauchy splitting region $G_\varepsilon$ with boundaries $S_0^\pm\simeq\partial G_\varepsilon$, so that
    \[
    	\mathscr{H}=\mathscr{H}_-\cup G_\varepsilon \cup \mathscr{H}_+.
    \]
    In GR the needed corner degrees of freedom can be taken to be
    (i) relative boost angles $\Gamma_0^\pm$ at $S_0^\pm$, and
    (ii) independent affine shifts $\Upsilon_0^\pm$ of the two corners along the generators.
    These edge modes carry a nontrivial corner symplectic form, which we derive from first principles. The result is 
    \begin{align}
    \Omega_{\partial G} = \ &\frac{1}{8\pi}\int_{S_0} \left[ \delta \Upsilon_0^+
          \wedge \delta (\lie_\ell {\bm \mu}) - \delta \Gamma_0^+ \wedge
          \delta \Delta \bm{\mu}_+  + \delta \Upsilon_0^+ \wedge
          \delta \Gamma_0^+ \Theta \bm{\mu}  \right] - (+ \leftrightarrow -) \nonumber
        \\ &+ \frac{1}{8\pi}\int_{S_0} \left[ 
          \delta \Upsilon_0^{+} \wedge \delta \Upsilon_0^{-} \lie_\ell(\bm{\mu}\Theta) \right],
    	\label{eq:intro-corner-symplectic}
    \end{align}
    (with $\Delta\bm{\mu}_\pm$ a background-subtracted area element).  The last term is the key new feature: it couples the complementary translation edge modes and encodes the fact that the null constraints glue $\mathscr{H}_+$ and $\mathscr{H}_-$ across the split.  In particular, the half-sided null translation generator is necessarily two-sided in $(\Upsilon_0^+,\Upsilon_0^-)$.
    
    Including \Eqref{eq:intro-corner-symplectic} in the subregion symplectic form makes the half-sided generators integrable on the extended phase space.  The resulting canonical corner charges can be written as
    \begin{subequations}
    \begin{align}
    	\mathscr{A}_\beta
    	&= \frac{1}{8\pi}\left[\int_{S_0^+}\beta\,\bm{\mu}-\int_{\infty}\beta\,\bm{\mu}\right],
    	\\
    	\mathscr{P}_\alpha
    	&= -\frac{1}{8\pi}\int_{S_0^+}\alpha\,e^{\Gamma_0^+}
    	\Bigl[\lie_{\ell}\bm{\mu}-(\Upsilon_0^+-\Upsilon_0^-)\lie_{\ell}(\bm{\mu}\Theta)\Bigr],
    	\label{eq:intro-A-P-def}
    \end{align}
        \end{subequations}
    where $\mathscr{A}_\beta$ is the (angle-dependent) area operator and $\mathscr{P}_\alpha$ generates half-sided null translations of the subregion relative to its complement.

    \subsubsection*{Crossed product algebras and gravitational dressing}
    
    With the corner data having been made dynamical, locality of ``bulk'' observables has to be relational. So in order to define a genuine horizon subregion algebra, we dress operators to the edge modes.  Concretely, we anchor points in $\mathscr{H}_{>u_0}$ to the future corner $S_0^+$ by flowing along the generators,
    \begin{equation}
    	p=\exp(u\ell)p_0, \ p_0\in S_0^+,
    	\label{eq:intro-dressing}
    \end{equation}
    and let $\mathcal{A}_{\mathscr{H}_{>u_0}}$ denote the resulting algebra of dressed local observables $\mathscr{O}(p)$ supported in $\mathscr{H}_{>u_0}$.
    
    Because the dressing depends on the edge modes, the corner charges act nontrivially on dressed ``bulk'' observables.  In particular, their Poisson brackets take the geometric form
    \begin{subequations}
    \begin{align}
    	\{ \mathscr{P}_\alpha, \mathscr{O}(p)\}
    	&= -\alpha e^{\Gamma_0^+}\mathfrak{L}_{\hat{\ell}} \mathscr{O}(p),
    	\label{eq:intro-P-bracket}\\
    	\{ \mathscr{A}_\beta, \mathscr{O}(p)\}
    	&= -(u-u_0)\beta\mathfrak{L}_{\hat{\ell}} \mathscr{O}(p).
    	\label{eq:intro-A-bracket}
    \end{align}
        \end{subequations}
    So $\mathscr{P}_\alpha$ and $\mathscr{A}_\beta$ generate outer automorphisms of $\mathcal{A}_{\mathscr{H}_{>u_0}}$, and (crucially) the area operator is not central: it fails to commute with dressed local observables precisely because it acts on their dressing.
    
    This means the natural classical horizon subalgebra is a crossed product between the dressed ``bulk'' algebra and the automorphisms generated by the corner charges,
    \begin{equation}
    	\widehat{\mathcal{A}}_{\mathscr{H}_{>u_0}} \simeq
    	\mathcal{A}_{\mathscr{H}_{>u_0}}
    	\rtimes
    	\bigl( C^\infty_\beta(\mathbb{S}^{d-2})^\ast \rtimes C^\infty_\alpha(\mathbb{S}^{d-2})^\ast \bigr),
    	\label{eq:intro-classical-crossed}
    \end{equation}
    i.e.\ the subregion algebra is obtained by adjoining the corner boost/translation generators that move the subregion relative to its complement.

    \subsection{Quantum gravity: the generalized entropy of subregions}
    
    \subsubsection*{Canonical quantization of the extended horizon phase space}
    
    We next pass to perturbative quantum gravity by linearizing around a stationary black hole background with a bifurcate Killing horizon.  After integrating out the null constraints (in particular the Raychaudhuri constraint) one is left with a set of horizon ``bulk'' degrees of freedom (matter and gravitons) together with the corner edge modes.  Denoting the resulting smeared field operators collectively by $\hat{\Phi}(f)$, canonical quantization amounts to imposing the standard abstract $\ast$-algebra structure and commutation relations dictated by the extended symplectic form:\footnote{See also \cite{Ciambelli:2023mir} for closely related work on canonical quantization of gravity on null surfaces and the role of the Raychaudhuri equation, as well as \cite{Freidel:2023bnj, Donnelly:2022kfs} for work on quantization of edge modes / corner symmetry algebras of gravitational subregions.}
    \begin{subequations}
    \begin{align}
    	&\hat\Phi(af+bg) = a\,\hat\Phi(f) + b\,\hat\Phi(g), \label{eq:intro-linear}\\
    	&\hat\Phi(f)^\dagger = \hat\Phi(f^\ast), \label{eq:intro-adjoint}\\
    	&[\hat\Phi(f),\hat\Phi(g)] = i\,\widehat{\Omega}_{\mathscr{H}}(f,g)\,\hat{\mathbf{1}}.
    	\label{eq:intro-CCR}
    \end{align}
        \end{subequations}
    The key point for the introduction is that the structure of the resulting von Neumann algebra mirrors that of the classical story: the horizon subalgebra is again a crossed product of the dressed bulk algebra by the corner edge mode algebra:
    \begin{equation}
    	\widehat{\mathcal{A}}_{\mathscr{H}_{>u_0}}
    	\simeq
    	\Bigl(\mathcal{A}^{\rm grav}_{\mathscr{H}_{>u_0}} \otimes
    	\mathcal{A}^{\rm mat}_{\mathscr{H}_{>u_0}}\Bigr)
    	\rtimes \mathcal{A}_{\partial G_\varepsilon}[\hat\Gamma^+_0,\hat\Upsilon^+_0].
    	\label{eq:intro-quantum-crossed}
    \end{equation}
    In a GNS representation built from the Hartle--Hawking state, this corresponds to an extended Hilbert space in which the edge mode sector provides the additional degrees of freedom needed for a consistent Lorentzian description of subregions in perturbative quantum gravity. 
    
    \subsubsection*{Crossed products and Type II\texorpdfstring{$_{\infty}$}{∞} horizon subalgebras}
    
    An important consequence of adjoining the edge modes is that the half-sided null translation flow becomes unitarily implementable on the appropriate subregion algebra: the operator $U(\delta u)=e^{i\hat{\mathscr{P}}\delta u}$ live in the enlarged crossed product algebra $\widehat{\mathcal{A}}_{\mathscr{H}_{>u_0}}$ and acts as an inner automorphism there, whereas it would act only as an outer automorphism on the underlying Type~III ``bulk'' algebra alone.
    
    Moreover, in the minisuperspace ($\ell=0$) reduction of the edge mode sector, one can form an intermediate crossed product by the boost automorphism, with flow parameter $s$,
    \begin{align}
    	\widehat{\mathcal{M}}_{\mathscr{H}_{>u}}
    = \mathcal{A}_{\mathscr{H}_{>u}}\rtimes \mathbb{R}_s,
    \end{align}
    which is a Type~II$_\infty$ factor for each cut $u$ once one conditions on the corner location via the translation edge mode.  This is the algebra that naturally carries a semifinite trace and a well-defined von Neumann entropy.
    
    In algebraic QFT, a key (state-dependent) one-sided flow is the Connes cocycle (CC) flow $u_{\Psi|\Omega;u}(s)$ associated to a state $|\Psi\rangle$ relative to the vacuum $|\Omega\rangle$ on a one-sided algebra $\mathcal{A}_{\mathscr{H}_{>u}}$ \cite{Connes1979NonCommIntegration,ConnesTakesaki1977FlowOfWeights,Ceyhan:2018zfg, Bousso:2020yxi}. The salient point is the CC flow acts nontrivially on the one-sided algebra and trivially on its commutant, i.e. it is the canonical way to ``boost only one side''.
    
    In perturbative quantum gravity, we find that this same one-sided flow is implemented in the bulk by the area operator. Within expectation values, this takes the form
    \begin{equation}
    	\langle \Psi|u_{\Psi|\Omega;u}(s)\hat{\mathscr{O}}^{\pm}u^{\dagger}_{\Psi|\Omega;u}(s)|\Psi\rangle
    	=
    	\langle \Psi|e^{i\beta\hat{\mathscr{A}}(u)s}\hat{\mathscr{O}}^{\pm}e^{-i\beta\hat{\mathscr{A}}(u)s}|\Psi\rangle,
    	\label{eq:intro-CC-area}
    \end{equation}
    where $\hat{\mathscr{O}}^{+}$ denotes an operator in the one-sided horizon algebra and $\hat{\mathscr{O}}^{-}$ an operator in its commutant. Conceptually, the equality reflects background independence plus the fact that acting on the corner edge modes changes the dressing of one-sided observables in precisely the same way that the CC flow acts on the algebra. This is a concrete realization of the ``bulk CC flow = kink transform'' conjecture laid out in \cite{Bousso:2020yxi}.
    
    \subsubsection*{Generalized entropy as the von Neumann entropy of a horizon subalgebra}
    
    A Type~II$_\infty$ factor admits a canonical trace, and therefore yields a von Neumann entropy.  For a state $|\hat\psi\rangle$ we can associate a density matrix $\rho_{\hat\psi}(u)\in \widehat{\mathcal{M}}_{\mathscr{H}_{>u}}$ and define \cite{Witten:2021unn}
    \begin{equation}
    	S(\hat\psi; \widehat{\mathcal{M}}_{\mathscr{H}_{>u}})
    	= -\,\mathrm{tr}\!\left[\rho_{\hat{\psi}}(u)\,\log\rho_{\hat{\psi}}(u)\right].
    	\label{eq:intro-entropy}
    \end{equation}
    
    In perturbative quantum gravity, this entropy coincides (up to a state-independent constant) with the generalized entropy of the horizon cut, except the cut location is itself an edge mode degree of freedom so semiclassical states generally involve quantum fluctuations in that location.  Concretely, if the translation edge mode has wavefunction $g(\Delta u_0)$ and we condition on a classical cut position $u_0$, then the Type~II$_\infty$ entropy becomes,
    \begin{equation}
    	S(\rho_{\hat\psi}; \widehat{\mathcal{M}}_{\mathscr{H}_{>u}}) \approx \bar{S}_{\rm gen}(u, \hat{\psi}):=
    	\int_{-\infty}^{\infty}\! d\Delta u_0\,|g(\Delta u_0)|^2\,
    	S_{\rm gen}(u-\Delta u_0;\hat\psi),
    	\label{eq:intro-S-avg}
    \end{equation}
    where \begin{align}S_{\rm gen}(u;\hat\psi)=\langle A(u)\rangle_{\hat\psi}/(4G_N)+S_{\rm bulk}(u;\hat\psi)\end{align} is the usual generalized entropy evaluated at the cut. 
    
    The construction is local, so it extends beyond exactly stationary horizons by working in a sufficiently small neighborhood of a cut where a local Rindler approximation applies. In this more general setting, $\hat{\mathscr{A}}$ is the perturbative quantum gravity operator which implements the ``left stretch'' defined in \cite{Bousso:2019dxk}. We obtain analogous results for finite causal diamonds by applying our construction to expanding and contracting lightsheets $\mathscr{N}^- \cup \mathscr{N}^+$ intersecting at a bifurcation surface, using techniques from \cite{Chandrasekaran:2019ewn}.\footnote{See \cite{Jensen:2023yxy} for earlier work on von Neumann algebras of generic codimension-two gravitational subregions.}
    
    \subsubsection*{Generalized second law and quantum focusing conjecture}
    
    Finally, we use the algebraic structure of horizon subregions to derive various entropy inequalities in the perturbative quantum gravity regime, adapting key aspects of the QFT results in \cite{Borchers1996, Wiesbrock1993, Faulkner:2015csl, Ceyhan:2018zfg}. The (averaged) GSL is essentially a consequence of nesting of the cut algebras under future-directed null translations: for $\delta u\ge 0$ there is a unitary $U(\delta u)=e^{i\hat{\mathscr{P}}\delta u}$ such that
    \begin{align}
    	U(\delta u)\,\widehat{\mathcal{M}}_{\mathscr{H}_{>u}}\,U(-\delta u)
    	\subset
    	\widehat{\mathcal{M}}_{\mathscr{H}_{>u}}.
    \end{align}
    Since the trace respects nesting, the Type II$_\infty$ entropy \eqref{eq:intro-entropy} is monotone under $\partial_u$. Combining with \Eqref{eq:intro-S-avg} yields the (averaged) generalized second law:
    \begin{equation}
    	\partial_u \bar S_{\rm gen}(u;\hat\psi) \ge 0.
    	\label{eq:intro-GSL}
    \end{equation}
    Quantum focusing is stronger: it is not just a nesting statement. It requires a (gravitational) half-sided modular inclusion algebra relating the translation semigroup to the vacuum modular flow. With this extra ingredient, adapting the QNEC proof in \cite{Ceyhan:2018zfg}, we obtain a proof of quantum focusing in perturbative quantum gravity:
    \begin{align}
    	\partial_u^2 \Sgen(u;\psihat)\le 0.
    \end{align}

    \subsection{Relation to previous work}
    Our construction is closely related to that of \cite{Faulkner:2024gst}, which studies crossed product algebras for cuts of a Killing horizon and gives an algebraic formulation of the semi-classical GSL.  In that framework, the GSL is ultimately driven by the nesting of the horizon cut algebras under forward null translations.  Half-sided modular inclusions enter \cite{Faulkner:2024gst} in a complementary way: they are used to identify the corresponding translation generator with the horizon ANEC operator at the level of the fixed background QFT.  
    
    By contrast, in our setting the horizon subalgebras are extended by gravitational edge modes $(\Gamma^{\pm}_0, \Upsilon_0^{\pm})$ at the corner of the subregion, along with the conjugate corner charges $(\hat{\mathscr{A}}_{\beta}, \hat{\mathscr{P}}_{\alpha})$. The half-sided null translation generator arises purely from the gravitational corner symplectic form for the edge modes; it acts on the ``bulk'' QFT operators via gravitational dressing, leading to the crossed product structure of the subregion algebra. This lifts upon quantization to a unitary implementation on horizon subalgebras. The identification with light-ray operators results straightforwardly from the null gravitational constraint equations. Finally, the relevant half-sided modular inclusion algebra is itself gravitational in the sense that the implementing unitaries are generated by dynamical corner degrees of freedom, rather than by QFT operators defined solely within the underlying Type~III$_1$ ``bulk'' algebra.
    
    A second difference is that in our framework the Type~II$_{\infty}$ algebra relevant for entropy is fundamentally tied to conditioning on the $\Upsilon^+_0$ edge mode. In the minisuperspace reduction, the Type~II$_\infty$ factors arise as a one-parameter family
    \begin{equation}
    	\widehat{\mathcal{M}}_{\mathscr{H}_{>u}} \simeq \mathcal A_{\mathscr{H}_{>u}}\rtimes \mathbb R_s,
    \end{equation}
    obtained after non-selective projective measurement onto a sharply localized value of the translation edge mode (i.e.\ a classical cut location $u$), so that the canonical trace and von Neumann entropy are associated to the post-measurement subalgebra.  At the same time, the unconditioned algebra generated by the bulk QFT degrees of freedom together with the boost and translation edge modes (i.e.\ the full crossed product algebra acting on the extended Hilbert space) remains Type~III$_1$ and does not itself come equipped with a semifinite trace.  
    
    This perspective also clarifies the relation to the quantum reference frame interpretation of crossed products in \cite{DeVuyst:2024fxc, Kirklin:2024gyl}: while \cite{DeVuyst:2024fxc, Kirklin:2024gyl} emphasize auxiliary observer/clock reference systems as playing a key role in the emergence of relational Type II$_{\infty}$ algebras (just as in \cite{Chandrasekaran:2022cip}) and the local GSL, here the necessary ``reference frame'' data consists of intrinsic corner edge modes arising from the null gravitational constraints and the null initial value problem; the Type~II$_{\infty}$ algebra emerges specifically from conditioning on the translation edge mode sector rather than by adjoining an external reference system. In particular, it is essential in our construction that we have a \emph{doubled} pair of edge modes $(\Gamma^{\pm}_0, \Upsilon_0^{\pm})$ on the respective corners $S_0^{\pm}$. Moreover, in our case there's a clear throughline from classical phase space $\rightarrow$ corner edge modes $\rightarrow$ integrability + gravitational dressing $\rightarrow$ canonical quantization $\rightarrow$ Tomita-Takesaki theory / gravitational half-sided modular inclusions. This is what allows us to prove the QFC in perturbative quantum gravity on top of Killing horizon backgrounds.

    A complementary analysis of the complete symmetry group and associated edge mode structure will appear simultaneously in \cite{KlingerKudlerFlamSatishchandran2026GEvN2}.
	
	\subsection{Notational conventions}
	
	We work on a $d$-dimensional Lorentzian spacetime $(M,g_{ab})$. Early Roman letters
	$a,b,c,\dots$ represent abstract spacetime indices, later Roman letters $i,j,k,\dots$ denote
	indices intrinsic to a null hypersurface,
	and capital early Roman letters $A,B,\dots$ label indices on codimension-two spatial cuts
	$S \simeq \mathbb{S}^{d-2}$ of that hypersurface. Indices are raised and lowered
	with $g_{ab}$ unless stated otherwise.
	
	Null boundaries $\mathscr{N}$ (and in particular the event horizon $\mathscr{H}$) are
	generated by a future-directed null vector field $\ell^a$ tangent to the
	null generators; they have a corresponding null normal $\ell_a$. Equality restricted to the
	null boundary is indicated by $\hateq$. Affine parameters along $\ell^a$ are denoted by $u$. A cut at
	$u = u_0$ is written $S_0$, and we use the shorthand
	$\mathscr{H}_{>u_0}$ ($\mathscr{H}_{<u_0}$) for the portion of the horizon to the future
	(past) of $S_0$.
	
	The induced $(d-2)$-metric on $\mathscr{N}$ is written as $q_{ab}$. The expansion and shear of the null congruence are denoted by $\Theta$
	and $\sigma_{ab}$, respectively; in particular
	$\Theta = q^{ab} \nabla_a \ell_b$ and $\sigma_{ab}$ is the traceless part of
	$q_a{}^c q_b{}^d \nabla_{(c} \ell_{d)}$. The inaffinity $\kappa$ is defined by $\ell^b \nabla_b \ell^a = \kappa \,\ell^a$. The pullback map from $M$ to $\mathscr{N}$ will be denoted by $\Pi_*$. So for example the pullback $\Pi_* \omega$ of a 1-form $\omega_a$ is represented in index notation as $\omega_i = \Pi_i^a \omega_a$. We will sometimes make use of the induced derivative operator on $\mathscr{N}$, which we denote by $\widehat{\nabla}_i$.
	
	On a null hypersurface $\mathscr{N}$ we denote by $\bm{\eta}$ the induced volume
	$(d-1)$-form and by $\bm{\mu}$ the area $(d-2)$-form on its spatial cuts
	$S$. When convenient we factor out these volume forms and work with
	tensor densities: boldface symbols denote quantities of the form
	$\boldsymbol{\omega} = \eta\,\omega$ on $\mathscr{N}$, and similarly
	$\boldsymbol{\varpi} = \mu\,\varpi$ on cuts, where $\omega$ and
	$\varpi$ are tensors independent of the choice of volume form.
	
	Variations of the fields are described by the exterior derivative
	$\delta$ on configuration/phase space. We sometimes use two independent
	variations $\delta$ and $\delta'$ in order to define the symplectic
	current
	$\bm{\omega} = \delta \bm{\theta}' - \delta' \bm{\theta}$, where $\bm{\theta}$ is the
	presymplectic potential. The corresponding symplectic form on a
	subregion $\mathscr{S} \subseteq \mathscr{N}$ is $\Omega_{\mathscr{S}} = \int_{\mathscr{S}} \bm{\omega}$. We use $\wedge$ to denote the wedge product on phase space, and $\curlywedge$ for the wedge product on spacetime.
	
	Contractions with spacetime vector fields are denoted by $i_\xi$,
	whereas contractions with vector fields on phase space (such as the
	Hamiltonian flow $\hat{\xi}$ generated by $\xi^a$) are written
	$\mathfrak{i}_{\hat{\xi}}$. We use $\lie_\xi$ for the Lie derivative on
	spacetime and $\mathfrak{L}_{\hat{\xi}}$ for the Lie derivative acting on phase
	space functionals.
	
	Throughout, curly letters such as $\mathcal{A}$ denote algebras of
	observables (classical Poisson algebras or von Neumann algebras in the
	quantum theory), with subscripts indicating the relevant region
	(e.g.\ $\mathcal{A}_{\mathscr{H}_{>u_0}}$). Hats indicate operators after
	quantization, e.g.\ $\hat{\mathscr{A}}$ for the area operator. We also use hats on algebras to denote the associated crossed product algebras, i.e. $\widehat{\mathcal{A}}_{\mathscr{H}_{>u_0}}$. Similarly $\widehat{\mathcal{H}}$ indicates the extended Hilbert space obtained from the GNS construction applied to the crossed product algebra. 
	
	Finally, we use the term ``bulk'' in quotes to refer to the matter/graviton degrees of freedom living on the codimension-one subregion $\mathscr{H}_{>u_0}$, as opposed to the edge modes living on the corner $S_0$. 
	
	\section{Phase spaces of gravitational subregions: preamble}
	
	A fundamental question that underlies the results of this paper is how to define
	phase spaces associated with subregions of spacetime in a
	gravitational theory.  This issue has been discussed extensively from
	different points of view in
	Refs.\ \cite{Donnelly:2016auv,Speranza:2017gxd,Ciambelli:2021nmv,Carrozza:2022xut,Pulakkat:2025eid,Carrozza:2021gju,Francois:2025sic,Araujo-Regado:2024dpr,Freidel:2025ous}.
	Here we focus on subregions of null boundaries, unlike most earlier work (with the exception of Refs.\ \cite{Ciambelli:2024swv,Freidel:2025ous}).
	A full understanding of the phase space, symmetries and generators
	associated with such subregions requires incorporating dressed subregions and edge modes
	\cite{Donnelly:2016auv,Carrozza:2022xut,Pulakkat:2025eid,Carrozza:2021gju},
	and will be described in detail 
	in the present context in \cref{sec:corneredgemodes} below. In this
	section, as a warm up, we  
	review extant frameworks which instead attempt to directly define corner charges associated with
	subregions without defining subregion phase spaces. These approaches do not make use of edge modes or dressing\cite{Iyer:1994ys,WZ,CFP,Harlow:2019yfa,Chandrasekaran:2020wwn,Chandrasekaran:2021vyu}, and as a consequence have a number of shortcomings which we review.


	\subsection{Phase space definitions}
	\label{sec:phasespacedefs}
    
\begin{figure}[t]
		\centering
		\begin{tikzpicture}[
			scale=1.2,
			every node/.style={font=\small},
			decoration={zigzag,segment length=3pt,amplitude=1.2pt}
			]
			
			\EternalBHBase
			
			
			\coordinate (HcutR) at ($(Bif)!0.6!(SfR)$); 
			
			\coordinate (Xext) at (3.2,-0.8);
			
			
			
			\fill[blue!12, pattern=north west lines, pattern color=blue]
			(HcutR) -- (SfR) -- (i0R) -- (Xext) -- cycle;
			
			
			\draw (SfL) -- (Bif);
			
			\draw[very thick, dashed, red]  (Bif)   -- (HcutR);
			\draw[very thick, dashed, blue] (HcutR) -- (SfR);
			
			\draw[very thick, dashed, blue] (SfR) -- (i0R);
			
			
			\node[right] at (0.25,1.6) {$\mathscr{H}_{>u_0}$};
			
			\fill (HcutR) circle (1.2pt);
			\node[below right] at (HcutR) {$u_0$};
			
		\end{tikzpicture}
		\caption{Penrose diagram of a two-sided eternal black hole.  The right
			future event horizon $\mathscr{H}$ with bifurcation surface $\mathscr{B}$ is shown, together
			with a cut at affine parameter $u_0$ that splits the horizon into a past
			portion $\mathscr{H}_{<u_0}$ (dashed red line) and a future portion
			$\mathscr{H}_{>u_0}$.  The subregion whose algebra we study is
			$\mathscr{H}_{>u_0} \cup \mathscr{I}^+$ (dashed blue line),
			whose domain of dependence in the exterior region is shaded in blue.
		}
		\label{fig:penrose-eternal-exterior}
	\end{figure}
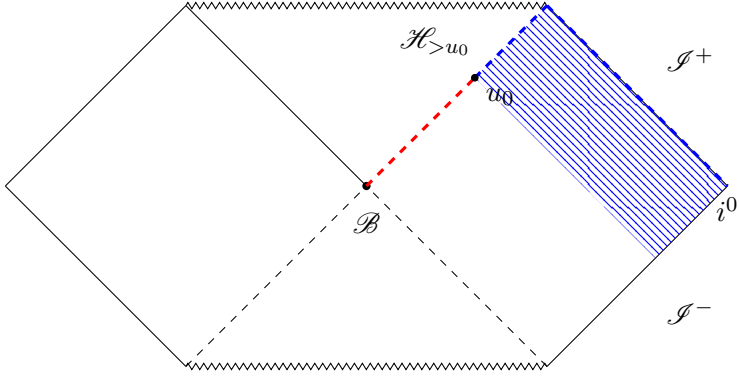

	We start by giving some examples of the types of gravitational phase
	spaces and subregion phase spaces that we would like to be able to define, to set 
	the context.  We consider null components $\mathscr{N}$ of the
	boundaries $\partial M$ of spacetimes $(M,g_{ab})$, focussing on
	boundaries at finite locations rather than asymptotic boundaries.
	The prototypical example is given by spacetimes obtained by
	perturbative excitations on top of the exterior region of two-sided,
	eternal black holes, illustrated in \cref{fig:penrose-eternal-exterior}. 
	Here we take $\mathscr{N}$ to be the right future horizon
	$\mathscr{H}$, which together with future null infinity
	$\mathscr{I}^+$ forms a Cauchy surface for the exterior region.
	
	We would like to define a gravitational phase space and an algebra of
	observables for the entire exterior region.
	In addition, we are interested in subregions.  Given a choice of
	affine parameter $u$ on ${\mathscr H}$, we can choose a cut $S_0$ of
	${\mathscr H}$ given by $u = u_0$, and we denote by $\mathscr{H}_{>u_0}$ the subregion of the
	horizon to the future of the cut.
	Then the surface $\mathscr{H}_{>u_0}\cup \mathscr{I}^+$ has a domain
	of dependence indicated by the blue shaded region.
	Again, we would like to define a subregion phase space and algebra of
	observables associated with this spacetime region.
	As mentioned above, such a definition in the gravitational
	case requires for consistency that the cut be dressed, that is, that its location be a functional
	of the field configuration
	\cite{Ciambelli:2021nmv,Carrozza:2022xut,Pulakkat:2025eid,Francois:2025sic,Araujo-Regado:2024dpr,Donnelly:2016rvo}, which is 
	associated with the existence of edge modes.
	In this section we neglect these modes.
	
\begin{figure}[t]
		\centering
		\begin{tikzpicture}[
			scale=1.2,
			every node/.style={font=\small},
			decoration={zigzag,segment length=3pt,amplitude=1.2pt}
			]
			
			\EternalBHBase
			
			\coordinate (HcutR) at ($(Bif)!0.6!(SfR)$);
			\coordinate (Q) at (0.4,2);
			
			
			\fill[red!12, pattern=north east lines, pattern color=red]
			(Bif) -- (SfL) -- (Q) -- (HcutR) -- cycle;
			
			\draw[very thick, dashed, red] (SfL) -- (Bif);
			\draw[very thick, dashed, red]  (Bif)   -- (HcutR);
			\draw[very thick, dashed, blue] (HcutR) -- (SfR);
			
			\node[left]  at (-1.3,1.0) {$\mathscr{H}'$};
			\node[right] at (0.5,0.3) {$\mathscr{H}_{<u_0}$};
			
			\fill (HcutR) circle (1.2pt);
			\node[below right] at (HcutR) {$u_0$};
			
		\end{tikzpicture}
		
		\caption{Interior version of the setup in
			\cref{fig:penrose-eternal-exterior}. Here the null Cauchy
			surface 
			$\mathscr{H}' \cup \mathscr{H}_{<u_0}$ (the dashed red
			line on the left and
			right future horizons) has  future domain of dependence shaded in
			red in the black hole interior, which terminates at the spacelike
			singularity, and gives rise to a gravitational subregion phase space and
			algebra of observables.}
		\label{fig:penrose-eternal}
	\end{figure}
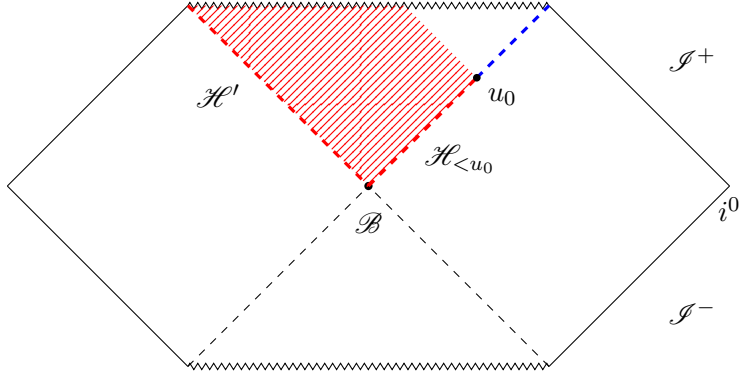

	In this eternal black hole example, we denote by $\mathscr{H}_{<u_0}$
	the region in ${\mathscr H}$ to the past of the cut $S_0$, the complement of
	$\mathscr{H}_{>u_0}$.  This region does not have a domain of
	dependence within the exterior region of the black hole (unlike the
	corresponding situation for spacelike boundaries analyzed in
	Ref.\ \cite{Pulakkat:2025eid}).  Instead, in this case it is natural
	to consider the global phase defined by the union of the right future
	horizon $\mathscr{H}$ and left future horizons $\mathscr{H}'$, and its
	future domain of dependence inside the black hole, illustrated in
	\cref{fig:penrose-eternal}.  Then the surface
	$\mathscr{H}' \cup \mathscr{H}_{<u_0}$ has the domain of dependence
	indicated by the red shaded area, corresponding to another gravitational
	subregion phase space.  In this case the spacetime region in question
	terminates at the singularity.
	
\begin{figure}[t]
		\centering
		\begin{tikzpicture}[scale=1.6,>=stealth,thick]
			
			\coordinate (im)  at (0,0);      
			\coordinate (P)   at (0,2);      
			\coordinate (Stop)at (0,3.6);    
			\coordinate (ip)  at (1.7,3.6);  
			\coordinate (i0)  at (2.6,2.0);  
			
			\coordinate (u0)  at ($(ip)!0.5!(i0)$);   
			\coordinate (B)   at ($(P)!0.65!(ip)$);   
			
			\path let
			\p1 = (B),
			\p2 = (i0),
			\p3 = (ip),
			\p4 = (Stop),
			\n1 = {(\y4-\y1)/(\y3-\y2)} 
			in
			coordinate (S0) at ($ (B) + \n1*(ip) - \n1*(i0) $);
			
			
			\begin{scope}
				\clip (P) -- (Stop) -- (ip) -- cycle; 
				\fill[pattern=north east lines,pattern color=red]
				(P) -- (B) -- (S0) -- (Stop) -- cycle;
			\end{scope}
			
			\begin{scope}
				\clip (im) -- (i0) -- (ip) -- (P) -- cycle;
				
				\coordinate (Qp) at ($(im)!0.302!(i0)$); 
				\coordinate (Qb) at ($(im)!0.756!(i0)$); 
				
				
				\fill[pattern=north east lines,pattern color=blue]
				(B) -- (ip) -- (i0) -- (Qb) -- cycle;
			\end{scope}
			
			\draw (im) -- (P) -- (Stop);
			
			\draw[decorate,decoration={zigzag,segment length=4pt,amplitude=1.2pt}]
			(Stop) -- (ip);
			
			\draw (im) -- (i0) node[midway,below right=2pt] {$\mathscr{I}^-$};   
			\draw (i0) -- (ip) node[midway,above right=2pt] {$\mathscr{I}^+$};   
			\draw (P)  -- (ip);                                                  
			
			\draw[very thick,red]  (P) -- (B);    
			\draw[very thick,blue] (B) -- (ip);   
			
			\draw[very thick,blue] (ip) -- (i0);
			
			
			\node[right=2pt]       at (i0) {$i^0$};
			\node[left=2pt]        at (P)  {$\mathcal{P}$};
			
			
			\node[left=1.5pt]  at (B) {$u_0$};
			
		\end{tikzpicture}
		\caption{
			Penrose diagram for a black hole collapse spacetime.  The
			event horizon forms at the point $\mathcal{P}$, and a cut at affine parameter
			$u_0$ on the horizon splits it into a future subregion $\mathscr{H}_{>u_0}$ and a
			past portion $\mathscr{H}_{<u_0}$ determined by the collapse geometry.
			As before, the cut is associated with both the subregion phase space of the
			region outside the horizon
			shaded in blue, and the subregion phase space of the region in the interior of the black
			hole shaded in red.
		}
		\label{fig:collapse}
	\end{figure}
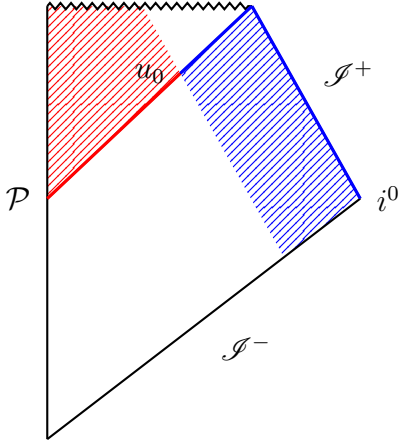

	Another example is the class of spacetimes generated by gravitational
	collapse, depicted in \cref{fig:collapse}.  Here again a cut of the
	future horizon corresponds both to a subregion phase space 
	corresponding to the region outside the horizon, shaded in blue,
	and to a different subregion phase space corresponding
	to the region inside the horizon, shaded in red.
	
	We next discuss definitions of the global phase spaces.  For simplicity we
	restrict to a single null boundary component ${\mathscr N}$,
	assumed to have topology 
	$\mathbb{R}\times \mathbb{S}^{d-2}$, where $d$ is the number of
	spacetime dimensions.
	Suppose now we are are given choice of a triple $(\ell_a, \ell^a,
	\kappa)$ on $\mathscr{N}$
	with $\ell^a \ell_a =0$ where $\ell_a$ is a choice of normal covector,
	defined up to the rescaling freedom \begin{align}\ell_a \to e^\Gamma \ell_a, \
	\ell^a \to e^\Gamma \ell^a, \ \kappa \to e^\Gamma (\kappa + \lie_\ell
	\Gamma),\end{align} where $\Gamma$ is any function on ${\mathscr{N}}$.
	Then the phase space defined in \cite{CFP} is the space of
	on-shell spacetimes $\mathcal{P}_{\mathscr{N}} =
	\left\{(M,g_{ab}) \mid \mathscr{N}\subset \partial M\right\}$
	for which $\ell^a = g^{ab} \ell_b$ and for which $\kappa$ is the
	inaffinity computed from the metric by
	$\ell^b \nabla_b \ell^a \hateq \kappa \ell^a$, where
	$\hateq$ denotes equality restricted to $\mathscr{N}$.
	When taking variations within this phase space, we will make use, throughout this paper, of the perturbative rescaling freedom to enforce $\delta \ell^i = 0$ as in \cite{CFP}, which implies that $\delta \kappa =0$.
	
    For a black hole horizon, the independent fields on the null
    surface for this phase space consist of the intrinsic normal
    $\ell^i$, inaffinity $\kappa$, conformal equivalence class of
    induced metrics $[q_{ij}]$, volume form $\left. \eta_{ijk}
    \right|_{S_0}$ on the fixed cut $S_0$, and any matter
    fields, all defined up to the rescaling freedom of the
    normal. One can solve the Raychaudhuri equation with 
    sources constructed from this data for the
    expansion $\Theta$ with the boundary condition that $\Theta
    \to 0$ in the far future.  One then solves $\lie_\ell
    \eta_{ijk} = \Theta \eta_{ijk}$ to obtain the volume form
    everywhere, which determines the choice of induced metric
    $q_{ij}$ from within the conformal equivalence class; see
    Sec.\ 3 of \cite{Chandrasekaran:2023vzb} and
    \cref{sec:canonicalquant} below for more details.

	The resulting boundary symmetry group is \cite{CFP}
	\be
	\label{symgroup}
	\mathcal{G} = \text{Diff}(\mathbb{S}^{d-2}) \rtimes
	\left(C^{\infty}_{\beta}(\mathbb{S}^{d-2})\rtimes
	C^{\infty}_{\alpha}(\mathbb{S}^{d-2})\right),
	\ee
	where the \emph{supertranslation} factor
	$C^{\infty}_{\beta}(\mathbb{S}^{d-2})\rtimes
	C^{\infty}_{\alpha}(\mathbb{S}^{d-2})$
	consists of vector fields on
	${\mathscr{N}}$ of the form $f \ell^a$ with
	\be
	\lie_\ell (\lie_\ell + \kappa) f = 0.
	\label{eq:symmetrydef}
	\ee
	If we specialize to a scaling of the normal for which $\kappa
	=0$ and consider an affine coordinate $u$ for which ${\vec \ell} =
	\partial_u$, then\footnote{In this paper we will for the most part use a general scaling of the
		normal, not restricted to $\kappa = 0$, except occasionally when it
		simplifies the presentation as here. None of the results in this paper will
		depend on the choice of scaling.}
	\begin{align}
		f(u,x^A) = \alpha(x^A) + u\beta(x^A),\label{eq:supertranslationgroup}
	\end{align}
	for some smooth functions\footnote{Note that the sign used here for
		$\beta$ is opposite to that used in Ref.\ \cite{CFP}.} $\alpha$ and $\beta$ on $\mathbb{S}^{d-2}$ 
	and where $x^A$ are coordinates on $\mathbb{S}^{d-2}$ \cite{CFP}.
	The symmetries
	parameterized by $\alpha$ and $\beta$ were called affine
	supertranslations and Killing supertranslations, respectively, in Ref.\ \cite{CFP}.
	Here we will instead call them angle-dependent translations and angle-dependent boosts, or
	for simplicity just translations and boosts; we will still refer to them collectively as supertranslations.
	
	In this paper we care only about the supertranslation factor in the
	symmetry group \eqref{symgroup}, so henceforth we
	ignore the $\text{Diff}(\mathbb{S}^{d-2})$ factor.  A closely related
	point is that the phase space $\mathcal{P}_\mathscr{N}$ is a
	restricted phase space which does not contain all the physical degrees
	of freedom of the Cauchy data on a null surface.  Larger phase spaces
	which do so are defined in Ref.\ \cite{Chandrasekaran:2023vzb}.  However the restricted
	phase space $\mathcal{P}_\mathscr{N}$ will be sufficient for our
	purposes.\footnote{The restriction $\delta \kappa=0$ which our phase space imposes
		is physically appropriate for perturbations about a background with a Killing horizon,
		since it corresponds to configurations with fixed temperature.
		Therefore $\mathcal{P}_{\mathscr{N}}$ is the appropriate phase space
		to consider when asking questions about black hole entropy, which is
		our aim.}

	\subsection{Corner charges of gravitational subregions}
	\label{sec:cornercharges}

	We now turn to describing the computation of corner charges associated
	with the null boundary symmetries, following the standard constructions
	\cite{Iyer:1994ys, WZ, CFP} and the extensions
	\cite{Harlow:2019yfa,Chandrasekaran:2020wwn,Chandrasekaran:2021vyu}. We
	decompose the pullback of the presymplectic potential $\bm{\theta}$ 
	to a boundary into a boundary term $\delta \bm{\alpha}$, a corner term
	$d \bm{\gamma}$ and a flux (or obstruction) term $\bm{\mathcal{E}}$
	\begin{align}
		\Pi_*\bm{\theta} = \delta \bm{\alpha} + d \bm{\gamma} +
		\bm{\mathcal{E}},
		\label{decompos1}
	\end{align}
	where $\Pi_*$ is the pullback
	map\footnote{These quantities were denoted $-\delta \bm{\ell}$
		and $d    \bm{\beta}$ in Ref.\ \cite{Chandrasekaran:2021vyu}.  In
		this paper
		we retain $\ell$ to refer to the null normal and $\beta$ to refer to
		a boost parameter}.
	We define the presymplectic current\footnote{Away from the boundaries
		we choose any smooth definition of $\bm{\gamma}$ (which need not be
		covariant) whose pullbacks to
		the boundaries agree with the decompositions \eqref{decompos1}.  All
		of the results depend only on the values of $\bm{\gamma}$ on the
		boundaries.  Similar remarks apply to expressions involving
		$\bm{\alpha}$ and $\bm{\mathcal{E}}$ away from the boundaries that occur below.}
	\be
	\bm{\omega} = \delta (\bm{\theta}- d \bm{\gamma}),
	\label{eq:omegadef}
	\ee
	and given any Cauchy surface $\Sigma$ we define the
	presymplectic form to be
	\be
	\Omega_{\Sigma} = \int_\Sigma \bm{\omega}.
	\label{OmegaSigma}
	\ee
	For example, for the spacetime region outside an eternal black hole
	associated with a cut $S_0$ of the horizon, illustrated in
	Fig.\ \ref{fig:penrose-eternal-exterior}, a slice $\Sigma$ extending
	from $S_0$ to spatial infinity would be a Cauchy slice. In the limit
	where $S_0$ is taken to the bifurcation twosphere  
	${\cal B}$, the symplectic form \eqref{OmegaSigma} becomes the standard symplectic
	form for the global phase space of the region outside the horizons.
	More generally, for cuts $S_0$ away from ${\cal B}$, the definition
	\eqref{OmegaSigma} is motivated by the idea of a subregion phase space, but it is
	nevertheless still a form defined on the global phase space.
	Next, defining $\mathscr{S}$ to be the
	subregion of the null boundary to the future of $S_0$ and specializing
	the Cauchy slice to be $\Sigma = \mathscr{S} \cup \scri^+$, we obtain\footnote{We are implicitly assuming here that there is no
		contribution from the limiting hyperbola at future timelike infinity
		\cite{1982RSPSA.381..323P,Compere:2023qoa}.
		In the context considered below where one of the two variations in $\bm{\omega}$ is evaluated on a symmetry,
		the validity of this
		assumption may require a relation between the choice of symmetry on
		the horizon and the choice of symmetry on future null infinity,
		analogous to the selection of the diagonal subgroup of ${\rm BMS}^+
		\times {\rm BMS}^-$ at spatial infinity \cite{Strominger:2013jfa,Prabhu_2019}.  This issue is beyond the
		scope of this paper and does not impact our results, since ultimately we 
		restrict attention to contributions to the symplectic form from the
		null surface.}
	\be
	\Omega_\Sigma = \int_\mathscr{S} \bm{\omega} + \int_{\scri^+}
	\bm{\omega} = \Omega_{\mathscr{S}} + \Omega_{\scri^+},
	\label{Omegadecomposition}
	\ee
	with separate contributions from the horizon and from null infinity.

	Consider now a vector field $\xi^a$ on spacetime which at each
	boundary reduces to a boundary symmetry.
	We denote by ${\hat \xi}$
	the corresponding vector field on phase space that maps solutions to
	linearized solutions via
	$\phi \to  \lie_\xi \phi$.  We denote by $\mathfrak{i}_{\hat{\xi}}$
	the contraction map on phase space differential forms  in
	$\Lambda^*\left(\mathcal{P}_{\mathscr{N}}\right)$, while 
	$i_{\xi}$ refers to the contraction map on spacetime differential forms
	in $\Lambda^*(M)$, following the notation of
	\cite{Donnelly:2016auv}. 
	On the global phase space the total symmetry generator
	corresponding 
	corresponding to ${\hat \xi}$
	is given by 
	\begin{align}
		\delta \mathcal{Q}^{\rm tot}_{\xi} = -\mathfrak{i}_{\hat{\xi}}\Omega_{\Sigma},
		\label{generatordef0}
	\end{align}
	and one can continue to use this formula to try to define
	$\mathcal{Q}^{\rm tot}_{\xi}$ more generally when $S_0 \ne {\cal B}$.
	Using the decomposition (\ref{Omegadecomposition}) 
	we can write $\delta \mathcal{Q}^{\rm tot}_{\xi} =
	\delta \mathcal{Q}_{\xi} + \delta \mathcal{Q}^{\scri^+}_{\xi}$
	with
	\begin{align}
		\delta \mathcal{Q}_{\xi} = -\mathfrak{i}_{\hat{\xi}}\Omega_{\mathscr{S}}
		\label{generatordef}
	\end{align}
	being the contribution from the horizon
	and $\delta \mathcal{Q}^{\scri^+}_{\xi} = -\mathfrak{i}_{\hat{\xi}}\Omega_{\scri^+}$.  In this paper we will focus exclusively 
	on the horizon contribution and neglect the contributions to charges
	from null infinity.

	We now make use of the identity
	\be
	\label{identity0}
	-  \mathfrak{i}_{\hat{\xi}} \bm{\omega} = d \delta \left[ \bm{Q}_\xi -
	i_\xi \bm{\alpha} -   \mathfrak{i}_{\hat{\xi}} \bm{\gamma} \right] -
	d i_\xi \bm{\mathcal{E}},
	\ee
	whose derivation we review below.
	Here $\bm{Q}_\xi$ is the Noether charge $d-2$ form defined by
	$d \bm{Q}_\xi  =   \mathfrak{i}_{\hat{\xi}} \bm{\theta} - i_\xi \bm{L}$ with
	$\bm{L}$ the Lagrangian.  Next we combine \crefrange{OmegaSigma}{generatordef} and 
	\Eqref{identity0}, integrate over $\mathscr{S}$ and make
	use of the fact that its boundary $\partial \mathscr{S}$ 
	consists of one component $S_0$ at $u=u_0$  and
	another component $S_\infty$ at $u = \infty$.  This gives
	\begin{align}
		\delta \mathcal{Q}_{\xi}= \ \delta \int_{S_0} \left[\bm{Q}_{\xi} -
		i_{\xi}\bm{\alpha}
		-   \mathfrak{i}_{\hat{\xi}} \bm{\gamma} 
		\right]
		- \delta \int_{S_\infty} \left[\bm{Q}_{\xi} -
		i_{\xi}\bm{\alpha}
		-   \mathfrak{i}_{\hat{\xi}} \bm{\gamma} 
		\right]
		-\int_{S_0}i_{\xi}\bm{\mathcal{E}}
		+\int_{S_\infty}i_{\xi}\bm{\mathcal{E}}.
		\label{eq:nonintegrablecharge0}
	\end{align}
	If we assume fall-off conditions on $\mathcal{P}_{\mathscr{N}}$ such
	that $i_{\xi} \bm{\mathcal{E}}\rightarrow 0$ sufficiently quickly as
	$u\rightarrow \infty$ \footnote{This will be true for event horizons
		and causal diamonds \cite{CFP, Chandrasekaran:2019ewn}.},
	then the last term vanishes and we obtain\footnote{When considering
		the total charge $\delta \mathcal{Q}_\xi^{\rm tot}$, the contribution
		from $S_\infty$ in \Eqref{eq:nonintegrablecharge} always cancels against a corresponding
		contribution coming from $\delta \mathcal{Q}_\xi^{\scri^+}$,
		as can be seen from integrating the identity \eqref{identity0} over a
		spatial Cauchy slice from $S_0$ to spatial infinity.  For this reason
		this term is often dropped from definitions of charges in the
		literature \cite{WZ,
			CFP,Harlow:2019yfa,Chandrasekaran:2020wwn,Chandrasekaran:2021vyu}
		(that is, a different splitting of the total charge into two pieces is
		used).
		We retain the term since we want to construct the actual
		generators of the symmetries on the horizon
		(\cref{sec:corneredgemodes} below).}
	\begin{align}
		\delta \mathcal{Q}_{\xi}= \ \delta \int_{S_0} \left[\bm{Q}_{\xi} -
		i_{\xi}\bm{\alpha}
		-   \mathfrak{i}_{\hat{\xi}} \bm{\gamma} 
		\right]
		- \delta \int_{S_\infty} \left[\bm{Q}_{\xi} -
		i_{\xi}\bm{\alpha}
		-   \mathfrak{i}_{\hat{\xi}} \bm{\gamma} 
		\right]
		-\int_{S_0}i_{\xi}\bm{\mathcal{E}}.
		\label{eq:nonintegrablecharge}
	\end{align}

	From the result \eqref{eq:nonintegrablecharge} we see that
	integrability of the charge is obstructed by the third term involving
	the flux $\bm{\mathcal{E}}$.  In the special case of the global phase space, i.e. $S_0 = {\cal
		B}$, the flux vanishes and the charge is integrable.  More generally
	however the charge is not integrable.
	The standard covariant phase space prescription used in \cite{CFP, WZ} amounts to
	dropping the obstruction term, yielding the integrable charge
	\begin{align}
		\mathcal{Q}_{\xi} =  \
		{\overset{\circ}{\mathcal{Q}}}_{\xi}[S_0]
		-   {\overset{\circ}{\mathcal{Q}}}_{\xi}[S_\infty],
		\label{WZcharges}
	\end{align}
	where
	\begin{align}
		\overset{\circ}{\mathcal{Q}}_{\xi}[S] =  \ \int_{S} \left[\bm{Q}_{\xi} -
		i_{\xi}\bm{\alpha}
		-   \mathfrak{i}_{\hat{\xi}} \bm{\gamma} 
		\right]
		\label{WZcorner}
	\end{align}
	is a ``corner charge''.

	Our goal in this paper is to construct subregion phase spaces and
	algebras, and to find symmetry generators on those spaces whose
	action under the Poisson bracket generates the symmetries, in order
	to allow quantization of the subregions.
	From this perspective, the standard covariant phase space prescription for calculating charges just described has
	two key shortcomings:
	
	\begin{itemize}
		\item One would like to have a definition of a subregion phase space on
		which the symmetries act.  Here one can try to define a subregion phase
		space to be the degrees of freedom on $\mathscr{N}$ to the
		future of the cut $S_0$.  However then the supertranslation symmetries do not
		preserve this phase space, since they can map degrees of freedom to the
		past of $S_0$ into the region to the future of $S_0$.  Thus no
		such definition seems to be possible.
		
		\item One would like to have the charges $\mathcal{Q}_\xi$
		generate the action of the symmetries under Poisson brackets.
		Here they fail to do so due to the obstruction term involving the
		flux in \Eqref{eq:nonintegrablecharge}.
		
	\end{itemize}
	In the following sections we will discuss how to modify the formalism
	to address these shortcomings, in several steps.  The final result
	will be a modification to the derivation of and context for the
	charges \eqref{WZcharges}, but the expressions for the charges themselves will
	be unaltered.
	
	Finally, we now discuss the identity \eqref{identity0}. Its validity requires
	that the symmetries $\xi^a$ be field-independent, and that all
	quantities be covariant, that is, do not depend on any non-dynamical
	background structures. The identity is a special case of
	Eq.\ (B2) of Ref.\ \cite{Chandrasekaran:2021vyu}, here we give a
	simpler and more direct derivation.
    
	From the decomposition \eqref{decompos1} we obtain
	\be
	- \mathfrak{i}_{\hat \xi} \bm{\omega} =
	- \mathfrak{i}_{\hat \xi} \delta \bm{\mathcal{E}}
	= \delta \mathfrak{i}_{\hat \xi}  \bm{\mathcal{E}} - \mathfrak{L}_{\hat{\xi}} \bm{\mathcal{E}},
	\ee
	where we have used Cartan's magic formula on phase space,
	$\mathfrak{L}_{\hat{\xi}} = \delta \mathfrak{i}_{\hat \xi} + \mathfrak{i}_{\hat \xi} \delta$.
	Using covariance we can replace the phase space Lie derivative with a
	spacetime Lie derivative $\lie_{\xi}$, and rewriting this using
	Cartan's magic formula on spacetime gives
	\be
	- \mathfrak{i}_{\hat \xi} \bm{\omega} 
	= \delta \mathfrak{i}_{\hat \xi}  \bm{\mathcal{E}} - d i_\xi
	\bm{\mathcal{E}} - i_\xi d \bm{\mathcal{E}}.
	\label{ans1}
	\ee
	Next we substitute the decomposition \eqref{decompos1}
	into the standard on-shell identity\cite{Iyer:1994ys}
	\be
	\delta \mathfrak{i}_{\hat{\xi}}\bm{\theta} = \delta d\bm{Q}_{\xi} +
	i_\xi d \bm{\theta},
	\ee
	and use the result to eliminate the first and third terms in
	\Eqref{ans1}, giving
	\be
	- \mathfrak{i}_{\hat \xi} \bm{\omega} 
	= d \delta \left[ \bm{Q}_\xi - \mathfrak{i}_{\hat \xi} \bm{\gamma} \right]
	- d i_\xi \bm{\mathcal{E}} - \delta \mathfrak{i}_{\hat \xi} \delta \bm{\alpha}
	+ i_\xi \delta d \bm{\alpha}.
	\label{ans2}
	\ee
	Next covariance implies that 
	$\mathfrak{L}_{\hat{\xi}} \delta \bm{\alpha} = \lie_\xi \delta \bm{\alpha}$.
	Expanding both sides using Cartan's magic formula, substituting into
	\Eqref{ans2} and using that $\delta$ and $i_\xi$ commute now
	yields the final result \eqref{identity0}.

	
	\subsection{Explicit corner charges}

	In this section we write down the explicit corner charges
	\eqref{WZcorner} for supertranslations for the phase space
	$\mathcal{P}_{\mathscr{N}}$ in general relativity.  We
	generalize slightly the treatment of \cite{CFP}
	which was specialized to vacuum general relativity by including a
	massless minimally coupled scalar field $\psi$, which will be useful
	later in the paper.
	
	We denote by $q_{ij}$ the induced metric and by $\eta_{ijk}$ the
	induced volume form on the null surface, where indices $i, j, \ldots$
	refer to tensors intrinsic to the surface.  We define $\mu_{ij} =
	\eta_{ijk} \ell^k$ and denote by $\Theta$ and $\sigma_{ij}$ the
	expansion and the shear.  The perturbation to the induced metric is
	$h_{ij} = \delta q_{ij}$, and its trace is $h = q^{ij} h_{ij}$, where
	$q^{ij}$ is any tensor with $q^{ij} q_{ik} q_{jl} = q_{kl}$.
	The pullback of the presymplectic potential to the boundary is
	\be
	\bm{\theta} = \frac{\bm{\eta}}{16 \pi} \left[ \lie_\ell h +
	\frac{1}{2} \Theta h + \sigma^{ij} h_{ij} \right] + \delta \psi
	\lie_\ell \psi \bm{\eta}.
    \label{eq:theta00}
	\ee
	In the decomposition \eqref{decompos1} we take the corner term
	$\bm{\gamma}$ to vanish, and the boundary and flux terms to be
	\begin{subequations}
          \label{decompos2}
          \begin{eqnarray}
            \label{alphaGR}
        \bm{\alpha} &=& \frac{1}{8 \pi} \Theta \bm{\eta}, \\
    	\label{fluxGR}
	\bm{\mathcal{E}} &=&
	\frac{\bm{\eta}}{16 \pi} \left[  
	- \frac{1}{2} \Theta h + \sigma^{ij} h_{ij} \right] + \delta \psi
	\lie_\ell \psi \bm{\eta}.
	\end{eqnarray}
        \end{subequations}
     These choices are uniquely determined by the Wald-Zoupas criteria \cite{WZ} as derived in Ref.\ \cite{CFP}.

     The symplectic form \eqref{eq:theta00} and decomposition \eqref{decompos2} obey
     the covariance assumption discussed in the previous subsection, that they do not depend on any
     non-dynamical background structures.  The requirement is not trivial since the choice of normal $\ell_a$
     is such a structure and violates the assumption.  However, all of the quantites are invariant under rescaling of the normal,
     and consequently are covariant, as described in Refs.\ \cite{CFP,Odak:2022ndm,Chandrasekaran:2023vzb}.

     Using the decomposition (\ref{decompos2}), we find 
     that the expressions Eqs.\ (6.6) and (6.27) of Ref.\ \cite{CFP} for the Noether charge and corner
	charge 
	are unmodified by the addition of the scalar
	field.  
	The resulting corner charge for supertranslations is
	\be
	\overset{\circ}{\mathcal{Q}}_\xi[S] = \frac{1}{8 \pi} \int_{S} \bm{\mu} \left[ \lie_\ell
	f + \kappa f - \Theta f \right].
	\label{eq:cornerchargest}
	\ee
	Defining $\alpha = f\big\lvert_{S}$ and $\beta = (\lie_\ell f +
	\kappa f)\big\lvert_{S}$ [cf. \Eqref{eq:supertranslationgroup} above], this can be
	written as
	\be
	\overset{\circ}{\mathcal{Q}}_\xi[S] = \overset{\ \ \circ}{\mathscr{A}}_{\beta}[S] +
	\overset{\circ}{ \mathscr{P}}_{\alpha}[S],
	\label{eq:APdefs0}
    \ee
	where
	\be \overset{\ \ \circ}{\mathscr{A}}_{\beta}[S]:= \frac{1}{8\pi}\int_{S}\beta\bm{\mu},
	\ \overset{\circ}{\mathscr{P}}_{\alpha} := -\frac{1}{8\pi}\int_{S}\alpha\Theta \bm{\mu}.
	\ee
	Here $\overset{\ \ \circ}{\mathscr{A}}_{\beta}[S]$ is the corner charge conjugate to
	angle-dependent 
	boosts (i.e.\ the area functional), while $\overset{\circ}{\mathscr{P}}_{\alpha}[S]$ is a kind of momentum corner charge
	conjugate to the angle-dependent translations.
	
	In the sections to follow we will show how to obtain actual integrable
	symmetry generators $\mathscr{A}_{\beta}$ and $\mathscr{P}_{\alpha}$ solving a version of \Eqref{eq:nonintegrablecharge}
	rather than just plain corner charges,
	which generate Hamiltonian flows associated to half-sided
	angle-dependent boosts and translations on horizon subregion phase spaces.
	These operators will play a key role throughout the rest of the paper.

	\section{Canonical half-sided supertranslations\label{sec:integrablegeneratorderiv}}

	In this section we take a first step towards defining subregion phase spaces and integrable symmetry generators on those phase spaces by considering half-sided supertranslations, that is, supertranslations
	which vanish to the past of the fixed cut $S_0$ but are nonzero to its future.  We will show that there is a modification of the covariant phase construction reviewed in the previous section which makes the generators integrable.  The construction will not be sufficient for our purposes, but it does give useful hints towards the subregion phase space definition that we arrive at in later sections.

	We start by writing the defining equation \eqref{generatordef} for a
	symmetry generator in a more explicit notation: 
	\begin{align}
		\delta \mathcal{Q}_{\xi} = \int_{\mathscr{N}} \ H(u-u_0)
		\bm{\omega}(\phi, \delta \phi, \mathfrak{i}_{\hat{\xi}} \delta \phi).
		\label{generatordefexplicit}
	\end{align}
	Here we have written the presymplectic current $\bm{\omega}$ as a
	function of two independent variations, written the dynamical fields
	collectively as $\phi$, and denoted the variation\footnote{This
		variation is often the Lie derivative $\lie_\xi \phi$, but for fields
		on a null surface there are additional terms when one uses a particular convention for gauge fixing the rescaling freedom of the null normal; see Appendix F of
		Ref.\ \cite{Chandrasekaran:2023vzb}.  We will use this gauge fixing
		convention throughout the paper.\label{footnote:variation}}
	in $\phi$ generated by
	the symmetry as $\mathfrak{i}_{\hat{\xi}} \delta\phi$.  
	The quantity $H(u-u_0)$ is the Heaviside step function which is unity
	for $u \ge u_0$ and vanishes otherwise.  It enforces that the integral
	be restricted to the subregion $\mathscr{S}$ of the null surface.
	We can interpret \Eqref{generatordefexplicit} as saying that we use
	the full, two-sided symmetry generator vector field on phase space\footnote{This expression for the phase space vector field for a diffeomorphism will acquire an extra term in \cref{sec:corneredgemodes} below, detailing how it acts on edge modes in the context of an extended phase space.},
    \be
	{\hat \xi} = \int_{\mathscr{N}} d^{d-1} y \, \mathfrak{i}_{\hat{\xi}} \delta \phi
	\frac{\delta }{\delta \phi(y)}, \label{twosidedphasespace}
	\ee
	where $y^i$ are coordinates on $\mathscr{N}$, but we use a
	truncated version of the presymplectic current, that is, we use $H
	\bm{\omega}$ instead of $\bm{\omega}$.

	As discussed in the previous section, the prescription
	\eqref{generatordefexplicit} is clearly somewhat ad hoc.  A modified
	prescription which is similarly ad hoc but which has the advantage of
	yielding integrable supertranslation charges is the following.
	We use the full presymplectic current $\bm{\omega}$, and instead use a
	truncated version of the vector field on phase space
	\be
	{\hat \xi}_{\rm T} = \int_{\mathscr{N}} d^{d-1} y \, H \, \mathfrak{i}_{\hat{\xi}} \delta \phi
	\frac{\delta }{\delta \phi(y)}.
	\label{phasespacetruncated}
	\ee
	[A third option, replacing the vector field $\xi^a$ on spacetime with
	its truncated version $H \xi^a$, will be discussed in
	\cref{sec:corneredgemodes} below.] 
	The choice \eqref{phasespacetruncated} yields the following
	replacement for the charge variation \eqref{generatordefexplicit}:
	\begin{align}
		\delta \mathcal{Q}_{\xi} = \int_{\mathscr{N}} \ 
		\bm{\omega}(\phi, \delta \phi, H \ \mathfrak{i}_{\hat{\xi}} \delta \phi).
		\label{generatordefexplicit1}
	\end{align}
	Note that the expressions \eqref{generatordefexplicit} and
	\eqref{generatordefexplicit1} differ because the presymplectic current
	depends on derivatives of the field variations.  This yields terms
	proportional to the derivative of $H$ and so proportional to delta functions
	localized on the cut $S_0$.  We also note that the phase space vector
	field \eqref{phasespacetruncated} does not correspond to any
	diffeomorphism on spacetime. Rather, it is defined only as a canonical transformation on phase space, which is why we refer to it as a ``canonical'' half-sided supertranslation.

	In this section, we give two different derivations of the
	integrability of the charges \eqref{generatordefexplicit1},
	showing that the new delta function terms exactly cancel the flux term
	that was an obstruction to integrability in the previous section. In \cref{sec:gen} we give a derivation for supertranslations for arbitrary theories of
	gravity for which the symplectic form obeys certain conditions.
	We verify those conditions are satisfied in general relativity in
	\cref{sec:grsymp}. In \cref{app:nonint} we provide an independent explicit calculation in general
	relativity using the results of \cref{sec:grsymp}, but without invoking the results of \cref{sec:gen}, and show that the supertranslations are integrable, but
	that half-sided $\text{Diff}(\mathbb{S}^{d-2})$ generators are not.

	\subsection{General theories of gravity \label{sec:gen}}

	Consider a general theory of gravity with a symplectic form
	$\Omega_{\mathscr{H}}$ defined on a future event horizon $\mathscr{H}$
	with null generator $\ell^i$. Assume that we can put the symplectic form
	into the following form:\footnote{The symplectic form
		$\Omega_{\mathscr{H}}$ includes contributions from matter fields
		$\psi$, not just the metric $g$.}  
	\begin{align}
		\Omega_{\mathscr{H}} = \sum_{\alpha}\int_{\mathscr{H}}\delta \bm{\Psi}_{(\alpha)}^{A_1\ldots A_{k_{\alpha}}} \wedge \delta \dot{{\Psi}}_{A_1\ldots A_{k_{\alpha}}}^{(\alpha)}, \label{eq:sympform}
	\end{align}
	where $\bm{\Psi}_{(\alpha)}^{A_1\ldots A_{k_{\alpha}}}\equiv \bm{\eta} \Psi_{(\alpha)}^{A_1\ldots A_{k_{\alpha}}}$ is a configuration space variable and
	\begin{align}
		\dot{{\Psi}}_{{A_1\ldots A_{k_{\alpha}}}}^{(\alpha)} \equiv c_\alpha \,
		q_{A_1B_1} \ldots q_{A_{k_{\alpha}} B_{k_{\alpha}}} \, \star\lie_{\ell}\bm{\Psi}^{{B_1\ldots B_{k_{\alpha}}}}_{(\alpha)}
		\label{conjugatemomentum}
	\end{align}
	is its conjugate momentum. Here $\alpha$ sums over the different types
	of fields in the theory, $c_\alpha$ is a constant and $\star$ is the
	Hodge dual operator with respect to $\bm{\eta}$.
	Capital Roman indices $A_1, A_2$ etc. in the down position refer to tensor fields intrinsic
	to ${\mathscr H}$ which have vanishing contraction with $\ell^i$, and in
	the up position refer to the dual of this space \cite{CFP}.
	Henceforth we
	suppress the explicit bookkeeping of the label $\alpha$ for notational
	simplicity; it doesn't change the mechanics of the calculations
	below.\footnote{More generally, we take the symplectic form to be a linear combination of terms of the form $\sum_{\alpha} \delta \bm{\Psi}_{(\alpha)}^{A_1\ldots A_{k_{\alpha}}} \wedge \delta \dot{{\Psi}}_{A_1\ldots A_{k_{\alpha}}}^{(\alpha)} + \sum_{\beta}\delta \Psi_{(\beta)}^{A_1\ldots A_{k_{\alpha}}} \wedge \delta \dot{\bm{\Psi}}_{A_1\ldots A_{k_{\beta}}}^{(\beta)}$ where in the latter terms the placement of $\bm{\eta}$ has switched from the configuration space variable to the conjugate momentum. This accommodates even more general classes of field theories, including e.g. the massless scalar field. The calculations to follow are identical for both types of terms, so we just stick to the former terms to avoid clutter.} 
	
	We want to compute $\mathfrak{i}_{\hat{\xi}_{\rm T}}\Omega_{\mathscr{H}}$ for
	the half-sided supertranslation \eqref{phasespacetruncated}. In
	general, for an arbitrary theory of 
	gravity with phase space $\mathcal{P}_{\mathscr{H}}$, this
	supertranslation won't
	necessarily be a symmetry of the phase space. But it still defines an
	admissible flow on phase space. We now show how one can get an
	integrable generator associated with this flow, even if the horizon is
	non-stationary.  
	The only assumptions on $\mathcal{P}_{\mathscr{H}}$ that we
	will make are that $\delta \ell^i \hateq 0$ and that solutions decay at
	late times to approach stationary black hole solutions, consistent
	with the no-hair theorem.

	From the definition \eqref{phasespacetruncated} of the half-sided
	supertranslation we have
	\begin{align}
		\mathfrak{i}_{\hat{\xi}_{\rm T}}\delta \bm{\Psi}^{A_1\ldots A_k} = H(u-u_0)
		\mathfrak{L}_{\hat {\xi}} \bm{\Psi}^{A_1\ldots A_k},
		\label{eq:qflow}
	\end{align}
	where $u$ is a null parameter adapted to $\ell^a$ (not necessarily
	affine), $u_0$ defines the corner $S_0$ in $\mathscr{H}$, and
	$H(u-u_0)$ is the Heaviside step function\footnote{If we're at null
		infinity, then $\Psi^{A_1\ldots A_k}$ is the shear tensor $C^{AB}$. In
		this case, the transformation rule is slightly modified to
		$\mathfrak{i}_{\hat{\xi}}\delta C^{AB} = \left(f\partial_uC^{AB} -
		2\left[D^A D^B - \frac{1}{2}q^{AB}D^2\right] f\right)H(u-u_0)$. But
		otherwise all the results in this section apply directly to null
		infinity as well.}.
	The conjugate momentum \eqref{conjugatemomentum} then satisfies
	\begin{align}
		\mathfrak{i}_{\hat{\xi}_{\rm T}}\delta \dot{{\Psi}}_{A_1\ldots A_k}  =
		\mathfrak{L}_{\hat {\xi}} \dot{{\Psi}}_{A_1\ldots A_k}H(u-u_0)
		+ c \,
		q_{A_1B_1} \ldots q_{A_{k} B_{k}} \, 
		\star \mathfrak{L}_{\hat {\xi}} \bm{\Psi}^{B_1\ldots B_k}\delta(u-u_0),\label{eq:qdotflow}
	\end{align}
	where $\hat{\xi}$ is the phase space vector field \eqref{twosidedphasespace}
	corresponding to the usual full supertranslation. 
	Inserting the results \eqref{eq:qflow} and \eqref{eq:qdotflow} into
	the symplectic form \eqref{eq:sympform} and making use of the identity\footnote{The minus sign in this equation arises from our conventions for orientations which follow those of Ref.\ \cite{CFP} and are as follows.  The orientation of $\mathscr{N}$ like that for any Cauchy surface is induced from the orientation of the spacetime by taking it to be the boundary of the region to its past. The orientation of $S_0$ is the natural orientation on boundary $\partial \mathscr{N}$ induced from the orientation of $\mathscr{N}$.}
	\be
	\int_{\mathscr H} \bm{\varpi} \delta(u - u_0) = - \int_{S_0} i_\ell
	\bm{\varpi}
        \label{identity3}
	\ee
	for any $(d-1)$-form $\bm{\varpi}$ now gives
	\begin{align}
		- \mathfrak{i}_{\hat{\xi}_{\rm T}}\Omega_{\mathscr{H}} &= - \int_{\mathscr{H}}
		H(u-u_0) \mathfrak{i}_{\hat{\xi}} \bm{\omega}
		-
		\int_{S_0} i_\ell \bm{\Xi}
		\label{eq:chargevariation1}
	\end{align}
	with
	\be
	\bm{\Xi} =  c \, \delta \bm{\Psi}^{A_1\ldots A_k} \,
	q_{A_1B_1} \ldots q_{A_{k} B_{k}} \, 
	\star \mathfrak{L}_{\hat {\xi}}  \bm{\Psi}^{B_1\ldots B_k}.
	\label{eq:Xidef}
	\ee
	The first term in \Eqref{eq:chargevariation1} was computed in
	\cref{sec:cornercharges} above, yielding the result
	\eqref{eq:nonintegrablecharge}, which finally gives
	\begin{align}
		\delta  \mathcal{Q}_{\xi} =  \
		\delta {\overset{\circ}{\mathcal{Q}}}_{\xi}[S_0]
		-  \delta {\overset{\circ}{\mathcal{Q}}}_{\xi}[S_\infty]
		-
		\int_{S_0}\left[ i_\ell \bm{\Xi} + i_\xi \bm{\mathcal{E}} \right].
		\label{eq:integratedflowalmost}
	\end{align}
	Note that $\hat{\xi}$ needs to be a symmetry of the phase space of
	the theory in order for the observable \eqref{eq:integratedflowalmost} to be well-defined
	on said phase space. 
	We see from \Eqref{eq:integratedflowalmost} that the charge is
	integrable if
	\be
	 i_\xi \bm{\mathcal E} = - i_\ell \bm{\Xi},
	\label{eq:fluxform}
	\ee
	when pulled back to the corner $S_0$, and if also
	\begin{align}
		\lim_{u\rightarrow \infty}i_{\xi}\bm{\mathcal{E}} =0.
        \label{eq:falloff}
	\end{align}
	which was used in the previous section in the derivation of the result
	\eqref{eq:nonintegrablecharge}. We will show in \cref{sec:grsymp} and \cref{app:nonint} that these conditions are satisfied for supertranslations $\xi^a = f \ell^a$ for general relativity. The resulting symmetry generator $\mathcal{Q}_{\xi}$ is
	\begin{align}
		\mathcal{Q}_{\xi} =  \
		{\overset{\circ}{\mathcal{Q}}}_{\xi}[S_0]
		-   {\overset{\circ}{\mathcal{Q}}}_{\xi}[S_\infty].\label{eq:integratedflow}
	\end{align}
	Note that this coincides with the charge expression \eqref{WZcharges} obtained in the previous section
	using the methods of \cite{CFP, WZ}, but now 
	$\mathcal{Q}_{\xi}$ actually generates a flow in phase space via
	Eqs.\ \eqref{eq:qflow} and \eqref{eq:qdotflow}. 
	
	Fundamentally, the difference between the prescription in this section
	and that of standard covariant phase space approaches is we are not
	truncating the phase space to the subregion $\mathscr{H}_{>u_0}$, but
	rather integrating over all of $\mathscr{H}$ and truncating the flow
	in phase space to be half-sided.   The latter prescription yields an additional
	corner term at $u_0$ that cancels out the non-integrability we get
	in the usual expression for $\delta \mathcal{Q}_{\xi}$.

	\subsection{General relativity \label{sec:grsymp}}

	The derivation of the last subsection applies to an arbitrary theory of gravity, but
	requires the conditions \eqref{eq:sympform} and \eqref{eq:fluxform} to hold. We now
	verify that these conditions are satisfied in the context of general relativity to
	see how this is all realized.

    The presymplectic form for general relativity in the phase
  space $\mathcal{P}_{\mathscr{H}}$ can be written as (see below for the derivation)
	\be
	\omega_{ijk} = \frac{1}{16 \pi} \delta \eta_{ijk} \wedge \delta \Theta
	+ \frac{1}{16 \pi} \delta( q^{AB} \eta_{ijk}) \wedge \delta \sigma_{AB}.
        \label{eq:sympfin0}
	\ee
    This is of the required form \eqref{eq:sympform} with $\bm{\Psi} =
    \bm{\eta}$ for one term and $\bm{\Psi}^{AB} = q^{AB} \bm{\eta}$ for
    the other.  Next, evaluating the form $\bm{\Xi}$ from
    Eqs.\ \eqref{eq:Xidef}, \eqref{eq:sympform} and \eqref{eq:sympfin0} yields
    the expression \eqref{eq:Xiformula} of \cref{app:nonint} (evaluated
    there by a different method), which satisfies the required condition
    \eqref{eq:fluxform} with the flux expression \eqref{fluxGR} as shown
    in \cref{app:nonint}. Finally, we consider the falloff condition
    \eqref{eq:falloff} on the flux \eqref{fluxGR}. On a future
    event horizon, $\sigma_{AB} \sim u^{-p}, \ \Theta \sim u^{-p}, \ p > 1$ (as
    argued for in \cite{CFP}) so $\bm{\mathcal{E}} \rightarrow 0$
    sufficiently quickly as $u\rightarrow \infty$. 

    We now turn to the derivation of the presymplectic form
    expression \eqref{eq:sympfin0}.
    In the phase space $\mathcal{P}_{\mathscr{H}}$, variations satisfy the conditions
    $\delta \ell^a \hateq 0$, $\ell^a h_{ab} \hateq 0$, $\delta \kappa \hateq 0$ and the no-hair theorem fall-off conditions at future infinity.\footnote{We explicitly track only the pure gravity contribution to the symplectic form. The matter contribution just comes along for the ride, so we leave it implicit for now. It will play a more explicit role in the next section \cref{sec:gredgecorner}.} 	The pullback to $\mathscr{H}$ of the presymplectic potential in GR is 
	\begin{align}
		\theta_{ijk} = \frac{1}{16\pi}\eta_{ijk}\ell^f(\nabla_f h - \nabla_e h_f{}{}^e) \hateq \frac{1}{16\pi}\eta_{ijk}(\lie_{\ell}h + h^m{}{}_{\ell}\nabla_{m}\ell^{\ell}), \label{eq:presymp}
	\end{align}
	where the final equality follows from the boundary condition \begin{align}\nabla_c(\ell^a \ell^b h_{ab})\hateq 0,\label{eq:identitymetricpert}\end{align} which itself follows straightforwardly from $\delta \kappa \hateq 0$. Specifically, since $\ell^b \nabla_b \ell^a = \kappa \ell^a$, we have (making repeated use of $\ell^a h_{ab}\hateq 0$)
	\begin{align}
		\ell^a\delta \kappa  \hateq \lie_{\ell}(\ell^c h_c{}{}^a) - \frac{1}{2}\nabla^a\left(\ell^b \ell^c h_{bc}\right),
	\end{align}
	from which the claim follows, after utilizing $\ell^a h_{ab} \hateq 0$ once again.

	To see how \Eqref{eq:presymp} results from \Eqref{eq:identitymetricpert}, start with $\ell^f \nabla_e h_f{}{}^e = \nabla_e(\ell^f h^e{}{}_f) - h^e{}{}_f \nabla_e \ell^f$. But it must be the case that \begin{align}\nabla_a(\ell^f h^e{}{}_f) = \Gamma^e \ell_a,\end{align} for some $\Gamma^e$, since $\ell^a h_{ab}\hateq 0$. So \begin{align}\Gamma^e \ell_e \ell_a = \nabla_a (\ell^e \ell^f h_{ef}) \hateq 0,\end{align} which implies $\Gamma^e \ell_e \hateq 0$. It then follows that $\nabla_e(\ell^f h^e{}{}_f) = 0$. Hence, \begin{align}\ell^f \nabla_e h_f{}{}^e = - h^e{}{}_f \nabla_e \ell^f,\end{align} which yields \Eqref{eq:presymp} as desired.
	
	The symplectic current is defined as $\omega = \delta \theta' - \delta' \theta$. We then compute from \Eqref{eq:presymp}:
	\begin{align}
		\omega_{ijk} = \frac{1}{16\pi}\eta_{ijk}\left(\frac{1}{2}h \lie_{\ell}h' + h h'^{m}{}{}_{\ell}\nabla_m \ell^{\ell} + h'^{m}{}{}_{\ell}\delta(\nabla_m \ell^{\ell})\right) - (h\leftrightarrow h'),
	\end{align}
	where we've used that $\delta \bm{\eta} = \frac{1}{2}h \bm{\eta}$.
	
	We have 
	\begin{align}
		h'^{m}{}{}_{\ell}\delta(\nabla_m \ell^{\ell}) \hateq  \frac{1}{2}\ell^p h'^{m}{}{}_{\ell}(\nabla_m h_p{}{}^{\ell} + \nabla_p h_m{}{}^{\ell} - \nabla^{\ell} h_{mp}),
	\end{align}
	where we've used $\delta \ell^a \hateq 0$ and $h'^{m}{}{}_{\ell}\nabla_m \delta \ell^{\ell} \hateq 0$.

	We can simplify this via several manipulations. Firstly, 
	\begin{align}
		h'^m{}{}_{\ell}(\nabla_m h_p{}{}^{\ell} - \nabla^{\ell}h_{mp}) = h'^m{}{}_{\ell}\nabla_m h_p{}{}^{\ell} - h'^{\ell}{}{}_m\nabla^m h_{\ell p} = 0.
	\end{align}
	Moreover, 
    \begin{subequations}
	\begin{align}
		\ell^p h'^m{}{}_{\ell}\nabla_p h_m{}{}^{\ell} &= h'^m{}{}_{\ell}\lie_{\ell}h_m{}{}^{\ell} +h'^m{}{}_{\ell}h_m{}{}^p\nabla_p \ell^{\ell} - h'^m{}{}_{\ell}h_{p}{}{}^{\ell}\nabla_m \ell^p 
		\\ &= h'^m{}{}_{\ell}\lie_{\ell}h_m{}{}^{\ell} + 2h'^m{}{}_{\ell}h^{p\ell}\nabla_{[p} \ell_{m]},
	\end{align}
    \end{subequations}
	where the second line follows from the first line after some index gymnastics. Since $\ell_a$ is a null normal, it satisfies $\ell_{[a}\nabla_{b}\ell_{c]} \hateq 0$, which in turn implies that $\nabla_{[a}\ell_{b]} \hateq w_{[a}\ell_{b]}$ for some $w_a$. 
	
	It then follows that
	\begin{align}
		h'^m{}{}_{\ell}h^{p\ell}\nabla_{[p} \ell_{m]} \hateq h'^m{}{}_{\ell}h^{p\ell}w_{[p} \ell_{m]} \hateq 0,
	\end{align}
	where we've used that $\ell^a h_{ab}\hateq 0$. Therefore, the symplectic current is 
	\begin{align}
		\omega_{ijk} = \frac{1}{16\pi}\eta_{ijk}\left[\frac{1}{2}h\lie_{\ell}h' + \frac{1}{2}\lie_{\ell}h_m{}{}^{\ell} h'_{\ell}{}{}^m + \frac{1}{2}h h'_{\ell}{}{}^m \nabla_m\ell^{\ell}\right] - (h\leftrightarrow h').\label{eq:sympformorig}
	\end{align}
		This can be expanded out as 
	\begin{align}
		16\pi\omega_{ijk} =\ & 2(\delta \eta_{ijk} \delta' \Theta - \delta'\eta_{ijk}\delta \Theta) + \frac{1}{2}\eta_{ijk}(\lie_{\ell}h_m{}{}^{\ell}h'_{\ell}{}{}^m - \lie_{\ell}h'_m{}{}^{\ell}h_{\ell}{}{}^m) \nonumber \\ &+ \frac{1}{2}\eta_{ijk}(hh'_{\ell}{}{}^m - h'h_{\ell}{}{}^m)\nabla_m\ell^{\ell},
	\end{align}
	where we've used that $\delta \Theta = \frac{1}{2}\lie_{\ell}h$ and that $\delta \bm{\eta} = \frac{1}{2}h\bm{\eta}$.
	
	Next, 
	\begin{align}
		\frac{1}{2}\eta_{ijk}(hh'_{\ell}{}{}^m - h'h_{\ell}{}{}^m)\nabla_m\ell^{\ell} &\hateq \frac{1}{2}\eta_{ijk}(hh'^{\ell m} - h'h^{\ell m})\sigma_{m\ell},
	\end{align}
	where we've once again used that $\ell^a h_{ab}\hateq 0$ when decomposing $\nabla_m \ell_{\ell}$ in terms of geometric quantities on $\mathscr{H}$. Lastly, note that $h'^{m\ell}\lie_{\ell}h_{m\ell} = 2h'^{m\ell}\delta \sigma_{m\ell} + \Theta h'^{m\ell}h_{m\ell} + h' \delta \Theta$ and similarly for $h \leftrightarrow h'$. Hence, 
	\begin{align}
		\frac{1}{2}\eta_{ijk}(\lie_{\ell}h_m{}{}^{\ell}h'_{\ell}{}{}^m - \lie_{\ell}h'_m{}{}^{\ell}h_{\ell}{}{}^m) = \eta_{ijk}(h'^{m\ell}\delta \sigma_{m\ell}-h^{m\ell}\delta' \sigma_{m\ell}) + \delta'\eta_{ijk} \delta \Theta - \delta \eta_{ijk}\delta'\Theta. 
	\end{align}

	There's one last bit of manipulation we have to do. Write $h_{\ell m} = \delta q_{\ell m}$. Then, using $\delta(q^{m\ell}\sigma_{m\ell}) = 0$,
	\begin{align}
		\frac{1}{2}\eta_{ijk}(hh'^{\ell m} - h'h^{\ell m})\sigma_{m\ell} = \delta \eta_{ijk}q^{m\ell} \delta'\sigma_{m\ell} - \delta'\eta_{ijk}q^{m\ell}\delta \sigma_{m\ell},
	\end{align}
	where we've used that $\delta q^{ab} = -h^{ab}$.
		So in the end we have 
	\begin{align}
		\omega_{ijk} = \frac{1}{16\pi }\left[\delta(\eta_{ijk}q^{m\ell})\delta'\sigma_{m\ell}- \delta'(\eta_{ijk}q^{m\ell})\delta \sigma_{m\ell}+\delta \eta_{ijk} \delta' \Theta -\delta'\eta_{ijk}\delta \Theta  \right],\label{eq:sympfin}
	\end{align}
	in agreement with
          \Eqref{eq:sympfin0}. This expression also agrees
          with the variation of the presymplectic potential
          \eqref{eq:theta00}, which is a nice consistency check.\footnote{One of the main reasons we didn't just take this route to begin with is to avoid relying on a particular decomposition of $\theta_{ijk}$ into boundary, corner, and flux terms. This is akin to doing a coordinate-free calculation instead of using specific coordinates; the choice of decomposition is essentially a choice of coordinates on phase space. As a bonus, being able to calculate the symplectic form independently of any such choice and then comparing against the result obtained from a particular decomposition of $\theta_{ijk}$ also serves as a good sanity check of the final result. The other main reason is that some of the intermediate results obtained via the coordinate-free calculation, namely \Eqref{eq:sympformorig}, will be needed in \cref{sec:gredgecorner}.}

          \section{Half-sided diffeomorphisms and null shocks\label{sec:spacetimeSTs}}

        The results of the previous section, while straightforward and explicit, do not provide insight into where the integrability is fundamentally coming from. What we've learned thus far is that using the full symplectic form contracted with a half-sided phase space vector field yields a corner term that cancels out the non-integrability obstruction. 
	At face value this method of obtaining non-trivial integrable generators might just seem like a neat trick. But it's actually hinting at an answer to a much deeper question: how should we think about the gauge-invariant dynamics of open subsystems\footnote{The standard meaning of open system involves a subregion phase space which admits a notion of time evolution, for which the time evolution
          depends on degrees of freedom external to the subspace.  Here we are using a slightly more general meaning, involving a one parameter family of subregion phase spaces for which time evolution maps from one phase space to another, as will become explicit in \cref{sec:edgemodes}.} in classical and quantum gravity?

We will address this question in two stages.  First, in this section,
we will show that if we suitably enlarge the horizon phase space by allowing null shocks,
then it admits integrable phase space symmetries associated with 
half-sided supertranslation diffeomorphisms on spacetime.
Second, in \cref{sec:corneredgemodes} below we will argue that the global horizon phase space
of this section has gauge fixed some of the diffeomorphism degrees of freedom.  
Restoring those gauge degrees of freedom
following the methods of \cite{Donnelly:2016auv,Speranza:2017gxd,Speranza2018a,Gomes:2018dxs,2017NuPhB.924..312G,Ciambelli:2021nmv,Ciambelli:2021vnn,Freidel:2021dxw,Speranza:2022lxr,Klinger:2023tgi,Pulakkat:2025eid,Francois:2025sic,Carrozza:2021gju,Carrozza:2022xut,Araujo-Regado:2024dpr,Freidel:2025ous}
will then allow us to define subregion phase spaces for the $+$ and $-$ regions, each of which comes with a pair of boost and translation edge modes, which can be consistenly glued together to obtain the original global horizon phase space.

Our results on half-sided supertranslations as spacetime diffeomorphisms can be summarized as follows.
        Instead of the vector field
\eqref{phasespacetruncated} truncated on phase space, consider a vector field $\xi^a$ that is truncated on spacetime:
\be
\xi^a = \xi^a_0 H(u-u_0),
\label{eq:spacetimetruncated}
\ee
where $\xi^a_0$ is a normal two-sided supertranslation of the form
\eqref{eq:supertranslationgroup}.
We would now like to define a vector field of the form \eqref{twosidedphasespace} on phase space by
defining the action $\phi \to \phi +  \mathfrak{i}_{\hat \xi} \delta \phi$ of the symmetry on the bulk fields $\phi$,
based on the known tranformation properties of $\phi$ under smooth diffeomorphisms given in \cref{app:nonint}.
There are two choices we can make:
\begin{enumerate}
\item A straightforward application of the formulae in \cref{app:nonint} yields a field variation
$\mathfrak{i}_{\hat \xi} \delta \phi$ which does not lie in our phase space
$\mathcal{P}_{\mathscr{H}}$, because the vector field
  (\ref{eq:spacetimetruncated}) is not a boundary symmetry, from \Eqref{eq:symmetrydef}.  This
  would invalidate the use of the formula \eqref{eq:sympfin0} for the
  symplectic form.  

\item We can modify slightly the distributional components of the
  field variation at the cut
  by setting to zero the variation $\delta \kappa$ of the inaffinity,
  together with compensating null shocks in the stress energy tensor,
  to obtain a field variation
$\mathfrak{i}_{\hat \xi} \delta \phi$ which does lie in
  $\mathcal{P}_{\mathscr{H}}$, as detailed in \cref{app:larger}.

  \end{enumerate}
We define our field variation $\mathfrak{i}_{\hat \xi} \delta \phi$
using option 2. This variation no longer corresponds directly to a diffeomorphism at
  the cut, and as a consequence the general formula \eqref{identity0} derived in \cref{sec:cornercharges} above
  for the
symplectic current $\mathfrak{i}_{\hat \xi} \bm{\omega}$ for a
diffeomorphism symmetry is no longer valid.  Instead, that formula now acquires additional
distributional terms that are localized at the cut $S_0$, which we
compute in \cref{sec:gredgecorner}.
  
The charge variation corresponding to this symmetry is
	\begin{align}
		\delta \mathcal{Q}_{\xi} = \int_{\mathscr{N}} \ 
		\bm{\omega}(\phi, \delta \phi,  \mathfrak{i}_{ \widehat{H \, \xi}_0} \delta \phi),
		\label{generatordefexplicit2}
	\end{align}
with the appropriate interpretation of the hat symbol as just described.  This formula 
should be compared with the charge variations
\eqref{generatordefexplicit} and \eqref{generatordefexplicit1}.
In \cref{sec:integrable2} we show that with the distributional
correction terms included, the symmetry generator is again integrable
for half-sided supertranslations, with the same corner charges as
before.
We also show there that the action of the symmetry gives rise to null shocks in the solutions.
We interpret the results as saying that half sided supertranslations are integrable
symmetries on an extended horizon phase space in which null shocks are allowed.
We extend the derivation to
general diffeomorphism invariant theories of gravity in \cref{sec:genedgecorner}.

	\subsection{Distributional corrections to the symplectic form\label{sec:gredgecorner}}

The general formula (\ref{identity0}) for the symplectic current contracted into a boundary
symmetry can be written as
	\be
	\label{identity00}
	  \mathfrak{i}_{\hat{\xi}} \bm{\omega} = - d \left(
            \delta \bm{Q}_\xi -
	i_\xi \bm{\theta}  \right),
	\ee
where we've used that $\bm{\gamma}=0$ in the setting of general relativity. We now show that for half sided supertranslations of the form \eqref{eq:spacetimetruncated}
that are truncated in spacetime, this formula acquires distributional
correction terms:
	\begin{align}
	   \mathfrak{i}_{\hat{\xi}}\bm{\omega} =
-d(\delta \bm{Q}_{\xi}-i_{\xi}\bm{\theta}) +
          \frac{1}{8\pi} \delta(\beta\bm{\eta}-\alpha \lie_{\ell}\bm{\eta})\delta(u-u_0).\label{eq:finalcornerterm}
	\end{align}
        Here $u=u_0$ is the location of the cut $S_0$, we have written the associated ``two-sided'' supertranslation symmetry as
        $\xi_0^a = f_0 \ell^a$, and we've decomposed it into angle-dependent translations
        \be
        \alpha = f_0 \big\lvert_{u_0}
        \label{eq:alphadef}
        \ee
        and angle-dependent boosts
\be
\beta = \lie_{\ell}f_0 + \kappa f_0
\label{eq:betadef}
\ee
   [cf.\ \Eqref{eq:supertranslationgroup} above].  We note that there are additional distributional components at the cut
contained in the first term on the right hand side arising from the discontinuity in the vector field $\xi^a$.

    We now turn to the derivation of the result \eqref{eq:finalcornerterm}.
	In general relativity, the Noether charge 2-form is given by the general expression 
	\begin{align}
		Q_{\xi, ab} = -\frac{1}{16\pi}\varepsilon_{abcd}\nabla^c \xi^d.
	\end{align}
	Since $\bm{\varepsilon} = \bm{\eta}\curlywedge \ell$, the pullback of the Noether charge 2-form is just 
	\begin{align}
		Q_{\xi,ij} &\hateq   \frac{1}{8\pi}\eta_{ijk}\mathfrak{q}_{\xi}^k, \ \mathfrak{q}^k_{\xi}:=\xi^b \nabla_b \ell^k - \beta_{\xi}\ell^k,
	\end{align}
	where $\beta_{\xi}$ is defined by $\lie_{\xi}\ell^a \hateq \beta_{\xi}\ell^a$.
	
	Firstly, note that $[d, \delta] = 0$. And, 
	\begin{align}
		d\bm{Q}_{\xi} =   j_{\xi}\bm{\eta}, 
	\end{align}
	since the LHS is a top-form on the horizon. Therefore, 
	\begin{align}
		j_{\xi} = \frac{6}{8\pi}\frac{1}{\sqrt{q}}\eta^{ijk}\eta_{\ell [jk}\widehat{\nabla}_{i]}\mathfrak{q}_{\xi}^{\ell} = \frac{1}{8\pi}\widehat{\nabla}_{\ell}\mathfrak{q}^{\ell}_{\xi},
	\end{align}
	where $\widehat{\nabla}_i = \Pi^a_i \nabla_a$ is the induced derivative operator on $\mathscr{H}$ (c.f. \S 3 of \cite{CFP}). It is easy to show that \begin{align}\widehat{\nabla}_{\ell}\mathfrak{q}^{\ell}_{\xi} = \nabla_{\ell}\mathfrak{q}^{\ell}_{\xi} - \varpi, \ \lie_{\mathfrak{q}_{\xi}}\ell_a = \varpi \ell_a.\end{align} So we can write 
	\begin{align}
		j_{\xi}  = \frac{1}{8\pi}(\nabla_{\ell}\mathfrak{q}^{\ell}_{\xi} - \varpi).\label{eq:divchargeaspect}
	\end{align}
	
	Next, we compute 
	\begin{align}
		\nabla_{\ell}\mathfrak{q}_{\xi}^{\ell} = \nabla^{k}\xi^{\ell} \nabla_{\ell} \ell_{k} + \xi^{\ell}\nabla_{k}\nabla_{\ell}\ell^k - \lie_{\ell}\beta_{\xi}  - \beta_{\xi}(\Theta + \kappa).
	\end{align}
	But,
	\begin{align}
		\nabla_{k}\nabla_{\ell}\ell^k = \nabla_{\ell}\nabla_k \ell^k + R^{k}_{}{}_{\ell k m}\ell^{m} = \nabla_{\ell}(\Theta + \kappa) + R_{\ell m}\ell^m.
	\end{align}
	Moreover, $\lie_{\xi}\Theta - \beta_{\xi}\Theta = \lie_{\ell}\lie_{\xi}\log \sqrt{q}$. And so, 
	\begin{align}
		\nabla_{\ell}\mathfrak{q}_{\xi}^{\ell} = \lie_{\ell}\lie_{\xi}\log \sqrt{q} + \lie_{\xi}\kappa - \lie_{\ell}\beta_{\xi} - \beta_{\xi}\kappa + \nabla^{k}\xi^{\ell} \nabla_{\ell}\ell_k + R_{\ell m}\xi^{\ell}\ell^m.
	\end{align}
	
	Furthermore, $\lie_{\xi}\log \sqrt{q} = \mathfrak{i}_{\hat{\xi}}\delta \log \sqrt{q} = \frac{1}{2}\mathfrak{i}_{\hat{\xi}}h$. Therefore, 
	\begin{align}
		\nabla_{\ell}\mathfrak{q}^{\ell}_{\xi} = \frac{1}{2}\lie_{\ell}\mathfrak{i}_{\hat{\xi}}h +\lie_{\xi}\kappa - \lie_{\ell}\beta_{\xi} - \beta_{\xi}\kappa + \nabla^k \xi^{\ell}\nabla_{\ell}\ell_k + R_{\ell m}\xi^{\ell}\ell^m.
	\end{align}
	
	Recall that $\nabla_{[a}\ell_{b]} \hateq w_{[a}\ell_{b]}$. In order to simplify the calculations, we extend $\ell_a$ to a first-order neighborhood off of $\mathscr{H}$ such that $\nabla_b \ell^2 \hateq 2\kappa \ell_b$ for all points in phase space. This basically amounts to identifying the inaffinity parameter with the surface gravity of a timelike vector field in the neighborhood of $\mathscr{H}$, which is merely a gauge fixing. This extension implies $w^a \ell_a \hateq 0$, i.e. $w^a$ is intrinsic to $\mathscr{H}$. Then, 
	\begin{align}
		\nabla^k \xi^{\ell}\nabla_{[\ell}\ell_{k]} \hateq w^k \nabla_k (\xi^{\ell}\ell_{\ell}) = 0.
	\end{align}
	
	Therefore, $\nabla^k \xi^{\ell}\nabla_{\ell}\ell_k = \nabla_{(k}\xi_{\ell)}\nabla^{\ell}\ell^k = \frac{1}{2}\mathfrak{i}_{\hat{\xi}}h_{\ell}{}{}^{k}\nabla_k \ell^{\ell}$. So finally we have 
	\begin{align}
		\nabla_{\ell}\mathfrak{q}^{\ell}_{\xi} = \frac{1}{2}\lie_{\ell}\mathfrak{i}_{\hat{\xi}}h +\lie_{\xi}\kappa - \lie_{\ell}\beta_{\xi} - \beta_{\xi}\kappa + \frac{1}{2}\mathfrak{i}_{\hat{\xi}}h_{\ell}{}{}^{k}\nabla_k \ell^{\ell} + R_{\ell m}\xi^{\ell}\ell^m.
	\end{align}
	
	Now we want to compute $\delta d\bm{Q}_{\xi}$. First note that $\delta \mathfrak{i}_{\hat{\xi}}h = -\mathfrak{i}_{\hat{\xi}}\delta^2 h + \mathfrak{L}_{\hat{\xi}}h = \lie_{\xi}h$ since $h$ is local and covariant. Similarly, $\delta \mathfrak{i}_{\hat{\xi}}h_{\ell}{}{}^k = \lie_{\xi}h_{\ell}{}{}^k$. So we just have 
	\begin{align}
		\delta(\nabla_{\ell}\mathfrak{q}^{\ell}_{\xi}) = \frac{1}{2}\lie_{\ell}\lie_{\xi}h +  \frac{1}{2}\lie_{\xi}h^{\ell}{}{}_k \nabla_{\ell}\ell^k + \mathfrak{i}_{\hat{\xi}}h_{\ell}{}{}^{k}\delta (\nabla_{\ell}\ell^k) + \delta\left(R_{\ell m}\ell^m \xi^{\ell}\right).
	\end{align}
	We then compute the last term:
	\begin{align}
		\mathfrak{i}_{\hat{\xi}}h_{\ell}{}{}^k\delta(\nabla_{\ell}\ell^k) &\hateq \frac{1}{2}\ell^m \mathfrak{i}_{\hat{\xi}}h_{\ell}{}{}^k(\nabla_k h_m{}{}^{\ell} - \nabla^{\ell}h_{km} + \nabla_m h_k{}{}^{\ell})
		\\ &= \ell^m \mathfrak{i}_{\hat{\xi}}h^{k\ell}\nabla_{[k}h_{\ell]m} + \frac{1}{2}\mathfrak{i}_{\hat{\xi}}h_{\ell}{}{}^k\lie_{\ell}h_k{}{}^{\ell} + \frac{1}{2}\mathfrak{i}_{\hat{\xi}}h_{\ell}{}{}^k(h_m{}{}^{\ell}\nabla_k \ell^m - h_k{}{}^m\nabla_m \ell^{\ell})
		\\ &= \frac{1}{2}\mathfrak{i}_{\hat{\xi}}h_{\ell}{}{}^k\lie_{\ell}h_k{}{}^{\ell} + \underbrace{\mathfrak{i}_{\hat{\xi}}h^{\ell}{}{}_{k}h_{m}{}{}_{\ell}\nabla^{[k} \ell^{m]}}_{=\,0}, 
	\end{align}
	where in the last line we've used that $\mathfrak{i}_{\hat{\xi}}h^{\ell}{}{}_{k}h_{m}{}{}_{\ell}\nabla^{[k} \ell^{m]} \hateq \mathfrak{i}_{\hat{\xi}}h^{\ell}{}{}_{k}h_{m}{}{}_{\ell}w^{[k} \ell^{m]}\hateq 0$.
	
	Hence, 
	\begin{align}
		\delta(\nabla_{\ell}\mathfrak{q}^{\ell}_{\xi}) = \frac{1}{2}\lie_{\ell}\lie_{\xi}h +  \frac{1}{2}\lie_{\xi}h_{\ell}{}{}^k \nabla_{\ell}\ell^k + \frac{1}{2}\mathfrak{i}_{\hat{\xi}}h_{\ell}{}{}^k\lie_{\ell}h_k{}{}^{\ell} + \delta\left(R_{\ell m}\ell^m \xi^{\ell}\right)\label{eq:divchargeaspectvar}.
	\end{align}
	
	At this stage, instead of putting together all the pieces of $\delta d\bm{Q}_{\xi}$ it will actually be easier to first compute the $di_{\xi}\bm{\theta}$ contribution. Note $di_{\xi}\bm{\theta} = \lie_{\xi}\bm{\theta}$ since $\bm{\theta}$ is a top-form on $\mathscr{H}$. Using \Eqref{eq:presymp}, we have 
	\begin{align}
		\lie_{\xi}\theta_{ijk} =\ & \frac{1}{16\pi}\eta_{ijk}\bigg(\frac{1}{2}\mathfrak{i}_{\hat{\xi}}h \lie_{\ell}h + \frac{1}{2}\mathfrak{i}_{\hat{\xi}}h h^k{}{}_{\ell}\nabla_k \ell^{\ell} + \lie_{\xi}\lie_{\ell}h \nonumber \\&+ \lie_{\xi}h^k{}{}_{\ell}\nabla_k{}{}\ell^{\ell} + \frac{1}{2}h^k{}{}_{\ell}\lie_{\ell}\mathfrak{i}_{\hat{\xi}}h_{k}{}{}^{\ell} + \frac{1}{2}\beta_{\xi}h^k{}{}_{\ell}\nabla_k\ell^{\ell}\bigg).
	\end{align}
	
	Thus far all our calculations have only been explicitly using the pure gravity piece of the presymplectic potential. But there's also a matter component $\bm{\theta}^{\psi}$ that we've kept implicit. We now need to make it explicit as well. Recall that 
	\begin{align}
		\bm{J}_{\psi} = \mathfrak{i}_{\hat{\xi}}\bm{\theta}^{\psi} - i_{\xi}\bm{L}^{\psi}.
	\end{align}
	Therefore, 
	\begin{align}
		\delta \bm{J}^{\psi} &= \delta \mathfrak{i}_{\hat{\xi}}\bm{\theta}^{\psi} - i_{\xi}d\bm{\theta}^{\psi} = d(i_{\xi}\bm{\theta}^{\psi}) - \mathfrak{i}_{\hat{\xi}}\bm{\omega}^{\psi},
	\end{align}
	where we've used Cartan's magic formula in phase space and in spacetime, along with the fact that $\mathfrak{L}_{\hat{\xi}}\bm{\theta}^{\psi} = \lie_{\xi}\bm{\theta}^{\psi}$ due to covariance. But at the same time, a standard calculation yields \cite{Iyer:1996ky, Jacobson:2018ahi},\footnote{For minimally coupled scalar field theories this is the complete result, but for gauge theories there will in general also be a corner improvement term involving the gauge connection and field strength. Since this additional term will not play a non-trivial role in our analysis, we just absorb it into $\bm{\omega}^{\psi}$ using the corner ambiguity in the symplectic current.}
	\begin{align}
		\Pi_*\bm{J}^{\psi} \hateq \bm{\eta}\  T_{\ell m}\xi^{\ell} \ell^m.
	\end{align}
	Hence,
	\begin{align}
		\Pi_*d(i_{\xi}\bm{\theta}^{\psi}) \hateq \Pi_*\mathfrak{i}_{\hat{\xi}}\bm{\omega}^{\psi} + \delta( \bm{\eta}\ T_{\ell m}\xi^{\ell} \ell^m) .
	\end{align}
	Additionally, from \Eqref{eq:divchargeaspectvar} above we have 
	\begin{align}
		\frac{1}{8\pi}\delta (\eta_{ijk}\nabla_{\ell}\mathfrak{q}^{\ell}_{\xi}) \supset \frac{1}{8\pi}\delta\left(\eta_{ijk}R_{\ell m}\ell^m \xi^{\ell}\right) = \delta\left(\eta_{ijk}T_{\ell m}\ell^m \xi^{\ell}\right),
	\end{align}
	where we've used the Einstein equation. Making the matter contribution to the presymplectic potential explicit $\theta_{ijk}\rightarrow \theta_{ijk} + \theta^{\psi}_{ijk}$, we find 
	\begin{align}
		\frac{1}{8\pi}\delta(\eta_{ijk}\nabla_{\ell}\mathfrak{q}^{\ell}_{\xi}) - \lie_{\xi}\theta_{ijk} \rightarrow \frac{1}{8\pi}\delta(\eta_{ijk}\nabla_{\ell}\mathfrak{q}^{\ell}_{\xi}) - \lie_{\xi}\theta_{ijk} -  \mathfrak{i}_{\hat{\xi}}\omega^{\psi}_{ijk},
	\end{align}
	i.e. the stress tensor terms cancel out. So we see that 
	\begin{align}
		\frac{1}{8\pi}\delta(\eta_{ijk}\nabla_{\ell}\mathfrak{q}^{\ell}_{\xi}) - \lie_{\xi}\theta_{ijk} =\  &\frac{1}{16\pi}\eta_{ijk}\bigg[ \frac{1}{2}\mathfrak{i}_{\hat{\xi}}h_{\ell}{}{}^k\lie_{\ell}h_k{}{}^{\ell} - \frac{1}{2}h^k{}{}_{\ell}\lie_{\ell}\mathfrak{i}_{\hat{\xi}}h_{k}{}{}^{\ell}+ \frac{1}{2}h\mathfrak{i}_{\hat{\xi}}h^k{}{}_{\ell}\nabla_k \ell^{\ell} \nonumber \\& - \frac{1}{2}\mathfrak{i}_{\hat{\xi}}h h^k{}{}_{\ell}\nabla_k \ell^{\ell} + \frac{1}{2}h\lie_{\ell}\mathfrak{i}_{\hat{\xi}}h- \frac{1}{2}\mathfrak{i}_{\hat{\xi}}h \lie_{\ell}h \nonumber \\ & + h(\lie_{\xi}\kappa - \lie_{\ell}\beta_{\xi} - \beta_{\xi}\kappa)\bigg] - \mathfrak{i}_{\hat{\xi}}\omega^{\psi}_{ijk},
	\end{align}
	where we've used the fact that $\bm{\theta}$ is local and covariant to go between $\mathfrak{L}_{\hat{\xi}}\bm{\theta}$ and $\lie_{\xi}\bm{\theta}$.
	
	Comparing to \Eqref{eq:sympformorig}, this means 
	\begin{align}
		\frac{1}{8\pi}\delta(\eta_{ijk}\nabla_{\ell}\mathfrak{q}^{\ell}_{\xi}) - \lie_{\xi}\theta_{ijk} = -\mathfrak{i}_{\hat{\xi}}\omega_{ijk} + \frac{1}{16\pi}\eta_{ijk}h(\lie_{\xi}\kappa - \lie_{\ell}\beta_{\xi} - \beta_{\xi}\kappa),
	\end{align}
	where we've made explicit the matter contribution to the symplectic current; that is, we rewrite $\omega_{ijk} \rightarrow \omega_{ijk} + \omega^{\psi}_{ijk}$.
	
	Let's now specialize to the class of vector fields $\xi^a = f \ell^a, \ f = f_0 H(u-u_0)$. We can do so safely at this point in the calculation, whereas if we had made this specialization at the start we would've incorrectly missed the $\delta(\nabla_{\ell}\mathfrak{q}^{\ell}_{\xi})$ term with no way of knowing, at that stage of the calculation, whether or not this term contributes distributional corrections to $\mathfrak{i}_{\hat \xi}\bm{\omega}$. Now $\mathfrak{q}_{\xi}^k = (\lie_{\ell}f + \kappa f)\ell^k$. In particular, this means $\delta \mathfrak{q}_{\xi}^k = 0$. Coming back to the $\varpi$ piece in
        \Eqref{eq:divchargeaspect}, this in turn implies $\delta
        \varpi = 0$. So all we have left to compute is the $\varpi \delta \bm{\eta}$ term:
	\begin{align}
		\lie_{\mathfrak{q}_{\xi}}\ell_a \hateq 2\mathfrak{q}^{b}_{\xi}\nabla_{[b}\ell_{a]}+\nabla_{a}(\ell_{b}\mathfrak{q}_{\xi}^{b}) = \nabla_{a}(\ell_{b}\mathfrak{q}_{\xi}^{b}),
	\end{align}
	where we've used that $w^a \ell_a \hateq 0$ for our particular extension of $\ell_a$ off of $\mathscr{H}$. In order to evaluate the remaining term, we also need an extension of $\mathfrak{q}^{\ell}_{\xi}$ to a first-order neighborhood off of $\mathscr{H}$. Since this is yet again just a gauge choice, and we only need to choose a gauge for the background spacetime, it suffices to extend $\mathfrak{q}_{\xi}^{\ell}$ such that $\nabla_{a}(\ell_{b}\mathfrak{q}_{\xi}^{b}) \hateq 0$.
	
	So in the end, we have the following result:
	\begin{align}
		d(\delta \bm{Q}_{\xi}-i_{\xi}\bm{\theta}) + \mathfrak{i}_{\hat{\xi}}\bm{\omega} = \frac{1}{16\pi}\bm{\eta}h(\lie_{\xi}\kappa - \lie_{\ell}\beta_{\xi}-\beta_{\xi}\kappa).\label{eq:violationcurrent}
	\end{align}
	For use below, recall that 
	\begin{align}
	  \mathfrak{i}_{\hat{\xi}}\delta \kappa = \lie_{\xi}\kappa - \lie_{\ell}\beta_{\xi} - \beta_{\xi}\kappa.
          \label{eq:kappavariation}
	\end{align}
		If we have an ordinary supertranslation $\xi_0^a = f_0 \ell^a$ then $\mathfrak{i}_{\hat{\xi}_0}\delta \kappa = 0$ since this is a symmetry of the phase space $\mathcal{P}_{\mathscr{H}}$, yielding $d(\delta \bm{Q}_{\xi_0}-i_{\xi_0}\bm{\theta}) + \mathfrak{i}_{\hat{\xi}_0}\bm{\omega} = 0$ as expected. But since we're doing a half-sided supertranslation, we actually get
	\begin{align}
		\bm{\eta}h\mathfrak{i}_{\hat{\xi}}\delta \kappa =
                \bm{\eta}h(\lie_{\ell}f_0 + \kappa f_0) \delta(u-u_0)
                - f_0\lie_{\ell}(\bm{\eta}h)\delta(u-u_0),
                \label{eq:kappadisc}
	\end{align}
	where we've integrated by parts on $\lie_{\ell}\delta(u-u_0)$. Making use of the decomposition \eqref{eq:alphadef} and \eqref{eq:betadef} and combining with Eqs.\ \eqref{eq:violationcurrent} and \eqref{eq:kappavariation} finally yields the result \eqref{eq:finalcornerterm}.


\subsection{Integrability and the null gravitational constraints}
\label{sec:integrable2}

In this section we show that the half-sided
  supertranslation vector field \eqref{eq:spacetimetruncated} gives rise to
  an integrable symmetry generator, by integrating the symplectic current
  \eqref{eq:finalcornerterm} over the entire null surface $\mathscr{H}$. We also demonstrate how this integrability goes hand in hand with connectedness of spacetime across the corner. 

  But in order to do so, we have to be careful about what prescription
  we are using to integrate over $u$ when we insert the symplectic
  current \eqref{eq:finalcornerterm} into the charge variation
  \eqref{generatordefexplicit2}, since the integrand contains
  distributional terms.  
    Whether or not we get an integrable symmetry generator comes down to a subtle order of operations. 
        
	One choice of prescription follows from introducing a region $G_{\varepsilon} = S_{0} \times [u_0-\varepsilon, u_0+\varepsilon]$. This is an infinitesimal tube around the true corner $S_0$. We then excise the region $G_{\varepsilon}$, apply Stokes' theorem treating $\partial G_{\varepsilon} = S_0^- \cup S_0^+$ as an internal boundary, and then take the limit $\varepsilon \rightarrow 0$ at the very end. We can think of $G_{\varepsilon}$ as a Cauchy splitting region since we're breaking up $\mathscr{H}$ into future/past pieces $\mathscr{H}_+ \cup \mathscr{H}_- = \mathscr{H} \setminus G_{\varepsilon}$ and acting on the state solely to the future.
        This prescription is equivalent to computing the symplectic form using the Cauchy principal value 
    \begin{align}
     \text{p.v.}\int du \ (\ldots) \  := \lim_{\varepsilon\rightarrow 0}\int_{\mathbb{R}\setminus[-\varepsilon,\varepsilon]}du \ (\ldots) .
        \label{eq:cpv1}
    \end{align}
    Thus we get 
	\begin{align}
		-\mathfrak{i}_{\hat{\xi}}\Omega_{\mathscr{H}}^{\text{p.v}} = \text{p.v.}\int_{\mathscr{H}\setminus G_{\varepsilon}}d(\delta \bm{Q}_{\xi}-i_{\xi}\bm{\theta}),
           \label{eq:cpv2}
	\end{align}
	since the principal value of the delta function in
        \Eqref{eq:finalcornerterm} is zero.  The prescription also
        omits the distributional contributions to the integrand in
        \Eqref{eq:cpv2}.

        In general this prescription will not yield an integrable symmetry generator, since it is equivalent to the standard covariant phase space prescription reviewed in \cref{sec:cornercharges} above. Essentially it misses corner degrees of freedom living in the Cauchy splitting region $G_{\varepsilon}$.  
	
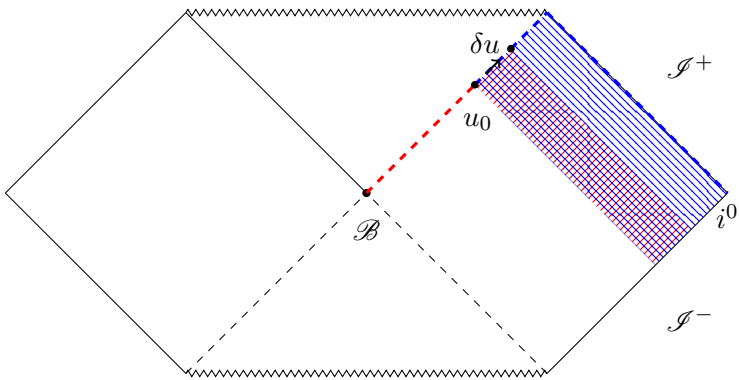
\begin{figure}[t]
		\centering
		\begin{tikzpicture}[
			scale=1.2,
			every node/.style={font=\small},
			decoration={zigzag,segment length=3pt,amplitude=1.2pt}
			]
			
			\EternalBHBase
			
			\coordinate (HcutR)    at ($(Bif)!0.6!(SfR)$);
			\coordinate (HcutRnew) at ($(Bif)!0.8!(SfR)$);
			
			\coordinate (Xold) at ($(HcutR)+(SpR)-(Bif)$);
			\coordinate (Xnew) at ($(HcutRnew)+(SpR)-(Bif)$);
			
			
			\fill[blue!12, pattern=north west lines, pattern color=blue]
			(HcutRnew) -- (SfR) -- (i0R) -- (Xnew) -- cycle;
			
			\fill[red!12, pattern=north east lines, pattern color=red]
			(HcutR) -- (HcutRnew) -- (Xnew) -- (Xold) -- cycle;
			\fill[blue!12, pattern=north west lines, pattern color=blue]
			(HcutR) -- (HcutRnew) -- (Xnew) -- (Xold) -- cycle;
			
			
			
			\draw (SfL) -- (Bif);
			
			\draw[very thick, dashed, red]  (Bif)      -- (HcutR);
			\draw[very thick, dashed, red]  (HcutR)    -- (HcutRnew);
			\draw[very thick, dashed, blue] (HcutR)    -- (HcutRnew);
			\draw[very thick, dashed, blue] (HcutRnew) -- (SfR);
			
			\draw[very thick, dashed, blue] (SfR) -- (i0R);
			
			
			\fill (HcutR) circle (1.2pt);
			\node[below right] at ($(HcutR)+(-0.25,-0.20)$) {$u_0$};
			
			\fill (HcutRnew) circle (1.2pt);
			
			\draw[->, thick]
			($(HcutR)+(0.12,0.12)$) -- ($(HcutRnew)+(-0.12,-0.12)$)
			node[midway, above, xshift=-3pt, yshift=1pt] {$\delta u$};
			
		\end{tikzpicture}
		\caption{Half-sided null translation generated by the corner charge $\mathscr{P}_\alpha$.  A cut $S_0$ at affine parameter $u_0$ on the future horizon is shifted to $u_0 + \delta u$, moving the subregion $\mathscr{H}_{>u_0}$ relative to its complement. The deformation can be visualized as inserting an impulsive null shock at $S_0$.}
		\label{fig:penrose-eternal-deformed}
	\end{figure}

	The other choice of prescription is to integrate with respect
        to $u$ along the entire domain $(-\infty, \infty)$, including
        the distributional components.  We will refer to this as the
        ``on-shell'' prescription for reasons that will become clear
        below. This prescription kills the contribution at $S_0$ from
        the exact term in the symplectic current
    \eqref{eq:finalcornerterm}, but it now picks up the delta function.
    Using that $i_{\xi}\bm{\mathcal{E}} \rightarrow 0$ as $u \rightarrow \infty$ and the identity \eqref{identity3}, we are left with
	\begin{align}
	  &-\mathfrak{i}_{\hat{\xi}}\Omega_{\mathscr{H}} = \delta \left(\mathscr{A}_{\beta} + \mathscr{P}_{\alpha}\right),\label{eq:cornertermintegrable} \end{align}
        where
        \begin{align}
        \label{eq:APdefs1}
        &\mathscr{A}_{\beta}:=
          \frac{1}{8\pi}\left[\int_{S_0}\beta\bm{\mu} -
            \int_{S_\infty}\beta \bm{\mu}\right], \ \mathscr{P}_{\alpha} := -\frac{1}{8\pi}\int_{S_0}\alpha\lie_{\ell}\bm{\mu}.
	\end{align}
	Here $\mathscr{A}_{\beta}$ is in fact the area operator, as the notation suggests. Indeed, from its definition it is clear that $\beta$ is just the rapidity parameter at the corner $S_0$. Similarly, $\mathscr{P}_{\alpha}$ is the translation generator at $S_0$. See \cref{fig:penrose-eternal-deformed} for a depiction of how the half-sided null translation generator $\mathscr{P}_{\alpha}$ acts on the horizon subregion.

Note that the charges (\ref{eq:APdefs1}) coincide with the
  charges \eqref{WZcharges}, \eqref{eq:APdefs0} and
  \eqref{eq:integratedflow} obtained previously.  However the method
  of derivation is now very different: the charges arise from the
  distributional corrections in the identity \eqref{eq:finalcornerterm}, rather than as
  boundary terms coming from the exact term in that identity.

   We can get more insight into the difference between the on-shell and principal value
        prescriptions by considering the constraint equations on the
        null surface, given by Eqs.\ \eqref{eq:constraints} below. 
        As we show there, acting on a state with half-sided spacetime
        supertranslation of the form \eqref{eq:spacetimetruncated}
        while continuing to satisfy the constraints gives rise to a
        state with integrable distributional shocks, with delta
        function contributions to the stress energy tensor and Weyl
        tensor given by Eqs.\ \eqref{eq:shocks}. 
        However, the principal value prescription misses these shocks,
which is why it gives rise to non-integrable charges.  With this
prescription, the global phase space cannot be obtained from a pair of
complementary subregion phase spaces that can be consistently glued
back together.

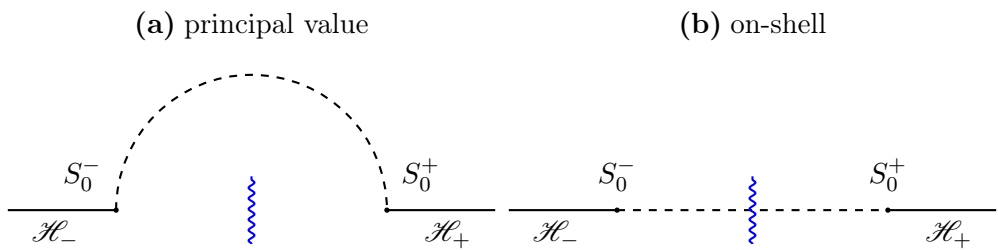
\begin{figure}[t]
    \centering
    \begin{tikzpicture}[
      scale=0.9,
      every node/.style={font=\small},
      shock/.style={blue,thick,decorate,decoration={snake,segment length=4pt,amplitude=1.1pt}},
    ]
    
    \begin{scope}
      \coordinate (L)      at (-3.6,0);
      \coordinate (R)      at ( 3.6,0);
      \coordinate (Sminus) at (-2,0);
      \coordinate (Splus)  at ( 2,0);
      \coordinate (u0)     at (0,0);
    
      \draw[thick] (L) -- (Sminus);
      \draw[thick] (Splus) -- (R);
    
      \node[below] at (-2.9,0) {$\mathscr{H}_{-}$};
      \node[below] at ( 2.9,0) {$\mathscr{H}_{+}$};
    
      \fill (Sminus) circle (1.1pt);
      \fill (Splus)  circle (1.1pt);
      \node[above=3pt] at (-2.5, 0) {$S_0^{-}$};
      \node[above=3pt] at (2.5,0)  {$S_0^{+}$};
    
      \draw[dashed,thick] (Sminus) arc[start angle=180,end angle=0,radius=2];
    
      \draw[shock] (u0) ++(0,-0.5) -- ++(0,1.0);
    
    
      \node[above] at (0,2.35) {\textbf{(a)}\;principal value};
    \end{scope}
    
    \begin{scope}[xshift=7.4cm]
      \coordinate (L)      at (-3.6,0);
      \coordinate (R)      at ( 3.6,0);
      \coordinate (Sminus) at (-2,0);
      \coordinate (Splus)  at ( 2,0);
      \coordinate (u0)     at (0,0);
    
      \draw[thick] (L) -- (Sminus);
      \draw[thick] (Splus) -- (R);
    
      \node[below] at (-2.9,0) {$\mathscr{H}_{-}$};
      \node[below] at ( 2.9,0) {$\mathscr{H}_{+}$};
    
      \fill (Sminus) circle (1.1pt);
      \fill (Splus)  circle (1.1pt);
      \node[above=3pt] at (Sminus) {$S_0^{-}$};
      \node[above=3pt] at (Splus)  {$S_0^{+}$};
    
      \draw[dashed,thick] (Sminus) -- (Splus);
    
      \draw[shock] (u0) ++(0,-0.5) -- ++(0,1.0);
    
    
      \node[above] at (0,2.35) {\textbf{(b)}\;on-shell};
    \end{scope}
    
    \end{tikzpicture}
    \caption{(a) Cauchy principal value vs.\ (b) on-shell prescriptions for computing the full symplectic form. The principal value prescription misses the shock (blue wiggly line) resulting from the null gravitational constraint equations. The on-shell prescription accounts for the constraints, thus allowing for consistent 
    definitions of subregion phase spaces for the two complementary subregions while retaining connectedness of spacetime across the corner.}
    \label{fig:pv-vs-onshell-prescriptions}
    \end{figure}

The ``on-shell'' prescription, on the other hand, is
a necessary ingredient for our goal of being able to define complementary subregion phase spaces,
and being able to consistently combine them together to get the full horizon phase space.
[We will argue in \cref{sec:corneredgemodes} below that additional ingredients are also necessary, so the on-shell prescription is necessary but not sufficient.]
    See
    \cref{fig:pv-vs-onshell-prescriptions} for a depiction of the two
    prescriptions. The discussion thus far can be summarized pithily
    as follows: the global horizon phase space admits half-sided null translation generators iff the constraint equations are satisfied across the Cauchy splitting region. Intuitively, this lends itself to the claim that in order for half-sided null translations to correspond to Hamiltonian time evolution in perturbative quantum gravity, spacetime must be connected across the corner. In \Eqref{eq:nulltranslationgenerator} below, we compute the half-sided null translation generator in terms of gravitational edge modes on both sides of $G_{\varepsilon}$ and show that this physical picture is realized explicitly.
    
	\section{Subregion phase spaces and gravitational edge modes\label{sec:corneredgemodes}}

    \subsection{Why gauge?}

    In \cref{sec:spacetimeSTs} we showed how interpreting the half-sided supertranslation as an actual diffeomorphism requires extending the global phase space by null shocks. While this is sufficient for the purposes of obtaining integrable half-sided boost and translation symmetry generators, it doesn't result in the fundamental construction we care about: the gauge-invariant dynamics of open subsystems in classical and quantum gravity. In order to achieve this, we need to first understand how to define gravitational phase spaces for subregions the horizon which are related to one another by intrinsic null time evolution.

    \cref{sec:spacetimeSTs} does, however, hint at what the construction entails. The existence of integrable half-sided boost and translation generators is a necessary step towards the construction because without these in hand, one cannot algebraically implement intrinsic null time evolution of a putative gravitational subregion phase space purely using degrees of freedom in that phase space. But it's also clear that this cannot be done purely using the ``bulk'' degrees of freedom of the subregion, on account of the null shocks we found in the previous section; there are corner degrees of freedom we must include. At this stage, there are two perspectives one can take in figuring out what the necessary corner degrees of freedom are. 
    
    The first of these, presented in \cref{sec:edgemodederiv}, is what we refer to as the ``bottom up'' approach. In order to even write down a well-defined gravitational subregion phase space, one has to extend the subregion field theory by gravitational edge modes at the corner. The role of these edge modes is to cancel out the anomalous transformation of ``bulk'' fields due to the breaking of diffeomorphism invariance we incurred by specifying some arbitrary subregion of the horizon. The edge modes carry their own gravitational action and corner symplectic form. What's non-trivial is that the edge modes obtained in this manner are also the corner degrees of freedom which yield integrable half-sided boost and translation generators. The bottom up nature of this approach is that we directly start with the subregion and ask how to write down a consistent self-contained phase space for it (and it alone). In the bottom up perspective, the gravitational edge modes are genuinely new physical degrees of freedom with respect to the ``bulk'' degrees of freedom of the associated subregion.

     The second approach, presented in
    \crefrange{sec:edgemodederivation2}{app:symform}, can be thought
    of as a ``top down'' one. The reasoning behind this approach is as
    follows. In \cref{sec:spacetimeSTs} we've essentially worked with
    a gauge-fixed description of the extended global horizon phase
    space, since we've chosen an arbitrary cut $S_0$ at some value of
    affine parameter $u_0$ that is fixed under field variations.
    The second approach undoes the gauge fixing and restores all of
    the gauge degrees of freedom, which gives rise to the edge modes\footnote{Thus the edge modes appear as gauge degrees of freedom in the top down  approach, but as physical degrees of freedom in the bottom up approach.  This apparent contradiction is resolved in \cref{sec:edgemodederivation2} below; see Table \ref{tab:table1}.}.

    Why are these extra gauge degrees of freedom necessary to include?  In
    most circumstances it is optional whether or not to use a gauge-fixed
    approach.  Here the gauge-fixed approach is a perfectly consistent
    way to analyze the full horizon phase space, and even a perfectly
    consistent way to split it into complementary subregion phase
    spaces. So why are the edge modes necessary?
    The answer can be found
    in a beautiful paper by Rovelli \cite{Rovelli:2013fga}. 
    
    Rovelli’s point can be phrased in our setting as follows.  Gauge redundancy is not merely
    a redundancy; it is the bookkeeping that is needed to relationally describe subsystems in a gauge-invariant
    system.  When we split the horizon into complementary subregions across a cut,
    we are trying to treat $\mathscr H^\pm$ as subsystems whose observables and dynamics can be defined
    without explicit reference to the complementary regions $\mathscr H^\mp$.  But the very act of forgetting the complement removes
    precisely the relational information that tells us how the intrinsic description of $\mathscr H^\pm$
    is embedded into (and glued to) the full spacetime. The subsystem therefore cannot
    be specified solely by the bulk fields restricted to $\mathscr H^\pm$: it must also include additional data living at the corner, encoding how $\mathscr H^\pm$ is to be glued to its complement.
    Introducing gravitational edge modes is a way of parameterizing this missing relational data. In other words, gauging is needed in order to consistently glue the two subregions back together, not to split them apart in the first place.

    For completeness, in \cref{app:horizondeformation} we discuss the role that fluctuations in the horizon location (due to gravitational dressing) play in the construction.

\color{black}
	\subsection{Subregion phase spaces: bottom up approach\label{sec:edgemodederiv}}

        We now show that the results of the \cref{sec:spacetimeSTs} can be
        reinterpreted in terms of a set of gravitational edge modes at the corner $S_0$ that encode the dynamics of half-sided boosts and translations. At a high level, it is easy to see why one might expect such an interpretation to exist. Recall that in \Eqref{eq:cornertermintegrable} we obtained integrability of $\mathfrak{i}_{\hat{\xi}}\Omega_{\mathscr{H}}$ \emph{directly} from the corner term that was calculated in \Eqref{eq:violationcurrent}. We never had to do the explicit manipulation/calculation of the symplectic form that was done in \cref{sec:gen}. 
        
        So it would be natural for \Eqref{eq:cornertermintegrable} to result from an extended symplectic form that extends the standard ``bulk'' symplectic form on the subregion $\mathscr{H}_{\pm}$ by a corner symplectic form on $\partial G_{\varepsilon} = S_0^- \cup S_0^+$ (the interior region of $G_{\varepsilon}$ is vanishingly small so only its boundaries matter). To this aim, let $\Gamma^{\pm}_0$ parameterize the relative boost angle at $S^{\pm}_0$. It corresponds to the ``internal'' gauge freedom $\ell^a_{0,\pm} \rightarrow e^{\Gamma^{\pm}_0}\ell^a_{0,\pm}$ in the choice of normal frame at $S_0^{\pm}$.\footnote{See \cite{Chandrasekaran:2023vzb} for a detailed construction of the relative boost angle in the context of horizon phase spaces and the characteristic null initial value problem.} Similarly, let $\Upsilon^{\pm}_0$ parameterize shifts in the location $u^{\pm}_0$ of $S^{\pm}_0$. 
        
        More precisely, choose coordinates $(u_{\pm},x_{\pm}^A)$ on $\mathscr{H}_{\pm}$ adapted to the null generator
	\begin{align}
		&\lie_{\ell_{\pm}} x^A_\pm = 0, \ \ell^a_{\pm} \nabla_a u_{\pm} = 1 ,
	\end{align}
	so that $u_{\pm}$ is an affine parameter along each generator and the $x^A_{\pm}$ label the
	generators. This choice is not unique. Any other adapted affine frame $(u'_{\pm},\ell'^a_{\pm})$ with the
	same generator labels $x^A_{\pm}$ is related to $(u_{\pm},\ell^a_{\pm})$ by an $\rm{Aff}(1)$
	transformation on each generator:
    \begin{subequations}
	\begin{align}\label{eq:aff1_transformation}
		&\ell'^a_{\pm} = e^{-\Gamma^{\pm}(x_\pm)}\ell^a_{\pm}, \\ 
		&u'_{\pm} = e^{\Gamma^{\pm}(x_\pm)}(u_{\pm} - u_0^{\pm}) + \Upsilon^{\pm}(x_\pm),\label{eq:affparamvar} \\ 
		&\lie_{\ell_{\pm}} \Gamma^{\pm} = \lie_{\ell_{\pm}} \Upsilon^{\pm} = 0 .
	\end{align}
        \end{subequations}
	Here $\Gamma^{\pm}(x^A_\pm)$ is an angle-dependent boost (rescaling of the affine frame)
	and $\Upsilon^{\pm}(x^A_\pm)$ is an angle-dependent translation (shift of the affine
	origin). \Eqref{eq:aff1_transformation} follows directly from preserving the conditions
	$\lie_{\ell'_{\pm}}x^A_\pm=0$ and $\ell'^a_{\pm}\nabla_a u'_{\pm}=1$.
    
    While the interpretation of $\Upsilon^{\pm}$ is straightforward, that of $\Gamma^{\pm}$ might still feel a bit abstract. We can make it even more concrete as follows. Let $n^{\pm}_a := -\nabla_a u_{\pm}$ be an auxiliary null normal, so that $(\ell^a_{\pm}, n_a^{\pm})$ form a null dyad on the normal bundle to $\mathscr{H}$. The \emph{spin connection} on the normal bundle is \cite{Ashtekar:2001jb} \begin{align}\omega^{\pm}_i := -\Pi_i^a n^{\pm}_b \nabla_a \ell^b_{\pm}.\end{align} Under the $\text{SO}(1,1)$ gauge transformation contained in \Eqref{eq:aff1_transformation} acting on the null dyad, we have \begin{align}\omega^{\pm}_i \rightarrow \omega^{\pm}_i + \widehat{\nabla}_i \Gamma^{\pm}.\end{align} Hence $\Gamma^{\pm}$ is exactly analogous to the $\text{U}(1)$ gauge parameter in electromagnetism.
    
    Thus far we've just characterized the gauge freedom in choosing a set of coordinates on $\mathscr{H}_{\pm}$. We haven't yet written down any edge modes, which would correspond to actual dynamical degrees of freedom which arise from gauge transformations. To that aim, consider the naive subregion symplectic form
    \begin{align}
       \Omega_{\mathscr{H}_\pm}= \int_{\mathscr{H}_{\pm}} \delta \bm{\theta}[g_{ab}, \psi]. 
    \end{align}
    If we aren't gauge-fixing anything, then under a regular two-sided supertranslation $\xi^a_g$ it should be the case that 
    \begin{align}
        \mathfrak{i}_{\hat{\xi}_g}\Omega_{\mathscr{H}_{\pm}} = 0,\label{eq:sympdegenalways}
    \end{align}
    where the subscript on $\xi^a_g$ is to indicate that it ought to correspond to a pure gauge transformation. But it is clear from \Eqref{eq:nonintegrablecharge} that this will not be satisfied for the standard expression \eqref{eq:theta00} in GR, on the phase space $\mathcal{P}_{\mathscr{H}_{\pm}}$, as this will just yield the usual $d[\delta \bm{Q}_{\xi_g}-i_{\xi_g}\bm{\theta}]$ result. This is because \Eqref{eq:nonintegrablecharge} and \Eqref{eq:theta00} assume a gauge-fixed choice of the $\text{Aff}(1)$ reference frame $(u^0_{\pm}, \ell^a_{0,\pm})$ at $S_0^{\pm}$; the behavior of $(u_{\pm}, \ell^a_{\pm})$ in the ``bulk'' region $\mathscr{H}_\pm$ doesn't matter since $\mathfrak{i}_{\hat{\xi_g}}\Omega_{\mathscr{H}_\pm}$ is a pure corner term. 

    Therefore, since $S_0^{\pm} = S_0^{\pm}[u^{\pm}_0]$ and $\bm{\theta} = \bm{\theta}[\ell_{0,\pm}^a]$, a truly gauge-invariant description that satisfies \Eqref{eq:sympdegenalways} requires the $\text{Aff}(1)$ corner frame $(u^{\pm}_0, \ell^a_{0,\pm})$ itself to be dynamical. This means we have to promote $(\Gamma_0^{\pm}, \Upsilon_0^{\pm})$ to gravitational edge modes with the following transformation under supertranslations:
    \begin{align}
    \mathfrak{i}_{\hat{\xi_g}}\delta \Gamma_0^{\pm} = -\alpha, \ \mathfrak{i}_{\hat{\xi_g}}\delta \Upsilon_0^{\pm} = -\beta,
    \end{align}
    where we emphasize that the phase space vector field $\hat{\xi}_g$ associated to $\xi^a_g$ now acts on the extended phase space of ``bulk'' fields and edge modes. This transformation corresponds to keeping the frame $(u^0_{\pm}, \ell^a_{0,\pm})$ fixed under the simultaneous action of a supertranslation on the spacetime manifold and an $\text{Aff}(1)$ transformation of the corner reference frame. 
    
	The action of a half-sided supertranslation on the edge modes is then defined as follows:
	\begin{align}
		\mathfrak{i}_{\hat{\xi}}\delta \Gamma^{+}_0 = -\beta, \ \mathfrak{i}_{\hat{\xi}}\delta \Upsilon^{+}_0 = -\alpha;\ 
		\mathfrak{i}_{\hat{\xi}}\delta \Gamma^{-}_0 = 0, \ \mathfrak{i}_{\hat{\xi}}\delta \Upsilon^{-}_0 = 0,
		\label{eq:actiononedgemodes}
	\end{align}
        along with the following matching conditions across $G_{\varepsilon}$:
    \begin{subequations}
    \begin{align}
    &\mathfrak{i}_{\hat{\xi}}h_{0,ij}^- = \mathfrak{i}_{\hat{\xi}}h_{0,ij}^+, \label{eq:continuity3} \\ &\mathfrak{i}_{\hat{\xi}}\left(\lie_{\ell}h_{ij}^-\right)_0 = \mathfrak{i}_{\hat{\xi}}\left(\lie_{\ell}h_{ij}^+\right)_0.\label{eq:continuity4} \\ 
    &\mathfrak{i}_{\hat{\xi}}\delta \psi^-_0 = \mathfrak{i}_{\hat{\xi}}\delta \psi^+_0,\label{eq:continuity5}
    \end{align}
    \end{subequations}
	In other words, under the action of $\xi^a$, we're transforming the horizon metric perturbation and conjugate momentum $\left(h_{0,ij}^{\pm}, \left(\lie_{\ell}h_{ij}^{\pm}\right)_0\right)$ in the normal (i.e.\ smooth) way under a diffeomorphism but we're transforming $\left(\Upsilon^{\pm}_0, \Gamma^{\pm}_0\right)$ discontinuously. This corresponds to a \emph{physical} deformation of observables on $\mathscr{H}_+$, holding fixed the observables on $\mathscr{H}_-$, by introducing a shock at the corner. We compute this shock in \crefrange{eq:stresstensorshock}{eq:weylshock} below.
	
	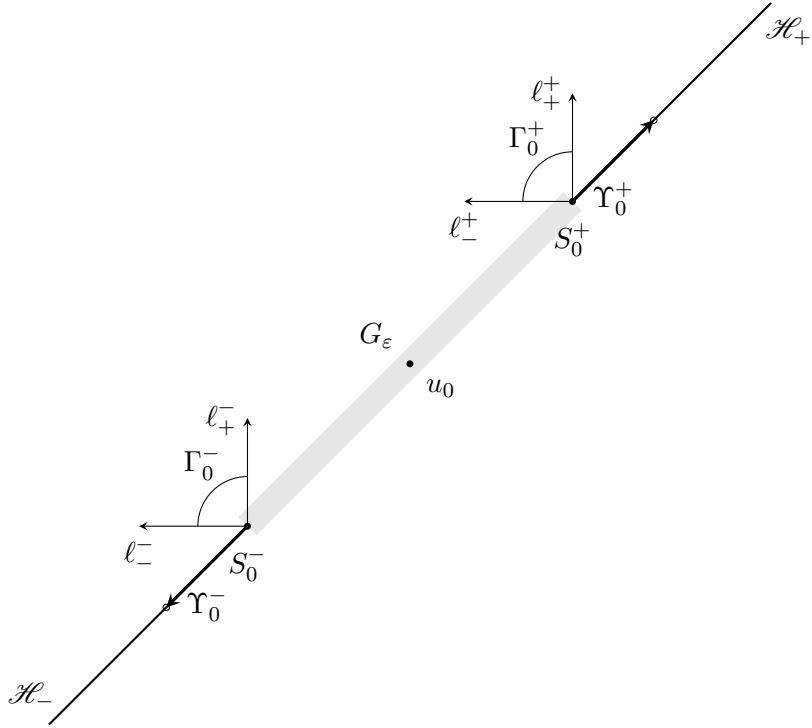
\begin{figure}[t]
		\centering
		\begin{tikzpicture}[
			scale=1.2,
			every node/.style={font=\small},
			decoration={zigzag,segment length=3pt,amplitude=1.2pt},
			>=stealth
			]
			
			\coordinate (HminusL) at (-4,-4);
			\coordinate (HplusR)  at ( 4, 4);
			
			\coordinate (Sminus) at (-1.8,-1.8); 
			\coordinate (u0)     at ( 0.0, 0.0); 
			\coordinate (Splus)  at ( 1.8, 1.8); 
			
			\coordinate (offset) at (-0.1,0.1);
			\coordinate (-offset) at (0.1,-0.1);
			
			\coordinate (G1) at ($(Sminus)+(offset)$);
			\coordinate (G2) at ($(Splus) + (offset)$);
			\coordinate (G3) at ($(Splus) +(-offset)$);
			\coordinate (G4) at ($(Sminus)+(-offset)$);
			
			\draw[thick] (HminusL) -- (HplusR);
			
			\fill[gray!20] (G1) -- (G2) -- (G3) -- (G4) -- cycle;
			
			\node[below left=4pt]  at (-3.7,-3.3) {$\mathscr{H}_{-}$};
			\node[above right=4pt] at ( 3.7, 3.3) {$\mathscr{H}_{+}$};
			
			\fill (u0) circle (1.1pt);
			\node[below right=2pt] at (u0) {$u_0$};
			\node[above left=2pt]  at (u0) {$G_{\varepsilon}$};
			
			\fill (Sminus) circle (1.1pt);
			\fill (Splus)  circle (1.1pt);
			\node[below=4pt] at (Sminus) {$S_0^{-}$};
			\node[below=4pt] at (Splus)  {$S_0^{+}$};
			
			\draw[->] (Sminus) -- ++(-1.2,0) node[below] {$\ell_-^{-}$};
			\draw[->] (Sminus) -- ++(0,1.2)  node[left]  {$\ell_+^{-}$};
			
			\draw (Sminus) ++(-0.55,0) arc[start angle=180,end angle=90,radius=0.55];
			\node at ($(Sminus)+(-0.5,0.7)$) {$\Gamma_0^{-}$};
			
			\draw[->] (Splus) -- ++(-1.2,0) node[below] {$\ell_-^{+}$};
			\draw[->] (Splus) -- ++(0,1.2)  node[left]  {$\ell_+^{+}$};
			
			\draw (Splus) ++(-0.55,0) arc[start angle=180,end angle=90,radius=0.55];
			\node at ($(Splus)+(-0.5,0.7)$) {$\Gamma_0^{+}$};
			
			\coordinate (SminusShift) at ($(Sminus)+(-0.9,-0.9)$); 
			\coordinate (SplusShift)  at ($(Splus) +(0.9,0.9)$);   
			
			\draw (SminusShift) circle (1.0pt);
			\draw (SplusShift)  circle (1.0pt);
			
			\draw[->, very thick] (Sminus) -- (SminusShift);
			\draw[->, very thick] (Splus)  -- (SplusShift);
			
			\node[below=4pt] at ($(Sminus)!0.5!(SminusShift)$) {$\Upsilon_0^{-}$};
			\node[below=4pt] at ($(Splus)!0.5!(SplusShift)$)  {$\Upsilon_0^{+}$};
			
		\end{tikzpicture}
		\caption{Splitting the horizon across a thin tube $G_\varepsilon$ around
			a cut $S_0$.  The horizon is decomposed into a past ``bulk'' portion $\mathscr{H}^{-}$ and
			a future ``bulk'' portion $\mathscr{H}^{+}$ separated by the Cauchy splitting region
			$G_\varepsilon$ with boundaries $S_0^-$ and $S_0^+$.  Each side carries
			its own null normals $\ell_\pm^a$ living on the normal plane and associated edge mode data: the
			relative boost angles $\Gamma_0^\pm$ and affine shifts $\Upsilon_0^\pm$
			along the generators.  These edge modes, along with their canonically conjugate corner charges, comprise the corner symplectic form that encodes the fluctuations of the subregions as a result of gravitational dressing.}
		\label{fig:splitting-region}
	\end{figure}
	
	Having now promoted $(\Gamma_0^{\pm}, \Upsilon_0^{\pm})$ to dynamical degrees of freedom, the only way for \Eqref{eq:sympdegenalways} to be satisfied is by extending the ``bulk'' symplectic form to include a corner term for the edge modes:
	\begin{align}
		\widehat{\Omega}_{\mathscr{H}} = \Omega_{\mathscr{H}_-} + \Omega_{\mathscr{H}_+}+ \Omega_{\partial G},\ \Omega_{\partial G} := \lim_{\varepsilon\rightarrow 0}\Omega_{\partial G_{\varepsilon}}.\label{eq:areacornersymp}
	\end{align}
    The corner term $\Omega_{\partial G}$ is non-trivial to derive; we compute it in \cref{app:symform} below. For now we just quote the result:
    \begin{align}
    \Omega_{\partial G} = \ &\frac{1}{8\pi}\int_{S_0} \left[ \delta \Upsilon_0^+
          \wedge \delta (\lie_\ell {\bm \mu}) - \delta \Gamma_0^+ \wedge
          \delta \Delta \bm{\mu}_+  + \delta \Upsilon_0^+ \wedge
          \delta \Gamma_0^+ \Theta \bm{\mu}  \right] - (+ \leftrightarrow -) \nonumber
        \\ &+ \frac{1}{8\pi}\int_{S_0} \left[ 
          \delta \Upsilon_0^{+} \wedge \delta \Upsilon_0^{-} \lie_\ell(\bm{\mu}\Theta) \right],\label{eq:cornersympformterm}
    \end{align}
    where $\Delta \bm{\mu}_{\pm} := \bm{\mu} - \bm{\mu}_{\pm\infty}$ is a background subtracted area element on $S_0$.\footnote{Note that we don't have edge mode contributions at $u\rightarrow \infty$. There is no excitable translation edge mode at future infinity because $\Theta \rightarrow 0$ there. And we don't have an independent boost edge mode at the future boundary since $\lie_{\ell_+}\Gamma_+ = 0$ on $S^+_0 \cup \mathscr{H}_+$. This is a realization of the fact that the non-trivial action of the area operator is contained entirely in the relative boost angle at the corner [see \Eqref{eq:halfsidedoperators} below].}    Note that this symplectic form is specialized to the background fields being continuous at the corner $S_0$, so that $\Upsilon_0^- = \Upsilon_0^+$, $\Gamma_0^- = \Gamma_0^+$ on the background spacetime, and $S_0^- = S_0^+ = S_0$ on the background spacetime as well, although the variations are allowed to be discontinuous.  This special case will be sufficient for the calculations in the remaining sections of the paper, since we will be specializing to linearized perturbations about a given background.
	
	Since $\Omega_{\mathscr{H}_{\pm}}$ only depends on the bulk fields, it is continuous in the limit $\varepsilon \rightarrow 0$ under the matching conditions above. Therefore, if we use perturbations satisfying \crefrange{eq:continuity3}{eq:continuity5}, we get
	\begin{align}
		\mathfrak{i}_{\hat{\xi}}\Omega_{\mathscr{H}_-} + \mathfrak{i}_{\hat{\xi}}\Omega_{\mathscr{H}_+} = 0.\label{eq:bulkhamconstraint}
	\end{align}
	
	On the other hand, applying the matching conditions to the corner symplectic form $\Omega_{\partial G}$ yields
	\begin{align}
		-\mathfrak{i}_{\hat{\xi}}\Omega_{\partial G} = \delta\left(\mathscr{A}_{\beta} + \mathscr{P}_{\alpha}\right).\label{eq:cornertimegenerators}
	\end{align}
    Thus, we recover \Eqref{eq:cornertermintegrable} as desired. In terms of the edge mode data (and their conjugate momenta), the area operator and null translation operator are (once again quoting the result from \cref{sec:edgemodederivation2})
    \begin{subequations}
    \label{eq:APformulae}
    \begin{align}
         &\mathscr{A}_{\beta}= \frac{1}{8\pi}\left[\int_{S_0^+}\beta\bm{{\mu}} - \int_{S_\infty}\beta
           \bm{{ \mu}}\right], \\ &\mathscr{P}_{\alpha} =
         -\frac{1}{8\pi}\int_{S_0^+}\alpha e^{ \Gamma_0^+} \left[\lie_{{
             \ell}}\bm{{\mu}} - \left(\Upsilon_0^+ - \Upsilon_0^-\right)\lie_{\ell}(\bm{\mu}\Theta)\right].\label{eq:nulltranslationgenerator}
    \end{align}
    \end{subequations}
    The extended symplectic form identifies the area operator and null translation operator with the following phase space flows
    \begin{subequations}
	\begin{align}
		\hat{\xi}_{\beta} &= \int_{S_0^+} d^{d-2}x \ \beta(x^A)\frac{\delta }{\delta \Gamma_0^+}, \\\hat{\xi}_{\alpha} &=\int_{S_0^+} d^{d-2}x \ \alpha(x^A)e^{\Gamma_0^+(x^A)}\frac{\delta }{\delta \Upsilon_0^+}.\label{eq:halfsidedoperators}
	\end{align}
    \end{subequations}
    Using \Eqref{eq:areacornersymp}, it is easy to show that
    \begin{subequations}
	\begin{align}
		&\left\{\mathscr{A}_{\beta}, \mathscr{P}_{\alpha}\right\} = \mathscr{P}_{-\alpha\beta}, \\ &\left\{\mathscr{A}_{\beta}, \mathscr{A}_{\beta'}\right\} = 0, \\
        &\left\{\mathscr{P}_{\alpha}, \mathscr{P}_{\alpha'}\right\} = 0.
	\end{align}
    \end{subequations}
	So we get the expected corner algebra between half-sided boosts and half-sided translations. 
    
    A key feature of \Eqref{eq:nulltranslationgenerator} is that the half-sided null translation generator $\mathscr P_\alpha$ is necessarily two-sided in the complementary translation edge modes: although the
    corresponding Hamiltonian flow $\hat\xi_\alpha$ acts only on the future translation mode
    $\Upsilon_0^+$, the generator itself depends on the relative displacement
    $\Upsilon_0^+-\Upsilon_0^-$. This two-sided coupling, as manifested in the final line of \Eqref{eq:cornersympformterm}, is the explicit realization of the discussion at the end of \cref{sec:integrable2}.
    A half-sided translation shifts $\mathscr H_+$ relative to $\mathscr H_-$, but the null constraint equations must continue to hold across the Cauchy splitting region. Imposing the constraints across $G_\epsilon$ forces an impulsive shock localized at $S_0$, as we will see below. In the edge mode description, the same physics is encoded by the fact that the split requires two independent translation edge modes $\Upsilon_0^\pm$, which enter through a bilocal coupling in the corner symplectic form.

    There's one last step needed in order for the construction above to be self-consistent. Since we want to the field configurations in the extended phase space to actually correspond to on-shell states, we have to check that the linearized constraint equations on $\mathscr{H}$ are satisfied under such perturbations. The linearized constraint equations are
    \begin{subequations}
      \label{eq:constraints}
\begin{align}
    &\lie_{\ell}\delta \Theta =  \Theta\delta \kappa  + \kappa \delta \Theta-\Theta\delta \Theta - 2\sigma\delta \sigma - 8\pi \ell^i \ell^j\delta T_{ij},\label{eq:raychauduri} \\ 
    &\lie_{\ell}\delta \sigma_{ij} = \sigma_{ij}\delta\kappa  + \kappa \delta \sigma_{ij} + 2q_{ij}\sigma^{\ell m}\delta \sigma_{\ell m}  + \sigma^2 h_{ij} - \ell^{\ell}\ell^m \delta C_{i \ell j m}.\label{eq:shearevolution}
\end{align}
    \end{subequations}
Contracting into $\hat{\xi}$, integrating over the region $G_{\varepsilon}$, and computing the discontinuity using the matching conditions \crefrange{eq:continuity3}{eq:continuity5} yields
    \begin{subequations}
            \begin{align}
    &\int_{u_0-\varepsilon}^{u_0 + \varepsilon} du\ \Theta\  \mathfrak{i}_{\hat{\xi}}\delta \kappa =  8\pi\int_{u_0-\varepsilon}^{u_0 + \varepsilon}du \ \ell^i \ell^j \mathfrak{i}_{\hat{\xi}}\delta T_{ij}, \\ 
     &\int_{u_0-\varepsilon}^{u_0 + \varepsilon}du \ \sigma_{ij}\ \mathfrak{i}_{\hat{\xi}}\delta \kappa =  \int_{u_0 - \varepsilon}^{u_0 + \varepsilon} du \ \ell^{\ell}\ell^m \mathfrak{i}_{\hat{\xi}}\delta C_{i \ell j m}.
        \end{align}
    \end{subequations} 
A simple calculation using Eqs.\ \eqref{eq:alphadef}, \eqref{eq:betadef} and \eqref{eq:kappavariation} specialized to $\kappa = 0$ and using $\beta_\xi = - \beta$ results in \begin{align}\mathfrak{i}_{\hat{\xi}}\delta \kappa = \left[\alpha + (u-u_0)\beta\right]\partial_u \delta(u-u_0) + 2\beta \delta(u-u_0),\end{align} hence it follows that 
\begin{subequations}
\begin{align}
&\lim_{\varepsilon\rightarrow 0}\int_{u_0-\varepsilon}^{u_0 + \varepsilon} du\ \Theta\  \mathfrak{i}_{\hat{\xi}}\delta \kappa = \left. \left(-\alpha\partial_u \Theta + \beta \Theta\right) \right|_{u=u_0},\\
&\lim_{\varepsilon\rightarrow 0}\int_{u_0-\varepsilon}^{u_0 + \varepsilon} du\ \sigma_{ij}\  \mathfrak{i}_{\hat{\xi}}\delta \kappa = \left. \left(-\alpha\partial_u \sigma_{ij} + \beta \sigma_{ij}\right) \right|_{u=u_0},
\end{align}
\end{subequations}
where we've integrated by parts on the $\partial_u \delta(u-u_0)$ term.

This means the perturbed spacetime needs to have an impulsive null matter shell and an impulsive gravitational wave in order to be on-shell:
    \begin{subequations}
      \label{eq:shocks}
   \begin{align}
    &\ell^i \ell^j\mathfrak{i}_{\hat{\xi}}\delta T_{ij}(u) = -\frac{1}{8\pi}\Bigl[\alpha \partial_u\Theta(u) -\beta \Theta(u)\Bigr]\delta(u-u_0)\label{eq:stresstensorshock}, \\ 
  &\ell^{\ell}\ell^m\mathfrak{i}_{\hat{\xi}}\delta C_{i \ell j m}(u) = -\Bigl[\alpha \partial_u\sigma_{ij}(u) - \beta \sigma_{ij}(u)\Bigr]\delta(u-u_0)\label{eq:weylshock}.
    \end{align}
   \end{subequations}
We can ask when the stress tensor shock satisfies the null energy condition. It will be violated if the following is true: 
\begin{align}
    \beta \Theta(u_0) \leq \alpha \partial_u \Theta(u_0).
\end{align}
If the background matter field satisfies the null energy condition, then the classical focusing theorem holds $\partial_u \Theta \leq 0$, and on the event horizon the classical area theorem also holds $\Theta \geq 0$. So if $\alpha,\beta > 0$, then the inequality above cannot hold on the event horizon. Thus it must be the case that 
\begin{align}
\ell^i \ell^j \mathfrak{i}_{\hat{\xi}}\delta T_{ij}(u) \geq 0,
\end{align}
as desired. The constraint $\alpha > 0$ just means we translate to the future, while the constraint $\beta > 0$ imposes the boost vector field be future-directed on $\mathscr{H}_{>u_0}$ (i.e.\ that it blueshift instead of redshift).

	In \cref{sec:edgemodes} below, we explore how the results we've obtained thus far can be interpreted in terms of classical crossed product algebras associated with horizon subregions.

	\subsection{Subregion phase spaces: top down approach}
	\label{sec:edgemodederivation2}
	
In this section we develop the alternative, top down approach to defining the phase spaces of the subregions $\mathscr{H}_+$ and $\mathscr{H}_-$, starting from the global horizon phase space of \cref{sec:spacetimeSTs}, extended to include shocks. The key idea is that we want to restore some of the gauge degrees of freedom (by which we mean not just diffeomorphisms but degeneracy directions of the symplectic form) which have been fixed in that phase space.

That gauge fixing can be understood as follows.
        Suppose that we consider dressed cuts of the horizon, that is, functionals
        $S = S[\phi]$ of the dynamical fields which transform
        covariantly under diffeomorphisms.  (These will naturally allow
        dressed observables, dressed subregions and dressed subregion phase
        spaces.)  It is always possible to make a field dependent
        diffeomorphism to map such a dressed cut $S[\phi]$ onto a
        fixed cut $S_0$, and then to consider only gauges which
        preserve $S_0$.  This is what the construction of \cref{sec:spacetimeSTs}
        effectively does, since the cut $S_0$ is fixed there.  We
        would
        like to undo this gauge fixing and to allow arbitrary dressed
        cuts $S[\phi]$.

Our starting point is the covariant framework for dressed subregion phase spaces and edge modes
	in gravitational theories that has been developed over the past
	several years, starting with the seminal work of Donnelley and Freidel
	\cite{Donnelly:2016auv} and with many developments and extensions	\cite{Speranza:2017gxd,Speranza2018a,Gomes:2018dxs,2017NuPhB.924..312G,Ciambelli:2021nmv,Ciambelli:2021vnn,Freidel:2021dxw,Speranza:2022lxr,Klinger:2023tgi,Pulakkat:2025eid,Francois:2025sic,Freidel:2025ous},
        particularly the work by Hoehn and collaborators
        \cite{Carrozza:2021gju,Carrozza:2022xut,Araujo-Regado:2024dpr}.
        Our
        application of the formalism
        will yield the two sets of edge modes
        $\left(\Gamma_0^{\pm}, \Upsilon_0^{\pm}\right)$ at the corners
        $S^{\pm}_0$ introduced in the previous subsection.
        Our approach
        differs from much of the literature in that we introduce
	two sets of edge modes rather than a single set, following
        Donnelley and Freidel \cite{Donnelly:2016auv}, which will be
        critical for our results.  The derivation will yield
                  the corner
          contribution \eqref{eq:areacornersymp} 
          to the symplectic form, and the generators \eqref{eq:APformulae}
          of half
        sided supertranslations.

	        We now briefly summarize the formalism and then apply it to our present context.
    The formalism introduces a
	reference spacetime ${\bar M}$ in addition to the physical spacetime
	$(M,g_{ab})$, and also an embedding map $X : {\bar M} \to M$.  We treat $X$ as a dynamical variable in the theory and consider an extended phase space consisting of pairs $(\phi, X)$, where $\phi$ are the original dynamical fields (a metric and matter fields). We define the pullback of the dynamical fields to the reference manifold as
	\be
	\pi = X_* \phi = \phi \circ X,
	\ee
	and we can use either $(\phi, X)$ or $(\pi,X)$ as coordinates on the extended phase space.
	We introduce a fixed null boundary ${\bar {\mathscr H}}$ and fixed corner ${\bar
		S}_0$ on the reference manifold, and define the corresponding
	objects on the physical manifold by mapping with the embedding map:
	\be
	\mathscr{H} = X( {\bar {\mathscr H}}), \ 
	S_0 = X( {\bar S}_0 ).
	\ee
	In this way the corner $S_0$ becomes field dependent or dressed.
	We define an action principle and Lagrangian on the reference
	manifold, in a way that depends only on $\pi$ and not $X$
	\cite{Carrozza:2021gju}, and then lift it to the physical manifold.
	One then finds that the dynamics of the embedding map is gauge (a
	degeneracy direction of the symplectic form) except at spacetime
	boundaries where it gives rise to edge modes.  
	Essentially the construction uses the Stueckelberg trick to restore
	covariance that is broken by the presence of non-dynamical structures
	(the boundary and cut).

        In the present context, the fields $\phi$ on $\mathscr{H}$
        consist of the various quantities we have defined,
        $\ell^i, \eta_{ijk}, \mu_{ij}, \Theta, q_{AB}$ and
        $\sigma_{AB}$, together with any matter fields.  On the
        reference surface ${\bar {\mathscr{H}}}$ the fields $\pi$
        consist of barred versions of these quantities, ${\bar
          \ell}^i, {\bar \eta}_{ijk}, {\bar \mu}_{ij}, {\bar \Theta},
        {\bar q}_{AB}$ and
        ${\bar \sigma}_{AB}$, related to the unbarred versions by pullbacks.
        
	A variation in the embedding map $X$ can be parameterized in terms of
	a vector field ${\vec \chi}$ on the physical spacetime
	\cite{Donnelly:2016auv}, defined so that the pullback of the perturbed
	embedding $X + \delta X$ is given by
	\be
	X_*^{-1} [X+ \delta X]_* = 1 + \lie_{\chi} + O(\delta X^2).
	\label{chiXdef0}
	\ee
        The vector field ${\vec \chi}$ will parameterize the edge modes.
	We will restrict to embeddings $X$ for which ${\vec \chi}$ evaluated on the null surface lies along the null generators:
	\be
	   {\vec \chi} = \chi {\vec \ell}.
           \label{eq:edgerestriction}
	\ee
	In effect we are considering just the subset of the full set of
	gravitational edge modes associated with supertranslations, which will
	be sufficient for our purposes.  The structure of the full set of edge
	modes is discussed in Ref.\ \cite{Ciambelli:2021vnn}.
        Because of the restriction \eqref{eq:edgerestriction}
        we can take the location of the null surface $\mathscr{H}$ to be fixed,
        and consider the embedding map to be a map $X : \bar{\mathscr{H}} \to \mathscr{H}$.
        More general embedding maps that shift the location of the
        horizon are discussed in \cref{app:horizondeformation}.

        It is natural to further restrict the embedding maps as
        follows. We fix a set of fields $(\bar{\ell}^a, {\bar \kappa},
        \bar{\ell}_a)$ on $\bar{\mathscr{H}}$, defined up to the rescaling freedom.  We assume\footnote{This assumption is not really necessary since
the additional degrees of freedom which it excludes turn out to be degeneracy directions of the symplectic form, i.e. pure gauge.  However imposing this condition here is convenient since it simplifies the calculations.}
        that the embedding map $X$ maps the equivalence class $[\bar{\ell}^a, {\bar \kappa},
          \bar{\ell}_a]$ onto the corresponding equivalence class
        $[\ell^a, \kappa,
          \ell_a]$ on $\mathscr{H}$ which defines the phase space $\mathcal{P}_{\mathscr{H}}$.

	We now discuss in more detail how to define a subregion phase space
	associated with the region ${\bar {\mathscr{H}}}_+$ of
	${\bar {\mathscr{H}}}$
	to the future of the cut ${\bar S}_0$.  A key point is that we
        allow the embedding map to be discontinuous at ${\bar S}_0$,
        giving rise to two independent set of edge modes \cite{Donnelly:2016auv}.
        In more detail, we consider two independent embedding maps
        \be
        X_- : \bar{\mathscr{H}}_- \to \mathscr{H}, \  X_+ : \bar{\mathscr{H}}_+ \to \mathscr{H},
        \ee
        and we define $S^\pm_{0} = X_{\pm}({\bar S}_0)$.
        We will eventually require that $S^+_0$ and $S^-_0$ coincide.
        We next fix an affine coordinate ${\bar u}$ on ${\bar
          {\mathscr{H}}}$ for which ${\bar \kappa} = 0$ and for which
        ${\bar S}_0$ is at ${\bar u} = 0$.  We also fix an affine
        coordinate $u$ on $\mathscr{H}$.  The edge modes 
        $\left(\Gamma_0^{\pm}, \Upsilon_0^{\pm}\right)$ are now
        defined in terms of the following parameterization of the maps
        $X_{\pm}$:
        \be
        u = \Upsilon_0^\pm + e^{\Gamma_0^\pm} {\bar u}.
        	\label{eq:edgemodedefs}
        \ee
        These quantities do depend on the choices of coordinates
        ${\bar u}$ and $u$, which have the freedom ${\bar u} \to {\bar
          b}
        {\bar u}$ and $u \to a + b u$, but the variation $\delta
        \Gamma_0^{\pm}$
        is invariant under these transformations, while the variation
        $\delta \Upsilon_0^{\pm}$ is invariant under ${\bar b}$ and
        $a$, and depends on $b$ in such a way that $\delta
        \Upsilon_0^{\pm} \partial_u$ is invariant.
        The modes $\Upsilon_0^\pm$ parameterize the location of the
        cuts $S_0^\pm$ which are now dressed, ie field dependent.
        
        We can now compute the vector field ${\vec \chi}$ that
        parameterizes variations of the embedding map, by combining
        Eqs.\ \eqref{chiXdef0}, \eqref{eq:edgerestriction} and
        \eqref{eq:edgemodedefs}.
        This gives
        \be
        \chi = \left[ \delta \Upsilon_0^+ + \delta \Gamma_0^+ ( u -
          \Upsilon_0^+) \right] H_+
        + \left[ \delta \Upsilon_0^- + \delta \Gamma_0^- ( u -
          \Upsilon_0^-) \right] H_-
        \label{eq:chiformula}
        \ee
        where
        \be
        H_+ = H( u - \Upsilon_0^+), \ H_- = H(-u +
        \Upsilon_0^-),
        \ee
        and $H$ itself is the Heaviside step function as earlier.
        We restrict the phase space by the assumption that
        $\Upsilon_0^+ > \Upsilon_0^-$, ensuring that the two terms in
        \Eqref{eq:chiformula} do not overlap.\footnote{This constraint is equivalent to requiring the null energy condition to be satisfied under half-sided null translations; see \Eqref{eq:stresstensorshock}.}
        It follows that the variations in the modes can be written
        directly in terms of $\chi$:
       \begin{subequations}
        \begin{align}
        \delta \Upsilon_0^\pm &= \left. \chi \right|_{S_0^{\pm}},
        \\       \delta \Gamma_0^\pm &= \left. (\lie_\ell +
        \kappa) \chi \right|_{S_0^{\pm}},
        \end{align}
                \end{subequations}
        where we have reverted to a general (non-affine) choice of
        normal ${\vec \ell}$.

        Given these edge modes, we now define the phase space 
        $\mathcal{P}_{\mathscr{H}_+}$
	to consist of the bulk fields $\phi$ defined on
        $\mathscr{H}_+$, together with the edge modes $\Gamma_0^+$ and
        $\Upsilon_0^+$. We similarly define 
        $\mathcal{P}_{\mathscr{H}_-}$.  Note that these are not
        subspaces of the full phase space $\mathcal{P}_{\mathscr{H}}$. 
        Instead, as explained by Donnelley and Freidel \cite{Donnelly:2016auv},
$\mathcal{P}_{\mathscr{H}}$ is obtained from
        $\mathcal{P}_{\mathscr{H}_+} \times
        \mathcal{P}_{\mathscr{H}_-}$ by imposing certain continuity or
        gluing conditions at the corner, and by performing a
        symplectic reduction with respect to the diagonal subgroup of
        the product of the two surface symmetry groups\footnote{So far we have treated the embedding maps $X_\pm$ as independent of the dynamical fields $\phi$.
          This is natural from the point of view of view of the phase space  $\mathcal{P}_{\mathscr{H}_+}$.
          However ultimately to define subregion phase spaces starting from the global horizon phase space,
          the embedding maps should be taken to be functionals of the dynamical fields $\phi$, yielding dressed subregion phase spaces.
          We will assume that this dressing is extrinsic, that is, $X_+$ is a functional of the fields on ${\mathscr H}_-$, for the reasons outlined in Refs.\ \cite{Pulakkat:2025eid,Araujo-Regado:2024dpr}; intrinsic dressing does not give rise to non-trivial corner symmetries.  Throughout the rest of the paper we will 
          continue to regard the embedding maps as independent fields, the specific choice of dressing will not play any role.}.

        We define the symplectic form for the full phase space in
        terms of the fields on the reference manifold.  We integrate
        over the entire null surface ${\bar {\mathscr{H}}}$, with no
        corner terms\footnote{Our symplectic form depends on the choice of splitting
        \eqref{decompos1} of the symplectic potential given by \Eqref{fluxGR}.
        That choice is uniquely determined by the Wald-Zoupas criteria \cite{WZ}
        which are physically reasonable \cite{CFP}.  Other choices would lead to different charges and to a different expression for the symplectic form, but we expect that the charges would still be integrable.}:
	\be
	\Omega_{\bar{\mathscr{H}}} = \int_{{\bar {\mathscr{H}}}} \delta \bar{ \bm{\mathcal{E}}}.
	\label{eq:omegaext}
	\ee
        Here $\delta \bm{\mathcal{E}}$ is the
        expression (\ref{eq:sympfin0}) for the sympletic current for general
        relativity, but with the fields replaced by their barred
        versions.  In \cref{app:symform} below we rewrite this
        symplectic form in terms of fields on the physical manifold
        $M$ and the edge modes.  The symplectic form is a function of
        background fields and of their variations.  For simplicity, in
        computing the symplectic form, we restrict attention to
        configurations of the background fields where the embedding
        map is continuous.  That is, we impose $\Upsilon_0^+ =
        \Upsilon_0^-$ and $\Gamma_0^+ = \Gamma_0^-$ on the background
        fields, but not on their variations.  This special case will
        be sufficient for the applications in the rest of the paper,
        since we will focus on linearized perturbations around a given
        background.  With this restriction we need not differentiate
        between the two cuts $S_0^+$ and $S_0^-$.
        The result obtained in \cref{app:symform} is
        \be
        \widehat{\Omega}_{\mathscr{H}} = \int_{{\mathscr{H}_-}} \delta  \bm{\mathcal{E}} + \int_{{\mathscr{H}_+}} \delta  \bm{\mathcal{E}} +
        \Omega_{\partial G},
	\label{eq:omegaext1}
        \ee
        where
        \begin{align}
        \Omega_{\partial G} =\ & \frac{1}{8\pi}\int_{S_0} \left[ \delta \Upsilon_0^+
          \wedge \delta (\lie_\ell {\bm \mu}) - \delta \Gamma_0^+ \wedge
          \delta \Delta \bm{\mu}_++ \delta \Upsilon_0^+ \wedge
          \delta \Gamma_0^+ \Theta \bm{\mu}  \right] - (+ \leftrightarrow -) \nonumber \\
        & + \frac{1}{8\pi}\int_{S_0} \left[ 
          \delta \Upsilon_0^+ \wedge \delta \Upsilon_0^- \lie_\ell(\bm{\mu}\Theta) \right].
        \label{eq:cornerterm1}
        \end{align}
        Here $\Delta \bm{\mu}_{\pm} := \bm{\mu} - \bm{\mu}(u={\pm\infty})$ is a background subtracted area element on $S_0$.
        Note that the last term couples together the two phase spaces 
        $\mathcal{P}_{\mathscr{H}_+}$ and
        $\mathcal{P}_{\mathscr{H}_-}$
        except in the special case when $\Theta$ and $\lie_\ell
        \Theta$ vanish on $S_0$.
        
        We next discuss two different kinds of diffeomorphism symmetries 
        \cite{Donnelly:2016auv,Speranza:2017gxd,Speranza2018a,Carrozza:2021gju,Freidel:2025ous}. 
        Consider first diffeomorphisms $Y : M \to M$ on the physical
        manifold, which can be field dependent.  These act on the
        fields on $M$ as $(\phi, X) \to (\phi \circ Y, Y^{-1} \circ
        X)$.  On the reference manifold they act as $(\pi, X) \to
        (\pi, Y^{-1} \circ X)$.  In particular the bulk fields $\pi$
        on the reference manifold are invariant under these
        transformations.
        They are therefore degeneracy directions of the symplectic
        form, with zero charges, since the symplectic form
	\eqref{eq:omegaext}
        on the
        reference manifold depends only on $\pi$ and not on $X$.
        For linearized supertranslations of the form ${\vec \xi_{\rm g}} = f
        {\vec \ell}$, the fields transform as given by
        \eqref{eq:symtransfs},
        while the embedding variation ${\vec \chi}$ transforms as \cite{Speranza:2017gxd}
	\be
	\mathfrak{i}_{\hat \xi_{\rm g}} {\vec \chi} = - \vec{\xi}_{\rm g}.
	\ee
        Here the subscript ``g'' denotes gauge, to distinguish these
        transformations from the second class discussed below which
        are not gauge.
        Writing ${\vec \xi_{\rm g}} = (\alpha_{\rm g} + \beta_{\rm g} u )
        \partial_u$ as in \Eqref{eq:supertranslationgroup} and
        comparing with \Eqref{eq:chiformula} now gives that
        \begin{subequations}
        \label{eq:transform11}
	\begin{align}
	&\mathfrak{i}_{{\hat \xi_{\rm g}}} \delta \Upsilon_0^{+} = - 
          \alpha_{\rm g}, \ \mathfrak{i}_{{\hat \xi_{\rm g}}} \delta \Upsilon_0^{-} = - 
          \alpha_{\rm g}, \\
	  &\mathfrak{i}_{{\hat \xi_{\rm g}}} \delta \Gamma_0^{+} = - \beta_{\rm g}, \ 
          	\mathfrak{i}_{{\hat \xi_{\rm g}}} \delta \Gamma_0^{-} = - \beta_{\rm g}.
	\end{align}
        \end{subequations}
	Substituting the transformations \eqref{eq:transform11} and 
        \eqref{eq:symtransfs} into the symplectic form
	\eqref{eq:omegaext1} now gives
        \be
        \mathfrak{i}_{{\hat \xi_{\rm g}}} \widehat{\Omega}_{\mathscr{H}} =
        0
        \ee
        as expected, which is a useful consistency check of the
        formula \eqref{eq:cornerterm1}. We note that our application of the formalism differs
          from that of
          Refs.\ \cite{Ciambelli:2021nmv,Speranza:2022lxr,Freidel:2021dxw}, 
        who choose a different symplectic form and as a consequence obtain nonzero integrable charges for these
        transformations.  We consider it preferable for these
        transformations to be exact gauge symmetries, following
        Refs.\ \cite{Carrozza:2022xut,Carrozza:2021gju}.

        The second kind of diffeomorphism symmetry consists of maps $Z
        : {\bar M} \to {\bar M}$ from the reference manifold to
        itself.  Under these transformations the fields on the
        physical manifold transform as $(\phi, X) \to (\phi, X \circ
        Z)$ while those on the reference manifold transform as $(\pi,
        X) \to (\pi \circ Z, X \circ Z)$.  We first discuss the
        perspective of the reference manifold.
        We specialize to a half sided supertranslation of the form
        \be
        {\vec \xi} = ( \alpha + \beta {\bar u}) H({\bar u})
        \frac{\partial}{\partial {\bar u}}.
        \label{eq:stref}
        \ee
        Combining this transformation with the symplectic form
        (\ref{eq:omegaext}) yields exactly the same calculation as was performed
        in \cref{sec:integrable2} above, but reinterpreted to apply to
        the reference manifold rather than the physical one.
        We conclude that the corresponding charge is integrable, and
        given by \Eqref{eq:cornertermintegrable} rewritten in terms of barred fields:
        \be
        -\mathfrak{i}_{\hat{\xi}}\Omega_{\bar{\mathscr{H}}} = \delta (\mathscr{A}_{\beta}
        + \mathscr{P}_{\alpha})
        \label{eq:genform}
        \ee
        with
        \be
         \mathscr{A}_{\beta}= \frac{1}{8\pi}\left[\int_{{\bar
               S}_0}\beta\bm{\bar {\mu}} - \int_{{\bar S}_\infty}\beta
           \bm{{\bar \mu}}\right], \  \mathscr{P}_{\alpha} =
         -\frac{1}{8\pi}\int_{{\bar S}_0}\alpha\lie_{{\bar
             \ell}}\bm{{\bar \mu}}.
         \label{eq:chargesref}
	\ee

        Consider next the perspective of the physical manifold for
        this symmetry transformation.  The bulk fields $\phi$ do not
        transform, while the edge modes transform as $X \to X \circ
        Z$, yielding from Eqs.\ \eqref{eq:edgemodedefs} and
        \eqref{eq:stref} that
        \begin{subequations}
        \begin{align}
		&\mathfrak{i}_{\hat{\xi}} \delta \Gamma^{+}_0 = \beta,
                \  \mathfrak{i}_{\hat{\xi}}\delta \Upsilon^{+}_0 =
                \alpha e^{\Gamma_0^+}, \\
		&\mathfrak{i}_{\hat{\xi}}\delta \Gamma^{-}_0 = 0,
                \ \mathfrak{i}_{\hat{\xi}}\delta \Upsilon^{-}_0 = 0.
		\label{eq:actiononedgemodes1}
	\end{align}
            \end{subequations}
        Substituting these transformations into the symplectic form
        \eqref{eq:cornerterm1} again yields an integrable charge of
        the form \eqref{eq:genform}, where now\footnote{An alternative expression for $\mathscr{P}_{\alpha}$ is \be\displaystyle \mathscr{P}_{\alpha} =
         -\frac{1}{8\pi}\int_{S_0^-}\alpha e^{ \Gamma_0^+} \lie_{{
               \ell}}\bm{{\mu}}\ee
         where the integral is evaluated over $S_0^-$ instead of $S_0^+$.}
\begin{subequations}
\begin{align}
         &\mathscr{A}_{\beta}= \frac{1}{8\pi}\left[\int_{S_0^+}\beta\bm{{\mu}} - \int_{S_\infty}\beta
           \bm{{ \mu}}\right], \\ &\mathscr{P}_{\alpha} =
         -\frac{1}{8\pi}\int_{S_0^+}\alpha e^{ \Gamma_0^+} \left[\lie_{{
             \ell}}\bm{{\mu}} - \left(\Upsilon_0^+ - \Upsilon_0^-\right)\lie_{\ell}(\bm{\mu}\Theta)\right].
                 \label{eq:chargesref1}
	\end{align}
                \end{subequations}
        This result is compatible with the formulae
        \eqref{eq:chargesref}, noting that (i) the factor $e^{-\Gamma_0^+}$ in $\mathscr{P}_{\alpha}$
        arises from taking the pullback from the reference
        manifold and (ii) the appearance of the $\Upsilon^+_0-\Upsilon_0^-$ piece of $\mathscr{P}_{\alpha}$ results from the $\pm$ coupling term in \Eqref{eq:cornerterm1}, where we drop terms that are $\mathcal{O}\left((\Upsilon_0^+ - \Upsilon_0^-)^2\right)$ for the reason discussed before \Eqref{eq:omegaext1}

        We summarize some of the features of the covariant dressed phase space formalism in Table \ref{tab:table1}.

\begin{table*}[t]
\centering
\footnotesize
\begin{tabular}{| c || c | c | c | c |}
\hline
&Reference manifold description & Physical manifold description\\ 
\hhline{| = # = | = | }

\hline
Nature of cut & Cut ${\bar S}_0$ of horizon is fixed  & Cut $S_0$ of horizon is dressed, its  \\
 & & location depends on the \\
&& field configuration \\
\hline
Symplectic & Symplectic form
\eqref{eq:omegaext} is independent of   & Symplectic form
\eqref{eq:omegaext1}
depends on edge  \\ form &  edge modes  &  modes \\
\hline
Nature of & Varying $\Upsilon_0^\pm, \Gamma_0^\pm$ at fixed    & Varying $\Upsilon_0^\pm, \Gamma_0^\pm$ at fixed
  \\ edge modes &  bulk fields $\pi$ is gauge. &  bulk fields $\phi$ is not gauge \\
  \hline
  Physical & Diffeomorphism & No diffeomorphism\\
  symmetries & Bulk fields $\pi$ transform under \eqref{eq:stref}    & Bulk fields $\phi$ do not transform
  \\  &  Edge modes transform via \eqref{eq:actiononedgemodes1} &  Edge modes transform via \eqref{eq:actiononedgemodes1}  \\
  \hline
  True gauge & No diffeomorphism & Diffeomorphism\\
  symmetries & Symmetries \eqref{eq:transform11} act only on edge    & Symmetries \eqref{eq:transform11} act on both 
  \\  &  modes; charges are zero &  bulk fields \& edge modes; charges zero \\
\hline
\end{tabular}
\caption{
  The covariant dressed phase space formalism makes use of two
  manifolds, the physical manifold $M$ and a reference manifold ${\bar
    M}$.  The two descriptions are mathematically equivalent, but
  many features of the phase space and symmetries appear different in the two domains.
  This table summarizes some of the differences.}
  \label{tab:table1}
\end{table*}

\normalsize

        \subsection{Derivation of edge mode contributions to the
          symplectic form\label{app:symform}}

        In this subsection we derive the edge mode contributions
        \eqref{eq:cornerterm1} to the symplectic form
        \eqref{eq:omegaext1} on the extended phase space.

        We start with the integral \eqref{eq:omegaext} over the
        reference null surface $\bar{\mathscr{H}}$, and rewrite it as
        an integral over $\mathscr{H}$ by using the embedding map $X$.
        Because the embedding
        map has a variation we obtain 	\cite{Speranza:2017gxd,Speranza2018a}
        \be
        \Omega = \int_{\mathscr{H}} \delta \bm{\mathcal{E}}( \phi,
        \delta_1 \phi + \mathfrak{i}_{\hat \chi_1} \delta \phi,
        \delta_2 \phi + \mathfrak{i}_{\hat \chi_2} \delta \phi),
        \ee
        where we have written explicitly the dependence on two independent variations.  Another notation for this is
        \be
        \Omega = \int_{\mathscr{H}}  \left[ \delta \bm{\mathcal{E}}
+ \mathfrak{i}_{\hat \chi} \delta \bm {\mathcal{E}} + \frac{1}{2} 
\mathfrak{i}_{\hat \chi} \mathfrak{i}_{\hat \chi}
\delta \bm {\mathcal{E}} \right].
        \label{omega11}
        \ee
        Here we have assumed that the background embedding map $X$ is
        continuous, but its variation parameterized by ${\vec \chi}$
        can have discontinuities at the cut ${\bar S}_0$.
        We now define the quantity
        \be
        \bm{Y}_\chi =  - \mathfrak{i}_{\hat \chi} \delta \bm
           {\mathcal{E}} - \frac{1}{2}  \mathfrak{i}_{\hat \chi}
           \mathfrak{i}_{\hat \chi} \delta \bm {\mathcal{E}}  + d
           \bm{\sigma}, 
           \label{eq:Ychidef}
        \ee
        where
	\be
	\bm{\sigma} = i_\chi \bm{\mathcal E} + \delta \bm{h}_\chi
	+ \lie_\chi \bm{h}_\chi
	+ \frac{1}{2} i_\chi i_\chi (\bm{L} - d \bm{\alpha})
	\label{eq:sigmadef}
	\ee
	and $\bm{h}_\chi = \bm{Q}_{\xi} -
	i_{\xi}\bm{\alpha}
	-   \mathfrak{i}_{\hat{\xi}} \bm{\gamma} $
	is the integrand of the corner charge \eqref{WZcorner}.
        Combining
        Eqs.\ \eqref{omega11}, \eqref{eq:Ychidef} and
        and using that the boundary of
        $\mathscr{H}$ consists of the surfaces $S_{\pm \infty}$ at $u
        = \pm \infty$ now gives
        \be
        \Omega = \int_{\mathscr{H}} \delta \bm{\mathcal{E}}
        + \int_{S_{-\infty}} \bm{\sigma} - \int_{S_{\infty}}
        \bm{\sigma}         - \int_{\mathscr{H}} \bm{Y}_\chi.
                \label{omega12}
        \ee

        We will first restrict attention to the case when
        ${\vec \chi}$ is continuous, in other words when the $+$ and
        $-$ edge modes coincide for the variations as well as for the
        background.  In this case we have the identity
        \be
        \bm{Y}_\chi = 0
        \label{eq:Yidentity}
        \ee
        which kills the last term in \Eqref{omega12}.
        This identity is Eq.\ (4.12) of Speranza \cite{Speranza:2017gxd}
        modified by the substitutions
        \begin{subequations}
          \begin{align}
        &{\bm L} \to {\bm L} - d \bm{\gamma}, \\
         &\bm{\theta} \to \bm{\theta} - \delta \bm{\alpha} - d
         \bm{\gamma} = \bm{\mathcal{E}}, \\
         &\bm{Q}_\chi \to  \bm{Q}_\chi-	i_{\xi}\bm{\alpha}
	-   \mathfrak{i}_{\hat{\xi}} \bm{\gamma} = \bm{h}_\chi,
          \end{align}
        \end{subequations}
        that make use of the ``gauge freedom'' described in
        Ref.\ \cite{Chandrasekaran:2021vyu}.
        Next, in the expression \eqref{eq:sigmadef} for $\bm{\sigma}$ we make use of the
        fact that $\bm{L}$, $\bm{\alpha}$ and $\bm{\mathcal{E}}$ all
        vanish at $S_{\pm \infty}$, since the shear and expansion
        vanish there, making use of Eqs.\ \eqref{eq:theta00} and \eqref{fluxGR}.
        Using the expression \eqref{eq:cornerchargest} for
        $\bm{h}_\chi$ and replacing $f$ there with the expression
        \eqref{eq:chiformula} for $\chi$ yields from \Eqref{omega12}
                \be
        \Omega = \int_{\mathscr{H}} \delta \bm{\mathcal{E}}
        + \int_{S_{\infty}} \delta \Upsilon_0^+ \wedge \delta \bm{\mu}
                - \int_{S_{-\infty}} \delta \Upsilon_0^+ \wedge \delta \bm{\mu}.
        \label{omega13}
        \ee
        Thus there are no contributions from the surface $S_0$ in this
        case.
        
        We now turn to the more general case where $\chi$ is has
        discontinuities, as in the expression (\ref{eq:chiformula})
        where the edge mode variations are unconstrained.
        In this case the identity \eqref{eq:Yidentity} is no longer
        valid, and $\bm{Y}_\chi$ acquires distributional
        corrections localized to the corner $S_0$ due to the
        discontinuities, similar to the
        distributional corrections in Eq.\ (\ref{eq:finalcornerterm}).
        These corrections then give an additional contribution to the
        symplectic form coming from the fourth term in \Eqref{omega12},
        which we will show reproduces the corner term 
        \eqref{eq:cornerterm1}. 

        Before turning to the explicit evaluation of $\bm{Y}_\chi$, we
        first derive some properties of the edge mode field $\chi$.
        We specialize to null vectors
        ${\vec \ell}$ for which $\kappa =0$ for simplicity, and choose
        $u$ with ${\vec \ell} = \partial_u$.
        Taking a variation of \Eqref{eq:chiformula} and also computing
        $\chi \wedge \lie_\ell \chi$ gives
        \be
        \delta \chi = - \chi \wedge \lie_\ell \chi + \varpi
        \label{eq:identity33}
        \ee
        where
        \be
        \label{eq:varpiformula}
        \varpi = - \delta \Upsilon_0^+ \wedge \delta \Upsilon_0^- \,
        \delta( u - \Upsilon_0).
        \ee
        Here we have assumed that $\Upsilon_0^+ = \Upsilon_0^-$ and
        $\Gamma_0^+ = \Gamma_0^-$ for the background quantities as
        discussed before \Eqref{eq:omegaext1}.  The result
        \eqref{eq:identity33} with $\varpi = 0$ is a special case of a
        general identity
        that is valid when the embedding map is continuous [Eq.\ (2.8)
          of Speranza \cite{Speranza:2017gxd}], which can be interpreted as imposing the
        flatness of a connection defined by ${\vec \chi}$ on the space of solutions.
                Here we see that there are
        distributional corrections to the identity that arise from the
        discontinuities in the embedding map.
        
        We now turn to evaluating the expression 
        for $\bm{Y}_\chi$ given by Eqs.\ \eqref{eq:Ychidef} and
        \eqref{eq:sigmadef}.
        Rather than use the specific
        expression (\ref{eq:chiformula}) for $\chi$, we will for the
        moment allow $\chi$ to be arbitrary, and make use of the
        expressions \eqref{fluxGR} for $\bm{\mathcal{E}}$ and
        \eqref{eq:cornerchargest} for $\bm{h}_\chi$.
        From \Eqref{eq:Yidentity} the resulting expression for
        $\bm{Y}_\chi$ must have
        the property that it 
        vanishes identically when $\chi$ corresponds to a perturbation
        in the phase space $\mathcal{P}_{\mathscr{H}}$, which requires
        [cf.\ \Eqref{eq:symmetrydef}]
        \be
           {\ddot \chi} = 0,
           \ee
           where dots denote derivatives with respect to $u$,
        as well as the identity (\ref{eq:identity33}) with $\varpi = 0$.
        Our $\chi$ violates these equations due to its discontinuity
        which is what generates the distributional corrections.
        Therefore we can proceed by replacing $\delta \chi$ everywhere
        with
        $-\chi \wedge {\dot \chi} + \varpi$, and by retaining only
        terms which contain derivatives of $\chi$ of order two or
        higher, or which contain $\varpi$.  All of the remaining terms
        must cancel each other.

        The only terms in Eqs.\ \eqref{eq:Ychidef} and
        \eqref{eq:sigmadef} which generate 
        terms proportional to ${\ddot \chi}$ or $\varpi$ are 
        $\delta \bm{h}_\chi$ and $\lie_\chi \bm{h}_\chi$.
        Starting with the expression \eqref{eq:cornerchargest} with
        $\kappa =0$ gives $8 \pi \bm{h}_\chi =  ({\dot \chi} - \Theta
        \chi) \bm{\mu}$, and taking a variation and using
        \Eqref{eq:identity33} yields  
        \be
        \delta \bm{h}_\chi + \lie_\chi \bm{h}_\chi = \frac{1}{8 \pi} \left[ {\dot \varpi} - \Theta \varpi + \Theta \chi \wedge {\dot \chi} - \delta \Theta \wedge \chi + h \wedge ({\dot\chi} - \Theta\chi)/2 \right] \bm{\mu}.
        \ee
        Next taking an exterior derivative using the identity (B.4h)
        of Ref.\ \cite{Chandrasekaran:2023vzb} gives
        \be
        d(\delta \bm{h}_\chi + \lie_\chi \bm{h}_\chi) = \frac{1}{8
          \pi} (\lie_\ell + \Theta) \left[ {\dot \varpi} - \Theta \varpi + \Theta \chi \wedge {\dot \chi} - \delta \Theta \wedge \chi + h \wedge ({\dot\chi} - \Theta\chi)/2 \right] \bm{\eta}.
        \ee
        Now the terms involving $\varpi$ can be seen to comprise a
        total derivative which vanishes upon intergrating over
        $\mathscr{H}$, since $\varpi$ is localized to $S_0$ from
        \Eqref{eq:varpiformula}. So we can drop these terms. Expanding
        out the expression, dropping terms 
        according to the prescription described above
        and combining with Eqs.\ \eqref{eq:Ychidef} and \eqref{eq:sigmadef}
        finally yields
        \be
        \bm{Y}_\chi = \frac{1}{16
          \pi}  \left[2 \Theta \chi \wedge {\ddot \chi} + h \wedge
          {\ddot \chi}   \right] \bm{\eta}.
        \ee
        We now insert the expression \eqref{eq:chiformula} for $\chi$
        and evaluate at $\Upsilon_0^+ = \Upsilon_0^- = \Upsilon_0$,
        which generates terms proportional to $\delta(u -
        \Upsilon_0)$  and $\delta'(u - \Upsilon_0)$.  We integrate by
        parts to eliminate the $\delta'$ terms, insert into the fourth
        term in \Eqref{omega12} and use the expression \eqref{omega13}
        for the remaining three terms
        to finally obtain \Eqref{eq:omegaext1}.

	\section{Crossed product algebra and canonical quantization \label{sec:edgemodes}}
	
	Having identified the relevant corner charges and their action on horizon observables, we now package the classical horizon phase space into an algebraic structure that is tailored for quantization. The main point of this section is that the combination of bulk horizon observables with the nontrivial action of the edge mode charges naturally organizes itself into a crossed product algebra.
	
	We first describe the classical crossed product generated by gravitationally dressed local observables on horizon subregions and the outer automorphisms induced by half-sided boosts and translations. We then quantize this extended phase space, promoting the edge modes to operators and constructing the corresponding abstract $\ast$-algebra and GNS Hilbert space. This sets the stage for identifying a Type II$_{\infty}$ factor at each cut and for interpreting its von Neumann entropy as a generalized entropy.
	
	\subsection{Classical construction \label{sec:classicalalgebra}}
	Consider the horizon subalgebra $\mathcal{A}_{\mathscr{H}_{>u_0}} = \left\{\mathscr{O}: \mathcal{P}_{\mathscr{H}_{>u_0}}\mapsto \mathbb{R}\right\}$ consisting of local phase space observables $\mathscr{O}(p)$ for $p\in \mathscr{H}_{+}$. In order to have a gauge-invariant algebra of local observables, we need all such $p$ to be gravitationally dressed.
	
	To that aim, we can get to any $p \in \mathscr{H}_{+}$ by applying the exponential map to a suitable choice of $p_0 \in S^+_0$: 
	\begin{align}
		p = \text{exp}(u \ell)p_0.\label{eq:dressing}
	\end{align}
	So we can gravitationally dress any $p\in \mathscr{H}_{+}$ to $S^+_0$ in this way, rendering the algebra $\mathcal{A}_{\mathscr{H}_{+}}$ gauge-invariant. Since $S_0^+$ is itself gauge-invariantly specified via the edge modes $\Gamma_0^+$ and $\Upsilon_0^+$, we should view the prescription \eqref{eq:dressing} as dressing $\mathscr{O}(p)$ to the frame $\left(\Gamma_0^+, \Upsilon_0^+\right)$.

    More precisely, in \Eqref{eq:dressing} the symbol $\exp(u\ell)$ should be understood as the spacetime flow along the
    horizon generator $\ell^a$. Concretely, let $\varphi_u$ denote the one-parameter family of
    diffeomorphisms obtained by integrating $\ell^a$ along $\mathscr{H}_+$, i.e.
    \begin{equation}
    \frac{d}{du}\,\varphi_u(p_0)=\ell\big|_{\varphi_u(p_0)}, \ \varphi_{u=0}=\mathrm{id},
    \end{equation}
    so that \Eqref{eq:dressing} is simply $p=\varphi_u(p_0)$.
    
    Given any (local, diffeomorphism-invariant) phase space observable $\mathscr{O}(x)$, its
    gravitationally dressed version relative to the corner frame $(\Gamma_0^+,\Upsilon_0^+)$ is then
    the composition
    \begin{equation}
    \mathscr{O}(u,x^A)\equiv\mathscr{O}\!\left(p(u,x^A)\right)
    =\mathscr{O}\!\left(\varphi_u\!\left(p_0(x^A)\right)\right),
    \ p_0(x^A)\in S_0^+,
    \end{equation}
    where $x^A$ labels the generator (held fixed under the flow) and $u$ is the affine parameter
    distance from $S_0^+$ along that generator. In this sense, a dressed operator is simply an
    ordinary local operator evaluated at a relationally specified point $p(u,x^A)$ determined
    by the edge mode data.
    
    Let's first focus on the null translation generator. The only thing we need is that the edge mode $\Upsilon_0^+(x^A)$ shifts the affine origin on each generator
    at the corner $S_0^+$.
    Infinitesimally, changing $\Upsilon_0^+(y^A)$ moves the base point along the corresponding null ray by an affine
    translation generated by $\ell^a$,
    \begin{equation}
    p_0(x^A) \rightarrow \exp\!\big(\delta\Upsilon_0^+(y^A)\delta(x^A - y^A)\,\ell \big)\,p_0(x^A).
    \end{equation}
    If we keep the physical point $p$ fixed while varying $\Upsilon_0^+$, the relation $p=\exp(u\ell)p_0$ then forces a
    compensating change of the affine parameter $u$ appearing in \Eqref{eq:dressing}. Explicitly, given
    \begin{align}
    p=\exp(u\ell)p_0(x^A)
    =\exp\!\big((u+\delta u)\ell\big)\,
       \exp\!\big(\delta\Upsilon_0^+(y^A)\delta(x^A-y^A)\,\ell\big)\,p_0(x^A),
    \end{align}
    we must have that 
    \begin{align}
        \delta u = -\,\delta\Upsilon_0^+(y^A)\,\delta(x^A-y^A).\label{eq:chainruleedgemodetransform}
    \end{align}

	The generator $\mathscr{P}_{\alpha}$ acts on an observable $\mathscr{O}(p) \in \mathcal{A}_{\mathscr{H}_{>u_0}}$ as 
	\begin{align}
		\left\{\mathscr{P}_{\alpha}, \mathscr{O}(p)\right\} = \frac{1}{16\pi}\int_{S^+_0} d^{d-2}y \frac{\delta \mathscr{O}(u(u_0), x^A)}{\delta \Upsilon^+_0(y^A)}\left\{\mathscr{P}_{\alpha}, \Upsilon_0^+(y^A)\right\}, 
	\end{align}
	which follows from the Poisson bracket chain rule. But $\left\{\mathscr{P}_{\alpha}, \Upsilon^+_0(y^A)\right\} = \alpha(y^A) e^{\Gamma_0^+(y^A)}$ and 
	\begin{align}
		\frac{\delta\mathscr{O}(u,x^A)}{\delta \Upsilon_0^+(y^A)} = \mathfrak{L}_{\hat{\ell}}\mathscr{O}(u,x^A)\frac{\delta u}{\delta \Upsilon_0^+(y^A)} = -\delta(x^A - y^A)\mathfrak{L}_{\hat{\ell}}\mathscr{O}(u, x^A),\label{eq:funcderivdressing}
	\end{align}
	where we've made use of \Eqref{eq:chainruleedgemodetransform} and the identity
    \begin{align}
        \frac{\delta\mathscr{O}(u,x^A)}{\delta u} = \mathfrak{i}_{\hat{\ell}}\delta\mathscr{O}(u,x^A) = \mathfrak{L}_{\hat{\ell}}\mathscr{O}(u,x^A).
    \end{align}
    Putting it all together, we arrive at
	\begin{align}
		\left\{\mathscr{P}_{\alpha}, \mathscr{O}(p)\right\}  = -\alpha e^{\Gamma_0^+}\mathfrak{L}_{\hat{\ell}}\mathscr{O}.\label{eq:classicaltranslationact}
	\end{align}
	The factor of $e^{\Gamma_0^+}$ is a boost weight which guarantees the RHS is boost-invariant, since $\mathfrak{L}_{\hat{\ell}}$ has boost weight $-1$.
    
\begin{figure}[t]
		\centering
		\begin{tikzpicture}[
			scale=0.8,
			every node/.style={font=\small},
			decoration={zigzag,segment length=3pt,amplitude=1.2pt},
			>=stealth
			]
			
			\begin{scope}
				\coordinate (HminusL) at (-3,-3);
				\coordinate (HplusR)  at ( 3, 3);
				
				\coordinate (Sminus) at (-1.8,-1.8); 
				\coordinate (u0)     at ( 0.0, 0.0); 
				\coordinate (Splus)  at ( 1.8, 1.8); 
				
				\coordinate (offset) at (-0.1,0.1);
				\coordinate (-offset) at (0.1,-0.1);
				
				\coordinate (G1) at ($(Sminus)+(offset)$);
				\coordinate (G2) at ($(Splus) + (offset)$);
				\coordinate (G3) at ($(Splus) +(-offset)$);
				\coordinate (G4) at ($(Sminus)+(-offset)$);
				
				\draw[thick] (HminusL) -- (HplusR);
				
				\fill[gray!20] (G1) -- (G2) -- (G3) -- (G4) -- cycle;
				
				\node[below left=4pt]  at (-2.7,-2.3) {$\mathscr{H}_{-}$};
				\node[above right=4pt] at ( 2.7, 2.3) {$\mathscr{H}_{+}$};
				
				\fill (u0) circle (1.1pt);
				\node[below right=2pt] at (u0) {$u_0$};
				\node[above left=2pt]  at (u0) {$G_{\varepsilon}$};
				
				\fill (Sminus) circle (1.1pt);
				\fill (Splus)  circle (1.1pt);
				
				\node[below=4pt] at (Sminus) {$S_0^{-}$};
				\node[below=4pt] at (Splus)  {$S_0^{+}$};
				
				\node[above right=4pt] at (1.75, 1.0) {$(\mathscr{A},\mathscr{P})$};
				
				\coordinate (ExcTop1) at (2.2, 2.8);
				\coordinate (ExcBot1) at (2.8, 2.2);
				
				\draw[decorate,->,thick,red] (ExcTop1) -- (ExcBot1);
				
				\coordinate (ExcPt1) at ($(ExcTop1)!0.5!(ExcBot1)$);
				
				\draw[
				decorate,
				decoration={snake,segment length=4pt,amplitude=1.0pt},
				->,
				thick,
				blue
				]
				(ExcPt1) to[out=-120,in=75] (Splus);
			\end{scope}
			
			\draw[->,thick] (1.5,0) -- (4.5,0)
			node[midway,above=4pt] {};
			
			\begin{scope}[xshift=6cm]
				\coordinate (HminusL) at (-3,-3);
				\coordinate (HplusR)  at ( 3, 3);
				
				\coordinate (Sminus) at (-1.8,-1.8); 
				\coordinate (u0)     at ( 0.0, 0.0); 
				\coordinate (Splus)  at ( 1.8, 1.8); 
				
				\coordinate (offset) at (-0.1,0.1);
				\coordinate (-offset) at (0.1,-0.1);
				
				\coordinate (G1) at ($(Sminus)+(offset)$);
				\coordinate (G2) at ($(Splus) + (offset)$);
				\coordinate (G3) at ($(Splus) +(-offset)$);
				\coordinate (G4) at ($(Sminus)+(-offset)$);
				
				\draw[thick] (HminusL) -- (HplusR);
				
				\fill[gray!20] (G1) -- (G2) -- (G3) -- (G4) -- cycle;
				
				\node[below left=4pt]  at (-2.7,-2.3) {$\mathscr{H}_{-}$};
				\node[above right=4pt] at ( 2.7, 2.3) {$\mathscr{H}_{+}$};
				
				\fill (u0) circle (1.1pt);
				\node[below right=2pt] at (u0) {$u_0+\delta u$};
				\node[above left=2pt]  at (u0) {$G_{\varepsilon}$};
				
				\fill (Sminus) circle (1.1pt);
				\fill (Splus)  circle (1.1pt);
				
				\node[below=4pt] at (Sminus) {$S_0^{-}$};
				\node[below=4pt] at (Splus)  {$S_0^{+}$};
				
				\node[above right=4pt] at (1.75, 1.0)
				{$(\mathscr{A}+\delta\mathscr{A},\,\mathscr{P}+\delta\mathscr{P})$};
				
				\coordinate (ExcTop2) at (-2.2,-2.8);
				\coordinate (ExcBot2) at (-2.8,-2.2);
				
				\draw[decorate,->,thick,red] (ExcBot2) -- (ExcTop2);
				
				\coordinate (ExcPt2) at ($(ExcTop2)!0.5!(ExcBot2)$);
				
				\draw[
				decorate,
				decoration={snake,segment length=4pt,amplitude=1.0pt},
				->,
				thick,
				blue
				]
				(ExcPt2) to[out=-120,in=75] (Sminus);
				
			\end{scope}
			
		\end{tikzpicture}
		\caption{Effect of half-sided translations on gravitational subregions in the presence of excitations. The left panel shows the split horizon $\mathscr{H}^- \cup G_\varepsilon
			\cup \mathscr{H}^+$ with a cut at $u_0$ and associated corner charges
			$(\mathscr{A},\mathscr{P})$ on $S_0^+$. The red squiggle represents an excitation, and the blue squiggle denotes gravitational dressing of the excitation to its respective corner. Under a half-sided null translation generated by $\mathscr{P}$, the cut is moved to $u_0 + \delta u$ (right panel), and the corner charges are shifted to $(\mathscr{A} + \delta \mathscr{A},\, \mathscr{P} + \delta \mathscr{P})$.  The bulk fields remain smooth across the Cauchy splitting region, while only the edge
			modes on $S_0^+$ transform non-trivially, illustrating how the subregion is moved
			relative to its complement purely through the corner degrees of freedom. The change in the charges due to the excitation leaving the subregion under the deformation is what leads to integrability of the null translation generator; this is a consequence of gravitational constraints coupling the ``bulk'' excitations to the edge modes.}
		\label{fig:dressed-excitation-two-panels}
	\end{figure}
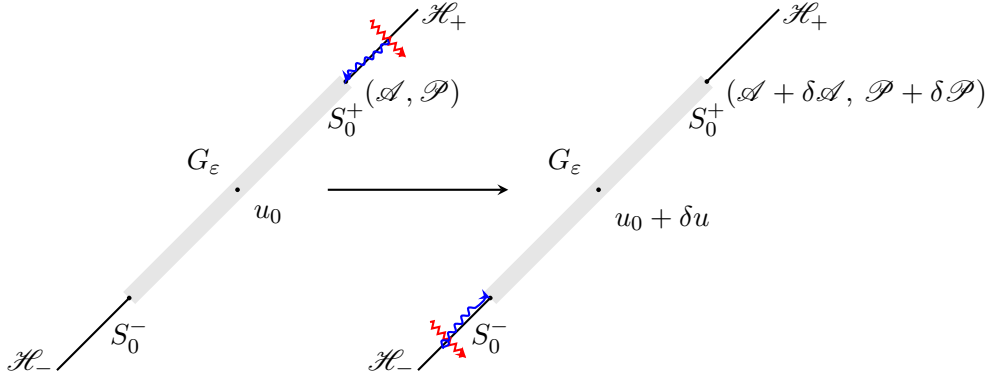

    Recall that under the matching conditions \crefrange{eq:continuity3}{eq:continuity5}, only the edge modes transform non-trivially, while ``bulk'' operators transform as pure gauge. So from the perspective of the point $p$, the cut $S_0^+$ is moving closer under a half-sided translation $\alpha$, meaning the distance \eqref{eq:dressing} shrinks in units of affine parameter. Hence the negative sign; see \cref{fig:dressed-excitation-two-panels}.
	
	Importantly, $\left\{\mathscr{P}_{\alpha}, \mathscr{O}(p)\right\} \in \mathcal{A}_{\mathscr{H}_{>u_0}}$. But $\mathscr{P}_{\alpha}$ is itself clearly not in $\mathcal{A}_{\mathscr{H}_{>u_0}}$. So it generates an \emph{outer automorphism} of $\mathcal{A}_{\mathscr{H}_{>u_0}}$. 
    
    We come now to the area operator. Under an infinitesimal rescaling of $u$, we have 
    \begin{equation}
    p_0(x^A) \rightarrow \exp\!\big((u-u_0)\delta\Gamma_0^+(y^A)\delta(x^A - y^A)\,\ell \big)\,p_0(x^A).
    \end{equation}
    Analogously to before, holding the physical point fixed while varying $\Gamma_0^+$ implies 
    \begin{align}
        \delta u = -(u-u_0)\delta(x^A - y^A).
    \end{align}
    Then, repeating the series of calculations from earlier but now using $\{\mathscr{A}_{\beta},\Gamma_0^+ \} = \beta$ yields
	\begin{align}
		\left\{\mathscr{A}_{\beta}, \mathscr{O}(p)\right\} = -(u-u_0)\beta \mathfrak{L}_{\hat{\ell}}\mathscr{O}.\label{eq:classicalboostact}
	\end{align}
	This too generates an outer automorphism of $\mathcal{A}_{\mathscr{H}_{>u_0}}$. It has no $e^{\Gamma_0^+}$ factor since it is already boost-invariant, due to the $(u-u_0)$ weight. 
    
    This means we can define a ``crossed product'' algebra 
	\begin{align}
		\widehat{\mathcal{A}}_{\mathscr{H}_{> u_0}} = \mathcal{A}_{\mathscr{H}_{>u_0}}\rtimes \left(C^{\infty}_{\beta}(\mathbb{S}^{d-2})^* \rtimes C^{\infty}_{\alpha}(\mathbb{S}^{d-2})^*\right),\label{eq:classicalcrossedprod} 
	\end{align}
	where $C^{\infty}_{\alpha}(\mathbb{S}^{d-2})^*$ is the dual (i.e.\ the space of generators $\mathscr{P}_{\alpha}$). Similarly, $C^{\infty}_{\beta}(\mathbb{S}^{d-2})^*$ is the space of generators $\mathscr{A}_{\beta}$. This is the classical analogue of the usual crossed product construction in the context of von Neumann algebras.

    The calculation above uses only the following structural feature of the dressing: points in $\mathscr{H}_+$ are identified by flowing along $\ell^a$ from $S_0^+$, meaning varying the corner data moves the
    dressed point only along $\ell^a$ at fixed generator label $x^A$. Any alternative ``one-sided'' $S_0^+$ anchored dressing that preserves this property (e.g.\ composing \Eqref{eq:dressing}
    with a diffeomorphism supported in $\mathscr{H}_+$ that is the identity at $S_0^+$, or using a different but
    $\Upsilon_0^+$ independent means of transporting tensors along the generator) yields the same
    functional derivative \Eqref{eq:funcderivdressing} and hence the same Poisson brackets \crefrange{eq:classicaltranslationact}{eq:classicalboostact}.
    
    Of course, what can change with the dressing is the concrete identification of a given operator with an
    element of $\mathcal{A}_{\mathscr{H}_{>u_0}}$: different dressings correspond to different (field-dependent) embeddings of the
    abstract subregion Poisson algebra into the full algebra, related by automorphisms. Within the class just described,
    this reshuffles representatives inside $\mathcal{A}_{\mathscr{H}_{>u_0}}$ but does not alter the geometric action of
    $\mathscr{P}_\alpha$ as null translations along $\ell^a$. In particular, the crossed product structure \Eqref{eq:classicalcrossedprod} remains preserved.

    The nearly unique choice of gravitational dressing \eqref{eq:dressing} is due to the fact that we cannot consider ``extrinsic'' dressings, i.e. dressings which depend on degrees of freedom not computable solely from fields on $\mathscr{H}_{>u_0}$ (such as shooting a spacelike geodesic out to the asymptotic boundary).\footnote{When we say ``extrinsic dressing'' we're referring to the dressing of ``bulk'' operators, not of the edge modes themselves. The terminology is sometimes used to refer to the latter in the literature, see e.g. \cite{Pulakkat:2025eid, Araujo-Regado:2024dpr}.} The extended symplectic form $\widehat{\Omega}_{\mathscr{H}_{>u_0}}$ only pairs variations of the one-sided ``bulk'' variables and ''two-sided'' edge modes $\left(\Phi_+;\Gamma_0^\pm,\Upsilon_0^\pm\right)$. Therefore the Poisson bracket $\{F,G\}_+$ is well-defined and closed only for functionals
    $F,G$ on the extended phase space $\widehat{\mathcal P}_{\mathscr{H}_{>u_0}}$.
    If $\mathscr O$ is extrinsically dressed then its functional derivatives include
    components along variations of fields in $\mathscr H_-$ and/or off $\mathscr H$ altogether,
    so $\mathscr O$ would not be a function on $\widehat{\mathcal P}_{\mathscr{H}_{>u_0}}$.
    It follows that extrinsic dressings do not define elements of $\mathcal A_{\mathscr H_{>u_0}}$:
    they would belong to a strictly larger algebra whose definition requires enlarging the phase space
    (and hence the symplectic form) beyond the null initial data of the one-sided subregion. Equivalently stated: the point of introducing the corner edge mode sector is that it supplies the
    entire relational reference frame data needed to construct gauge-invariant observables
    localized to $\mathscr H_{>u_0}$, without adjoining an auxiliary/external system.
	
	An extremely important point here is that $\mathscr{A}_{\beta}$ is \emph{not} in the center of $\mathcal{A}_{\mathscr{H}_{<u_0}}\cup \mathcal{A}_{\mathscr{H}_{>u_0}}$, as is evident from the brackets \crefrange{eq:classicaltranslationact}{eq:classicalboostact}. This might seem confusing at first, because one typically thinks of the area operator as acting purely on the relative boost angle edge mode in the Cauchy splitting region, i.e. $\left\{\mathscr{A}_{\beta}, \Gamma^+_0\right\} = \beta$. In the standard quantum error correction story (which does not know about diffeomorphism invariance), the area operator indeed lives in the center \cite{Harlow:2016vwg}. 
	
	But the calculations above show that $\mathscr{A}_{\beta}$ nevertheless acts non-trivially on the ``bulk'' algebra $\mathcal{A}_{\mathscr{H}_{>u_0}}$ precisely because of the dressing \eqref{eq:dressing}. The same goes for $\mathscr{P}_{\alpha}$ (which there is no analogue of in the quantum error correction picture). So the fact that we get a \emph{crossed product} as opposed to just a trivial tensor product is precisely because of gravitational dressing, which only shows up when treating gravity dynamically. This is also the reason we get an integrable null translation generator despite the non-stationarity. See \cref{fig:dressed-excitation-two-panels}. 
	
	We should therefore think of the edge modes on $\partial G_{\varepsilon}$ as playing the role of an observer, except we didn't have to put in an observer by hand; it just falls out of treating gravitational subsystems dynamically.
	
	On an eternal black hole background one can instead anchor the split at the bifurcation
	surface $\mathscr{B}$ and study the algebra of (say) the right exterior / right horizon degrees of
	freedom.  In that stationary setting, the modular flow of the exterior algebra is geometric:
	it is generated by boosts about $\mathscr{B}$, and adjoining a timeshift variable that implements
	this flow as an inner automorphism leads to a crossed product (Type II$_{\infty}$) algebra with a
	canonical trace, as emphasized in \cite{Chandrasekaran:2022eqq}. In our language,
	this timeshift is precisely the boost edge mode $\Gamma^+_0$ living on $\partial G_\epsilon$,
	with conjugate generator given by $\mathscr{A}_{\beta}$.  A purely Lorentzian covariant phase space derivation of this canonical corner pair was given in \cite{Chandrasekaran:2023vzb}, and can be viewed as the horizon analogue
	of the crossed product degree of freedom.
	
	The present construction both clarifies and generalizes this bifurcation surface picture.
	First, it makes explicit why $\mathscr{A}_{\beta}$ fails to be central once we treat the split region
	dynamically: because dressed bulk operators depend on the corner data \eqref{eq:classicalboostact}, the
	corner charges act nontrivially on the ``bulk'' algebra.  Second, away from a bifurcation
	surface (or in non-stationary situations) one must also adjoin the null translation edge
	mode $\Upsilon^+_0$ and its conjugate generator $\mathscr{P}_{\alpha}$ in order to implement half-sided
	translations as inner automorphisms; this is the additional ingredient behind the
	crossed product structure \eqref{eq:classicalcrossedprod} beyond the bifurcation surface
	story.
	
	Previous work \cite{Ciambelli:2021nmv} writes down integrable symmetry generators in gravitational theories by making use of gravitational dressing but as \cite{Speranza:2022lxr,Carrozza:2022xut,Carrozza:2021gju} make it clear, said result corresponds to moving both the cut and the dynamical fields in a way that cancels out any non-trivial action on observables. In other words, the resulting symmetry generators can always be made to vanish on all solutions by exploiting the corner ambiguity in the symplectic form. They differ from our $\left(\mathscr{A}_{\beta}, \mathscr{P}_{\alpha}\right)$, which clearly act non-trivially on phase space observables, and cannot be rendered trivial via corner ambiguities.
	
	The main difference between these works and our approach is that by considering the extended subregion phase spaces of both $\mathscr{H}_+$ and its complement $\mathscr{H}'_+$, we obtain integrable non-trivial half-sided flows that move $\mathscr{H}_+$ relative to $\mathscr{H}'_+$. Transforming the edge modes on $S^+_0$ while keeping ``bulk'' fields on $\mathscr{H}$ and the edge modes on $S_0^-$ fixed constitutes the key step.
	
	Lastly, we can now formalize the statement behind \Eqref{eq:bulkhamconstraint} by defining the ``bulk'' Hamiltonian
	$H_{\rm bulk}[\xi]$ via
	\begin{equation}\label{eq:Hbulk-def}
		\delta H_{\rm bulk}[\xi] =
		\mathfrak{i}_{\hat{\xi}}\Omega_{\mathscr{H}_-} + \mathfrak{i}_{\hat{\xi}}\Omega_{\mathscr{H}_+}.
	\end{equation}
	Then \Eqref{eq:bulkhamconstraint} is precisely the statement that the right-hand side vanishes on the allowed phase space:
	\begin{equation}\label{eq:Hbulk-vanishes}
		\delta H_{\rm bulk}[\xi] = 0\Rightarrow
		H_{\rm bulk}[\xi] = \text{const.}\equiv 0\,,
	\end{equation}
	where in the last step we fix the additive constant by a choice of zero-point energy. Equivalently,
	$\xi$ lies in the kernel of the bulk presymplectic form, so it is a pure-gauge direction
	of the bulk sector.
	
	Consequently, for any observable $\mathscr{O}(p) \in \mathcal{A}_{\mathscr{H}_{>u_0}}$, we have the Dirac constraint
	\begin{equation}\label{eq:bulk-constraint-bracket}
		\{H_{\rm bulk}[\xi],\,\mathscr{O}(p)\} = 0\,.
	\end{equation}
	
	Non-trivial boundary dynamics of observables in $\mathcal{A}_{\mathscr{H}_{>u_0}}$ is instead generated by the corner term:
	\begin{equation}\label{eq:Hcorner-def}
		\delta H_{\rm corner}[\xi] =
		-\mathfrak{i}_{\hat{\xi}}\Omega_{\partial G}
		=
		\delta\!\left(\mathscr{A}_\beta + \mathscr{P}_\alpha\right).
	\end{equation}
	
	In particular, gauge-invariant dressed bulk operators $\mathscr{O}(p)$ do transform
	non-trivially, but only through their dressing dependence on the corner degrees of freedom as in \crefrange{eq:classicaltranslationact}{eq:classicalboostact}. Thus half-sided time evolution of horizon subregions is localized entirely to the corner: the ``bulk'' Hamiltonian is a constraint (trivial on bulk observables), while the physical time evolution arises from the corner charges through gravitational dressing.
	
	\subsection{Canonical quantization \label{sec:canonicalquant}}
	We'd like to quantize the algebra $\widehat{\mathcal{A}}_{\mathscr{H}_{> u_0}}$, and construct the associated GNS Hilbert space. To this aim, we follow the method in \cite{Prabhu:2022zcr}. We now briefly review the construction but cast it in the present context. 
	
	To start with, the construction only works if we have a linear theory, because it needs the phase space to have a vector space or affine space structure. So let's fix a background solution $g$ and consider the ``bulk'' phase space of linearized solutions about $g$, which we denote $\delta \mathcal{P}_{\mathscr{H}_{>u_0}} := T_g\mathcal{P}_{\mathscr{H}_{>u_0}}$. 
	
	Since $\delta \mathcal{P}_{\mathscr{H}_{>u_0}}$ is infinite dimensional, the symplectic form is typically weakly nondegenerate. This means we cannot globally invert the symplectic form. We instead have to define the Poisson bracket directly through the symplectic form itself. This can always be done on a restricted class of observables, one that is nevertheless sufficiently general enough to capture the types of operators we normally care about.
	
	Suppose we have an observable $\mathscr{O}(p)$ for which there exists a vector field $\hat{X}_{\mathscr{O}}$ on $\delta \mathcal{P}_{\mathscr{H}_{>u_0}}$ such that 
	\begin{align}
		\delta \mathscr{O} = \Omega(\cdot, \hat{X}_{\mathscr{O}}).\label{eq:observablecond}
	\end{align}
	In other words, $\mathscr{O}(p)$ generates a flow on phase space; it goes without saying that the flow need not have anything to do with a diffeomorphism of spacetime. Given two such observables, the Poisson bracket between them is just 
	\begin{align}
		\left\{\mathscr{O}_1(p), \mathscr{O}_2(p)\right\} = -\Omega(\hat{X}_{\mathscr{O}_1}, \hat{X}_{\mathscr{O}_2}).\label{eq:poissonbracketdef}
	\end{align}
	
	We therefore define an observable as a function $\mathscr{O}\colon \delta \mathcal{P}_{\mathscr{H}_{>u_0}}\mapsto \mathbb{R}$ that satisfies \Eqref{eq:observablecond}. The associated algebra $\mathcal{A}_{\mathscr{H}_{>u_0}}$ is then equipped with the product \eqref{eq:poissonbracketdef}. The obvious question is what class of observables we actually obtain this way. Let's tackle this question now.
	
	Recall as shown in \cref{sec:grsymp}, the ``bulk'' symplectic form on $\mathscr{H}_{>u_0}$ takes the form 
	\begin{align}
		\Omega_{\mathscr{H}_+} = \frac{1}{16\pi}\int_{\mathscr{H}_{+}}\left[\delta(\bm{\eta}q^{m\ell})\delta' \sigma_{m\ell} + \delta\bm{\eta}\delta' \Theta - (\delta \leftrightarrow \delta')\right].
	\end{align}
	But these fields are of course not completely independent, because of the linearized Raychaudhuri equation for $\delta \Theta$. So before proceeding, we need to integrate this out in the symplectic form. 
	
	For the next few sections we restrict to (non-stationary) linearization around a bifurcate Killing horizon. This furnishes a canonical choice of cyclic and separating vacuum state which satisfies the KMS condition, namely the Hartle-Hawking state \cite{Kay:1988mu}. KMS states have the well-known property that vacuum modular flow generates a local geometric boost about the corner \cite{Witten:2018zxz}. This identification is necessary in order to relate the crossed product algebra we've constructed to a Type II$_{\infty}$ von Neumann algebra for which entropies can be defined. But in \cref{sec:genbackentropy} we generalize the results to non-stationary linearization around a non-stationary event horizon.
	
	Before moving ahead, one last point worth emphasizing is that we don't strictly need a global timelike Killing field for the above. We could also linearize around an isolated horizon, and all the calculations would go through the same way. Recall that an isolated horizon is just one which has $\Theta = \sigma = 0$, even if there's no global timelike Killing field. It has a local timelike Killing field, in the neighborhood of the horizon. Physically, it's a black hole in equilibrium with its environment across the event horizon, even if there's dynamics happening inside/outside (e.g.\ emission of Hawking radiation inside of a reflecting box). In such a case the linearized phase space would arise from non-stationary excitations of the event horizon from such sources. This is a good description of the late time dynamics of an astrophysical black hole formed from collapse.
	
	The linearized Raychaudhuri equation just reads 
	\begin{align}
		\partial_u \delta \Theta = -8\pi G_N T_{uu}.
	\end{align}
	An important note: we're expanding the metric in powers of $G_N$ about a fixed background, such that $h_{ab}$ (which is dynamical) appears at order $\sqrt{G}_N$. The matter stress tensor itself does not have a $G_N$ counting; it can be of any order. It just corresponds to some free field theory on the fixed background. But $T_{uu} = T_{uu}(\delta g, \psi)$ should be interpreted as a function of a 1-form on phase space, where $\psi$ is the matter field.
	
	We can solve this using the retarded boundary condition $\lim_{u\rightarrow \infty}\delta \Theta \rightarrow 0$ (which is necessarily a property of teleological event horizons). This yields 
	\begin{align}
		\delta \Theta(u) = 8\pi G_N\int_{u}^{\infty}ds \  T_{uu}(s).\label{eq:deltathetalin}
	\end{align}
	Since $\delta \Theta = \frac{1}{2}\lie_{\ell}h$, it follows that 
	\begin{align}
		\Delta h(u) = 16\pi G_N \int_{u}^{\infty}ds \ (s-u) T_{uu}(s).\label{eq:deltaarealin}
	\end{align}
	
	Therefore, 
	\begin{align}
		\frac{1}{2}\Delta h(u) \wedge \delta \Theta(u) = 8\pi G_N^2\int_{u}^{\infty}ds\int_{u}^{\infty}ds' \ (s-u) T_{uu}(s)\wedge  T_{uu}(s').
	\end{align}
	Upon inserting this into the symplectic form, we compute 
	\begin{align}
		\frac{1}{2}\int_{u_0}^{\infty}du \left[\int_u^{\infty}ds\int_{u}^{\infty}ds'\ (s-u)\left( T'_{uu}(s) T_{uu}(s') - T_{uu}(s) T'_{uu}(s')\right)\right]
	\end{align}
	We can relabel $s \rightarrow s'$ in the second term and do the $u$ integral. This simplifies the expression to 
	\begin{align}
		\frac{1}{2} \left[\int_{u_0}^{\infty}ds\int_{u_0}^{\infty}ds'\ \left(s-s'\right)\left(\text{min}(s,s')-u_0\right)T_{uu}(s) T'_{uu}(s')\right].
	\end{align}
	
	Let $\mathcal{K}_{u_0}(s,s') := \left(s-s'\right)\left(\text{min}(s,s')-u_0\right)$ denote the kernel. And let $\bar{h}_{ab}$ represent the trace-free part of the metric perturbation. Putting it all together, the ``bulk'' symplectic form becomes (putting back in the factors of $G_N$)
	\begin{align}
		\Omega_{\mathscr{H}_+} = &\frac{1}{16\pi G_N}\int_{\mathscr{H}_{+}}\bm{\eta}\ \bar{h}^{m\ell}\wedge\delta\sigma_{m\ell} \nonumber \\ &+ \frac{G_N}{32\pi}\int_{S_0^+}\bm{\mu}\int_{u^+_0}^{\infty}ds\int_{u^+_0}^{\infty}ds'\  T_{uu}(s)\mathcal{K}_{u^+_0}(s,s') T'_{uu}(s').
	\end{align}
	Now everything is written in terms of independent fields. The shear $\sigma_{ab}$ is free data which determines the Weyl tensor through \Eqref{eq:shearevolution}. It enters into the symplectic form in the usual way as the radiative gravitational data on the horizon. The contribution to the symplectic form from the expansion turns into a bilinear form on the space of radiative matter data, with a non-trivial kernel. (One can also evaluate the corner symplectic form $\Omega_{\partial G}$ in terms of \crefrange{eq:deltathetalin}{eq:deltaarealin} but we won't write it down explicitly in order to avoid clutter, as there's no added insight in doing so.)
    
    We can now write down the area operator and half-sided translation generator by making use of \crefrange{eq:deltathetalin}{eq:deltaarealin}:
    \begin{subequations}
	\begin{align}
		&\mathscr{A}_{\beta}= \int_{u^+_0}^{\infty}du\int_{S^+_0}d^{d-2}x \sqrt{q}\ \beta(x^A)(u-u^+_0)T_{uu}(u)- \mathscr{A}_{\beta}(\infty)\label{eq:linmodham}, \\ &\mathscr{P}_{\alpha} = -\int_{u_0^+}^{\infty}du\int_{S^+_0}d^{d-2}x  \sqrt{q}\ \alpha(x^A) T_{uu}(u) \label{eq:linanec}.
	\end{align}
        \end{subequations}
	Notice that $\mathscr{A}_{\beta}$ and $\mathscr{P}_{\alpha}$ are just the half-sided vacuum modular Hamiltonian and half-sided ANEC operator of the matter theory, respectively; recall that these operators satisfy a half-sided modular inclusion algebra. The equivalence is just a consequence of linearizing around a Killing horizon. But as we've shown, these are the operators that actually implement half-sided boosts and translations of horizon subalgebras in gravity. This will be relevant for \crefrange{sec:gravmodham}{sec:gravmodinc} below. This is a nice marriage of covariant phase space and Tomita-Takesaki theory. Also note that both $\mathscr{A}_{\beta}$ and $\mathscr{P}_{\alpha}$ are $\mathcal{O}(1)$ in $G_N$ counting. So that means despite working perturbatively in $G_N$, we can nevertheless generate $\mathcal{O}(1)$ changes to the relative boost angle at the corner and to the location of the corner.
	
	We can now explicitly write down the algebra of observables $\mathcal{A}_{\mathscr{H}_{>u_0}}$. The basic observables are smeared versions of $\left(\delta\sigma_{ij}, \psi, \Gamma_0^+, \Upsilon^+_0\right)$, where $\psi$ is the matter field (taken to be a free scalar field for simplicity). The smeared observables take the following form:
    \begin{subequations}
	\begin{align}
		&\hat{\sigma}(f) = \int_{\mathscr{H}_+}f^{ij}\delta\sigma_{ij}, \\
		&\hat{\Gamma}_0^+(f) = \int_{S_0^+}f \Gamma_0^+, \\ 
		&\hat{\Upsilon}_0^+(f) = \int_{S_0^+}f\Upsilon_0^+, \\ 
		&\hat{\psi}(f) = \int_{\mathscr{H}_+}f\psi,
	\end{align}
        \end{subequations}
	where $f_{ij}$ is a smearing tensor, and $f$ is a scalar smearing function. The (smeared) linearized metric perturbation can be obtained from the relation $\delta \sigma_{ij} = \frac{1}{2}\lie_{\ell}\bar{h}_{ij}$. In this sense, it is a memory observable:
	\begin{align}
		\hat{h} = \int du \ \hat{\sigma}.
	\end{align}
	That these observables satisfy \Eqref{eq:observablecond} is easy to check. The constraint it places is that the smearing $f$ has to be in the same function class as the fields themselves, i.e. that it has the right fall-off conditions. Let's collectively denote the smeared observables by $\hat{\Psi}(f)$.
	
	We then impose the following canonical quantization conditions:
    \begin{subequations}
	\begin{align}
		&\text{\textbf{Linearity:}}\  \hat{\Phi}(af + bg) = a\hat{\Phi}(f) + b\hat{\Phi}(g),\label{eq:linearitycond} \\
		&\text{\textbf{Self-adjointness:}} \ \hat{\Phi}(f)^{\dagger} = \hat{\Phi}(f^*),\label{eq:selfadjointcond} \\ 
		&\text{\textbf{Canonical commutation relations:}} \ [\hat{\Phi}(f), \hat{\Phi}(g)] = i\widehat{\Omega}_{\mathscr{H}}(X_{\hat{\Phi}_f}, X_{\hat{\Phi}_g})\hat{\boldsymbol{1}},\label{eq:ccrcond}
	\end{align}
        \end{subequations}
	where $\widehat{\Omega}_{\mathscr{H}}$ is given by \Eqref{eq:areacornersymp}. 
	
	The abstract $*$-algebra $\widehat{\mathcal{A}}_{\mathscr{H}_{> u_0}}$ is then generated by arbitrary polynomials of $\hat{\Phi}(f)$, subject to the conditions above. More explicitly, we define the abstract polynomial canonical commutation relation (CCR) $*$-algebra
    $\widehat{\mathcal{A}}_{\mathscr{H}_{>u_0}}$ as the universal unital $*$-algebra generated by formal symbols
    $\hat{\Phi}(f)$, modulo the canonical quantization conditions above. Concretely, one may formally write the algebra as a quotient
    \begin{equation}
    \widehat{\mathcal{A}}_{\mathscr{H}_{>u_0}}
    =
    \mathbb{C}\langle \hat{\Phi}(f)\ \colon f \ \text{admissible}\rangle / \langle \mathcal{R}\rangle_{*\text{-ideal}}
    \end{equation}
    where $\mathbb{C}\langle \cdots\rangle$ denotes the free associative unital algebra, $\langle \ldots \rangle_{*\text{-ideal}}$ denotes the two-sided $*$-ideal, and $\mathcal{R}$ is the set of relations \crefrange{eq:linearitycond}{eq:ccrcond}.
    Equivalently, every element $\hat{\mathscr{O}}\in \widehat{\mathcal{A}}_{\mathscr{H}_{>u_0}}$ can be represented as a
    finite $\mathbb{C}$-linear combination of monomials
    \begin{equation}
    \hat{\mathscr{O}}
    =
    \sum_{n=0}^{N}\ \sum_{I} c_{I}\,
    \hat{\Phi}(f_{i_1})\cdots \hat{\Phi}(f_{i_n}),\ c_I\in\mathbb{C},
    \end{equation}
    with the understanding that different representatives are identified using the quotient relations above.
    The involution is fixed by $\hat{\mathbf 1}^\dagger=\hat{\mathbf 1}$ and
    $(\hat{\mathscr{O}}_1\hat{\mathscr{O}}_2)^\dagger=\hat{\mathscr{O}}^\dagger_2\hat{\mathscr{O}}^\dagger_1$.
    
    The construction above uses the polynomial CCR algebra, in which the generators $\hat{\Phi}( f)$
    are formally unbounded. For operator algebra constructions (commutants/bicommutants, weak closures, etc.) it is often
    preferable to work with bounded exponentials/Weyl operators. The corresponding Weyl $C^*$-algebra is generated by unitary
    symbols $\hat W(f)$ satisfying the Weyl relations
    \begin{subequations}
    \begin{align}
    &\hat W(f)\,\hat W(g)
    =
    \exp\left({-\frac{i}{2}\widehat{\Omega}_{\mathscr H}(X_{\hat{\Phi}_{ f}}, X_{\hat{\Phi}_{ g}})}\right),\\ &\hat W(f+g)
    \hat W(f)^\dagger=\hat W(- f),\\
    &\hat W(0)=\hat{\mathbf 1}.
    \end{align}
        \end{subequations}
    In any regular representation one may identify $\hat W(f)=e^{i\hat{\Phi}( f)}$ and recover the smeared fields
    as the self-adjoint generators of the one-parameter groups $t\mapsto W(t f)$, i.e.
    \begin{equation}
    \hat{\Phi}(f)=-i\,\frac{d}{dt}\,\hat W(t f)\Big|_{t=0}.
    \end{equation}
    Thus the polynomial and Weyl description differ primarily by the choice of generators/completion:
    the Weyl algebra is a bounded (hence $C^*$) completion of the same CCR data, while the polynomial algebra is a convenient
    dense $*$-subalgebra for algebraic manipulations. In particular, whenever we later form commutants/bicommutants, the intended
    meaning is the von Neumann algebra generated in the chosen GNS representation by bounded functionals of the smeared fields rather than the bare polynomial algebra of unbounded generators.

    Forging ahead, the ``bulk'' operators are dressed to $S_0^+$ in the same way as in the classical construction of \cref{sec:classicalalgebra}. And based on the results therein, $\widehat{\mathcal{A}}_{\mathscr{H}_{> u_0}}$ is a crossed product algebra of the form
	\begin{align}
		\widehat{\mathcal{A}}_{\mathscr{H}_{> u_0}} = \left(\mathcal{A}^{\text{grav}}_{\mathscr{H}_{> u_0}}[\hat{\sigma}]\otimes\mathcal{A}^{\text{mat}}_{\mathscr{H}_{> u_0}}[\hat{\psi}]\right) \rtimes \mathcal{A}_{\partial G}[\hat{\Gamma}_0^+, \hat{\Upsilon}_0^+],
	\end{align}
	where $\mathcal{A}_{\partial G}[\hat{\Gamma}_0^+, \hat{\Upsilon}_0^+] = C^{\infty}_{\beta}(\mathbb{S}^{d-2})\rtimes C^{\infty}_{\alpha}(\mathbb{S}^{d-2})$, corresponding to the automorphisms generated by $\hat{\mathscr{P}_{\alpha}}$ and $\hat{\mathscr{A}}_{\beta}$ respectively. As indicated, these generators are now formally operators. 
    
    Lastly, a convenient choice of dressing for the operators in the complement algebra $\widehat{\mathcal{A}}_{\mathscr{H}'_{> u_0}}$ can be obtained by fixing the reference point in the exponential map \eqref{eq:dressing} to be an inherently gauge invariant point in the complement region, such as the bifurcation surface, the asymptotic limit along the left horizon, or the point from which the event horizon of a collapse black hole forms. In perturbative quantum gravity, this ensures that $\widehat{\mathcal{A}}_{\mathscr{H}'_{> u_0}}$ is also the commutant algebra $\widehat{\mathcal{A}}'_{\mathscr{H}_{> u_0}}$.\footnote{In full quantum gravity there will be chaotic scrambling dynamics, so gravitational dressing alone would be insufficient in constructing the commutant. But that is outside the scope of this work.} This will prove important for the proof of the QFC in \cref{sec:quantfocus}. 
	
	We can now construct the extended Hilbert space. But we have to be careful about which extended Hilbert space we're referring to. In the approach taken by \cite{Chandrasekaran:2022eqq}, one only ever works with the Hilbert space of the entire region, i.e. the entire bulk Cauchy slice of the two-sided black hole, which we can write as $\mathscr{H}'\cup \mathscr{H}$ by boosting the spacelike Cauchy slice. Writing the slice this way is better suited for the present setting. 
	
	This is because in a Type III$_1$ QFT (which is what the ``bulk'' fields correspond to) the Hilbert space does not have a tensor product structure across a codimension-two entangling surface, due to the infinite amount of entanglement between modes in the vacuum on either side of the surface \cite{Witten:2018zxz}. This naively prevents one from constructing a Hilbert space of the subregion.
	
	However the subregion algebra is perfectly well-defined, as we've just seen. So the extended Hilbert space follows directly from the GNS construction applied to the subregion algebra $\widehat{\mathcal{A}}_{\mathscr{H}_{> u_0}}$, given a cyclic and separating vacuum state $|\Omega\rangle$. For the two-sided black hole, $|\Omega\rangle$ is just the Hartle-Hawking state. Let $\mathcal{H}$ be the Fock space of the ``bulk'' fields $\hat{\sigma}$ and $\hat{\psi}$. Then the extended GNS Hilbert space obtained from the crossed product algebra is just
	\begin{align}
		\widehat{\mathcal{H}} = \mathcal{H}\otimes L^2\left(\mathcal{G}\right), \ \mathcal{G} := C^{\infty}_{\beta}(\mathbb{S}^{d-2})\rtimes C^{\infty}_{\alpha}(\mathbb{S}^{d-2}).\label{eq:extendedhilbertspace}
	\end{align}
	States of the subregion correspond instead to density matrices in the associated algebra. This can be done if we have a Type II or Type I algebra, since they are equipped with a notion of trace. For the purposes of computing entropies, this is sufficient. Of course, since $\mathcal{G}$ is infinite-dimensional, it is not clear how to define the measure on $L^2(\mathcal{G})$. Soon we will specialize to a mini-superspace approximation in which we only consider the $\ell = 0$ sector of the edge mode phase space, so that we reduce down to the finite-dimensional group $\mathbb{R}\rtimes \mathbb{R}$. But in \cref{app:angle-dep-ST} we describe how to make sense of the general setting in \Eqref{eq:extendedhilbertspace} using an inductive limit (in the weak operator topology) of a truncated basis of spherical harmonics.
	
	Furthermore, in \cref{app:directintegral} we show that the extended GNS Hilbert space can be written as a direct integral over edge mode configurations, which upon choosing a trivialization takes the form of a tensor product between a “hard mode’’ Hilbert space and an $L^2$ space of edge mode wavefunctions. We also explain that this tensor product split is not canonical: because the full algebra is a crossed product, the edge unitaries act by outer automorphisms on the hard algebra, so any separation into “hard’’ versus “edge’’ degrees of freedom depends on a choice of dressing/trivialization rather than being fixed by the algebra itself. As noted therein, this constitutes an algebraic realization of the background independence of perturbative quantum gravity (see also \cite{Witten:2023xze}).
	
	With all that being said, we can actually go a bit further in our current construction compared to that of \cite{Chandrasekaran:2022cip}. Recall that in QFT we can still define Hilbert spaces for subregions by introducing a small codimension-one region of size $\varepsilon$ around the entangling surface, splitting the full region across the entangling surface; this breaks vacuum entanglement between modes below this length scale, thus allowing the full Hilbert space to factorize into subregion Hilbert spaces. We can think of this regularization as a brick wall boundary condition.\footnote{Though of course, if we're in a gauge theory there will be edge modes conjugate to the gauge charges living on this inner boundary, as emphasized by \cite{Donnelly:2016auv}.}
	
	We can also do this in gravity, except we have to introduce gravitational edge modes at the boundary of the excised region. That is, the inner boundary has to be treated dynamically in gravity, as opposed to just being a brick wall (see \cite{Jafferis:2019wkd} for a discussion of this point in the context of Euclidean path integrals in JT gravity). This is exactly the split $\mathscr{H} = \mathscr{H}_- \cup G_{\varepsilon} \cup \mathscr{H}_+$ that we've already constructed in previous sections. The abstract crossed product degrees of freedom that allow us to obtain subregion traces and entropies correspond precisely to the edge modes on $\partial G_{\varepsilon}$. Then,\footnote{In algebraic language, the split property of algebraic QFT guarantees the existence of a Type I von Neumann algebra $\mathcal{N}$ satisfying $\mathcal{A}_{\mathscr{H}_{+}}\subset \mathcal{N} \subset \left(\mathcal{A}'_{\mathscr{H}_{+}}\right)'$ \cite{Witten:2018zxz}.}
	\begin{align}
		\widehat{\mathcal{H}} = \lim_{\varepsilon \rightarrow 0}\left(\mathcal{H}_{\mathscr{H'}\cup\mathscr{H}_-} \otimes \mathcal{H}_{\mathscr{H}_+}\otimes L^2(\mathcal{G})\right).\label{eq:decomphilbertspace}
	\end{align}
	As far as this construction of one-sided Hilbert spaces is concerned, the key differences between (perturbative) quantum gravity and QFT are (i) the former has a natural scale $\varepsilon \sim \ell_p$ and (ii) one-sided vacuum gravitational modular flow is just the Connes cocycle flow of the QFT. 
    
    Regarding (ii), recall that under our matching conditions \crefrange{eq:continuity3}{eq:continuity5}, ``bulk'' fields (matter + graviton) transform trivially under boosts/translations. But the edge modes at the corner transform as half-sided flows. So intuitively, the half-sided modular inclusion in QFT is directly implemented by the generators conjugate to the gravitational edge modes. We make this slightly more explicit in \cref{sec:gravmodham}. In other words, the non-trivial commutators are just
    \begin{subequations}
	\begin{align}
		[\hat{\mathscr{P}}_{\alpha}, \hat{\mathscr{O}}(p)] &= -i\alpha \partial_u \hat{\mathscr{O}}(p),\label{eq:translationcommutator} \\ [\hat{\mathscr{A}}_{\beta}, \hat{\mathscr{O}}(p)] &= -i(u-u_0)\beta \partial_u \hat{\mathscr{O}}(p)\label{eq:boostcommutator}.
	\end{align}
        \end{subequations}
	This is the content of \crefrange{eq:linmodham}{eq:linanec} combined with \crefrange{eq:classicaltranslationact}{eq:classicalboostact}.\footnote{At the classical level, since $\mathscr{O}(p)$ is gauge-invariant, we can write $\mathfrak{L}_{\hat{\ell}}\mathscr{O}(p) = \lie_{\ell}\mathscr{O}(p)$. By using a choice of affine parameter $u$ in the background spacetime which absorbs the $e^{\Gamma_0^+}$ background boost weight, we can then write $e^{\Gamma_0^+}\lie_{\ell}\mathscr{O}(p) = \partial_u \mathscr{O}(p)$. Finally, mapping $\mathscr{O}(p)\rightarrow \hat{\mathscr{O}}(p)$ yields \crefrange{eq:translationcommutator}{eq:boostcommutator}.} As we've seen, the action of these generators simply inserts a finite energy shock at the corner\footnote{Actually it follows from \crefrange{eq:stresstensorshock}{eq:weylshock} that when perturbing around a Killing horizon, the stress tensor and Weyl shocks actually vanish at linear order.}, so we get a well-defined state in perturbative quantum gravity. This is why \Eqref{eq:decomphilbertspace} holds.
	
	Let's now get a sense for the structure of the crossed product algebra and extended Hilbert space. For simplicity, let's smear all operators by smearing functions that projects onto the $\ell = 0$ mode. This allows us to ignore the angle-dependence. We can then rewrite the algebra and Hilbert space as follows:
    \begin{subequations}
	\begin{align}
		&\widehat{\mathcal{A}}_{\mathscr{H}_{> u_0}} = \left(\mathcal{A}_{\mathscr{H}_{> u_0}}\rtimes \mathbb{R}_{s}\right)\rtimes \mathbb{R}_{u}.\label{eq:simplifiedextalgebra}\\ 
		&\widehat{\mathcal{H}} = \mathcal{H}\otimes L^2(\mathbb{R}_s) \otimes L^2(\mathbb{R}_u),\label{eq:simplifiedexthilbertspace}
	\end{align}
        \end{subequations}
	where $\mathbb{R}_s$ is the automorphism group generated by the uniform half-sided boost generator $\hat{\mathscr{A}}$, and $\mathbb{R}_u$ is the automorphism group generated by the uniform half-sided translation generator $\hat{\mathscr{P}}$. This is just a finite-dimensional mini-superspace approximation of the edge mode sector of the horizon subregion phase space. See \cref{app:angle-dep-ST} for a discussion on how one might try to generalize to the full infinite dimensional case.
	
	Forging ahead, we consider states $|\hat{\Psi}\rangle \in \widehat{\mathcal{H}}$ of the form
	\begin{align}
		|\hat{\Psi}(u_0)\rangle = \int_{-\infty}^{\infty}dx \int_{-\infty}^{\infty}du \ f(x)g(u-u_0)|\Psi\rangle |x\rangle |u\rangle,\label{eq:quantumgravstate}
	\end{align}
	where $f, g$ are square-integrable functions. We've chosen a basis where $|x\rangle$ represents eigenstates of $\hat{\mathscr{A}}$, while $|u\rangle$ represents eigenstates of the edge mode $\hat{\Upsilon}_0^+$. The conjugate momentum acting on $L^2(\mathbb{R}_s)$ is therefore the relative boost angle edge mode $\hat{\Gamma}_0^+$, whereas on $L^2(\mathbb{R}_u)$ it is $\hat{\mathscr{P}}$. The reason for the choice of $|x\rangle$ basis is the same as in \cite{Chandrasekaran:2022eqq}, while the reason for the choice of $|u\rangle$ basis is to make the computation of $\partial_{u} S_{\text{gen}}$ very natural. 
	
	Notice that $|\Psi\rangle$ does not depend on $u$ or $x$ explicitly. In the perturbative quantum gravity regime, ``bulk'' QFT states should not depend on the edge modes because we need to have a smooth $G_N \rightarrow 0$ limit of the states in $\widehat{\mathcal{H}}$. Rather, the coupling between ``bulk'' fields and edge modes is contained entirely in gravitational dressing of ``bulk'' operators after integrating out the null gravitational constraints. Also note in \Eqref{eq:quantumgravstate}, the $u_0$ parameterizing $|\hat{\Psi}(u_0)\rangle$ is a c-number and just corresponds to the classical location of the corner. The RHS contains a wavefunction $g(u-u_0)$ peaked around the corner $u_0$, corresponding to a superposition over different corners in the neighborhood of $u_0$. In order to remain in the perturbative quantum gravity regime, i.e. to make sense of ``localized'' horizon subregions, we need to limit ourselves to states for which $\Delta \hat{\Upsilon}_0^+ = \mathcal{O}(\epsilon)$ for some $\epsilon \ll 1$.
	
	Similarly, we need to restrict to states for which $\Delta \hat{\Gamma}_0^+ = \mathcal{O}(\epsilon)$ so that we're not in an arbitrary superposition over relative boost angles. This is the same condition imposed in \cite{Chandrasekaran:2022eqq}. It follows that $\Delta \hat{\mathscr{A}} = \mathcal{O}(1/\epsilon)$, and similarly $\Delta \hat{\mathscr{P}} = \mathcal{O}(1/\epsilon)$, by the Heisenberg uncertainty principle. Importantly, the superposition over edge mode configurations is a feature of the state; the algebra knows nothing about it. From the perspective of the algebra, $s$ and $u$ are just flow parameters. But in a typical state the edge modes $\hat{\Gamma}_0^+$ and $\hat{\Upsilon}_0^+$ (which are the dynamical versions of $s$ and $u$) will fluctuate.
	
	Let's work in the Heisenberg picture under the time evolution generated by $\hat{\mathscr{P}}$ along the horizon. An operator $\hat{\mathscr{O}}(p) \in \widehat{\mathcal{A}}_{\mathscr{H}_{>u_0}}$ can be written
	\begin{align}
		\hat{\mathscr{O}}(u) = \int_{-\infty}^{\infty} du_0 \ \hat{\mathscr{O}}_0(u_0)e^{i\hat{\mathscr{P}} (u-u_0)}.
	\end{align}
	For any crossed product algebra, there exists a canonical faithful normal conditional expectation $E_{u_0}\colon \widehat{\mathcal{A}}_{\mathscr{H}_{>u_0}}\mapsto \widehat{\mathcal{M}}_{\mathscr{H}_{>u_0}}$ which satisfies 
	\begin{align}
		E_{u_0}[\hat{\mathscr{O}}(u)] = \hat{\mathscr{O}}_0(u) \in \widehat{\mathcal{M}}_{\mathscr{H}_{>u_0}}.\label{eq:conditionalexpect}
	\end{align}
	This defines localization onto a cut algebra $\widehat{\mathcal{M}}_{\mathscr{H}_{>u_0}}$.
	
	The GNS construction yields a representation $\Pi\colon \widehat{\mathcal{M}}_{\mathscr{H}_{>u_0}}\mapsto \mathcal{B}(\widehat{\mathcal{H}})$ (the set of bounded linear operators on $\widehat{\mathcal{H}}$). Using this we can define a one-parameter family of time translated algebras
	\begin{align}
		\widehat{\mathcal{M}}_{\mathscr{H}_{>u}} = \text{Ad} \ U(\delta u)\left(\Pi\left(\widehat{\mathcal{M}}_{\mathscr{H}_{>u_0}}\right)\right), \ U(\delta u) := e^{i\hat{\mathscr{P}} \delta u},
	\end{align}
	where $u := u_0 + \delta u$ and $\text{Ad} \ U(\delta u) )\Pi\left(\widehat{\mathcal{M}}_{\mathscr{H}_{>u_0}}\right)= U(\delta u)\Pi\left(\widehat{\mathcal{M}}_{\mathscr{H}_{>u_0}}\right)U^{\dagger}(\delta u)$ is just shorthand for conjugation by the half-sided translation unitary. This is a covariant $*$-automorphism, hence the adjoint notation.
	
	Putting these two ingredients together, given any operator $\hat{\mathscr{O}}(p)\in \mathcal{B}(\widehat{\mathcal{H}})$, we can define a projection onto $\widehat{\mathcal{M}}_{\mathscr{H}_{>u}}$ by
	\begin{align}
		\hat{\mathscr{O}}(u)= \text{Ad} \ U(\delta u)\left(E_{u_0}[\hat{\mathscr{O}}(u_0)]\right).
	\end{align}
	This allows us to talk about operators along the one-parameter family of algebras. Let's compute the expectation value of $\hat{\mathscr{O}}(u)$ in state $|\hat{\Psi}(u_0)\rangle$.
	
	We first compute 
	\begin{align}
		\langle u''|\hat{\mathscr{O}}(u)|u'\rangle &= \langle u''|U(\delta u)\Pi(\hat{\mathscr{O}}_0(u_0))U(-\delta u)|u'\rangle  \\ 
		&= \langle u'' - \delta u|\Pi(\hat{\mathscr{O}}_0(u_0))|u' - \delta u\rangle 
		\\ &= \delta(u'' - u')\pi(\varphi_{\delta u - \Delta u_0}(\hat{\mathscr{O}}(u_0))),
	\end{align}
	where $\varphi_{\delta u}$ is the automorphism generated by $\hat{\mathscr{P}}$, i.e. the preimage of $\text{Ad} \ U(\delta u)$ under $\Pi$, and we've defined the representation $\pi\colon \widehat{\mathcal{M}}_{\mathscr{H}_{>u_0}} \mapsto \mathcal{B}(\mathcal{H}\otimes L^2(\mathbb{R}_s))$. Lastly, $\Delta u_0 = u' - u_0$ is the spread in measurements of the corner location due to fluctuations in $\hat{\Upsilon}^+_0$. For notational simplicity, let $|\hat{\psi}\rangle = \int dx \ f(x) |\Psi\rangle |x\rangle$ denote a state in the ``base'' Hilbert space $\widehat{\mathcal{H}}_{\mathcal{M}} := \mathcal{H}\otimes L^2(\mathbb{R}_s)$. Then,
	\begin{align}
		\langle \hat{\Psi}(u_0)|\hat{\mathscr{O}}(u)|\hat{\Psi}(u_0)\rangle = \int_{-\infty}^{\infty}d\Delta u_0 \ |g(\Delta u_0)|^2  \langle \hat{\psi}|\pi(\varphi_{\delta u-\Delta u_0}(\hat{\mathscr{O}}(u_0))) |\hat{\psi}\rangle.
	\end{align}
	
	To get some intuition for this expression, recall that in the perturbative quantum gravity regime we want $g(u'-u_0)$ to be highly peaked around $u_0$, with width $\Delta u_0 \ll 1$ (not to be confused with $\delta u'$). A natural class of such wavefunctions is a Gaussian. Then, to leading order in $\Delta u_0$,
	\begin{align}
		\langle \hat{\Psi}(u_0)|\hat{\mathscr{O}}(u)|\hat{\Psi}(u_0)\rangle \approx &\langle \hat{\psi}|\pi(\varphi_{\delta u}(\hat{\mathscr{O}}(u_0)))|\hat{\psi}\rangle \nonumber \\ &+ \frac{(\Delta u_0)^2}{2} \frac{d^2}{d\Delta u_0^{2}}\langle \hat{\psi}|\pi(\varphi_{\delta u - \Delta u_0}(\hat{\mathscr{O}}(u_0)))|\hat{\psi}\rangle\Big\lvert_{\Delta u_0 = 0}.\label{eq:translationexpectation}
	\end{align}
	So to leading order, the expectation value of an operator in $\widehat{\mathcal{M}}_{\mathscr{H}_{>u}}$ on the full Hilbert space $\widehat{\mathcal{H}}$ is just the expectation value of the time translated operator on the ``base'' Hilbert space $\widehat{\mathcal{H}}_{\mathcal{M}}$. The subleading corrections account for the gravitational fluctuations in the location of the corner.
	
	We now have all the ingredients we need in order to compute the generalized entropy of a horizon subalgebra.
	
	\section{Generalized entropy of a horizon subalgebra\label{sec:genentropyderiv}}
	
	With the horizon subalgebras and their GNS representations in hand, we can finally answer one of our central questions: what is the entropy of a horizon subregion? In this section we show that, for each cut $u \geq u_0$ of the horizon, the crossed product algebra $\widehat{\mathcal{M}}_{\mathscr{H}_{>u}}$ is a Type II$_{\infty}$ factor admitting a canonical trace, and that the associated von Neumann entropy coincides (up to a state-independent constant and a small smearing in $u$) with the generalized entropy of the horizon at that cut.
	
	The strategy is to relate the area operator $\hat{\mathscr{A}}$ to the Connes cocycle flow of the underlying Type III$_1$ horizon algebra and to use the crossed product trace to build a density matrix $\rho_{\hat{\Psi}}(u) \in \widehat{\mathcal{M}}_{\mathscr{H}_{>u}}$. We then show that the algebraic von Neumann entropy reproduces the usual generalized entropy formula, but evaluated in a quantum superposition of cut locations determined by the translation edge mode. Finally, we show how the nesting property of the family $\widehat{\mathcal{M}}_{\mathscr{H}_{>u}}$ implies a generalized second law.
	
	Throughout this section we will be slightly imprecise in the standard way familiar from large-$N$
	effective algebra discussions \cite{Chandrasekaran:2022eqq}. Concretely, let
	$\widehat{\mathcal{A}}_{\mathscr{H}_{>u}}$ denote the unital $*$-algebra generated (in the GNS representation on
	$\widehat{\mathcal{H}}$) by bounded functions of the smeared matter/graviton fields together with
	the edge mode unitaries (e.g.\ Weyl operators) implementing the horizon symmetries.\footnote{We
		avoid unbounded generators themselves by working with bounded functional calculus / exponentials, so
		that commutants are well-defined inside $B(\widehat{\mathcal{H}})$.}
	We then define the associated horizon von Neumann algebra as the double commutant
	\begin{equation}
		\mathcal{A}_{\mathscr{H}_{>u}}''\subset B(\widehat{\mathcal{H}}),
	\end{equation}
	where $(\cdot)'$ denotes the commutant taken inside $B(\widehat{\mathcal{H}})$.
	By the bicommutant theorem, $\mathcal{A}_{\mathscr{H}_{>u}}''$ coincides with the weak/strong-operator closure of
	$\mathcal{A}_{\mathscr{H}_{>u}}$; thus any use of $(\cdot)'$ or $(\cdot)''$ below should be read as a statement
	about these closures in the chosen representation, rather than about the bare polynomial algebra.
	
	A further caveat is that our dressing map (and hence the generators of $\mathcal{A}_{\mathscr{H}_{>u}}$) is defined
	perturbatively as a formal series in $G_N$.  At finite $G_N$ we therefore only control the algebra and
	its commutation relations order-by-order in this expansion, so identifications involving commutants or
	bicommutants are likewise perturbative and could receive nonperturbative corrections.  In particular,
	the clean von Neumann-algebraic picture is sharpest in the strict $G_N\to0$ limit (the analogue of the
	strict large-$N$ limit of \cite{Chandrasekaran:2022eqq}).
	
	\subsection{One-parameter family of Type II\texorpdfstring{$_{\infty}$}{∞} algebras\label{sec:gravmodham}}
	For any value of $u$, the algebra $\widehat{\mathcal{M}}_{\mathscr{H}_{>u}}$ is a Type II$_{\infty}$ algebra.
	
	To see this, let's first recall the definition of the Connes cocycle flow. Consider an excited state $|\Psi\rangle \in \mathcal{H}$. Take the vacuum state $|\Omega\rangle \in \mathcal{H}$ to be the Hartle-Hawking state as before. The relative Tomita operator is an antilinear operator defined via \cite{Witten:2018zxz} 
	\begin{align}
		S_{\Omega|\Psi;u} \hat{\mathscr{O}}|\Psi\rangle = \hat{\mathscr{O}}^{\dagger}|\Omega\rangle,
	\end{align}
	for all operators $\hat{\mathscr{O}} \in \mathcal{A}_{\mathscr{H}_{>u}}$. The relative modular operator is then defined as 
	\begin{align}
		\Delta_{\Omega|\Psi;u} = S_{\Omega|\Psi;u}^{\dagger}S_{\Omega|\Psi;u}.
	\end{align}
	The relative modular operator does not belong to $\mathcal{A}_{\mathscr{H}_{>u}}$ or its complement $\mathcal{A}'_{\mathscr{H}_{>u}}$; it acts non-trivially on operators in both algebras.
	
	Finally, the Connes cocycle (CC) flow is given by 
	\begin{align}
		u_{\Psi|\Omega;u}(s) = \Delta^{is}_{\Psi|\Omega; u}\Delta_{\Omega; u}^{-is} = \Delta_{\Psi; u}^{is}\Delta_{\Omega|\Psi; u}^{-is}.
	\end{align}
	In particular, $u_{\Psi|\Omega;u}(s)\in \mathcal{A}_{\mathscr{H}_{>u}}$ for all values of $s$. Here $\Delta_{\Omega;u} := \Delta_{\Omega|\Omega;u}$ is the vacuum modular operator.
	
	Since $|\Omega\rangle$ is the Hartle-Hawking state on a spacetime with bifurcate Killing horizon, and the Hartle-Hawking state satisfies the KMS condition with temperature $\beta$, it follows that $\Delta_{\Omega;u}^{is}$ generates two-sided boosts about $u$ by the Bisognano-Wichmann theorem \cite{BisognanoWichmann1975DualityScalar, BisognanoWichmann1976DualityQFT} (or generalizations thereof, c.f. \cite{Casini:2017roe}). Now, the CC flow has the following important property:
    \begin{subequations}
	\begin{align}
		&\langle \Psi|u_{\Psi|\Omega;u}(s)\hat{\mathscr{O}}u^{\dagger}_{\Psi|\Omega;u}(s)|\Psi\rangle = \langle \Psi|\Delta^{is}_{\Omega;u}\hat{\mathscr{O}}\Delta^{-is}_{\Omega;u}|\Psi\rangle, \ \hat{\mathscr{O}}\in \mathcal{A}_{\mathscr{H}_{>u}} \\ 
		&\langle \Psi|u_{\Psi|\Omega;u}(s)\hat{\mathscr{O}}'u^{\dagger}_{\Psi|\Omega;u}(s)|\Psi\rangle = \langle \Psi|\hat{\mathscr{O}}'|\Psi\rangle, \ \hat{\mathscr{O}}'\in \mathcal{A}'_{\mathscr{H}_{>u}}.
	\end{align}
        \end{subequations}
	In words, $u_{\Psi|\Omega;u}(s)$ generates physical one-sided boosts about $u$. Since $\mathcal{A}'_{\mathscr{H}_{>u}} \simeq \mathcal{A}_{\mathscr{H}_{<u}}$ we can also write $\hat{\mathscr{O}}'$ as $\hat{\mathscr{O}}^{-}$, while $\hat{\mathscr{O}}$ can be written as $\hat{\mathscr{O}}^{+}$. We will go back and forth between these two notations as needed.
	
	If we compare the CC flow to the action of the area operator $\hat{\mathscr{A}}(u) = U(\delta u)\hat{\mathscr{A}}(u_0)U(-\delta u)$, we see that
	\begin{align}
		\langle \Psi|u_{\Psi|\Omega;u}(s)\hat{\mathscr{O}}^{\pm}u^{\dagger}_{\Psi|\Omega;u}(s)|\Psi\rangle = \langle \Psi|e^{i\beta\hat{\mathscr{A}}(u)s}\hat{\mathscr{O}}^{\pm}e^{-i\beta\hat{\mathscr{A}}(u)s}|\Psi\rangle.\label{eq:ccisgrav}
	\end{align}
	Equation \eqref{eq:ccisgrav} is an elegant identity in and of itself. It says that in perturbative quantum gravity, the bulk Connes cocycle flow coincides with the action of the area operator on one-sided observables in excited states. This is basically a consequence of (perturbative) background independence: the area operator acts non-trivially on ``bulk'' observables through gravitational dressing by keeping the ``bulk'' fields fixed and changing the gravitational edge modes, whereas the CC flow directly acts on the bulk fields and knows nothing about the gravitational edge modes. See \cite{Bousso:2020yxi} for previous discussion of this point.
	
	However, CC flow and the area operator are not literally the same objects. CC flow is explicitly state-dependent, while the area operator corresponds to a manifestly state-independent action on states in $\mathcal{H}$. More concretely, combining Eqs.\ \eqref{eq:boostcommutator} and \eqref{eq:ccisgrav}, we see that $\hat{\mathscr{A}}(u)$ acts as the one-sided vacuum modular Hamiltonian on the algebra $\widehat{\mathcal{M}}_{\mathscr{H}_{>u}}$, so how can it also coincide with the CC flow of one-sided observables in excited states? This is because the area operator acts universally as a one-sided boost but also knows how the complementary regions $\mathscr{H}_{>u_0}$ and $\mathscr{H}_{<u_0}$ are glued back together gravitationally, and our matching conditions \crefrange{eq:continuity3}{eq:continuity5} correspond to physically transforming excited states (whereas $\Delta^{is}_{\Omega;u}$ is pure gauge). 
	
	Moving ahead, it is a standard result that the crossed product of the Type III$_1$ algebra $\mathcal{A}_{\mathscr{H}_{>u}}$ with its modular automorphism group is a type II algebra. The modular automorphism group generated by $\vartheta_{s}(\hat{\mathscr{O}}):= \Delta_{\Omega;u}^{is}\hat{\mathscr{O}}\Delta^{-is}_{\Omega;u}$ in this case is just $\mathbb{R}_s$, based on the discussion above, so $\widehat{\mathcal{M}}_{\mathscr{H}_{>u}}$ is indeed a Type II algebra.
	
	A Type II algebra has a well-defined notion of trace \cite{Witten:2021unn}, i.e. a positive linear functional $\text{tr}$ on operators in the algebra that satisfies 
	\begin{align}
		\text{tr}[\hat{\mathscr{O}}_1\hat{\mathscr{O}}_2] = \text{tr}[\hat{\mathscr{O}}_2\hat{\mathscr{O}}_1],
	\end{align}
	for any pair of operators $\hat{\mathscr{O}}_1, \hat{\mathscr{O}}_2 \in \widehat{\mathcal{M}}_{\mathscr{H}_{>u}}$. Specifically,
	\begin{align}
		\text{tr}[\hat{\mathscr{O}}] = \int_{-\infty}^{\infty}dx \ e^{\beta x}\langle \Psi|\hat{\mathscr{O}}(x)|\Psi\rangle.
	\end{align}
	Note that $\text{tr}(\hat{1}) = \infty$, so not all operators have a finite trace. This is what makes $\widehat{\mathcal{M}}_{\mathscr{H}_{>u}}$ a Type II$_{\infty}$ algebra. Moreover, the trace has no canonical normalization.
	
	Since each algebra $\widehat{\mathcal{M}}_{\mathscr{H}_{>u}}$ has a trace, we can define a one-parameter family of density matrices $\rho_{\hat{\psi}}(u) \in \widehat{\mathcal{M}}_{\mathscr{H}_{>u}}$ by 
	\begin{align}
		\text{tr}[\rho_{\hat{\psi}}(u)\hat{\mathscr{O}}(u)] = \langle \hat{\psi}|\hat{\mathscr{O}}(u)|\hat{\psi}\rangle.
	\end{align}
	This allows us to define a time-dependent von Neumann entropy:
	\begin{align}
		S(\hat{\psi} ;{\widehat{\mathcal{M}}_{\mathscr{H}_{>u}}}) = -\text{tr}[\rho_{\hat{\psi}}(u)\log \rho_{\hat{\psi}}(u)] = -\langle \hat{\psi}|\log \rho_{\hat{\psi}}(u)|\hat{\psi}\rangle.\label{eq:entropycut}
	\end{align}
	All that remains now is to relate this to the generalized entropy $S_{\text{gen}}(u)$ at the cut $u$.
	
	As an aside, note that the full algebra is still a Type III$_1$ algebra, because the $\mathbb{R}_u$ automorphism group rescales the Type II$_{\infty}$ trace by a $u$-dependent factor. So the full algebra has no meaningful notion of entropy, only the algebra localized to a cut does.
	
	\subsection{A generalized entropy formula and the generalized second law \label{sec:gravmodinc}}
	In \cite{Chandrasekaran:2022eqq}, it was shown that \Eqref{eq:entropycut} evaluated at the bifurcation surface is (the $\mathcal{O}(1)$ piece of) the generalized entropy of the black hole at the bifurcation surface. Using the machinery we've set up, it is straightforward to generalize this result to arbitrary cuts of the horizon.
	
	Before getting there, it is important to note that the cut algebra
	$\widehat{\mathcal{M}}_{\mathscr{H}_{>u}}$ admits states $\rho_{\hat{\Psi}}$
	that are reduced density matrices, obtained by tracing over the degrees of
	freedom that carry the record of the cut displacement $\Delta u_0$.
	Equivalently, for observables in $\widehat{\mathcal{M}}_{\mathscr{H}_{>u}}$
	one may view this reduction as evaluating an unconditional (non-selective)
	expectation value, i.e.\ averaging over $\Delta u_0$ rather than conditioning
	on a particular outcome. The induced classical weight
	$p(\Delta u_0)=|g(\Delta u_0)|^2$ is then just the Born distribution associated
	with the cut-position wavefunction $g(\Delta u_0)$ for the cut location
	$u_0+\Delta u_0$ in the state $|\hat{\Psi}\rangle\in\widehat{\mathcal{H}}$.
	
	The use of an unconditional expectation value reflects the fact that the semi-classical black
	hole dynamics localizes (decoheres) the cut location, so that we
	effectively average over the outcomes of a measurement of $\hat{\Upsilon}_0^{+}$
	rather than selecting a specific branch. The edge mode is part of the quantum
	reference frame (quantum mechanical observer), and $\hat{\Upsilon}_0^{+}$ is entangled
	with the semi-classical black hole dynamics through the null gravitational constraint equations;
	thus there is no external classical apparatus that would produce a fundamental
	``collapsed'' cut position---only an effective mixture in the reduced
	description.
	
	Using \Eqref{eq:translationexpectation} we then have
	\begin{align}
		S(\rho_{\hat{\psi}} ;{\widehat{\mathcal{M}}_{\mathscr{H}_{>u}}}) &= \int_{-\infty}^{\infty}d\Delta u_0\  |g(\Delta u_0)|^2\langle \hat{\psi}|U(\delta u - \Delta u_0)\log \rho_{\hat{\psi}}U(-\delta u_0 + \Delta u_0)|\hat{\psi}\rangle.\label{eq:entropyformulagrav}
	\end{align}
	
	Let $u_0$ be the location of the bifurcation surface. Following \cite{Chandrasekaran:2022eqq} but recast in terms of the construction in this paper, it can be shown that 
	\begin{align}
		\log \rho_{\hat{\psi}}(u_0) \approx -\beta \hat{\mathscr{A}}(\infty)+ h_{\Omega|\Psi}(u_0) - h_{\Omega}(\infty)- h_{\Psi}(u_0),\label{eq:densitymatrix}
	\end{align}
	where $h_{\Omega|\Psi} := -\log \Delta_{\Omega|\Psi}$, $h_{\Psi} = -\log \Delta_{\Psi},$ and $h_{\Omega} = -\log \Delta_{\Omega}$.
	
	It follows straightforwardly from the action of the half-sided translation operator $\hat{\mathscr{P}}$ that\footnote{When we move $u_0 \rightarrow u_0 + \delta u$ we are only moving the location of the choice of cut, not the bifurcation surface itself. The latter of course can't be moved.}
	\begin{align}
		U(\delta u)\log \rho_{\hat{\psi}}U(-\delta u) \approx -\beta \hat{\mathscr{A}}(\infty) + h_{\Omega|\Psi}(u) - h_{\Omega}(\infty)- h_{\Psi}(u).
	\end{align}
	
	Using the $G_{\varepsilon}$ splitting form of the extended Hilbert space \eqref{eq:decomphilbertspace}, we can actually write $h_{\Omega|\Psi}(u) = -\log \rho_{\Omega}(u) + \log\rho'_{\Psi}(u)$ for any non-zero $\varepsilon$. Here $\rho_{\Omega}(u)$ is the half-sided ``bulk'' density matrix of the vacuum state reduced to $\mathscr{H}_{>u}$, while $\rho'_{\Psi}$ is the half-sided ``bulk'' density matrix of the excited state $|\Psi\rangle$ reduced to the complementary region $\mathscr{H}'_{>u}$. Note that $K_{\Omega}(u) = -\log \rho_{\Omega}(u)$ is just the half-sided vacuum modular Hamiltonian of the ``bulk'' fields. Therefore, $h_{\Omega|\Psi}(u) = K_{\Omega}(u) + \log \rho_{\Psi}'$.
	
	Using \Eqref{eq:linmodham}, we therefore have 
	\begin{align}
		U(\delta u)\log \rho_{\hat{\psi}}U(-\delta u) \approx \beta \hat{\mathscr{A}}(u) + \log \rho_{\Psi}'(u) - h_{\Omega}(\infty)-h_{\Psi}(u).
	\end{align}
	
	Since $\langle \hat{\psi}|h_{\Psi}|\hat{\psi}\rangle = 0$, the expectation value reduces to 
	\begin{align}
		\langle \hat{\psi}|U(\delta u)\log \rho_{\hat{\psi}}U(-\delta u)|\hat{\psi}\rangle \approx \beta \langle \hat{\mathscr{A}}(u)\rangle_{\hat{\psi}} + S_{\text{bulk}}(u; \Psi) + \text{const.}, 
	\end{align}
	where the constant is just the infinite vacuum entanglement of the ``bulk'' fields, and we've used purity to relate the ``bulk'' entropy on $\mathscr{H}'_{>u}$ to the ``bulk'' entropy on $\mathscr{H}_{>u}$. Since the trace on a Type II$_{\infty}$ algebra is only defined up to an overall rescaling, the gravitational entropy obtained in this manner is only defined up to an overall (infinite state-independent) constant. So we've shown that 
	\begin{align}
		\langle \hat{\psi}|U(\delta u)\log \rho_{\hat{\psi}}U(-\delta u)|\hat{\psi}\rangle\approx S_{\text{gen}}(u; \hat{\psi}) + \text{const.}
	\end{align}
	
	Putting it all together, we've shown that the gravitational von Neumann entropy \eqref{eq:entropyformulagrav} is
	\begin{align}
		S(\hat{\psi} ;{\widehat{\mathcal{M}}_{\mathscr{H}_{>u}}}) \approx \int_{-\infty}^{\infty}d\Delta u_0 \ |g(\Delta u_0)|^2 S_{\text{gen}}(u - \Delta u_0; \hat{\psi}) + \text{const.}\label{eq:finalgraventropy}
	\end{align}
	The fluctuation $\Delta u_0$ is now just a classical random variable, with probability distribution $p(\Delta u_0) := |g(\Delta u_0)|^2$. We can therefore define a classical average of the generalized entropy as follows:
	\begin{align}
		\bar{S}_{\text{gen}}(u; \hat{\psi}) = \int_{-\infty}^{\infty}d\Delta u_0 \ p(\Delta u_0) S_{\text{gen}}(u |\Delta u_0; \hat{\psi}),
	\end{align}
	where each $S_{\text{gen}}(u | \Delta u_0)$ is the value of generalized entropy for an observed value of $\Delta u_0$, and hence is associated to a sharply defined cut $u - \Delta u_0$ i.e. $S_{\text{gen}}(u-\Delta u_0;\hat{\psi}) = S_{\text{gen}}(u|\Delta u_0; \hat{\psi})$. 
	
	Therefore, 
	\begin{align}
		S(\hat{\psi} ;{\widehat{\mathcal{M}}_{\mathscr{H}_{>u}}}) \approx \bar{S}_{\text{gen}}(u; \hat{\psi}) + \text{const}.\label{eq:vnentropyisgenentropy}
	\end{align}
	So the gravitational von Neumann entropy is actually the generalized entropy averaged over all observations of the $\hat{\Upsilon}^+_0$ edge mode.
	
    Next, the one-parameter family $\widehat{\mathcal{M}}_{\mathscr{H}_{>u}}$ satisfies a nesting property that follows from (i) translation covariance of the net and (ii) isotony (monotonicity).
    Let $U(\delta u)$ implement a null translation by affine parameter $\delta u$ along the
    null generator. Covariance corresponds to the statement
    \[
    U(\delta u)\widehat{\mathcal{M}}_{\mathscr{H}_{>{u_0}}}U(\delta u)^\dagger = \widehat{\mathcal{M}}_{\mathscr{H}_{>{u_0+\delta u}}},
    \]
    which relies on the fact that we've gravitationally dressed ``bulk'' operators to the edge modes. For $\delta u>0$ the null translation shifts the cut/region so that the translated
    region is geometrically nested, $\mathscr{H}_{>u_0+\delta u} \subseteq \mathscr{H}_{>u_0}$, and
    then isotony implies
    \[
    \widehat{\mathcal{M}}_{\mathscr{H}_{>{u_0+\delta u}}}\subseteq \widehat{\mathcal{M}}_{\mathscr{H}_{>{u_0}}}.
    \]
    Equivalently, the algebras are nested under future-directed null translations:\footnote{A possible confusion is that in the ambient crossed product algebra
$\widehat{\mathcal{A}}_{\mathscr{H}_{> u_0}} = \mathcal{A}_{\mathscr{H}_{> u_0}}\rtimes (\mathbb{R}_{s}\rtimes \mathbb{R}_{u})$,
boost and null translation unitaries obey the Borchers relation
$\Delta^{it}U(a)\Delta^{-it}=U(e^{-2\pi t}a)$, thus implying that
$U(a)\Delta^{it}U(-a)=\Delta^{it}U((1-e^{-2\pi t})a)$. In words, conjugating a boost by a translation generally produces a
``translation dressed'' boost.  If one tried to define the cut algebra at $u$ as the von Neumann algebra generated
by these translated unitaries inside the fixed $\widehat{\mathcal{A}}_{\mathscr{H}_{> u}}$, the resulting family need not be nested.
In this paper the Type $\mathrm{II}_\infty$ cut factor is instead defined fiberwise by conditioning on the
translation edge mode via the canonical conditional expectation $E_u\colon\widehat{\mathcal{A}}_{\mathscr{H}_{> u}}\mapsto\widehat{\mathcal{M}}_{\mathscr{H}_{> u}}$.  The expectations are covariant under translations,
$\mathrm{Ad}_{U(a)}\!\circ E_u=E_{u+a}\!\circ\mathrm{Ad}_{U(a)}$, so the above dressing is precisely compensated for by
shifting the conditioning map when the cut is moved. See the discussion around \Eqref{eq:covariant-expectation} for further details.} 
	\begin{align}
		U(\delta u)\Pi\left(\widehat{\mathcal{M}}_{\mathscr{H}_{>u_0}}\right)U(-\delta u) \subset \Pi\left(\widehat{\mathcal{M}}_{\mathscr{H}_{>u_0}}\right),
	\end{align}
	for any $u_0$. It follows that $\partial_u S(\hat{\psi} ;{\widehat{\mathcal{M}}_{\mathscr{H}_{>u}}}) \geq 0$ because $\text{tr}$ respects nesting of the algebra. See \cref{sec:montonocityentropy} for a proof of this statement.
	
	Applying this to \Eqref{eq:vnentropyisgenentropy}, we arrive at the generalized second law (GSL) in perturbative quantum gravity:
	\begin{align}
		\partial_u \bar{S}_{\text{gen}}(u;\hat{\psi}) \geq 0.
	\end{align}
	As one might expect, the GSL only holds on expectation, due to fluctuations in the corner location.
	
	The deviation from the mean can also be computed easily:
	\begin{align}
		S_{\text{gen}}(u|\Delta u_0; \hat{\psi}) - \bar{S}_{\text{gen}}(u; \hat{\psi}) = -\frac{(\Delta u_0)^2}{2}\partial_u^2 \bar{S}_{\text{gen}}(u; \hat{\psi}) + \mathcal{O}((\Delta u_0)^4).
	\end{align}
	This actually gives us a nice interpretation of quantum focusing: it controls the fluctuations in the generalized entropy arising from fluctuations in the location of the corner. Furthermore, the quantum focusing conjecture $\partial_u^2 \bar{S}_{\text{gen}}(u) \leq 0$ would then correspond to the statement that the fluctuations occur around a local minimum. It essentially ensures that there isn't an entropic pressure to run away from the classical location of the cut.

    \subsection{Non-stationary backgrounds\label{sec:genbackentropy}}
	Let's take stock of what we've constructed thus far:
	\begin{itemize}
		\item Everything up to and including \cref{sec:classicalalgebra} is completely general, applying to the full non-linear classical phase space.
		\item \cref{sec:canonicalquant} restricted to spacetimes with bifurcate Killing horizons. In this section we extend all of the results therein to linearization around a general non-stationary event horizon.
		\item The details of \cref{sec:genentropyderiv} thus far strongly relied on the existence of a global KMS state on the horizon. This cannot be generalized to non-stationary backgrounds. But in this section we make use of local Rindler frames to derive the results of \cref{sec:genentropyderiv} thus far in a small neighborhood of any cut of the horizon.
	\end{itemize}
	We start with the last point. Consider a cyclic and separating state $|\Omega\rangle$ on $\mathscr{H}'_{>u_0}\cup \mathscr{H}_{>u_0}$. There's no canonical choice of vacuum in a general spacetime, so we will arbitrarily refer to this state as the ``vacuum'' state. 
	
	By the equivalence principle, any smooth state should look like the Hartle-Hawking vacuum at length scales smaller than both the typical radius of curvature $R_0$ of the background spacetime and the typical length scale $\lambda_0$ of excitations of the background matter fields, and the metric itself should look locally like that of a Killing horizon.
	
	To make this a bit more precise, consider a cut $S_0$ of $\mathscr{H}$ at location $u_0$. Smoothness of the metric guarantees we can always find a coordinate system $(\tilde{u}, \tilde{v}, x^A)$ in a neighborhood of $S_0$ such that
	\begin{align}
		ds^2 = -&2d\tilde{u} d\tilde{v} - 2\tilde{v} \Omega_A \big\lvert_{\tilde{u}=\tilde{v}=0} d\tilde{u}dx^A +  q_{AB}(x^C)dx^A dx^B \nonumber \\&+ \left(\tilde{u}\partial_{\tilde{u}}q_{AB}(x^C, \tilde{u})\big\lvert_{\tilde{u}=0}+ \tilde{v}\partial_{\tilde{v}}q_{AB}(x^C, \tilde{v})\big\lvert_{\tilde{v} = 0}\right)dx^A dx^B  + \mathcal{O}(\tilde{u}^2, \tilde{v}^2, \tilde{u}\tilde{v}).\label{eq:localrindlermetric}
	\end{align}
	The boost vector field at $S_0$ takes the form $\chi = \kappa (\tilde{u} \partial_{\tilde{u}} - \tilde{v}\partial_{\tilde{v}}) + \mathcal{O}(\tilde{u}^2, \tilde{v}^2, \tilde{u}\tilde{v})$, and satisfies $\nabla_{(\mu}\chi_{\nu)} = 0 + \mathcal{O}(\tilde{u}, \tilde{v})$. This just follows from
	\begin{align}
		\nabla_{(\mu}\chi_{\nu)} &= \kappa \tilde{u}\partial_{\tilde{u}}g_{\mu\nu} - \kappa \tilde{v}\partial_{\tilde{v}}g_{\mu\nu} + \kappa g_{\mu \tilde{u}}\partial_{\nu}\tilde{u} - \kappa g_{\mu \tilde{v}}\partial_{\nu}\tilde{v}
		\\ &= \kappa g_{\tilde{u}\tilde{v}}\left[\delta^{\tilde{u}}_{(\nu}\delta^{\tilde{v}}_{\mu)} - \delta^{\tilde{v}}_{(\nu}\delta^{\tilde{u}}_{\mu)}\right] + \mathcal{O}(\tilde{u}, \tilde{v})
	\end{align}
	Now, \Eqref{eq:localrindlermetric} describes the local geometry but we also need to describe the local state. It is well-known \cite{Hollands:2014eia} that in a linearized theory on a (generically curved) globally hyperbolic spacetime, one can always find a state $|\Omega\rangle$ that is Hadamard and Gaussian. This means that the two-point function of all operators in the state has the same UV singularity structure as in Minkowski, and all higher-point functions are determined by the two-point function. 
	
	But recall we're considering a linearized theory with an ultralocal symplectic form $\Omega_{\mathscr{H}}$. Furthermore, since $\mathscr{H}$ is always a smooth characteristic initial value surface in a sufficiently small neighborhood of $S_0$, with a well-posed characteristic initial value problem, it follows that $[\mathscr{O}(\tilde{u}, x^A), \mathscr{O}(\tilde{u}, y^A)] = 0$ when $x^A \neq y^A$ \cite{Wall:2011hj}. So ultralocality combined with Hadamard + Gaussian actually tells us that $\langle \mathscr{O}_1(\tilde{u}_1, x_1^A)\ldots \mathscr{O}_n(\tilde{u}_n, x_n^A)\rangle_{\Omega_g} \sim \langle \mathscr{O}_1(\tilde{u}_1, x_1^A)\ldots \mathscr{O}_n(\tilde{u}_n, x_n^A)\rangle_{\Omega_ \eta}$ for all $n$ in the limit $|\tilde{u}_i - \tilde{u}_j| \ll R_0, \lambda_0$ where $\eta$ is the metric of some Killing horizon and $\Omega_{\eta}$ is the Hartle-Hawking vacuum. In particular, there's no condition on $|x^A_i - x^A_j|$. Hence, we refer to this construction as a local Rindler frame, since the horizon looks like a Killing horizon in the Hartle-Hawking state in the neighborhood of (the entirety of) $S_0$ in these coordinates.
	
	Moreover, as we've shown in \cref{sec:classicalalgebra}, for any metric $g$ and any cut $\tilde{u}$,
    \begin{subequations}
	\begin{align}
		&[\hat{\mathscr{A}}_g(\tilde{u}), \hat{\mathscr{O}}(p)] = -i(\tilde{u}_p - \tilde{u})\partial_{\tilde{u}}\hat{\mathscr{O}}(p), \\ 
		&[\hat{\mathscr{P}}_g, \hat{\mathscr{O}}(p)] = -i\partial_{\tilde{u}}\hat{\mathscr{O}}(p), 
	\end{align}
        \end{subequations}
	for any gauge-invariant operator $\hat{\mathscr{O}}(p)$ gravitationally dressed to $S_{\tilde{u}}$ as in \cref{sec:classicalalgebra}. So the local boost $\chi$ at $S_0$ can be identified with the action of the area operator $\hat{\mathscr{A}}_{\eta} := \hat{\mathscr{A}}_g\big\lvert_{\tilde{u}=0}$.\footnote{Note the shear and expansion don't vanish at $\tilde{u} = \tilde{v} = 0$ for the metric \eqref{eq:localrindlermetric}. So the local Rindler frame doesn't resemble a Killing horizon under first order shape deformations of $S_0$; in particular, $\hat{\mathscr{A}}_{\eta}$ is not given by the same expression as in the global Killing horizon case. But all we need for the existence of a local KMS state is the presence of a local boost Killing field.} Under the shape deformation $\tilde{u}$, we have that 
	\begin{align}
		\hat{\mathscr{A}}_g(\tilde{u}) = \hat{\mathscr{A}}_{\eta} + \tilde{u}\ [\hat{\mathscr{P}}_{g}, \hat{\mathscr{A}}_g]\Big \lvert_{g = \eta} + \mathcal{O}(\tilde{u}^2)
	\end{align}
	while holding the state fixed. Combining what we have so far, this means that at $S_0$ itself, $\Delta_{\Omega_{g}; \mathcal{A}_{>u_0}}$ can be identified with the boost generator $\hat{\mathscr{A}}_{\eta}(u_0)$. Ultimately we want to be able to at least say something about the physics to first order in deviation $\tilde{u} \ll 1$ away from $S_0$. We will soon calculate what $\hat{\mathscr{A}}_g(\tilde{u})$ is after solving the gravitational constraints in complete generality when doing perturbative quantum gravity on an arbitrary non-stationary background event horizon.
	
	But before getting there, note that by background independence we can equivalently view the deformation $\hat{\mathscr{A}}_g$ as a change to the state $|\Omega_{g}\rangle = |\Omega_{\eta}\rangle + \tilde{u}_0|\Delta \Omega\rangle + \mathcal{O}(\tilde{u}^2)$ while holding the background metric fixed.\footnote{This statement should strictly speaking be interpreted as holding inside of arbitrary $n$-point correlation functions.} In this latter perspective, the first law of entanglement tells us that 
	\begin{align}
		S_{\text{bulk}}(\tilde{u}; \Omega_g) - S_{\text{bulk}}(\tilde{u};\Omega_{\eta}) =  \langle \Omega_{\eta}|K_{\Omega_g}(\tilde{u})|\Omega_{\eta}\rangle + \mathcal{O}(\tilde{u}^2).
	\end{align}
	Therefore, 
	\begin{align}
		K_{\Omega_g}(\tilde{u}) = \tilde{u} [\hat{\mathscr{P}}_{g}, \hat{\mathscr{A}}_g]\big \lvert_{g = \eta} + \mathcal{O}(\tilde{u}^2).
	\end{align}
	Crucially, this means the action of $\Delta_{\Omega_g; \mathcal{A}_{>\tilde{u}}}$ on $\mathcal{A}_{>\tilde{u}}$ agrees with the action of $\hat{\mathscr{A}}_g(\tilde{u})$ to first order in $\tilde{u}$. Hence we've argued that in a first order neighborhood of (the entirety) of $S_0$, the ``vacuum'' modular Hamiltonian acts on gauge-invariant local operators in the future algebra as the local geometric flow generated by $\hat{\mathscr{A}}_g(\tilde{u})$.
	
	Putting this all together, we immediately see that all the results of \cref{sec:gravmodinc} directly carry over to a first order neighborhood of $S_0$. In particular, 
	\begin{align}
		&S(\hat{\psi}_g ;{\widehat{\mathcal{M}}_{\mathscr{H}_{>\tilde{u}}}}) \approx \bar{S}_{\text{gen}}(\tilde{u}; \hat{\psi}_g) + \text{const.} + \mathcal{O}(\tilde{u}^2)\label{eq:localsgen}.
	\end{align}
	Thus we've shown that in perturbative quantum gravity the generalized entropy is the von Neumann entropy of a Type II$_{\infty}$ algebra on an arbitrary non-stationary background event horizon to first order in the neighborhood of any cut of the horizon.\footnote{Technically it also follows that $\partial_{\tilde{u}} \bar{S}_{\text{gen}}(\tilde{u};\hat{\psi}_g)\lvert_{\tilde{u}=0} \ \geq 0\label{eq:localgsl}$ but for a non-stationary background this will be dominated by the classical area term at $\mathcal{O}(1/G_N)$, which will be true simply by the classical area theorem, so the GSL is only interesting when quantizing around a Killing horizon background.\label{footnote:gslnontrivial}}
	
	Since the original location $u_0$ was arbitrary, \Eqref{eq:localsgen} holds in the neighborhood of any cut of the horizon. However, we obviously can't ``patch'' them together to claim they hold globally on the horizon. To see this, let's say the non-stationarity of the black hole comes from an arbitrary series of shocks falling across the horizon. We can always approximate the black hole's non-stationarity by this picture, with the timescale between shocks as the characteristic timescale of change of the underlying dynamics; from the perspective of the shocks this is the characteristic thermalization time of the black hole. 
	
	Then, physically all that's happening is if we zoom in on a sufficiently small region of the horizon that doesn't contain a shock, the state and geometry will look approximately stationary. This local equilibrium is necessary if we want to talk about black hole thermodynamics. But when a shock crosses the horizon, there will be a sudden quench, so we can't talk about black hole thermodynamics in the vicinity of the shock. Roughly speaking, $|\tilde{u}| \sim 1/T$ where $T$ is the (local) temperature of the black hole. So \Eqref{eq:localsgen} is the best we can do locally. However, it's possible to derive an integrated GSL between the far past and far future of a black hole evolving under an arbitrary series of shocks. This was shown in \cite{Chandrasekaran:2022eqq}.
	
	All that remains is to actually calculate $\hat{\mathscr{A}}_g(u)$ and $\hat{\mathscr{P}}_{g}(u)$ after integrating out the gravitational constraints, namely the Raychaudhuri equation. This is straightforward; the computation is analogous to the one in \cref{sec:canonicalquant}. At the classical level, the linearized constraint is
	\begin{align}
		\partial_u \delta \Theta(u) + \Theta(u) \delta \Theta(u)  = -2\sigma(u) \delta \sigma(u) - 8\pi G T_{uu}(u). 
	\end{align}
	Let $\delta S(u) := -2\sigma(u) \delta \sigma(u) - 8\pi G_N T_{uu}(u)$ represent the linearized source for the perturbed expansion.
	
	Using the teleological boundary condition $\delta \Theta \rightarrow 0$ as $u\rightarrow \infty$, the solution is just 
	\begin{align}
		\delta \Theta(u) = -\frac{1}{\sqrt{q(u)}}\int_{u}^{\infty}ds \  \delta S(s)\ \sqrt{q(s)}.
	\end{align}
	Therefore, the area operator and half-sided translation operator in perturbative quantum gravity are given by
    \begin{subequations}
	\begin{align}
		&\hat{\mathscr{P}}_g = -\frac{1}{8\pi G_N}\frac{1}{\sqrt{q(u_0^+)}}\int_{u^+_0}^{\infty}du \ \int_{S_0^+}d^{d-2}x \ q(u) \ \hat{S}(u),\label{eq:genbacktrans} \\ 
		&\hat{\mathscr{A}}_g= \frac{1}{8\pi G_N}\int_{u^+_0}^{\infty}du\int_{S^+_0}d^{d-2}x \sqrt{q(u)}\ \left(U(u) - U(u_0)\right)\hat{S}(u)- \hat{\mathscr{A}}(\infty),\label{eq:genbackarea}
	\end{align}
        \end{subequations}
	where $\hat{S}(u)$ is the smeared operator obtained from $\delta S(u)$ and $U(u) = \int^u du'/\sqrt{q(u')}$ is the clock measured in terms of the expansion $\Theta(u)$.\footnote{To be maximally precise: the symmetry generators are integrated over the entire half-line but this assumes geodesic completeness of the horizon. Generically, caustics will form in finite affine time so we can't extend the integration range all the way to infinity. Instead, we should really view these operators as generating the local flows $[\hat{\mathscr{A}}_g, \hat{\mathscr{O}}(p)] = i(u-u_0)\partial_u \hat{\mathscr{O}}(p), \ [\hat{\mathscr{P}}_g, \hat{\mathscr{O}}(p)] = i\partial_u \hat{\mathscr{O}}(p)$ as long as $p$ lies in a convex normal neighborhood of $u^+_0$. Away from this neighborhood the identification breaks down. In other words we're only locally identifying the form of the symmetry generators with that in a global Rindler frame, solely in order to write down the local geometric action of the operators.} One can further use these expressions to compute the non-degenerate symplectic form if desired, though we won't do that here.
	
	So on a general non-stationary background, the area operator and half-sided translation operator not only include the usual stress tensor term but also a term linear in the graviton operator $\hat{\sigma}$ (multiplied by the background classical shear). In other words, the graviton operator $\hat{\sigma}$ is the pure gravity contribution to these generators.

    \section{Implications}

    A central result of this paper is that once corner edge modes are included, the half-sided null translation/boost symmetries become unitarily implemented as gravitational half-sided modular inclusions on a one-parameter family of Type~$\mathrm{II}_\infty$ factors $\widehat{\mathcal{M}}_{\mathscr{H}_{>u}}$. Generalized entropy then becomes amenable to the methods of Tomita--Takesaki theory.  In this section we exploit this structure as a calculational framework to derive several important aspects of perturbative quantum gravity purely from the bulk. The main technical step, as we've seen, is to rewrite $\bar S_{\rm gen}(u)$ as (minus) an Araki relative entropy on $\widehat{\mathcal{M}}_{\mathscr{H}_{>u}}$, so that its null derivatives are controlled by the null translation generator associated with the gravitational half-sided modular inclusion algebra.
    
    This perspective lets us straightforwardly adapt algebraic QFT methods essentially verbatim: the quantum expansion $\bm{\Theta}(u)=\partial_u \bar S_{\rm gen}(u)$ admits a variational representation as an infimum over purifications generated by Connes-cocycle flow in the commutant, and gravitational half-sided modular inclusion implies the commutants grow as the cut is pushed forward.  The enlarging minimization domain forces the infimum (and hence $\bm{\Theta}$) to be nonincreasing, yielding an algebraic proof of quantum focusing in exact parallel with the QNEC proof of Ceyhan-Faulkner \cite{Ceyhan:2018zfg}. We then apply Tomita-Takesaki technology to finite null segments (causal diamonds, to be precise) and to the question of bulk reconstruction from the corner algebra, where the corner translation generator plays the role of ``time evolution'' along a null Cauchy slice.

    	\subsection{Gravitational half-sided modular inclusions and edge modes\label{sec:hsm_inclusion_bridge}}
	
	Ordinary QFTs typically satisfy a property known as half-sided modular inclusions.
	Given two von Neumann algebras $\mathcal{N}\subset \mathcal{M}$ with common cyclic/separating vector $\Omega$, one says $(\mathcal{N}\subset \mathcal{M},\Omega)$ is a half-sided modular inclusion if the modular group of $\mathcal{M}$ leaves $\mathcal{N}$ invariant
	under flows along the time direction, e.g.
	\begin{equation}
		\Delta_{\mathcal{M}}^{it}\,\mathcal{N}\,\Delta_{\mathcal{M}}^{-it}\subset \mathcal{N},\ t\ge 0.
	\end{equation}
	A theorem of Borchers \cite{Borchers1996} and Wiesbrock \cite{Wiesbrock1993} then implies the existence of a unique one-parameter unitary group
	\begin{equation}
		U(a)=e^{iaG},\ a\in\mathbb{R},
	\end{equation}
	with positive generator
	\begin{equation}
		G\ge 0,
	\end{equation}
	such that $U(a)$ implements the half-sided translation semigroup on the net and satisfies the ax+b commutation relations with the modular flow.
	A convenient form of these relations is
	\begin{equation}
		\Delta_{\mathcal{M}}^{it}\,U(a)\,\Delta_{\mathcal{M}}^{-it}=U(e^{-2\pi t}a), \
		J_{\mathcal{M}}\,U(a)\,J_{\mathcal{M}}=U(-a), \
		[K_{\mathcal{M}},G]=iG,
		\label{eq:axb_borchers}
	\end{equation}
	where $K_{\mathcal{M}}:=-(\log \Delta_{\mathcal{M}})$ is the modular Hamiltonian of $(\mathcal{M},\Omega)$ and $J_{\mathcal{M}}$ the modular conjugation.
	
	In the null-plane vacuum of a relativistic QFT one can further identify $G$ with the ANEC operator:
	\begin{equation}
		G \;\propto\; \int d^{d-2}x \int_{-\infty}^{\infty} du\,T_{uu}(u,x^A).
	\end{equation}
	Then the modular Hamiltonians for translated cuts become affine linear in the cut position, so that for nested cuts
	\begin{equation}
		K_{u_1}-K_{u_2} = 2\pi (u_2-u_1)\,G\;\ge\;0,
		\ u_2>u_1.
		\label{eq:qft_positive_difference}
	\end{equation}
	
	The gravitational story is more or less the same, except that unlike in the case of QFT, the null translation generator $G$ now lives naturally in the crossed product algebra $\widehat{\mathcal{A}}_{\mathscr{H}_{>u}}$. This is because in gravity, or at least in our construction specifically, $G$ corresponds to a half-sided translation; it acts as an outer automorphism on the ``bulk'' subregion algebra $\mathcal{A}_{\mathscr{H}_{>u}}$ i.e. the QFT algebra, but in gravity it becomes an inner automorphism after including dynamical edge modes.
	
	In our gravitational construction, the uniform half-sided translation generator is the operator $\hat{\mathscr{P}}$
	(conjugate to the translation edge mode $\Upsilon_0^+$), and more generally the angle-dependent generators are $\hat{\mathscr{P}}_\alpha$.
	On gravitationally dressed bulk observables $\hat{\mathscr{O}}(p)$ (dressed to the cut as in Sec.~5.1) these act as genuine horizon translations:
	\begin{equation}
		[\hat{\mathscr{P}}_\alpha,\hat{\mathscr{O}}(p)] = -\,i\,\alpha^+(x^A)\,\partial_u \hat{\mathscr{O}}(p), \
		[\hat{\mathscr{A}}_\beta,\hat{\mathscr{O}}(p)] = -\,i\,(u-u_0)\,\beta^+(x^A)\,\partial_u \hat{\mathscr{O}}(p),
		\label{eq:half_sided_actions}
	\end{equation}
	while the nontrivial commutator between the generators is the half-sided ax+b algebra
	\begin{equation}
		[\hat{\mathscr{A}}_\beta,\hat{\mathscr{P}}_\alpha]=i\,\hat{\mathscr{P}}_{-\alpha\beta}, \
		[\hat{\mathscr{A}}_\beta,\hat{\mathscr{A}}_{\beta'}]=[\hat{\mathscr{P}}_\alpha,\hat{\mathscr{P}}_{\alpha'}]=0.
		\label{eq:corner_axb}
	\end{equation}
	Two points are essential for the present discussion:
	
	\begin{enumerate}
		\item \textbf{Inner vs.\ outer automorphism}: the half-sided translation automorphisms on the ``bulk'' horizon algebra $\mathcal{A}_{\mathscr{H}_{>u_0}}$ are, in general, outer, as we have already discussed. By contrast, once we pass to the crossed product algebra $\widehat{\mathcal{A}}_{\mathscr{H}_{>u_0}}$, the same half-sided translations are implemented unitarily as an inner automorphism on $\widehat{\mathcal{A}}_{\mathscr{H}_{>u_0}}$, and as an outer automorphism on $\widehat{\mathcal{M}}_{\mathscr{H}_{>u_0}}$,
		\begin{equation}
			U(\delta u)=e^{i\hat{\mathscr{P}}\,\delta u}\in \widehat{\mathcal{A}}_{\mathscr{H}_{>u_0}}, \
			\mathrm{Ad}(U(\delta u))\colon\widehat{\mathcal M}_{\mathscr{H}_{>u_0}}\mapsto  \widehat{\mathcal{M}}_{\mathscr{H}_{>u_0}}, \
			\delta u\ge 0.
			\label{eq:inner_translation}
		\end{equation}
		
		\item \textbf{Light ray operator representations:} In general, light ray operators are merely identifications of the group action, not the fundamental source of positivity. As we've shown, given a Killing horizon background one can solve the linearized constraints (Raychaudhuri) and identify
		$\hat{\mathscr{A}}$ and $\hat{\mathscr{P}}$ with the following half-sided light ray operators:
        \begin{subequations}
		\begin{align}
			&\hat{\mathscr{A}} \;\simeq\; \frac{1}{8\pi}\int_{u_0}^{\infty}\!du\int_{S_0^+}\!d^{d-2}x\,\sqrt q\,(u-u_0)\,\hat{T}_{uu}(u,x),\label{eq:half_anec_identification_1}\\
			&\hat{\mathscr{P}} \;\simeq\; -\frac{1}{8\pi}\int_{u_0}^{\infty}\!du\int_{S_0^+}\!d^{d-2}x\,\sqrt q\,\hat{T}_{uu}(u,x).
			\label{eq:half_anec_identification_2}
		\end{align}
                \end{subequations}
		(On a general nonstationary background $\hat{\mathscr{P}}$ and $\hat{\mathscr{A}}$ acquire additional pure gravity terms involving the graviton operator $\widehat\sigma$, see \cref{sec:genbackentropy} below). However, this should be read as local identifications of generators via their commutator with dressed observables (and within an appropriate domain), not as the method by which to prove (or disprove) spectral positivity of $\hat{\mathscr{P}}$. The reason is the same as in QFT: the half-sidedness refers to the net/inclusion and to the semigroup $\delta u\ge 0$,
		not to a manifestly positive kernel in the bulk integral.
	\end{enumerate}
	
	So what is actually positive in our context? Assume, as is standard on the null plane and as we do here, that the net $u\mapsto \mathcal{A}_{\mathscr{H}_{>u_0}}$ in the vacuum satisfies the half-sided modular inclusion property; then the same is true for the net of crossed product algebras $u\mapsto \widehat{\mathcal{M}}_{\mathscr{H}_{>u_0}}$, that is,
    \begin{equation}
        \mathrm{Ad}\big(U(\delta u)\big)\Big(\widehat{\mathcal{M}}_{\mathscr{H}_{>u_0}}\Big)\subset \widehat{\mathcal{M}}_{\mathscr{H}_{>u_0}}, \
        \delta u\ge 0.
        \label{eq:semigroup_inclusion}
    \end{equation}
    This is because the operators $\hat{\mathscr{A}}$ and $\hat{\mathscr{P}}$ coincide with the half-sided vacuum modular flow and half-sided null translation operators when acting on $\mathscr{O}\in \mathcal{A}_{\mathscr{H}_{>u}}$. This follows from \crefrange{eq:linmodham}{eq:linanec} as well as \crefrange{eq:translationcommutator}{eq:boostcommutator}. Under this assumption, the theorem of Borchers/Wiesbrock applies to the pair of nested factors
	$\widehat{\mathcal{M}}_{\mathscr{H}_{>u_2}}\subset \widehat{\mathcal{M}}_{\mathscr{H}_{>u_1}}$ with cyclic and separating vacuum vector $|\hat{\Omega}\rangle$:
	
	\begin{quote}
		\emph{There exists a unitary group $U(a)=e^{iaG}$ implementing the half-sided translation semigroup on the net,
			with a positive self-adjoint generator $G\ge 0$, satisfying the ax+b commutation relations with the modular flow.}
	\end{quote}

    But as we have noted, in our gravitational realization the same translation semigroup is implemented as an inner automorphism on $\widehat{\mathcal{A}}_{\mathscr{H}_{>u_0}}$ by $U(\delta u)=e^{i\hat{\mathscr{P}}\,\delta u}$. Therefore we may identify the generator $G$ from the abstract half-sided modular inclusion with the null translation generator $\hat{\mathscr{P}}$ conjugate to the edge mode $\Upsilon^+_0$:
	\begin{equation}
		G \equiv \hat{\mathscr{P}},\ \text{and more generally},\ 
		G_{\alpha} \equiv \hat{\mathscr{P}}_\alpha \;\;\text{for}\;\;\alpha(x^A)\ge 0.
		\label{eq:identify_positive_generator}
	\end{equation}
	$\hat{\mathscr{P}}$ is identified abstractly as the self-adjoint generator of the half-sided translation unitaries in the crossed product algebra $\widehat{\mathcal{A}}_{\mathscr{H}_{>u_0}}$. And $\hat{\mathscr{P}}$ is positive by modular inclusion. The action of $\hat{\mathscr{P}}$ on dressed ``bulk'' observables can be represented by the half-sided ANEC operator. This representation is not what determines positivity (or lack thereof). What happens in gravity that doesn't happen in QFT is the gravitational edge modes at the corner render the half-sided translations inner on the crossed product subregion algebra $\widehat{\mathcal{A}}_{\mathscr{H}_{>u_0}}$, whereas in QFT the analogous translations on the ``bulk'' subregion algebra $\mathcal{A}_{\mathscr{H}_{>u_0}}$ are outer.

    We conclude by elaborating on a subtlety regarding half-sided modular inclusions and Type II$_{\infty}$ algebras.\footnote{We thank Marc Klinger for raising this point.} One might naively worry that Type II$_{\infty}$ algebras cannot support half-sided modular inclusions due to a result of Wiesbrock's which ties half-sided modular inclusions to Type III$_1$ algebras \cite{Wiesbrock1993}. But it's important to distinguish a bare half-sided modular inclusion from a \emph{standard} half-sided modular inclusion wherein one imposes additional conditions on the reference vector beyond being cyclic and separating for the two nested algebras. Concretely, if \((\mathcal N\subset \mathcal M,\Omega)\) is a half-sided modular inclusion and furthermore one has a unique translation invariant vacuum vector (see \cite[Cor.~11]{Wiesbrock1993}),
    then one obtains the conclusion that \(\mathcal M\) must be of Type~III\(_1\)
    \cite[Thm.~12]{Wiesbrock1993}.\footnote{Technically Corollary 11 of \cite{Wiesbrock1993} follows from assuming cyclicity of $|\hat{\Omega}\rangle$ with respect to the relative commutant. As we will argue, uniqueness of the vacuum state $|\hat{\Omega}\rangle$ does not hold for our Type II$_{\infty}$ crossed product cut algebra. Therefore, it must ultimately be the case that any choice of cyclic and separating vacuum vector on $\widehat{\mathcal{M}}_{\mathscr{H}_{>u_0}}$ fails to be cyclic on the relative commutant. It is easy to show that this is indeed the case.}
    
    So what is ruled out is the coexistence of (i) half-sided modular inclusion of a Type II$_{\infty}$ algebra together with (ii) the strengthened standardness assumption that produces a unique translation invariant
    vacuum vector in the same representation.  But in fact the standardness assumption fails for our crossed product cut algebra $\widehat{\mathcal{M}}_{\mathscr{H}_{>u_0}}$, so there is no contradiction. The argument is as follows.

    In the covariant crossed product representation
    \(\widehat{\mathcal H}=\mathcal H\otimes L^2(\mathbb R_s)\otimes L^2(\mathbb R_u)\)
    from \Eqref{eq:simplifiedexthilbertspace}, we chose the \(\ket{u}\) basis so that the null translation edge mode
    \(\hat \Upsilon_0^+\) acts by multiplication on \(L^2(\mathbb R_u)\), and its conjugate generator is
    \(\hat{\mathscr P}\) (acting as \(-i\partial_u\) on the \(u\)-wavefunction). If \(U(a)\psi=\psi\) for all \(a\), then \(\psi(u)\) is (a.e.) constant.  On \(\mathbb R_u\) this implies
    \(\psi=0\) in \(L^2(\mathbb R_u)\). So the strengthened Wiesbrock assumption that yields a unique translation invariant vacuum vector
    cannot even be formulated as a vector statement in the non-compact crossed product translation sector.
    If one instead IR regulates the translation sector (e.g.\ by replacing \(L^2(\mathbb R_u)\) by \(L^2(S^1_L)\) with
    period \(L\), or imposing a box normalization on \(u\)), then the constant wavefunction \(\psi_0(u)=L^{-1/2}\)
    is a normalizable \(U(a)\)-invariant vector in the \(u\)-sector.  However, the space of invariant vectors in
    the full Hilbert space is then infinite dimensional.  In particular, there are infinitely many \(|\hat\Omega\rangle\) with
    \(\hat{\mathscr P}|\hat\Omega\rangle=0\); take any $\ket{\chi}\in \widehat{\mathcal{H}}_{\mathcal{M}}$ and set
    \(|\hat\Omega\rangle=\ket{\chi}\otimes\ket{\psi_0}\).
    
    In ordinary QFT on a Rindler horizon, the half-sided translation semigroup is implemented in the same Hilbert space representation that carries the vacuum. In other words, translations act directly on the ``bulk'' degrees of freedom (they move local excitations), so the only translation-invariant state is the empty state. In perturbative quantum gravity, by contrast, the cut location is a quantum degree of freedom with (translation) edge labels in the vacuum representation. So the uniqueness property of the vacuum fails here for a very intuitive reason: the null translation generator acts on an edge mode/quantum reference frame sector rather than acting directly on the ``bulk'' QFT degrees of freedom.\footnote{The argument in \cref{sec:quantfocus} below for quantum focusing uses only the nesting \(\widehat{\mathcal{M}}_{u+a}\subset \widehat{\mathcal{M}}_{u}\) under half-sided modular inclusions, together with a variational formula for the quantum expansion \(\bm{\Theta}(u)\) as an infimum over commutant unitaries/purifications \(V\in U(\widehat{\mathcal{M}}'_{u})\), adapted from the analogous result for relative entropy in flat space QFT \cite{Ceyhan:2018zfg}; at no point does the argument invoke the existence of a unique translation invariant vacuum vector.\label{footnote:qfcuniquevac}}

	\subsection{Quantum focusing \label{sec:quantfocus}}
	
	In the present paper we have restricted to a one-parameter family of horizon cuts labeled by a
	single affine parameter \(u\). In that setting,
	the (averaged over corner fluctuations) generalized entropy \(\Sgen(u;\psihat)\) defined in \cref{sec:gravmodinc}
	is a function of one variable, and the ``quantum focusing conjecture'' (QFC) reduces to the
	concavity statement
	\begin{equation}
		\partial_u^2 \Sgen(u;\psihat)\le 0.
		\label{eq:QFC_concavity}
	\end{equation}
	Equivalently, defining the (one-dimensional) quantum expansion
	\begin{equation}
		\bm{\Theta}(u) := \partial_u \Sgen(u;\psihat),
		\label{eq:Theta_def}
	\end{equation}
	the QFC is the monotonicity/focusing statement
	\begin{equation}
		\partial_u \bm{\Theta}(u)\le 0 \ \Leftrightarrow \ \bm{\Theta}(u+a)\le \bm{\Theta}(u) \ \ \text{for all }a\ge 0.
		\label{eq:Theta_monotone}
	\end{equation}
	Our goal in this section is to prove \Eqref{eq:Theta_monotone} (hence \Eqref{eq:QFC_concavity})
	via the same approach adopted by Ceyhan--Faulkner in \cite{Ceyhan:2018zfg} to prove the QNEC. Concretely, we will (i) rewrite \(\Sgen\) as (minus) a relative
	entropy on the one-parameter family of Type \(\mathrm{II}_\infty\) horizon algebras, (ii) invoke gravitational half-sided modular
	inclusion (HSMI) to obtain an expression for \(\bm{\Theta}\) as an infimum
	over purifications, and then (iii) use the nesting of commutants along the HSMI to show
	that this infimum is non-increasing in \(u\).
    
    Just as in \cref{sec:gravmodinc}, we restrict to a Killing horizon background (see \cref{footnote:gslnontrivial} as to why this is the non-trivial setting for entropy inequalities in semi-classical gravity). Recall that the family \(\{\widehat{\mathcal{M}}_u\}\) forms a half-sided modular inclusion.\footnote{In order to avoid notational clutter in this section, we use the short-hand $\mathcal{A}_u \equiv \mathcal{A}_{\mathscr{H}_{>u}}$ for any given algebra.} Concretely, there exists a strongly continuous one-parameter unitary group
	\begin{align}
		U(a)=e^{i a \Pnull},\  a\in \mathbb{R},
	\end{align}
	implementing the null translation automorphisms such that for all \(a\ge 0\),
	\begin{equation}
		U(a)\,\M_{u_0}\,U(-a) \subset \M_{u_0}.
		\label{eq:HSMI}
	\end{equation}
	We will also use the induced commutant nesting:
	\begin{equation}
		\M_{u_0}' \subset U(a)\,\M_{u_0}'\,U(-a)\equiv \M_{u_0+a}', \ a\ge 0,
		\label{eq:commutant_nesting}
	\end{equation}
	i.e.\ the commutants grow as we move the cut forward. This point deserves further elucidation. It is naively confusing that $\hat{\mathscr{P}}$ can act non-trivially on $\widehat{\mathcal{M}}_{u_0}'$ given that $\hat{\mathscr{P}} \in \widehat{\mathcal A}_{u_0}$, i.e. $[\hat{\mathscr{P}}, \mathscr{O}'] = 0, \ \forall \mathscr{O}' \in \widehat{\mathcal A}_{u_0}'$. But recall the algebra $\M_{u_0}$ was defined in terms of the conditional expectation \eqref{eq:conditionalexpect}
    \begin{equation}
    E_{u_0}\colon \widehat{\mathcal A}_{u_0} \mapsto \widehat{\mathcal M}_{u_0},\
    \hat{\mathscr O}_{u_0} = E_{u_0}(\hat{\mathscr O}) \in \mathcal \M_{u_0}.
    \end{equation}
    The one-parameter family of conditional expectations $\{E_u\}$ are covariant with respect to null translations, i.e.
    \begin{equation}\label{eq:covariant-expectation}
    \text{Ad}\ {U(a)}\circ E_{u_0}=E_{u_0+a}\circ \text{Ad}\ {U(a)}.
    \end{equation}
    Now take $\hat{\mathscr O}'\in \widehat{\mathcal A}'_{u_0}$. Since it commutes with $\hat{\mathscr{P}}$, we have
    \begin{equation}
    [\hat{\mathscr{P}},\hat{\mathscr O}']=0
    \Leftrightarrow
    \text{Ad} \ {U(a)}(\hat{\mathscr O}')=\hat{\mathscr O}',\ \forall a\,.
    \end{equation}
    Let $\hat{\mathscr O}'_{u_0}=E_{u_0}(\hat{\mathscr O}')$. Using \Eqref{eq:covariant-expectation} and $\text{Ad} \ U(a)(\hat{\mathscr O}')=\hat{\mathscr O}'$,
    we obtain the relation
    \begin{equation}\label{eq:sliding}
    \text{Ad} \ U(a)\big(\mathscr O'_{u_0}\big) =
    \text{Ad} \ U(a)\big(E_{u_0}(\hat{\mathscr O}')\big) =
    E_{u_0+a}\big(\text{Ad} \ U(a)(\hat{\mathscr O}')\big) =
    \hat{\mathscr O}'_{u_0+a}\,.
    \end{equation}
    Differentiating \Eqref{eq:sliding} at $a=0$ and using
    $\partial_a\text{Ad} \ U(a)(\hat{\mathscr{O}})|_{a=0}=i[\hat{\mathscr{P}},\hat{\mathscr{O}}]$ yields
    \begin{equation}\label{eq:shape-derivative}
    [\hat{ \mathscr{P}},\hat{\mathscr O}'_{u_0}]=
    -i\partial_{u_0}\hat{\mathscr{O}}'_{u_0}.
    \end{equation}
    Thus even when $[\hat{\mathscr{P}},\hat{\mathscr O}']=0$ in the full crossed product algebra,
    the conditional expectation $\hat{\mathscr O}'_{u_0}=E_{u_0}(\hat{\mathscr O}')$ of this operator upon collapsing onto a classical cut location need not commute with $\hat{\mathscr P}$. Instead, the commutator measures the $u_0$ dependence (i.e.\ shape derivative) induced by conditioning on the cut.

	Having clarified that crucial point, we can now proceed with the proof. Let \(\rho_{\HH}(u)\in \M_u\) be the density matrix associated to the uplift $|\hat{\Omega}\rangle$ of the Hartle-Hawking state to the extended Hilbert space. The Araki relative entropy is \cite{Witten:2018zxz}
	\begin{equation}
		\Srel(u):=-\langle \hat{\psi}|\log \Delta_{\hat{\psi}|\hat{\Omega};\M_u}|\hat{\psi}\rangle.
		\label{eq:Srel_u_def}
	\end{equation}
	It is easy to show, using the results of \cref{sec:gravmodinc}, that
	\begin{equation}
		\Sgen(u;\psihat) \approx - \Srel(u)+\text{const}.
		\label{eq:Sgen_minus_Srel}
	\end{equation}
	Consequently,
	\begin{equation}
		\bm{\Theta}(u)=\partial_u \Sgen(u;\psihat) \approx -\partial_u \Srel(u).
		\label{eq:Theta_vs_Srelprime}
	\end{equation}
	Thus, in the present setting, proving quantum focusing \(\partial_u\bm{\Theta}\le 0\) is equivalent to proving
	\begin{equation}
		\partial_u^2 \Srel(u)\ \ge\ 0,
		\label{eq:Srel_convex}
	\end{equation}
	i.e.\ convexity of the relative entropy along the half-sided inclusion parameter \(u\).
	This is the exact same structural reduction as in \cite{Ceyhan:2018zfg}: there the QNEC is recast
	as a convexity property of \(\Srel\) as a function of the null cut deformation.
	
	We now import the method of proof used by Ceyhan--Faulkner. The essential input is the HSMI structure, which gives a canonical way to compare the algebras at nearby cuts and to produce a distinguished family of ``optimal purifications'' generated by relative modular
	flow (Connes cocycle flow to be precise).
	
	Fix \(u\) and the reduced state on \(\M_u\), i.e.\ \(\rho_{\psihat}(u)\).
	A general purification that leaves \(\rho_{\psihat}(u)\) invariant is obtained by acting on \(|\psihat\rangle\)
	with a unitary in the commutant \(\M_u'\):
	\begin{equation}
		|\hat{\psi}_V\rangle = V\,|\psihat\rangle,\ V\in U(\M_u'),
		\label{eq:purification_V}
	\end{equation}
    where $U(\M_u')$ is the subalgebra of $\M_u'$ consisting of unitary operators. Indeed, for any \(\mathscr{O}\in \M_u\),
	\(\langle \hat{\psi}_V|\mathscr{O}|\hat{\psi}_V\rangle=\langle\psihat|V^\dagger \mathscr{O} V|\psihat\rangle=\langle\psihat|\mathscr{O}|\psihat\rangle\).
	
	In order to simplify notation, let \(\D_{\psihat|\HH;u}\) denote the relative modular operator of \((|\psihat\rangle,|\HH \rangle)\) for \(\M_u\),
	and \(\D_{\HH;u}\) the modular operator of \(|\HH \rangle\) for \(\M_u\).
	Recall from the previous section the definition of the Connes cocycle flow:
	\begin{equation}
		\cc_{\psihat|\HH;u}(s):=\D_{\psihat|\HH;u}^{\,is}\,\D_{\HH;u}^{-is},\ s\in\mathbb{R}.
		\label{eq:Connes_cocycle}
	\end{equation}
	This picks out the associated (distinguished) commutant unitaries
	\begin{equation}
		V_s := J_{\HH;u}\,\cc_{\psihat|\HH;u}(s)\,J_{\HH;u}\ \in\ U(\M_u'),
		\label{eq:Vs_from_J}
	\end{equation}
	where \(J_{\HH;u}\) is the modular conjugation for \((\M_u,|\HH \rangle)\).
	The corresponding purified states are
	\begin{equation}
		|\psihat_s\rangle:=V_s\,|\psihat\rangle.
		\label{eq:Psi_s_def}
	\end{equation}
	This is the exact analogue of the one-parameter family \(|\psi_s\rangle\), constructed from
	the Connes cocycle as in Ceyhan--Faulkner, that saturates the relevant infimum.
	
	The adaptation of the Ceyhan--Faulkner theorem to our present one-parameter situation is:\footnote{The proof of the theorem in \cite{Ceyhan:2018zfg} is completely general, as it works abstractly at the level of von Neumann algebras and half-sided modular inclusions. So the theorem applies verbatim to our construction.}
	
	\begin{theorem-non}[Variational formula for the null shape derivative]
		Assume \(\{\M_u\}\) forms a half-sided modular inclusion w.r.t.\ \(|\HH \rangle\), and let \(\Srel(u)\) be
		\Eqref{eq:Srel_u_def}. Then, for almost every \(u\),
		\begin{equation}
			-\frac{1}{2\pi}\,\partial_u \Srel(u)
			=\inf_{V\in U(\M_u')}\ \langle\psihat|V^\dagger\,\hat{\mathscr{P}}_u\,V|\psihat\rangle
			=\inf_{s\in\mathbb{R}}\ \langle\psihat_s|\hat{\mathscr{P}}_u|\psihat_s\rangle,
			\label{eq:variational_formula}
		\end{equation}
		where, as discussed in \cref{sec:hsm_inclusion_bridge}, the half-sided translation generator \(\hat{\mathscr{P}}\) is the same as the positive semi-definite operator $G$ implementing half-sided modular inclusions in the theorem of Borchers/Wiesbrock. The infimum is achieved (or approximated arbitrarily well) by the cocycle family \(|\psihat_s\rangle\).
	\end{theorem-non}
	We note that \Eqref{eq:variational_formula} is the direct analogue of Eq.\ (1.1) in
	Ceyhan--Faulkner, with \(\partial_u \Srel\) playing the role of the null shape derivative and \(\hat{\mathscr{P}}\) playing the role of the averaged null energy operator. The second equality (inf over \(s\)) is the statement that Connes-cocycle flow gives the minimizing purification family.
	
	Given \Eqref{eq:Theta_vs_Srelprime}, the variational formula immediately rewrites the quantum expansion
	\(\bm{\Theta}(u)\) as
	\begin{equation}
		\bm{\Theta}(u) \approx 2\pi \inf_{V\in U(\M_u')} \langle\psihat|V^\dagger\,\hat{\mathscr{P}}_u\,V|\psihat\rangle.
		\label{eq:Theta_as_inf}
	\end{equation}
	
	We now show that \(\bm{\Theta}(u)\) is nonincreasing in \(u\), i.e.\ \(\partial_u\bm{\Theta}\le 0\).
	The salient observation is that as we move the cut forward, the commutant grows
	(\Eqref{eq:commutant_nesting}), hence the minimization domain in
	\Eqref{eq:Theta_as_inf} becomes larger, which can only decrease the infimum.
	
	Fix \(a\ge 0\). By \Eqref{eq:HSMI}, \(\M_{u+a}\subset \M_u\), hence \(\M_{u+a}'\supset \M_u'\).
	Therefore,
	\begin{align}
		\inf_{V\in U(\M_{u+a}')} \langle\psihat|V^\dagger\,\hat{\mathscr{P}}_{u+a}\,V|\psihat\rangle
		&\le
		\inf_{V\in U(\M_{u}')} \langle\psihat|V^\dagger\,\hat{\mathscr{P}}_{u+a}\,V|\psihat\rangle.
		\label{eq:inf_domain_monotone}
	\end{align}
	At this stage we use the HSMI property to identify \(\hat{\mathscr{P}}_{u+a}\) with the translated operator:
	\begin{equation}
		\hat{\mathscr{P}}_{u+a} = U(a)\,\hat{\mathscr{P}}_u\,U(-a),
		\label{eq:E_covariance}
	\end{equation}
	Then for any \(V\in U(\M_u')\),
	\begin{align}
		\langle\psihat|V^\dagger\,\hat{\mathscr{P}}_{u+a}\,V|\psihat\rangle
		&=
		\langle\psihat|V^\dagger\,U(a)\hat{\mathscr{P}}_u U(-a)\,V|\psihat\rangle
		=
		\langle\psihat| \widetilde V^\dagger\,\hat{\mathscr{P}}_u\,\widetilde V|\psihat\rangle,
	\end{align}
	where \(\widetilde V:=U(-a)VU(a)\in U(\M_{u+a}')\) (since conjugation by \(U(a)\) maps \(\M_{u+a}'\) to \(\M_u'\)
	and vice versa). Thus the right-hand infimum in \Eqref{eq:inf_domain_monotone} can be re-expressed as
    \begin{align}
    \inf_{V\in U(\M_{u}')} \langle\psihat|V^\dagger\,\hat{\mathscr{P}}_{u+a}\,V|\psihat\rangle = \inf_{\widetilde{V}\in \text{Ad}(U(a))(U(\M_{u}'))} \langle\psihat|\widetilde{V}^\dagger\,\hat{\mathscr{P}}_{u}\,\widetilde{V}|\psihat\rangle = \inf_{V\in U(\M_{u}')} \langle\psihat|V^\dagger\,\hat{\mathscr{P}}_{u}\,V|\psihat\rangle,
    \end{align}
    where we've used that the infimum is preserved under unitary automorphisms. From this we conclude
	\begin{equation}
		\inf_{V\in U(\M_{u+a}')} \langle\psihat|V^\dagger\,\hat{\mathscr{P}}_{u+a}\,V|\psihat\rangle
		\le
		\inf_{V\in U(\M_{u}')} \langle\psihat|V^\dagger\,\hat{\mathscr{P}}_{u}\,V|\psihat\rangle.
		\label{eq:inf_decreases}
	\end{equation}
	Multiplying by \(2\pi\) and using \Eqref{eq:Theta_as_inf} at \(u\) and \(u+a\) gives
	\begin{equation}
		\bm{\Theta}(u+a)\le \bm{\Theta}(u),\ \forall\,a\ge 0.
		\label{eq:Theta_focusing}
	\end{equation}
	Taking \(a\to 0^+\) (assuming differentiability, which is the same regularity assumption already implicit
	in writing \(\partial_u \Sgen\) and \(\partial_u^2 \Sgen\) in \cref{sec:gravmodinc}) yields
	\begin{equation}
		\partial_u \bm{\Theta}(u)\le 0
		\ \Leftrightarrow \
		\partial_u^2 \Sgen(u;\psihat)\le 0.
	\end{equation}
	This constitutes a proof of the quantum focusing conjecture in perturbative quantum gravity.
	
	\subsection{Causal diamonds\label{sec:causaldiamonds}}
	
	Thus far we’ve focused entirely on horizon subalgebras. What if we consider instead sub-
	algebras associated to generic codimension-two subregions of spacetime? More specifically,
	let's consider a finite causal diamond in spacetime. In this final section we demonstrate that the same machinery can be adapted to finite causal diamonds. In that context, the relevant null surfaces are the lightsheets of the diamond boundary.
	
	We consider the double cone defined by a pair of timelike related points, use a relational prescription to dress the causal diamond, and repeat the construction comprising the previous sections. The resulting family of subalgebras along the contracting lightsheet again forms a one-parameter family of Type II$_{\infty}$ factors, and the associated von Neumann entropy of a cut of the lightsheet can be similarly identified with the (averaged) generalized entropy of that cut. 
	
	More precisely, given a spacetime \((M,g_{ab})\), consider two points $p^+, p^- \in M$ such that $p^+$ is in a convex normal neighborhood of $p^-$ and is in its chronological future, i.e., \(p^+\) is inside the future light cone of \(p^-\). The intersection of the chronological past of $p^+$ with the chronological future of $p^-$ defines a \emph{causal diamond} or a \emph{double cone}: 
	\begin{align}
		\mathscr{D}(p^-,p^+) := I^+(p^-) \cap I^-(p^+).
	\end{align}
	
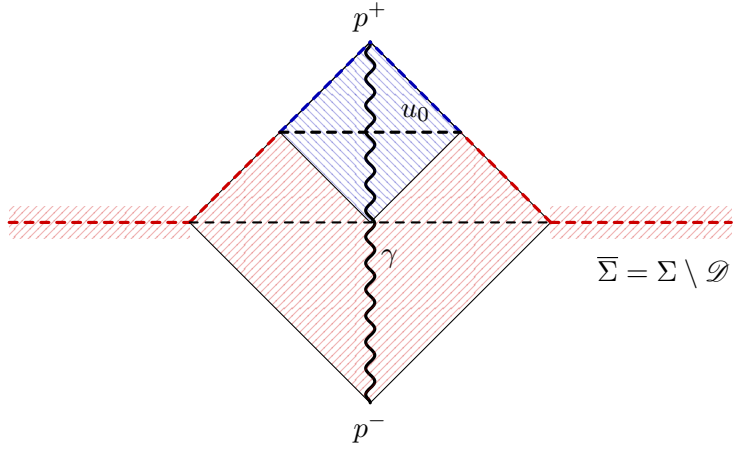
\begin{figure}[t]
		\centering
		\begin{tikzpicture}[scale=1.2, line cap=round, line join=round, font=\small]
			\colorlet{Hred}{red!80!black}
			\colorlet{Hblue}{blue!70!black}
			
			\tikzset{
				bdry/.style={black},
				redhatch/.style={pattern=north east lines, pattern color=Hred},
				bluehatch/.style={pattern=north west lines, pattern color=Hblue},
				futbdry/.style={Hblue, dashed, very thick},
				pastbdry/.style={Hred, dashed, very thick},
				cut/.style={Hblue, dashed, very thick},
				slice/.style={black, dashed, thick},
				worldline/.style={black, very thick},
			}
			
			\coordinate (pm) at (0,-2);   
			\coordinate (pp) at (0, 2);   
			\coordinate (L)  at (-2,0);   
			\coordinate (R)  at ( 2,0);   
			
			\coordinate (p0) at (0,0);    
			
			\draw[worldline, decorate, decoration={snake, amplitude=0.6mm, segment length=3.5mm}]
			(pm) -- (pp);
			\node[below] at (pm) {$p^-$};
			\node[above] at (pp) {$p^+$};
			\node[right] at ($(pm)!0.4!(pp)$) {$\gamma$};
			
			\coordinate (sL) at (-1,1);   
			\coordinate (sR) at ( 1,1);   
			
			\begin{scope}[opacity=0.55]
				\fill[redhatch, even odd rule]
				(pm) -- (L) -- (pp) -- (R) -- cycle
				(p0) -- (sL) -- (pp) -- (sR) -- cycle;
				
				\fill[bluehatch]
				(p0) -- (sL) -- (pp) -- (sR) -- cycle;
				
				\fill[redhatch] (-4, 0.18) -- (-2, 0.18) -- (-2,-0.18) -- (-4,-0.18) -- cycle;
				\fill[redhatch] ( 2, 0.18) -- ( 4, 0.18) -- ( 4,-0.18) -- ( 2,-0.18) -- cycle;
			\end{scope}
			
			\draw[bdry] (pm) -- (L) -- (pp) -- (R) -- cycle;
			\draw[bdry] (p0) -- (sL) -- (pp) -- (sR) -- cycle;
			
			\draw[pastbdry] (L) -- (sL);
			\draw[pastbdry] (R) -- (sR);
			\draw[futbdry]  (sL) -- (pp);
			\draw[futbdry]  (sR) -- (pp);
			
			\draw[black, dashed, very thick] (sL) -- (sR);
			\node[above] at ($(sL)!0.75!(sR)$) {$u_0$};
			
			\draw[pastbdry] (-4,0) -- (L);
			\draw[slice]    (L) -- (R);
			\draw[pastbdry] (R) -- (4,0);
			\node[below] at (3.25,-0.25) {$\overline{\Sigma} = \Sigma\setminus\mathscr{D}$};
			
			\node[Hblue] at (0,1.55) {};
			\node[Hred]  at (0,-1.05) {};
		\end{tikzpicture}
		\caption{Penrose diagram of the causal diamond $\mathscr{D}(p^-,p^+)=I^+(p^-)\cap I^-(p^+)$ with tips $p^\pm$. The null boundary $\mathscr{N}=\partial \mathscr{D}$ (solid) is generated by null rays from $p^\pm$ and meets at the bifurcation surface. The wiggly curve is a timelike geodesic $\gamma(p^-,p^+,g)$ connecting the tips; the dashed line $u=u_0$ denotes a cut that splits the diamond into ``above'' and ``below'' regions as defined by light signals emitted from $\gamma$ (blue/red shading). A Cauchy slice $\Sigma$ intersects the diamond, with its exterior complement $\bar\Sigma=\Sigma\setminus \mathscr{D}$ indicated by the horizontal strip.}
		
		\label{fig:causal-diamond}
	\end{figure}

	The fact that $p^{\pm}$ lie in a convex normal neighborhood means $\mathscr{D}(p^-, p^+)$ has to be ``sufficiently small'' so that conjugate points don't form. Then, the null generators emanating from \(p^\pm\) form smooth null surfaces \(\mathscr{N}^\pm\) respectively, which intersect at a smooth \(2\)-surface \(\mathscr{B}\), the bifurcation surface, which is topologically $\mathbb{S}^{d-2}$. Moreover, we can always find a timelike geodesic $\gamma(p^{-},p^+, g)$ connecting $p^-$ to $p^+$. See \cref{fig:causal-diamond}.
	
	We denote the null boundary of the causal diamond by 
	\begin{align}
		\mathscr{N} = \partial \mathscr{D}(p^-, p^+) = \mathscr{N}^+ \cup \mathscr{N}^-.
	\end{align}
	We adopt the same boundary conditions on field configuration space: $\delta \ell^a_{\pm} = 0, \delta \kappa_{\pm} = 0$. As a next step, in order to write down the boundary phase space $\mathcal{P}_{\mathscr{N}}$ of the causal diamond we need to know what the fall-off conditions on the metric are as we approach $p^{\pm}$. We can do this via a blowup construction that maps from $p^{\pm}$ to the projectivization of its normal bundle, effectively ``zooming in'' on the singularity to make it a smooth codimension-two manifold \cite{Choquet-Bruhat:2010vby, Choquet-Bruhat:2010bed, Chandrasekaran:2019ewn}. 
	
	We describe this construction for $p^+$ but it works identically for $p^-$. Consider the tangent space $Tp^+$ and introduce local coordinates $\left\{y^i\right\}$ on $Tp^+$ such that $y^i(p^+) = 0$ and $\left\{\partial_i\right\}$ is an orthonormal basis. The vector field $\partial_0$ is a future-directed timelike vector field. Then, the past lightcone in $Tp^+$, which describes $\mathscr{N}^+$ in the local neighborhood of $p^+$, is described by coordinates $(\tilde{u} = 0, r,\tilde{x}^A)$ where
	\begin{align}
		r^2 = (y^1)^2 + (y^2)^2 + (y^3)^2, \ \tilde{u} = y^0 - r,
	\end{align}
	and $\tilde{x}^A$ are coordinates on the space of past-directed null directions at $p^+$ isomorphic to the 2-sphere. 
	
	Indeed, there exists an exponential map from $Tp^+$ to a local neighborhood of $p^+$ which extends the coordinates $\tilde{u}, r, \tilde{x}^A)$ to this neighborhood. In these coordinates, the metric takes the form \cite{Choquet-Bruhat:2010vby, Choquet-Bruhat:2010bed, Chandrasekaran:2019ewn}
	\begin{align}
		ds^2 = &( 1 + \mathcal{O}(r^2)) d{\tilde{u}}^2 - 2 (1 + \mathcal{O}(r^4)) d\tilde{u} dr - 2 \left(\mathcal{O}(r^3)\right)_A d\tilde{u} dx^A \nonumber \\ &+ r^2(q_{AB}^0 + \mathcal{O}(r^2)) dx^A dx^B \label{eq:diamondtipmetric},
	\end{align}
	where $q_{AB}^0$ is the standard 2-sphere metric. As is evident, the fall-off conditions are such that the metric near $p^+$ behaves as the Minkowski metric at the tip of a light cone. It is easy to show that 
	\begin{align}
		\Theta = -\frac{2}{r} + \mathcal{O}(r^3), \ \sigma_{AB} = \mathcal{O}(r^3).
	\end{align}
	Note that as a consequence, 
	\begin{align}
		\lim_{r\rightarrow 0}\delta \Theta  = \lim_{r\rightarrow 0}\delta \sigma_{AB} = 0.\label{eq:perturbedmetricfalloff}
	\end{align}
	
	We can directly lift the symmetry vector fields $\xi^a$ as well as the flux term $\bm{\mathcal{E}}$ and charge expression $\bm{Q}_{\xi}-i_{\xi}\bm{\alpha}$ from \cref{sec:grsymp} and \cref{sec:gredgecorner}; they are valid for any null surface with the boundary conditions $\delta \ell^a = \delta \kappa = 0$ on field configuration space. Then, given a cut $S(r)$ of $\mathscr{N}^+$ in the neighborhood of $p^+$, a simple calculation yields
	\begin{align}
		\lim_{r\rightarrow 0}\int_{S(r)}i_{\xi}\bm{\mathcal{E}} = \lim_{r\rightarrow 0}\left(\bm{Q}_{\xi}[S(r)] - i_{\xi}\bm{\alpha}[S(r)]\right) = 0.\label{eq:chargefluxfalloff}
	\end{align}
	The same goes for $p^-$.
	
	One important point is that unlike the event horizon $\mathscr{H}$, $\mathscr{D}(p^-, p^+)$ is not gauge invariantly specified as defined. In order to make it gauge invariant, we can define $p^{\pm}$ relationally. In particular, given any choice of $p^{\pm}$ we can dress them relationally to one another by fixing 
	\begin{align}
		\delta \left(\int_{\tau^-}^{\tau^+} d\tau \sqrt{-g_{ab}T^a T^b}\right) = 0,\label{eq:dressingdiamond}
	\end{align}
	where $T^a = (d/d\tau)^a$ is the tangent vector along the timelike geodesic connecting the two points, and $\tau^{\pm} = \tau^{\pm}(p^{\pm})$. So this is the statement that we keep the proper time between the two points fixed in phase space.
	
	Since $T^b \nabla_b T^a = 0$, this yields the simple condition
	\begin{align}
		\delta (\tau^+ - \tau^-) = \frac{1}{2}\int_{\tau^-}^{\tau^+} d\tau \ h_{\tau\tau}(\tau),\label{eq:dressingconditiondiamond}
	\end{align}
	where we've chosen to parameterize $\tau$ such that $T^a T_a = -1$ in the background spacetime.
	
	By an argument similar to that of \cref{app:horizondeformation}, calculations reduce to ones wherein we just gauge fix $p^{\pm}$ when defining the phase space, i.e. $\delta p^{\pm} = 0$.
	
	Now, consider a cut $S_0$ of $\mathscr{N}^+$. Fix the tip $p^+$ at $u = 0$ in affine parameterization on $\mathscr{N}^+$. Given \Eqref{eq:perturbedmetricfalloff} and \Eqref{eq:chargefluxfalloff}, as well as the discussion immediately above, everything in \crefrange{sec:integrablegeneratorderiv}{sec:genentropyderiv} goes through exactly as before up to a few important modifications that we now discuss.
	
	Instead of a background Killing horizon we now have a ball-shaped causal diamond in a maximally symmetric spacetime. The causal diamond has a conformal Killing vector
	\begin{align}
		\zeta \hateq \frac{u}{u_0}(u-u_0)\partial_u,
	\end{align}
	regardless of whether the maximally symmetric spacetime is empty AdS, empty dS, or Minkowski spacetime. A simple calculation shows that
	\begin{align}
		\lie_{\zeta}g_{ij}(u) \hateq \Theta(u)f(u)g_{ij}, \ f(u) = \frac{u}{u_0}(u-u_0),\label{eq:ckvcondition}
        \end{align}
        hence why it's a conformal isometry. 
	
	For a ball-shaped region, $q(u) = q_0 u^4$. Adapting \Eqref{eq:genbackarea}, we can then compute the area operator
	\begin{align}
		\hat{\mathscr{A}}_g= -\frac{1}{8\pi G_N}\int_{0}^{u^+_0}du\int_{S^+_0}d^{d-2}x \  f(u)\hat{S}(u, x),\label{eq:causaldiamondareaop}
	\end{align}
	which is exactly the half-sided boost generator from \Eqref{eq:cornertermintegrable}.\footnote{It is also the vacuum modular Hamiltonian of a CFT on the causal diamond of a ball-shaped region \cite{Faulkner:2015csl}.} But now it is a conformal boost. And with that, everything in \crefrange{sec:integrablegeneratorderiv}{sec:edgemodes} goes through as before. 
	
	The main difference is in how we approach the analysis of \cref{sec:genentropyderiv}. The primary issue is that \Eqref{eq:ckvcondition} is a conformal isometry, not a true isometry. Even if the matter fields were CFTs, linearized gravity is not a CFT. So the global vacuum of the maximally symmetric spacetime will not satisfy KMS when reduced to $\mathscr{N}_{<u_0}$. Fundamentally this is because lightcones have non-stationary geometry. So instead we have to follow the approach in \cref{sec:genbackentropy}, by considering causal diamonds for which the proper time is much smaller than both $R_0$ and $\lambda_0$. Recall $R_0$ and $\lambda_0$ are the typical radius of curvature of background spacetime and length scale of background excitations, respectively.
	
	To that aim, we may as well just consider a causal diamond in a general non-stationary background spacetime. Then the calculations in \cref{sec:genbackentropy} go through in exactly the same way.\footnote{The crossed product algebra for a generic causal diamond is naturally Type II$_{\infty}$ with a semifinite trace that diverges on the identity. In special situations with a preferred observer and Hamiltonian (e.g.\ the de Sitter static patch), one can instead pick a finite-trace corner and obtain a Type II$_1$ factor, as in \cite{Chandrasekaran:2022cip}. We will not assume such extra structure here.} In particular, the result
	\begin{align}
		&S(\hat{\psi}_g ;{\widehat{\mathcal{M}}_{\mathscr{H}_{>\tilde{u}}}}) \approx \bar{S}_{\text{gen}}(\tilde{u}; \hat{\psi}_g) + \text{const.} + \mathcal{O}(\tilde{u}^2)\label{eq:localsgendiamond}
	\end{align}
    continues to hold for an arbitrary cut of a generic (smooth) causal diamond.

    \subsection{Cauchy slice holography at the corner}
	\label{sec:corner-cauchy-slice-holography}
	
	In this final section, we sketch out a construction of ``Cauchy slice holography'' that follows from the results of \crefrange{sec:corneredgemodes}{sec:edgemodes}, wherein the bulk exterior algebra can be reconstructed entirely from the algebra of observables at spatial infinity combined with the corner algebra associated with a given subregion of the event horizon. The construction we outline below is essentially a version of the arguments in \cite{Marolf:2008mf, Laddha:2020kvp}, but which applies not only to black hole spacetimes but also to (semi-infinite) portions of the event horizon.\footnote{We adopt the phrase ``Cauchy slice holography'' from \cite{Araujo-Regado:2022gvw}, which develops a very general notion of the concept.}
	
	Let $\mathcal{U}$ denote the exterior region of interest. We know that the horizon subregion $\mathscr{H}_{>u_0}$ together with future null infinity $\mathscr{I}^+$ is a characteristic Cauchy
	surface for $\mathcal{U}$, so that
	\begin{equation}
		\mathcal{U} = D^+\!\left(\mathscr{H}_{>u_0}\cup\mathscr{I}^+\right), \ \Sigma^+ = \mathscr{H}_{>u_0}\cup\mathscr{I}^+ .
	\end{equation}
	We also assume fall-off conditions at $i^+$ such that no $i^+$ hyperboloid
	terms contribute to the symplectic form.
	
	Let $\Psi^I$ denote the collection of linearized fields in the exterior effective theory
	(matter and/or graviton in a fixed gauge), with linearized equations \begin{align}E_{IJ}\Psi^J=0.\end{align}
	Let $\omega(\delta_1\Psi,\delta_2\Psi)$ be the corresponding covariant symplectic current,
	and write the symplectic form on $\Sigma^+$ as
	\begin{equation}
		\Omega_{\Sigma^+}(\delta_1\Psi,\delta_2\Psi) =
		\int_{\mathscr{H}_{>u_0}}\omega(\delta_1\Psi,\delta_2\Psi) +
		\int_{\mathscr{I}^+}\omega(\delta_1\Psi,\delta_2\Psi).
	\end{equation}
	For any bulk point $p\in \mathcal{U}$, let $G_p$ denote a (distributional) solution of the
	adjoint linearized equations with a delta function source at $p$.

    For any two (possibly distributional) configurations $\Phi^I$ and $\Psi^I$ one has (using the Lagrange identity)
    \begin{equation}\label{eq:GreenIdentitySympl}
    d\omega(\Phi,\Psi)
    = \Phi^I E_{IJ}\Psi^J \,\epsilon -
    \Psi^I E^\dagger_{IJ}\Phi^J \,\epsilon,
    \end{equation}
    where $E^\dagger$ is the formal adjoint (with respect to the spacetime volume form $\epsilon$).
    In particular, if $\Phi$ is a homogeneous solution $E\Phi=0$ and $\Psi$ solves the adjoint
    equations with a delta function source at $p$ for some fixed component $I_{\star}$
    \begin{equation}\label{eq:AdjointGreen}
    E^\dagger_{IJ}\Psi^J = \delta_{I}{}{}^{I_{\star}}\,\frac{\delta^{(d)}(\cdot,p)}{\sqrt{-g}},
    \end{equation}
    then \Eqref{eq:GreenIdentitySympl} reduces distributionally to
    \begin{equation}\label{eq:domegaDelta}
    d\omega(\Phi,\Psi) = \Phi^{I_\star}\delta^{(d)}(\cdot,p)\,d^dx.
    \end{equation}
    
    Now let $\mathcal V\subset \mathcal U$ be any region whose boundary consists of two characteristic
    surfaces $\Sigma_-$ and $\Sigma_+$ for $\mathcal U$, with $p\in \mathcal V$, and with orientations such that
    $\partial\mathcal V=\Sigma_+\cup (-\Sigma_-)$. Integrating \Eqref{eq:domegaDelta} over $\mathcal V$ and using Stokes' theorem gives
    \begin{align}
    \int_{\Sigma_+}\omega(\Phi,\Psi)-\int_{\Sigma_-}\omega(\Phi,\Psi)
    &= \int_{\mathcal V} d\omega(\Phi,\Psi)
    = \int_{\mathcal V}\Phi^{I_\star}\,\delta^{(d)}(\cdot,p)\,d^dx
    = \Phi^{I_\star}(p).
    \end{align}
    If we choose $\Psi$ to be a retarded adjoint Green's function sourced at $p$, then for a past
    surface $\Sigma_-$ lying entirely to the past of $p$ we have $\Psi|_{\Sigma_-}=0$, and hence
    $\int_{\Sigma_-}\omega(\Phi,\Psi)=0$. 
    
    Taking $\Sigma_+=H_{>u_0}\cup\mathcal I^+$, we can therefore write down the following characteristic reconstruction formula:
	\begin{equation}
		\Phi^I(p) =
		\Omega_{\Sigma^+}(\Phi,\,G_p^{I})
		=
		\int_{\mathscr{H}_{>u_0}}\omega(\Phi,\,G_p^{I})
		+
		\int_{\mathscr{I}^+}\omega(\Phi,\,G_p^{I}),
		\label{eq:null-cauchy-reconstruction}
	\end{equation}
    Because $\Sigma_+$ is a characteristic Cauchy surface for $\cal U$, the characteristic null
    initial value problem asserts that specifying a complete set of characteristic null initial data on
    $\mathscr H_{>u_0}$ and on $\mathscr I^+$ determines a unique bulk solution in $\cal U$. Denote such data by $\varphi^I_{\mathscr H_{>u_0}}(u,x^A), \ \varphi^I_{\mathscr I^+}(u,x^A)$ (for instance: suitable radiative components plus the data fixed by constraints). The map
    \begin{equation}
    (\varphi_{\mathscr H_{>u_0}},\varphi_{\mathscr I^+})\mapsto \Phi^I(p)
    \end{equation}
    is linear in the data for a linear field theory. Therefore, there exist kernels $\mathscr K^I_{H_{>u_0}}$
    and $\mathscr K^I_{\mathscr I^+}$ such that
    \begin{align}
    \Phi^I(p)
    = \ & \int_{\mathscr H_{>u_0}} du\,d^{d-2}x\;\mathscr K^I_{\mathscr H_{>u_0}}(p|u,x^A)\,\varphi_{\mathscr H_{>u_0}}(u,x^A)
    \nonumber\\
    &\quad+\int_{\mathscr I^+} du\,d^{d-2}x\;\mathscr K^I_{\mathscr I^+}(p|u,x^A)\,\varphi_{\mathscr I^+}(u,x^A),\label{eq:hkll-null}
    \end{align}
	
	Upon quantization, \crefrange{eq:null-cauchy-reconstruction}{eq:hkll-null} become operator
	identities in the exterior effective theory, showing that every dressed bulk operator
	$\widehat{\mathscr{O}}(p)$ localized at $p\in\mathcal{U}$ can be reconstructed from the von Neumann
	algebra generated by the boundary algebras on $\mathscr{H}_{>u_0}$ and $\mathscr{I}^+$.
	Concretely, writing the von Neumann join as
	\begin{equation}
		\mathcal{A}\,\vee\,\mathcal{B}
		=
		(\mathcal{A}\cup \mathcal{B})'',
	\end{equation}
	we obtain a form of null Cauchy slice reconstruction:
	\begin{equation}
		\mathcal{A}_{\rm ext}
		=
		\mathcal{A}_{\mathscr{H}_{>u_0}}\,\vee\,\mathcal{A}_{\mathscr{I}^+},
		\label{eq:cauchy-slice-holography}
	\end{equation}
	where $\mathcal{A}_{\rm ext}$ denotes the exterior bulk effective algebra generated by
	gauge-invariant, gravitationally dressed bulk operators supported in $\mathcal{U}$.
	
	What we've obtained thus far is nothing more than the standard HKLL reconstruction formula \cite{Hamilton:2006az} applied to null Cauchy surfaces. In order to derive something resembling a form of "Cauchy slice holography" we have to make use of the (unitary) null time evolution operator $U(\alpha) = \text{exp}(i \hat{\mathscr{P}}_{\alpha})$ on horizon subregions, which maps ``bulk'' operators in the horizon subalgebra onto operators in an arbitrarily small neighborhood of the corner.  To this aim, consider the corner $S_0$ at affine parameter $u=u_0$ associated with $\mathscr{H}_{>u_0}$. For any $\varepsilon>0$, define an arbitrarily thin ``corner strip''
	\begin{equation}
		G^{\varepsilon}_{u_0}
		=
		\{(u,x^A)\in \mathscr{H} \;|\; u_0 \le u < u_0+\varepsilon\}.
	\end{equation}
	As we've shown, the half-sided translation generator $\hat{\mathscr{P}}_\alpha$ (with $\alpha(x^A)\ge 0$) is a pure corner term whose action on gravitationally dressed observables is to Lie drag along the horizon generators:
	\begin{equation}
		[\mathscr{P}_\alpha,\hat{\mathscr{O}}(p)]
		=
		-i\alpha\partial_{u}\hat{\mathscr{O}}(p).
		\label{eq:corner-P-action}
	\end{equation}
	The corresponding unitary $U(\alpha):=e^{\,i \hat{\mathscr{P}}_\alpha}$ implements
	half-sided translations as inner automorphisms on the crossed product algebra:
	\begin{equation}
		U(\alpha)\,\hat{\mathscr{O}}(u,x^A)\,U(\alpha)^\dagger
		=
		\hat{\mathscr{O}}(u+\alpha,x^A),
		\ \alpha(x^A)\ge 0.
		\label{eq:corner-translation-inner}
	\end{equation}
	
	It follows immediately that the entire extended algebra is generated
	by an arbitrarily thin neighborhood of the corner together with the corner edge mode unitaries:
	\begin{equation}
		\widehat{\mathcal{M}}_{\mathscr{H}_{>u_0}}
		=
		\Big(
		\mathcal{A}\!\left(G^{\varepsilon}_{u_0}\right)
		\cup
		\{\,e^{\,i\hat{\mathscr{P}}_\alpha}:\alpha\ge 0\,\}
		\cup
		\{\,e^{\,i\hat{\mathscr{A}}_\beta}:\beta\ge 0\,\}
		\Big)'' ,\ \forall\,\varepsilon>0.
		\label{eq:horizon-generated-by-corner}
	\end{equation}
	In words, once the corner edge modes are included, any operator supported
	at finite $u-u_0>0$ is obtained by conjugating a near-corner operator by corner unitaries, and hence belongs to the
	von Neumann algebra generated by the corner strip.
	
	Exactly the same reasoning applies at null infinity. Fix a cut $C_0\subset\mathscr{I}^+$ at retarded time $v=v_0$,
	let $\mathscr{I}^+_{>v_0}$ denote the portion to its future, and define the thin strip
	\begin{equation}
		N^{\varepsilon}_{v_0} =
		\{(v,x^A)\in \mathscr{I}^+ \;|\; v_0 \le v < v_0+\varepsilon\}.
	\end{equation}
	After obtaining the analogous corner edge mode completion at $\mathscr{I}^+$, with BMS supertranslation charges
	$\hat{\mathscr{T}}_\alpha$ that generate $v\mapsto v+\alpha(x^A)$ on $\mathscr{I}^+$ as inner automorphisms,
	we have the counterpart of \Eqref{eq:horizon-generated-by-corner}:
	\begin{equation}
		\widehat{\mathcal{M}}_{\mathscr{I}^+_{>v_0}}
		=
		\Big(
		\mathcal{A}\!\left(N^{\varepsilon}_{v_0}\right)
		\cup
		\{\,e^{\,i\hat{\mathscr{T}}_\alpha}:\alpha\ge 0\,\}
		\Big)'' ,
		\ \forall\,\varepsilon>0.
		\label{eq:scri-generated-by-corner}
	\end{equation}
	The limit $v_0\rightarrow \infty$ corresponds to the analogous limit $C_0 \rightarrow i^0$ (more precisely, the codimension-one hyperboloid at $i^0$).
	
	Combining null Cauchy slice reconstruction \eqref{eq:cauchy-slice-holography} with corner generation of the horizon algebra \crefrange{eq:horizon-generated-by-corner}{eq:scri-generated-by-corner} yields a sharpened
	``corner holography'' result. Define the corner algebras
    \begin{subequations}
	\begin{align}
		\mathcal{A}_{\rm corner}(S_0)
		&:=   \Big(
		\mathcal{A}\!\left(G^{\varepsilon}_{u_0}\right)
		\cup
		\{\,e^{\,i\hat{\mathscr{P}}_\alpha}:\alpha\ge 0\,\}
		\cup
		\{\,e^{\,i\hat{\mathscr{A}}_\beta}:\beta\ge 0\,\}
		\Big)'' , \\
		\mathcal{A}_{\rm corner}(i^0)
		&:= \Big(
		\mathcal{A}\!\left(N^{\varepsilon}_{\infty}\right)
		\cup
		\{\,e^{\,i\hat{\mathscr{T}}_\alpha}:\alpha\ge 0\,\}
		\Big)'',
	\end{align}
        \end{subequations}
	for any fixed $\varepsilon>0$ (the resulting algebras are independent of $\varepsilon$ by the arguments above). Then the full exterior bulk effective algebra is generated by those of the two corners:
	\begin{equation}
		\mathcal{A}_{\rm ext} =
		\mathcal{A}_{\rm corner}(S_0)\,\vee\,\mathcal{A}_{\rm corner}(i^0).
		\label{eq:two-corner-holography}
	\end{equation}
	Equivalently, every gauge-invariant, gravitationally dressed bulk operator localized at $p\in\mathcal{U}$ obeys
	\begin{equation}
		\widehat{\mathscr{O}}(p)
		\;\in\;
		\mathcal{A}_{\rm corner}(S_0)\,\vee\,\mathcal{A}_{\rm corner}(i^0),\label{eq:corner_holography_statement}
	\end{equation}
	with an explicit constructive representation provided by the HKLL formula \eqref{eq:hkll-null},
	together with the fact that the $\mathscr{H}_{>u_0}$ and $\mathscr{I}^+$ operator algebras appearing there are themselves
	generated by arbitrarily small corner neighborhoods once the corner edge mode completion is included.\footnote{One might naively find this result in conflict with the conventional wisdom that non-perturbative in $1/G_N$ effects are needed in order to restore unitarity of black hole evaporation, since \Eqref{eq:corner_holography_statement} implies unitary evolution of gravitationally dressed ``bulk'' operators in the subregion algebra just from the corner edge mode unitaries. But recall that \Eqref{eq:corner_holography_statement} is only valid in perturbative quantum gravity. Namely, we can only evolve by $\text{exp}(\mathcal{O}(1))$ amounts along the horizon in $G_N$ counting under the action of edge mode unitaries, whereas to reach the Page time we need to evolve by $\text{exp}(\mathcal{O}(1/G_N))$ amounts, which is outside the regime of validity of our canonical quantization procedure.}

        \section*{Acknowledgements}
        For helpful discussions and/or comments on the manuscript we thank Laurent Freidel, Marc Klinger, Gautam Satischandran, and Pranav Pulakkat.  EF was supported in part by NSF grant PHY-2409350 and by a Simons foundation fellowship. GPT-5 was used as an aid for the calculations in this paper and for its writing.

	\appendix
    \crefalias{section}{appendix}
    \crefalias{subsection}{appendix}
	
	\section{Integrability of half-sided symmetry generators in GR\label{app:nonint}}

	In this appendix we show explicitly that the generators \eqref{generatordefexplicit1} are integrable
	for supertranslations in vacuum general relativity, but not for 
	${\rm diff}(\mathbb{S}^{2})$ generators.  The result 
	is also valid when matter fields are included, but we omit them here for brevity.
	
	Contracting the symplectic form \eqref{eq:sympfin0}
    with the symmetry \eqref{phasespacetruncated} gives
    \be
	\label{jjj}
	16 \pi \mathfrak{i}_{{\hat \xi}_{\rm T}} \omega_{ijk} =
	(\mathfrak{i}_{{\hat \xi}_{\rm T}}
	\delta \eta_{ijk})  \delta \Theta - (\mathfrak{i}_{{\hat \xi}_{\rm T}}
	\delta \Theta ) \delta \eta_{ijk}
	+ \mathfrak{i}_{{\hat \xi}_{\rm T}}
	\delta(q^{AB} \eta_{ijk})  \delta \sigma_{AB} - (\mathfrak{i}_{{\hat
			\xi}_{\rm T}}
	\delta \sigma_{AB} ) \delta (q^{AB} \eta_{ijk}).
	\ee
	Now the transformations of the fields under the symmetry $
	{\hat  \xi}$ given by \Eqref{twosidedphasespace}
	are (see Appendix F of Ref.\ \cite{Chandrasekaran:2023vzb})
	\begin{subequations}
		\begin{align}
			&\mathfrak{i}_{\hat{\xi}}\delta \ell^i = 0, \\ 
			&\mathfrak{i}_{\hat{\xi}}\delta \eta_{ijk} = (\widehat{\nabla}_m \xi^m + \beta_\xi) \eta_{ijk}, \\
			&\mathfrak{i}_{\hat{\xi}}\delta \Theta = \lie_{\xi} \Theta - \beta_\xi
			\Theta, \\
			&\mathfrak{i}_{\hat{\xi}}\delta q_{AB} = \lie_{\xi} q_{AB}, \\
			&\mathfrak{i}_{\hat{\xi}}\delta \sigma_{AB} = \lie_{\xi} \sigma_{AB} - \beta_\xi
			\sigma_{AB},
		\end{align}
	\end{subequations}
	where $\beta_\xi$ is defined by $\lie_{\xi}\ell^a = \beta_\xi \ell^a$.
    For the special case of supertranslations ${\vec \xi} = f
        {\vec \ell}$ these transformations reduce to
	\begin{subequations}
          \label{eq:symtransfs}
		\begin{align}
			&\mathfrak{i}_{\hat{\xi}}\delta \ell^i = 0, \\ 
			&\mathfrak{i}_{\hat{\xi}}\delta \eta_{ijk} =
                        f \Theta \eta_{ijj}, \\
			&\mathfrak{i}_{\hat{\xi}}\delta \Theta =
                        \lie_{\ell}(f  \Theta), \\
			&\mathfrak{i}_{\hat{\xi}}\delta q_{AB} = f
                        (\sigma_{AB} + 2 \Theta q_{AB}), \\
			&\mathfrak{i}_{\hat{\xi}}\delta \sigma_{AB} =
                        \lie_{\ell} (f \sigma_{AB}).
		\end{align}
	\end{subequations}

	For the transformations under the truncated phase vector field $\hat
	\xi_{\rm T}$, three of these just get
	multiplied by $H(u - u_0)$:
	\begin{subequations}
		\label{easyones}
		\begin{align}
			&\mathfrak{i}_{\hat{\xi}_{\rm T}}\delta \ell^i = 0, \\ 
			&\mathfrak{i}_{\hat{\xi}_{\rm T}}\delta \eta_{ijk} =  H (\widehat{\nabla}_m \xi^m + \beta_\xi) \eta_{ijk}, \\
			&\mathfrak{i}_{\hat{\xi}_{\rm T}}\delta q_{AB} = H \lie_{\xi} q_{AB}.
		\end{align}
	\end{subequations}
	The transformations of the remaining quantities $\Theta$ and
	$\sigma_{AB}$ can be expressed in terms of these three using the
	identities $\lie_\ell \eta_{ijk} = \Theta \eta_{ijk}$ and $\lie_\ell
	q_{AB} = \Theta q_{AB} + 2 \sigma_{AB}$, which yield after taking variations
	\begin{subequations}
		\begin{align}
			&\delta \Theta  =  \frac{1}{6} \eta^{ijk} (\lie_\ell - \Theta)
			\delta \eta_{ijk}, \\
			&\delta \sigma_{AB} =  \frac{1}{2}  (\lie_\ell - \Theta) \delta
			q_{AB}
			+ \frac{1}{4} q_{AB} (\Theta q^{CD} + 2 \sigma^{CD}) \delta q_{CD}
			- \frac{1}{4} q_{AB} q^{CD} \lie_\ell \delta q_{CD}.
		\end{align}
	\end{subequations}
	It follows that
	\begin{subequations}
		\label{hardones}
		\begin{align}
			&\mathfrak{i}_{\hat{\xi}_{\rm T}}\delta \Theta  = H \mathfrak{i}_{\hat{\xi}}\delta \Theta +
			\lie_\ell H (\widehat{\nabla}_m \xi^m + \beta_\xi), \\
			&\mathfrak{i}_{\hat{\xi}_{\rm T}}\delta \sigma_{AB}  = H \mathfrak{i}_{\hat{\xi}}\delta \sigma_{AB} +
			\frac{1}{2} \lie_\ell H \left(\delta^C_A \delta^D_B - \frac{1}{2} q_{AB}
			q^{CD} \right) \lie_{\xi} q_{CD}.
		\end{align}
	\end{subequations}
	
	Now inserting the results \eqref{easyones} and \eqref{hardones} in \Eqref{jjj} gives
	\begin{eqnarray}
		\label{jjj3}
		\mathfrak{i}_{{\hat \xi}_{\rm T}} \omega_{ijk} &=& H \mathfrak{i}_{{\hat
				\xi}} \omega_{ijk}  - \lie_\ell H \, \Xi_{ijk},
	\end{eqnarray}
	where
	\be
	16 \pi \Xi_{ijk} =  (\widehat{\nabla}_m \xi^m + \beta_\xi) \delta \eta_{ijk} +
	\frac{1}{4} (2 \delta^C_A \delta^D_B - q_{AB} q^{CD}) \lie_{\xi}
	q_{CD} \delta (q^{AB} \eta_{ijk}).
    \label{eq:Xiformula}
	\ee
	Using the identities $q^{AB} \lie_{\xi} q_{AB} = 2 (\widehat{\nabla}_m \xi^m +
	\beta_\xi)$ and $\delta \eta_{ijk} = h \eta_{ijk}/2$ with $h_{AB} = \delta
	q_{AB}$ and $h = q^{AB} h_{AB}$, this can be simplified to 
	\be
	\Xi_{ijk} = -\frac{1}{32 \pi} \left[ h^{AB} \lie_{\xi} q_{AB} - 2 h (\widehat{\nabla}_m
	\xi^m + \beta_\xi) \right] \eta_{ijk}.
	\ee

	Now comparing with \Eqref{eq:fluxform} the condition for integrability is
	\be
	\label{condt111}
	{\underline {i_{\xi} {\mathcal E}}} = -{\underline {i_\ell \Xi}},
	\ee
	where
	the underline denotes a pullback to the surface $u = u_0$
	and the flux 
	${\mathcal E}_{ijk}$ is given by the first term in \Eqref{fluxGR}.
	For supertranslations where ${\vec \xi} = f {\vec \ell}$, both sides
	of the condition \eqref{condt111} evaluate to
	\be
	\frac{1}{16 \pi} f\mu_{ij} h^{AB} (\sigma_{AB} - \Theta q_{AB}/2)
	\ee
	and the condition is satisfied.  For ${\rm diff}(\mathbb{S}^{2})$ transformations ${\vec
		\xi} = \xi^A(\theta^B) \partial_A$, the left hand side vanishes
	but the right hand side does not (as can be seen from taking $h_{AB}$
	to be traceless), so the symmetry is not integrable.

\section{An enlarged horizon phase space\label{app:larger}}

In the body of the paper we mostly used the definition of a global horizon phase space
given in \cref{sec:phasespacedefs}, in which the variation of the
inaffinity $\delta \kappa$ is constrained to vanish when we use the
convention for fixing the perturbative rescaling freedom $\delta
\ell^i = 0$ as described there.   In this appendix, we define a larger
phase space with $\delta \kappa \ne 0$.   We use this phase space to give an alternative
derivation of the distributional corrections
\eqref{eq:finalcornerterm} to the symplectic current, and to clarify
the definition of the symmetry corresponding to a half-sided
supertranslation in \cref{sec:spacetimeSTs}.  

The larger phase space ${\hat {\mathcal P}}_{\mathscr H}$ is described
in Appendix H of Ref.\ \cite{CFP} and in
Ref.\ \cite{Chandrasekaran:2023vzb}, where it was denoted ${\mathscr
  P}_{\mathfrak{q}}$. It consists of replacing the equivalence class
of fields $(\ell_a, \ell^a, \kappa)$ of \cref{sec:phasespacedefs}
with the smaller equivalence class $(\ell_a, \ell^a)$.  The
independent fields are the same as those listed in
\cref{sec:phasespacedefs},
except that there is now a larger set of pairs $(\ell^i,\kappa)$ to
choose from.  The expressions \eqref{eq:sympfin0} and \eqref{fluxGR} for the symplectic
form and flux acquire correction terms:
	\be
	\bm{\omega} = \frac{1}{16 \pi} \delta {\bm \eta} \wedge
        (\delta \Theta + 2 \delta \kappa)
	+ \frac{1}{16 \pi} \delta( q^{AB} {\bm \eta}) \wedge \delta
        \sigma_{AB} + \delta \psi \wedge \delta ( \bm{\eta} \lie_\ell \psi),
        \label{eq:sympfin01}
	\ee
        and
        \be
         \label{fluxGR1}
	\bm{\mathcal{E}} =
	\frac{\bm{\eta}}{16 \pi} \left[  2 \delta \kappa
	- \frac{1}{2} \Theta h + \sigma^{ij} h_{ij} \right] + \delta \psi
	\lie_\ell \psi \bm{\eta},
        \ee
        while the charge $\bm{h}_\xi$ is unaltered.  These quantities
        satisify the general identity \eqref{identity0} with no correction terms when
        the fields are smooth.

        We now describe the definition of the half-sided
        supertranslation symmetry, expanding on the discussion given
        in \cref{sec:spacetimeSTs}.  We consider a smoothed out
        symmetry of the form ${\vec \xi} = f(u) \partial_u$ with
        $f(u) = 0$ for $u \le - \varepsilon$, $f(u) = \alpha + \beta
        u$ for $u \ge \varepsilon$, and we choose some smooth
        interpolation in the splitting region
        $[-\varepsilon,\varepsilon]$.
        This is a boundary symmetry of the phase space
        ${\hat {\mathcal P}}_{\mathscr H}$.  The corresponding field
        variations are given by Eqs.\ \eqref{eq:symtransfs} together
        with $\delta \kappa = \lie_\ell \lie_\ell f$, where for
        simplicity we have taken $\kappa = 0$ in the background.
        However, this is not the symmetry we want, since from the
        identity \eqref{identity0} the symplectic current is exact,
        and the charge variation has no term localized on the cut.

        Instead we define the symmetry by adding a correction term:
        \be
        \mathfrak{i}_{\hat \xi} \delta \phi = \mathfrak{i}_{{\hat
            \xi}_{\rm diffeo}} \delta \phi +\mathfrak{i}_{{\hat
            \xi}_{\rm corr}} \delta \phi.
        \ee
        Here the first term is the diffeomorphism described in the
        previous paragraph, and the correction term is nonzero only in the
        splitting region.  It consists of two pieces, an inaffinity
        perturbation
        \be
        \label{eq:cancelkappa}
        \mathfrak{i}_{{\hat \xi}_{\rm corr}} \delta \kappa = -
        \lie_\ell \lie_\ell f
        \ee
        in order to cancel out the inaffinity variation from the first
        term and give a variation in our original phase space
        $\mathcal{P}_{\mathscr{H}}$,
        and a perturbation to the matter stress energy tensor in order
        to ensure that the constraint equations are still satisfied,
        described in \cref{sec:edgemodederiv}.  In more detail, the
        linearized Raychaudhuri equation can be written as
        \be
        (\partial_u \partial_u + \Theta \partial_u) \mathfrak{i}_{{\hat
          \xi}_{\rm corr}} h= - 2 \Theta
        \partial_u \partial_u f - 8 \pi \partial_u \psi \partial_u (\mathfrak{i}_{{\hat
            \xi}_{\rm corr}} \delta \psi),
        \ee
        and we can choose $\mathfrak{i}_{{\hat
            \xi}_{\rm corr}} \delta \psi$ within the splitting region so that the solution vanishes
        outside the splitting region in the limit $\varepsilon \to 0$. 

        In order to compute the charge variation localized to the cut
        for this symmetry, we can focus on ${\hat \xi}_{\rm corr}$,
        since the contribution from ${\hat \xi}_{\rm diffeo}$
        vanishes. Inserting the field variations $\mathfrak{i}_{{\hat
            \xi}_{\rm corr}} \delta \phi$ into the symplectic current
        \eqref{eq:sympfin01}
        and integrating over $\mathscr{H}$, the only term that gives a
        nonvanishing contribution in the limit $\varepsilon \to 0$ is
        the $\delta \kappa$ term.  Thus we obtain from \Eqref{eq:cancelkappa}
        \be
        \delta {\mathcal Q}_\xi =  - \mathfrak{i}_{{\hat \xi}_{\rm
            corr}} \Omega_{\mathscr{H}} = - \frac{1}{8 \pi} \int_{\mathscr{H}}
        \delta \bm{\eta}\ \partial_u \partial_u f.
        \ee
        Thus we have recovered the right hand side of \Eqref{eq:violationcurrent}, and 
        the rest of the calculation then proceeds as in that section.

    \section{Half-sided supertranslation generators beyond GR\label{sec:genedgecorner}}
	In this appendix we derive the main results of \crefrange{sec:gredgecorner}{sec:edgemodederiv} for a general diffeomorphism invariant theory of gravity.
	
	To start with, for a half-sided supertranslation $\xi^a = f_0 H(u-u_0)\ell^a$,
	\begin{align}
		\lie_{\xi}\bm{\theta} = H(u-u_0)\lie_{\xi_0}\bm{\theta} + f_0\bm{\theta} \delta(u-u_0).
	\end{align}
	Moreover, as shown in \cref{sec:gen}, 
	\begin{align}
		-\mathfrak{i}_{\hat{\xi}}\bm{\omega} = H(u-u_0)\left(d\delta \bm{Q}_{\xi_0} - \lie_{\xi_0}\bm{\theta}\right) - \delta(u-u_0)f_0\bm{\mathcal{E}}.
	\end{align}
	Putting these two together, and decomposing $\bm{\theta} = \delta \bm{\alpha} + \bm{\mathcal{E}}$, 
	\begin{align}
		\lie_{\xi}\bm{\theta} = \delta(u-u_0)\delta(f_0 \bm{\alpha}) + H(u-u_0)\delta d\bm{Q}_{\xi_0}.
	\end{align}
	Therefore,
	\begin{align}
		d[\delta \bm{Q}_{\xi} - i_{\xi}\bm{\theta}] + \mathfrak{i}_{\hat{\xi}}\bm{\omega} = \delta\left[ d \bm{Q}_{\xi}-H(u-u_0) d\bm{Q}_{\xi_0}\right] - \delta(u-u_0)\delta( f_0\bm{\alpha}).
	\end{align}
	Integrating by parts on the $H(u-u_0) d\bm{Q}_{\xi_0}$ term allows us to write 
	\begin{align}
		d[\delta \bm{Q}_{\xi} - i_{\xi}\bm{\theta}] + \mathfrak{i}_{\hat{\xi}}\bm{\omega} = d\delta\left[ \bm{Q}_{\xi}-H(u-u_0) \bm{Q}_{\xi_0}\right] + \delta(u-u_0)\delta\left[\bm{Q}_{\xi_0} - f_0i_{\ell}\bm{\alpha}\right]\label{eq:gencontactterm}.
	\end{align}
	
	As shown by Wald-Iyer \cite{Iyer:1994ys}, the Noether charge 2-form can be written in general as
	\begin{align}
		\bm{Q}_{\xi, ab} = \bm{W}_{abc}\xi^c - \bm{E}_{ab}{}{}^{cd}\nabla_{[c}\xi_{d]},\label{eq:waldiyer}
	\end{align}
	where 
    \begin{subequations}
	\begin{align}
		\bm{E}_{ab}{}{}^{cd} &= \varepsilon_{abef}E^{efcd}, \\ 
		E_{abcd} &= \frac{\partial L}{\partial R_{abcd}} - \nabla_{a_1}\frac{\partial L}{\partial \nabla_{a_1}R_{abcd}} + \ldots + (-1)^m\nabla_{(a_1} \ldots \nabla_{a_m)}\frac{\partial L}{\partial \nabla_{(a_1}\ldots \nabla_{a_m)}R_{abcd}}.
	\end{align}
	    \end{subequations}
	For the second term in \Eqref{eq:waldiyer}, the pullback can be written 
	\begin{align}
		\Pi^a_i \Pi^b_j\bm{E}_{ab}{}{}^{cd}\nabla_{[c}\xi_{d]} = \bm{\eta}_{ijk}\mathfrak{q}^k_{\xi}, \ \mathfrak{q}_{\xi}^a := E^{abcd}\ell_b \nabla_{[c} \xi_{d]}.\label{eq:gendecomp}
	\end{align}
	Note that $\mathfrak{q}^a_{\xi}$ is intrinsic to $\mathscr{H}$ because $\mathfrak{q}_{\xi}^a \ell_a \hateq 0$, which just follows from the usual antisymmetry of $R_{abcd}$.
	
	For convenience, introduce an auxiliary null normal $n_a$ normalized by $n_a \ell^a = -1$ and a basis $e^A_a$ for the corner satisfying $\ell^a e^A_a = n^a e^A_a = 0, e^A_a e^a _B = \delta^A_B$. We extend the basis away from the corner via parallel transport: $n^b \nabla_b e^a_A = \ell^b \nabla_b e^a_A = 0$. 
	
	We can write 
	\begin{align}
		\Pi^a_i \Pi^b_j\bm{E}_{ab}{}{}^{cd}\nabla_{[c}\xi_{d]} = \bm{\mu}_{ij}E^{abcd}n_a \ell_b \nabla_{[c}\xi_{d]}.
	\end{align}
	By antisymmetry of $E^{abcd}$ under $c\leftrightarrow d$, we know that any diagonal component of $\nabla_{c}\xi_{d}$ vanishes after contraction. So we only care about $e^c_A n^d \nabla_{c}\xi_{d}$, $e^c_A \ell^d \nabla_{c}\xi_{d}$, and $\ell^c n^d \nabla_{c}\xi_{d}$. The other components are obtained via the exchange $c \leftrightarrow d$. 
	
	We have 
	\begin{align}
		e^c_A \ell^d \nabla_c \xi_d \hateq 0,
	\end{align}
	where we've used that the derivative is along $\mathscr{H}$ to plug in $\xi^a = f\ell^a$ and we've made repeated use of the parallel transport condition $\ell^b\nabla_b e^a_A = 0$. Similarly, 
	\begin{align}
		e^c_A n^d \nabla_c \xi_d \hateq -f \omega_A - \nabla_A f,
	\end{align}
	where $\omega_A := -n^d e_A^c \nabla_c \ell_d$ is the spatial projection of the spin connection $\omega_i$ defined in the previous section. And lastly, 
	\begin{align}
		\ell^c n^d\nabla_c \xi_d \hateq -\beta,
	\end{align}
	where we've used that $\nabla_{[a}\ell_{b]} \hateq w_{[a}\ell_{b]}$ and that $\ell^a w_a \hateq 0$.
	
	So putting it together, 
    \begin{subequations}
	\begin{align}
		&\Pi^a_i \Pi^b_j\bm{E}_{ab}{}{}^{cd}\nabla_{[c}\xi_{d]} = -\bm{\mu}_{ij}\left(\beta E + E^A \alpha_A\right), \\ &E:= E^{abcd}n_a \ell_b n_c \ell_d, \ E^A := E^{abcd}n_a \ell_b e^A_c \ell_d, \\ &D_A \alpha := \nabla_A \alpha + \alpha \ \omega_A,
	\end{align}
        \end{subequations}
	where we've decomposed $f$ into an angle-dependent translation piece $\alpha$ and an angle-dependent boost piece $\beta$, just as in the previous section. We can think of $D_A$ as a gauge covariant derivative on the normal bundle.
	
	In the end, 
	\begin{align}
		\bm{Q}_{\xi_0} - f_0i_{\ell}\bm{\alpha} = \bm{\mu}\left[\beta E + E^A D_A\alpha + \alpha \left(W -K\right)\right],
	\end{align}
	where we've written $i_{\xi}\bm{W} = \bm{\mu}\alpha W$ and $i_{\ell}\bm{\alpha} = \bm{\mu}K$. Hence, if we use the ``on-shell'' prescription defined in \cref{sec:integrable2}, \Eqref{eq:gencontactterm} just reduces to 
	\begin{align}
		-\mathfrak{i}_{\hat{\xi}}\Omega_{\mathscr{H}}= \delta\left(\int_{S_0}\bm{\mu}\left[-\beta E - E^A D_A\alpha + \alpha \left(W -K\right)\right]\right) + \delta \int_{\infty}\bm{\mu} \ \beta E.
	\end{align}
	
	We can then immediately write down the half-sided boost and translation generators for a general gravitational theory:\footnote{Only in GR does the half-sided boost generator $\mathscr{K}_{\beta}$ coincide with the area operator $\mathscr{A}_{\beta}$.}
	\begin{align}
		\mathscr{K}_{\beta} = -\frac{1}{8\pi}\left[\int_{S_0}\bm{\mu} \beta E -\int_{S_\infty}\bm{\mu} \beta E \right], \ \mathscr{P}_{\alpha} = \frac{1}{8\pi}\int_{S_0}\bm{\mu} \left[\alpha(W-K) + E^AD_A\alpha\right].
	\end{align}
	
	In the case of GR, it is easy to check that $E = -1$ and $E^A = 0$ as well as $W = 0$ and $K = \Theta$. So we recover the results of the previous section. But in that section we derived this result from scratch, without ever invoking \cref{sec:gen}. So this acts as a non-trivial consistency check when restricted to GR.
    
	The main body of the paper restricts to the setting of GR for simplicity, but the results above allow one to straightforwardly obtain the analogous results for general diffeomorphism invariant theories of gravity.

	\section{Transverse deformations of the horizon\label{app:horizondeformation}}
	One point we've glossed over in the main body of the paper is the fact that the horizon itself is embedded in spacetime in a metric-dependent manner. Given a spacetime $(M,g_{ab})$, recall the global definition of the event horizon:
	\begin{align}
		\mathscr{H}[g] = \partial J^-[g](\mathscr{J}^+).\label{eq:eventhorizondef}
	\end{align}
	Therefore when the metric fluctuates, so too does the location of the horizon in spacetime. Let's briefly discuss how this folds into our calculations.
	
	As a convenient representation, define a scalar field $\mathscr{V}$ such that 
    \begin{subequations}
	\begin{align}
		\mathscr{H}[g] &= \left\{x\in M \mid \mathscr{V}[g](x)=0\right\}, \\
		\ell_a &\hateq \nabla_a \mathscr{V}[g].
	\end{align}
        \end{subequations}
	Additionally, let $X[g]\colon M_0\mapsto M$ be an embedding of the horizon from a fixed reference manifold $M_0$ into the actual spacetime $(M,g_{ab})$. Then $\mathscr{V}[g](X[g](y)) = 0$ specifies the (dressed) location of $\mathscr{H}$. We can think of it as a (smooth) metric-dependent diffeomorphism that moves $\mathscr{H}[g]$ relative to fixed reference horizon $\mathscr{H}_0$ while satisfying the dressing condition \eqref{eq:eventhorizondef}.
	
	Under a metric perturbation, 
	\begin{align}
		\delta \left(\mathscr{V}[g](X[g](y))\right) \hateq 0 \Rightarrow \delta \mathscr{V}[g] \hateq -\ell_a \delta X^a.
	\end{align}
	Moreover, since $\delta \ell_a \hateq \nabla_a \delta \mathscr{V}$,
	\begin{align}
		\delta(g^{ab}\ell_a \ell_b) \hateq 0 \Rightarrow \ell^a \nabla_a (\ell_b \delta X^b) = -\frac{1}{2}h_{\ell\ell}.  
	\end{align}
	In terms of the affine parametrization $u$ of $\ell^a$, the dressing \eqref{eq:eventhorizondef} implies $\ell_a\delta X^a \rightarrow 0$ as $u\rightarrow \infty$. In words, the location of the horizon at future infinity, which we denote $\mathscr{H}^+_+$, does not fluctuate since the horizon always approaches a stationary vacuum solution in that limit. Hence, 
	\begin{align}
		\left(\ell_a\delta X^a\right)(u) \hateq \frac{1}{2}\int_{u}^{\infty}du \ h_{uu}.
	\end{align}
	This completely determines the change in the (dressed) location of the horizon under a metric perturbation.
	
	Now, for any smooth differential form $\bm{\omega}$ on field configuration space,
	\begin{align}
		\delta(X(\bm{\omega})) = X_{*}(\delta\bm{\omega} + \lie_{\chi}\bm{\omega})\label{eq:dressingvariation}
	\end{align}
	where $X^{-1}_{*}\left(X + \delta X\right)_* = 1 + \lie_{\chi} + \mathcal{O}(\delta X^2)$ defines the infinitesimal generator $\chi^a$ of the diffeomorphism. Note that $\chi^a$ is a 1-form on field space because the diffeomorphism depends on the metric.

    The vector field $\chi^a$ satisfies $\chi^a \ell_a = \gamma$, i.e. it deforms the horizon location by some amount $\mathscr{V} \rightarrow \mathscr{V} + \gamma$. We denote by \begin{align}\bm{\mathcal{Q}}_{\chi} = \bm{Q}_{\chi} - i_{\chi}\bm{\alpha}^{\perp}\end{align} the generator of horizon deformations (or more precisely, the associated density). 

    We emphasize that the quantity which enters into the generator is the pullback $\Pi^a_i \Pi^b_j \left(\chi^c\theta_{abc}\right)$ as opposed to $\chi^k\left(\Pi^a_i \Pi^b_j \Pi^c_k \theta_{abc}\right)$ even though in standard covariant phase space calculations one typically works with the latter. This is because standard treatments assume a vector field tangent to the boundary, whereas in our case $\chi^k \hateq 0$. So we instead have \begin{align}\Pi^a_i \Pi^b_j \left(\chi^c\theta_{abc}\right) = \delta\left(i_{\chi} \alpha^{\perp}\right)_{ij} + \left(i_{\chi}\mathcal{E}^{\perp}\right)_{ij}\end{align} as defining the transverse boundary term $\bm{\alpha}^{\perp}$ and transverse flux term $\bm{\mathcal{E}}^{\perp}$. That this actually leads to an (integrable) generator on phase space just follows from the fact that $i_{\chi}\bm{\mathcal{E}}^{\perp}\rightarrow 0$ at $\partial \mathscr{H}$ due to the fall-off conditions satisfied by $\gamma$. 
	
	Now, a straightforward calculation implies \cite{Speranza:2022lxr}
	\begin{align}
		\Omega_{\mathscr{H}} = \delta\int_{\mathscr{H}}\bm{\theta} = \int_{\mathscr{H}}\left(\delta \bm{\theta} +\lie_{\chi}\bm{\theta}\right) + \int_{\partial \mathscr{H}}\left(\delta \bm{\mathcal{Q}}_{\chi} + \lie_{\chi}\bm{\mathcal{Q}}_{\chi}\right). 
	\end{align}
	Using Cartan's magic formula along with the fact that $\bm{\theta}$ is a top-form on $\mathscr{H}$,
	\begin{align}
		\int_{\mathscr{H}}\lie_{\chi}\bm{\theta} = \int_{\mathscr{H}}d(i_{\chi}\bm{\theta}) = \int_{\partial \mathscr{H}}i_{\chi}\bm{\theta}.
	\end{align}
    Moreover, using that $\partial \partial \mathscr{H} = \varnothing$, we also have 
    \begin{align}
        \int_{\partial \mathscr{H}}\lie_{\chi}\bm{\mathcal{Q}}_{\chi} = \int_{\partial \mathscr{H}}i_{\chi}d\bm{\mathcal{Q}}_{\chi}.\label{eq:dqchiterm}
    \end{align}
	The two cases of interest are when $\partial \mathscr{H} = \mathscr{H}^+_+ \cup \mathscr{B}$ where $\mathscr{B}$ is the bifurcation surface of a Killing horizon, and when $\partial \mathscr{H} = \mathscr{H}^+_+ \cup p_0$ where $p_0$ is the tip of the lightcone from which the horizon of a collapse black hole emanates. We already know $\chi^a \rightarrow 0$ at $\mathscr{H}^+_+$. In the former case, $\chi^a\big\lvert_{\mathscr{B}} = 0$ as well because the bifurcation surface remains fixed under first order metric perturbations. In the latter case, $\int_{p_0}i_{\chi}\bm{\theta} = 0$ since $p_0$ has zero area and $\bm{\theta}$ is smooth everywhere. The same argument implies \Eqref{eq:dqchiterm} vanishes as well.
	
	The $i_{\chi}\bm{\alpha}^{\perp}$ part of $\bm{\mathcal{Q}}_{\chi}$ clearly vanishes at $\partial \mathscr{H}$ because $\chi^a$ goes to zero there and $\bm{\alpha}^{\perp}$ is smooth. The Noether charge piece $\delta \bm{Q}_{\chi}$ is a bit more non-trivial. Let's introduce an auxiliary null normal $n^a$ normalized by $n^a \ell_a \hateq -1$. We can always take $\chi^a = \gamma n^a$ in this frame. We then calculate,
	\begin{align}
		Q_{\chi, ij} &= -\frac{1}{8\pi}\mu_{ij}n_c \ell_d\nabla^{[c} \chi^{d]} = -\frac{1}{16\pi}\mu_{ij}\left(\lie_n \gamma + \chi_c\lie_{\ell}n^c\right).
	\end{align}
	The second term clearly vanishes at $\partial \mathscr{H}$ given the fall-off conditions on $\chi^a$, so we're just left with
	\begin{align}
		\int_{\partial \mathscr{H}}\bm{\mathcal{Q}}_{\chi} = -\frac{1}{16\pi}\int_{\partial \mathscr{H}} \bm{\mu}\ \lie_n \gamma.
	\end{align}
	
	But we also know that 
	\begin{align}
		\mathfrak{i}_{\hat{\chi}}\delta n^a = \lie_{\chi}n^a \hateq -n^a \lie_n \gamma,
	\end{align}
	where we've extended $n^a$ to all of phase space by demanding that it transform covariantly. Since $\delta \ell_a \hateq 0$ and $\delta(\ell_a n^a) \hateq 0$, we have that $\ell_a \delta n^a \hateq 0$, hence it follows that $\lie_n \gamma \hateq 0$ identically on $\mathscr{H}$. 
    
    As a consistency check of this result, we compute \begin{align}\delta \kappa \hateq -\frac{1}{2}\ell^b \ell^c n^a\left(\nabla_b h_{ca} + \nabla_c h_{ba} - \nabla_a h_{bc}\right)= \frac{1}{2}\lie_n h_{\ell\ell},\end{align}
    where we've repeatedly made use of the fact that $\ell^a h_{ab}\hateq 0$. Therefore, \begin{align}\mathfrak{i}_{\hat{\chi}}\delta \kappa \hateq \gamma \ \mathfrak{i}_{\hat{n}}\delta \kappa - \left(\lie_{\ell}+\kappa\right) \lie_n \gamma,\end{align} where we've used that \begin{subequations}\begin{align}&\ell^a \lie_n g_{ab} = \mathfrak{i}_{\hat{n}}\left(\ell^a h_{ab}\right) \hateq 0, \\ &\lie_n\left(\ell^a \ell^b \lie_n g_{ab}\right) = \mathfrak{i}_{\hat{n}}\lie_n h_{\ell\ell} \hateq \mathfrak{i}_{\hat{n}}\delta \kappa, \\ &\ell^b \nabla_b n^a \hateq -\kappa n^a + \Pi^a_i v^i,\end{align}\end{subequations} for some $v^i$ tangent to $\mathscr{H}$. But recall that we're working in a phase space $\mathcal{P}_{\mathscr{H}}$ where $\delta \kappa \hateq 0$ for all smooth variations. In particular, this means $\mathfrak{i}_{\hat{n}}\delta \kappa \hateq 0$. Combining this with the fact that $\lie_n \gamma \hateq 0$, we have $\mathfrak{i}_{\hat{\chi}}\delta \kappa \hateq 0$ meaning $\chi^a$ is indeed an admissible perturbation.
	
	We therefore conclude that 
	\begin{align}
		\int_{\partial \mathscr{H}}\delta \bm{\mathcal{Q}}_{\chi} = 0.
    \end{align}
	
	Therefore, despite the horizon being gravitationally dressed, we find that
	\begin{align}
		\Omega_{\mathscr{H}} = \int_{\mathscr{H}}\delta \bm{\theta},
	\end{align}
	Hence, calculations reduce to ones wherein we just gauge fix the location of the horizon in spacetime when defining the phase space, i.e. $\ell_a \delta X^a \hateq 0$.
	
	Another way to think about this is as follows. If instead of an explicit gravitational dressing we promoted the embedding fields evaluated at the boundaries $X\big\lvert_{\partial\mathscr{ H}}$ to putative dynamical edge modes $\mathscr{X}$ on $\partial \mathscr{H}$, then by the calculation above the symplectic form acquires corner terms of the form
	\begin{align}
		\left(\delta \mathscr{X} \wedge \delta \bm{\mathcal{Q}}_{\chi}\right)\Big\lvert_{\partial \mathscr{H}}.\label{eq:horizondeformcoupling}
	\end{align}
    
	But as we've just seen, $\delta \bm{\mathcal{Q}}_{\chi}\big\lvert_{\partial \mathscr{H}} = 0$. That is, the (putative) transverse horizon deformation edge modes $\mathscr{X}$ are not dynamical after all, but rather pure gauge artifacts. This is a non-trivial property of the phase space $\mathcal{P}_{\mathscr{H}}$ that we work with in this paper. And this conclusion agrees with the one we arrived at via the direct gravitational dressing approach.

	\section{Direct integral structure and non-factorization \label{app:directintegral}}
	
	In this we spell out in a bit more detail what is meant by the extended GNS Hilbert
	space
	\begin{equation}
		\label{eq:extended-H-space-appendix}
		\widehat{\mathcal{H}} \;\cong\; \mathcal{H} \,\otimes L^2(\mathcal{G}), \ 
		\mathcal{G} := C^\infty_\beta(\mathbb{S}^{d-2}) \rtimes C^\infty_\alpha(\mathbb{S}^{d-2})\,,
	\end{equation}
	and how this arises from the crossed product structure of the subregion algebra
	\begin{equation}
		\widehat{\mathcal{A}}_{\mathscr{H}_{>u_0}} 
		= 
		\Bigl(\mathcal{A}^{\text{grav}}_{\mathscr{H}_{>u_0}}[\hat{\sigma}]\otimes \mathcal{A}^{\text{mat}}_{\mathscr{H}_{>u_0}}[\hat{\psi}]\Bigr)
		\rtimes \mathcal{A}_{\partial G_\varepsilon}[\hat{\Gamma}^+_0,\hat{\Upsilon}^+_0]\,.
	\end{equation}
	We will also explain in what sense the ``hard mode $\otimes$ edge mode'' tensor product structure is not
	canonical once the crossed product has been taken.
	
	\subsection{GNS construction for the crossed product}
	
	Let
	\begin{equation}
		\mathcal{A}_0 
		= 
		\mathcal{A}^{\text{grav}}_{\mathscr{H}_{>u_0}}[\hat{\sigma}]\otimes \mathcal{A}^{\text{mat}}_{\mathscr{H}_{>u_0}}[\hat{\psi}]
	\end{equation}
	denote the von Neumann algebra generated by the dressed “bulk’’ operators on the portion
	of the horizon to the future of the cut $u>u_0$. This algebra acts on the Fock space
	$\mathcal{H}$ of the linearized fields $(\hat{\sigma},\hat{\psi})$ with cyclic and separating
	Hartle–Hawking state $|\Omega\rangle\in\mathcal{H}$.
	
	The edge mode algebra $\mathcal{A}_{\partial G_\varepsilon}[\hat{\Gamma}^+_0,\hat{\Upsilon}^+_0]$ is, by
	construction, the group algebra of the infinite-dimensional group
	\begin{equation}
		\mathcal{G} = C^\infty_\beta(\mathbb{S}^{d-2}) \rtimes C^\infty_\alpha(\mathbb{S}^{d-2})\,,
	\end{equation}
	where the two factors are generated by the area operator and half-sided translation operator
	$\hat{\mathscr{A}}_\beta$ and $\hat{\mathscr{P}}_\alpha$ and act as automorphisms of $\mathcal{A}_0$.  More
	precisely, for each $(\beta,\alpha)\in \mathcal{G}$ we have an automorphism
	$\vartheta_{(\beta,\alpha)}:\mathcal{A}_0 \to \mathcal{A}_0$ implemented on $\mathcal{H}$ by a
	unitary $U(\beta,\alpha)$,
	\begin{equation}
		\vartheta_{(\beta,\alpha)}(\hat{\mathscr{O}}) 
		= U(\beta,\alpha)\,\hat{\mathscr{O}}\,U(\beta,\alpha)^{-1}, \
		\hat{\mathscr{O}} \in \mathcal{A}_0\,,
	\end{equation}
	generated infinitesimally by the commutators with $\hat{\mathscr{A}}_\beta$ and $\hat{\mathscr{P}}_\alpha$ as
	in \crefrange{eq:linmodham}{eq:linanec} and \crefrange{eq:translationcommutator}{eq:boostcommutator}.
	
	The crossed product algebra $\widehat{\mathcal{A}}_{\mathscr{H}_{>u_0}}$ is then, by definition, the
	algebra generated by $\mathcal{A}_0$ and an additional set of unitaries $\lambda(g)$,
	$g\in \mathcal{G}$, subject to the covariance relations
	\begin{equation}
		\lambda(g)\,\hat{\mathscr{O}}\,\lambda(g)^{-1} 
		= \vartheta_g(\hat{\mathscr{O}}), \
		\hat{\mathscr{O}}\in\mathcal{A}_0,\; g\in \mathcal{G}\,.
	\end{equation}
	The canonical GNS representation of this crossed product is naturally constructed on a
	Hilbert space of the form
	\begin{equation}
		\widehat{\mathcal{H}} 
		= L^2(\mathcal{G},\mathrm{d}\mu_{\mathcal{G}};\mathcal{H}) \;\cong\; \mathcal{H}\otimes L^2(\mathcal{G})\,,
	\end{equation}
	where $\mathrm{d}\mu_{\mathcal{G}}$ is a left-invariant measure on $\mathcal{G}$, and
	$L^2(\mathcal{G},\mathrm{d}\mu_{\mathcal{G}};\mathcal{H})$ is the space of square-integrable $\mathcal{H}$-valued
	functions on $\mathcal{G}$.  Concretely:
	\begin{itemize}
		\item A vector $|\widehat{\Psi}\rangle\in\widehat{\mathcal{H}}$ is a map
		$g\mapsto |\Psi(g)\rangle\in\mathcal{H}$ such that
		$\int_{\mathcal{G}} \mathrm{d}\mu_{\mathcal{G}}(g)\,\|\Psi(g)\|^2 < \infty$.
		\item The action of $\hat{\mathscr{O}}\in\mathcal{A}_0$ is given fiberwise by
		\begin{equation}
			\bigl(\widehat{\pi}(\hat{\mathscr{O}})\Psi\bigr)(g) 
			= 
			\vartheta_{g^{-1}}(\hat{\mathscr{O}})\,|\Psi(g)\rangle\,.
			\label{eq:bulk-action-fiberwise}
		\end{equation}
		In other words, $\hat{\mathscr{O}}$ acts on the fiber at $g$ via the automorphism
		$\vartheta_{g^{-1}}$ of $\mathcal{A}_0$.
		\item The edge unitaries $\lambda(h)$ act by the left-regular representation of $\mathcal{G}$:
		\begin{equation}
			\bigl(\widehat{\pi}(\lambda(h))\Psi\bigr)(g) 
			= 
			|\Psi(h^{-1}g)\rangle\,.
			\label{eq:edge-left-regular}
		\end{equation}
	\end{itemize}
	One checks that the relations of the crossed product are satisfied:
	\begin{equation}
		\widehat{\pi}(\lambda(h))\,\widehat{\pi}(\hat{\mathscr{O}})\,\widehat{\pi}(\lambda(h))^{-1}
		= \widehat{\pi}\bigl(\vartheta_h(\hat{\mathscr{O}})\bigr)\,,
	\end{equation}
	and that $\widehat{\mathcal{A}}_{\mathscr{H}_{>u_0}}$ is represented faithfully on $\widehat{\mathcal{H}}$.
	This is the precise meaning of \Eqref{eq:extendedhilbertspace}.
	
	In practice, it is convenient to pick a reference configuration $g=e$ (the identity element of $\mathcal{G}$) and
	identify each fiber $\mathcal{H}_g$ with the original Fock space $\mathcal{H}$ via the unitary
	\begin{equation}
		W_g:\mathcal{H}\to\mathcal{H}_g, \
		W_g |\Psi\rangle = \vartheta_g(|\Psi\rangle)\,,
	\end{equation}
	where $\vartheta_g$ is now regarded as acting on states rather than operators. Choosing
	these identifications for all $g\in \mathcal{G}$ trivializes the bundle of fibers over $\mathcal{G}$ and induces the
	explicit tensor product identification
	\begin{equation}
		\widehat{\mathcal{H}} \;\cong\; \mathcal{H}\otimes L^2(\mathcal{G})\,.
	\end{equation}
	Note, however, that this step depends on the choice of trivialization $\{W_g\}$ and therefore
	is not canonical. This will be important below.
	
	\subsection{Direct integral decomposition and edge mode wavefunctions}
	
	The description above in terms of $L^2(\mathcal{G},\mathcal{H})$ is naturally phrased as a direct
	integral decomposition of the extended Hilbert space:
	\begin{equation}
		\widehat{\mathcal{H}} 
		= \int_{\mathcal{G}}^{\oplus} \mathrm{d}\mu_{\mathcal{G}}(g)\,\mathcal{H}_g, \
		\mathcal{H}_g \cong \mathcal{H}\,.
	\end{equation}
	Each fiber $\mathcal{H}_g$ is a copy of the ``hard mode'' Hilbert space associated to a given value
	of the edge data $(\hat{\Gamma}^+_0,\hat{\Upsilon}^+_0)$, and a generic state is a square-integrable
	superposition
	\begin{equation}
		|\widehat{\Psi}\rangle 
		\;\sim\; \Bigl\{ g\mapsto |\Psi(g)\rangle\in\mathcal{H}\Bigr\}\,.
	\end{equation}
	In this language, the edge mode degrees of freedom are encoded in the dependence of the
	wavefunction on $g$, while the hard degrees of freedom live in the fiber Hilbert space
	$\mathcal{H}_g$.
	
	As mentioned in the main text, it is simpler to work in a reduced model where we keep only the $\ell=0$ spherical
	harmonics of the generators, so that the group $\mathcal{G}$ collapses to a finite-dimensional
	semidirect product
	\begin{equation}
		\mathcal{G} \;\cong\; \mathbb{R}_s \rtimes \mathbb{R}_u\,,
	\end{equation}
	generated by the uniform half-sided boost and translation parameters $s$ and $u$ appearing
	in \crefrange{eq:simplifiedextalgebra}{eq:simplifiedexthilbertspace}. In this case,
	\begin{equation}
		\widehat{\mathcal{H}} \;\cong\; \mathcal{H}\otimes L^2(\mathbb{R}_s)\otimes L^2(\mathbb{R}_u)\,,
	\end{equation}
	and it is natural to represent the conjugate edge operators as differential operators on
	$L^2(\mathbb{R}_s)\otimes L^2(\mathbb{R}_u)$, while the edge configurations
	$(\hat{\Gamma}^+_0,\hat{\Upsilon}^+_0)$ act as multiplicative operators. In this representation the
	Heisenberg-picture action of the generators $\hat{\mathscr{A}}$ and $\hat{\mathscr{P}}$ on the fiber $\mathcal{H}$
	is implemented by shifts in $(s,u)$ acting on the edge mode wavefunctions, in agreement
	with the commutation relations written in \crefrange{eq:translationcommutator}{eq:boostcommutator}.
	
	\subsection{Non-canonical nature of the hard/edge tensor product}
	
	Although \Eqref{eq:extended-H-space-appendix} suggests a simple tensor product
	factorization into ``hard'' and ``edge'' sectors,
	\begin{equation}
		\widehat{\mathcal{H}} \stackrel{?}{\cong} 
		\bigl(\mathcal{H}_{\text{hard}}\bigr)\otimes
		\bigl(\mathcal{H}_{\text{edge}}\bigr)\,,
	\end{equation}
	the crossed product structure implies that there is no canonical way to identify such a
	factorization at the level of the algebra.
	
	The key point is that the bulk algebra $\mathcal{A}_0$ does not act on
	$\widehat{\mathcal{H}}$ as $\mathcal{A}_0\otimes \mathbf{1}_{L^2(\mathcal{G})}$, but rather in the
	twisted, fiberwise fashion of \Eqref{eq:bulk-action-fiberwise}:
	\begin{equation}
		\bigl(\widehat{\pi}(\hat{\mathscr{O}})\Psi\bigr)(g) 
		= \vartheta_{g^{-1}}(\hat{\mathscr{O}})\,|\Psi(g)\rangle\,.
	\end{equation}
	In other words, the “same’’ operator $\hat{\mathscr{O}}\in\mathcal{A}_0$ is represented differently in each
	fiber $\mathcal{H}_g$, related by the automorphisms $\vartheta_g$.  The edge unitaries
	$\lambda(h)$, on the other hand, act by shifting the label $g$ as in
	\Eqref{eq:edge-left-regular}.  The crossed product relations precisely express the fact
	that the edge sector does not commute with the hard sector: it acts by conjugation on
	$\mathcal{A}_0$.
	
	From the algebraic point of view, $\widehat{\mathcal{A}}_{\mathscr{H}_{>u_0}}$ is not isomorphic to a
	simple tensor product $\mathcal{A}_0\otimes\mathcal{A}_{\text{edge}}$.  Instead, it is a
	semidirect product in which the edge algebra implements outer automorphisms of the hard
	algebra.  Consequently, there is no distinguished subalgebra of
	$\widehat{\mathcal{A}}_{\mathscr{H}_{>u_0}}$ that can be identified as “pure hard modes’’ and that
	commutes with a “pure edge’’ algebra.  Any such split requires a choice of trivialization
	$\{W_g\}$ of the direct integral and is therefore representation-dependent.
	
	In particular, writing
	\begin{equation}
		\widehat{\mathcal{H}} \;\cong\; \mathcal{H}\otimes L^2(\mathcal{G})
	\end{equation}
	amounts to choosing a specific identification of each fiber $\mathcal{H}_g$ with a fixed copy
	of $\mathcal{H}$, and hence a specific way of labeling excitations as “hard’’ versus “edge’’.
	Different choices of dressing of the bulk operators to the corner $S_0$ correspond to
	different trivializations and therefore to different, but equivalent, tensor product
	decompositions. What is invariant is the crossed product algebra
	$\widehat{\mathcal{A}}_{\mathscr{H}_{>u_0}}$ and its representation as a direct integral over edge
	configurations.
	
	This is the sense in which \Eqref{eq:extendedhilbertspace} should be understood: the extended Hilbert space
	carries a canonical direct integral representation over the edge data, and once a choice of
	trivialization is made this representation can be written as a tensor product
	$\mathcal{H}\otimes L^2(\mathcal{G})$.  However, the induced split into “hard’’ and “edge’’ factors is
	not canonical at the algebraic level, precisely because the crossed product structure ties
	together the action of the half-sided generators and the corner edge modes.
	
	This is the mathematical codification of background independence in perturbative quantum gravity.
	
	\section{Supertranslations and Type II\texorpdfstring{$_{\infty}$}{∞} algebras}
	\label{app:angle-dep-ST}
	
	In the previous parts of this appendix we considered a reduced finite-dimensional mini-superspace of edge modes, associated with the two-parameter affine group
	\begin{equation}
		\mathcal{G} \;\cong\; \mathbb{R}_s \rtimes \mathbb{R}_u,
	\end{equation}
	where \(s\) generates half-sided boosts and \(u\) generates half-sided translations along the horizon. The corresponding crossed product algebra
	\begin{equation}
		\widehat{\mathcal{A}}_{\mathscr{H}_{>u_0}} = \mathcal{A}_{\mathscr{H}_{>u_0}} \rtimes G
	\end{equation}
	is a Type~II\(_\infty\) factor with a canonical semifinite trace. Here \(\mathcal{A}\) denotes the ``bulk'' Type~III horizon algebra associated with the cut \(u=u_0\).
	
	In the full theory the relevant symmetry is the infinite-dimensional group of angle-dependent supertranslations, which act independently on each null generator of the horizon. In this subsection we sketch how the Type II$_\infty$ trace and algebraic von Neumann entropy constructions extend to this infinite-dimensional setting. The construction is slightly delicate because the supertranslation group is no longer finite dimensional or locally compact, so the crossed product and its trace must be defined via angular regulators and inductive limits.

	\subsection{Angle-dependent supertranslation group}
	
	Let \(x^A\) denote angular coordinates on the horizon cross-sections \(\mathbb{S}^{d-2}\). A general angle-dependent supertranslation is specified by a pair of functions
	\begin{equation}
		\beta(x^A),\; \alpha(x^A) \in C^\infty(\mathbb{S}^{d-2}),
	\end{equation}
	corresponding to angle-dependent boosts and translations
	\begin{equation}
		u \;\mapsto\; \mathrm{e}^{\beta(x^A)} u + \alpha(x^A)
	\end{equation}
	along each null generator. The associated group can be written as the semi-direct product
	\begin{equation}
		\mathcal{G}_{(\beta,\alpha)} = C^\infty_{\beta}(\mathbb{S}^{d-2}) \rtimes C^\infty_{\alpha}(\mathbb{S}^{d-2}),
	\end{equation}
	with group law given pointwise by the finite-dimensional affine structure. At each fixed angle \(x^A\), the pair \(\big(\beta(x^A),\alpha(x^A)\big)\) furnishes a copy of the original \(\mathcal{G} \cong \mathbb{R}_s\rtimes\mathbb{R}_u\) group action.
	
	On the quantum side, it is convenient to introduce local edge mode generators
    \begin{subequations}
	\begin{align}
		&\hat{\mathscr{A}}(u_0,f) = \int_{\mathbb{S}^{d-2}} d^{d-2}x\, f(x^A)\,\hat{\mu}(u_0,x^A),\\
		&\hat{\mathscr{P}}(u_0, g) = \int_{\mathbb{S}^{d-2}} d^{d-2}x\, g(x^A)\,\hat{\Pi}_q(u_0,x^A),
	\end{align}
        \end{subequations}
	where $\hat{\mu}(u_0,x^A)$ is the area density operator on the cut,
	\begin{equation}
		\hat{\mathscr{A}}_{\mathcal{S}(u_0)} = \int_\Sigma d^{d-2}x\,\hat{\mu}(u_0,x^A)
	\end{equation}
	is the area operator of a patch \(\mathcal{S} \subset \mathbb{S}^{d-2}\), and \(\hat{\Pi}_{\mu}\) is the canonical conjugate to $\mu$ (the null expansion operator in the linearized theory). The smearing functions \(f,g\) play the role of angle-dependent boost/translation parameters; for instance, smearing with \(f=\beta\) and \(g=\alpha\) gives
    \begin{subequations}
	\begin{align}
		&\hat{\mathscr{A}}(\beta) = \int d^{d-2}x\,\beta(x^A)\,\hat{\mu}(u_0,x^A),\\
		&\hat{\mathscr{P}}(\alpha) = \int d^{d-2}x\,\alpha(x^A)\,\hat{\Pi}_{\mu}(u_0,x^A),
	\end{align}
        \end{subequations}
	Infinitesimally, these operators generate automorphisms of the bulk algebra \(\mathcal{A}\),
	\begin{equation}
		\vartheta_{(\beta,\alpha)}(\mathscr{O})
		=
		\mathrm{e}^{i \left(\hat{\mathscr{A}}(\beta) + \hat{\mathscr{P}}(\alpha)\right)}\,\mathscr{O}\,
		\mathrm{e}^{-i\left(\hat{\mathscr{A}}(\beta) + \hat{\mathscr{P}}(\alpha)\right)},\ \mathscr{O}\in\mathcal{A},
	\end{equation}
	which generalize the action of \(\mathcal{G}\).
	
	In the main text we often write $\hat{\mathscr{A}}(u)$ for the (suitably normalized) area operator of the entire cross-section at affine parameter \(u\). In the angle-dependent setting the more fundamental object is the local area density \(\hat{\mu}(u,x^A)\). The various area operators \(\hat{\mathscr{A}}(f)\) or \(\hat{\mathscr{A}}_{\mathcal{S}}(u)\) are obtained by smearing this density with test functions or integrating over regions \(\mathcal{S}\). When we speak of an ``entropy density'' below we mean the integrand that appears when one writes \(S_{\text{gen}}(u;\mathcal{S})\) as an integral over \(\mathcal{S}\).
	
	Formally, the crossed product algebra generated by the bulk degrees of freedom and the angle-dependent supertranslations is
	\begin{equation}
		\widehat{\mathcal{A}}_{\mathscr{H}_{>u_0}}
		=
		\mathcal{A} \rtimes_\vartheta \mathcal{G}_{(\beta,\alpha)}.
	\end{equation}
	It is the von Neumann algebra generated by \(\mathcal{A}\) together with unitaries \(\lambda(g)\), \(g\in \mathcal{G}_{(\beta,\alpha)}\), subject to
	\begin{equation}
		\lambda(g)\,\mathscr{O}\,\lambda(g)^{-1} = \vartheta_g(\mathscr{O}),\ \mathscr{O}\in\mathcal{A}.
	\end{equation}
	In direct analogy with the finite-dimensional case, one expects a GNS representation on an extended Hilbert space
	\begin{equation}
		\widehat{\mathcal{H}}_{\partial G}
		\;\cong\;
		L^2\!\big(\mathcal{G}_{(\beta,\alpha)},d\mu_{\mathcal{G}_{(\beta,\alpha)}};\mathcal{H}\big),
	\end{equation}
	where \(\mathcal{H}\) carries the original representation of \(\mathcal{A}\) and \(\mu_{\mathcal{G}_{(\beta,\alpha)}}\) is a formal generalization of the Haar measure on \(\mathcal{G}_{(\beta,\alpha)}\). However, \(\mathcal{G}_{(\beta,\alpha)}\) is infinite dimensional and not locally compact, so there is no honest Haar measure. To make this construction precise we introduce an angular ultraviolet regulator and then pass to an inductive limit.

	\subsection{Angular mode cutoff and inductive limit}
	
	Let \(\{Y_{\ell m}(x^A)\}\) be an orthonormal basis of spherical harmonics on \(\mathbb{S}^{d-2}\). For a fixed cutoff \(\ell_{\max}\), define the finite-dimensional subspace
	\begin{equation}
		\mathcal{V}_{\ell_{\max}}
		=
		\mathrm{span}\big\{Y_{\ell m} \,\big|\, 0\le\ell\le\ell_{\max}\big\}
		\;\subset\; C^\infty(\mathbb{S}^{d-2}).
	\end{equation}
	Restricting the parameters \(\beta(x^A),\alpha(x^A)\) to \(\mathcal{V}_{\ell_{\max}}\) gives a finite-dimensional approximation of the supertranslation group,
	\begin{equation}
		\mathcal{G}_{(\beta,\alpha)}^{(\ell_{\max})}
		=
		\mathcal{V}_{\ell_{\max}}^{(s)} \rtimes \mathcal{V}_{\ell_{\max}}^{(u)}
		\;\cong\;
		(\mathbb{R}_s \rtimes \mathbb{R}_u)^{N_{\ell_{\max}}},
	\end{equation}
	where
	\begin{equation}
		N_{\ell_{\max}} := \sum_{\ell=0}^{\ell_{\max}} (2\ell+1)
	\end{equation}
	is the number of angular modes. At this level \(\mathcal{G}_{(\beta,\alpha)}^{(\ell_{\max})}\) is a finite-dimensional, locally compact Lie group with a well-defined Haar measure \(d\mu_{\ell_{\max}}\).
	
	We may then form the finite-mode crossed product
	\begin{equation}
		\widehat{\mathcal{A}}_{\mathscr{H}_{>u_0}}^{(\ell_{\max})}
		=
		\mathcal{A} \rtimes_\vartheta \mathcal{G}_{(\beta,\alpha)}^{(\ell_{\max})},
	\end{equation}
	represented on
	\begin{equation}
		\widehat{\mathcal{H}}^{(\ell_{\max})}
		=
		L^2\!\big(\mathcal{G}_{(\beta,\alpha)}^{(\ell_{\max})}, d\mu_{\ell_{\max}}; \mathcal{H}\big)
		\;\cong\;
		\bigotimes_{n=1}^{N_{\ell_{\max}}} L^2\!\big(\mathbb{R}_s \rtimes \mathbb{R}_u, d\mu_{ax+b}; \mathcal{H}\big),
	\end{equation}
	where \(d\mu_{ax+b}\) is the Haar measure on the single-mode affine group \(\mathbb{R}_s \rtimes \mathbb{R}_u\). From the discussion earlier in the appendix, each factor yields a Type~II\(_\infty\) algebra with its own semifinite trace, and the tensor product carries a canonical semifinite trace
	\begin{equation}
		\mathrm{tr}_{\ell_{\max}} : \left(\widehat{\mathcal{A}}_{\mathscr{H}_{>u_0}}^{(\ell_{\max})}\right)_+ \to [0,\infty]
	\end{equation}
	given as the product of the single-mode traces. Here $\left(\widehat{\mathcal{A}}_{\mathscr{H}_{>u_0}}^{(\ell_{\max})}\right)_+$ refers to the positive cone of $\widehat{\mathcal{A}}_{\mathscr{H}_{>u_0}}^{(\ell_{\max})}$.
	
	The full angle-dependent algebra is obtained as the inductive limit
	\begin{equation}
		\widehat{\mathcal{A}}_{\mathscr{H}_{>u_0}}
		=
		\overline{\bigcup_{\ell_{\max}} \widehat{\mathcal{A}}_{\mathscr{H}_{>u_0}}^{(\ell_{\max})}}^{\;\mathrm{WOT}},
	\end{equation}
	taken in the weak operator topology (WOT). For any operator \(\hat{\mathscr{O}}\) that involves only finitely many angular modes, the expectation values and traces stabilize at sufficiently large \(\ell_{\max}\), and the limit
	\begin{equation}
		\mathrm{tr}(\hat{\mathscr{O}})
		=
		\lim_{\ell_{\max}\to\infty} \mathrm{tr}_{\ell_{\max}}(\hat{\mathscr{O}})
	\end{equation}
	exists and is independent of the details of the regulator. This defines a canonical semifinite trace on a dense $*$-subalgebra of $\widehat{\mathcal{A}}_{\mathscr{H}_{>u_0}}$.
	
	In addition to the ultraviolet cutoff in \(\ell\), it is physically natural to localize in angle. Let \(\mathcal{S} \subset \mathbb{S}^{d-2}\) be a measurable subset of the cross-section with finite area, as before. We define:
	
	\begin{itemize}
		\item The localized bulk algebra \(\mathcal{A}(\mathcal{S})\) as the von Neumann algebra generated by dressed bulk operators supported on \(\mathscr{H}_{>u_0}\) and smeared with test functions that vanish outside \(\mathcal{S}\).
		
		\item The localized supertranslation subgroup \(\mathcal{G}_{(\beta, \alpha)}(\mathcal{S})\subset \mathcal{G}_{(\beta,\alpha)}\) consisting of pairs \((\beta,\alpha)\) with \(\beta(x^A)=\alpha(x^A)=0\) for \(x^A \notin \mathcal{S}\).
	\end{itemize}
	
	We then form the localized crossed product
	\begin{equation}
		\widehat{\mathcal{A}}_{\mathscr{H}_{>u_0}}(\mathcal{S})
		=
		\mathcal{A}(\mathcal{S}) \rtimes_\vartheta \mathcal{G}_{(\beta,\alpha)}(\mathcal{S}).
	\end{equation}
	With the angular mode cutoff in place, $\widehat{\mathcal{A}}_{\mathscr{H}_{>u_0}}(\mathcal{S})$ is a crossed product by a finite-dimensional group and is thus a Type~II\(_\infty\) factor with a canonical semifinite trace \(\mathrm{tr}_{\ell_{\max},\mathcal{S}}\). Passing to the inductive limit defines a semifinite trace
	\begin{equation}
		\mathrm{tr}_{\mathcal{S}} : \big(\widehat{\mathcal{A}}_{\mathscr{H}_{>u_0}}(\mathcal{S})\big)_+ \to [0,\infty]
	\end{equation}
	on the localized algebra.
	
	\subsection{Entropy and generalized entropy density}
	
	Given a normal state \(\omega\) on $\widehat{\mathcal{A}}_{\mathscr{H}_{>u_0}}(\mathcal{S})$, the trace \(\mathrm{tr}_\mathcal{S}\) defines a density operator $\hat{\rho}_\omega(u;\mathcal{ S})\in\widehat{\mathcal{A}}_{\mathscr{H}_{>u_0}}(\mathcal{S})$ by
	\begin{equation}
		\mathrm{tr}_\mathcal{S}\!\big(\hat{\rho}_\omega(u;\mathcal{S})\,\hat{\mathscr{O}}\big)
		= \omega(\hat{\mathscr{O}}),\ \hat{\mathscr{O}}\in\widehat{\mathcal{A}}_{\mathscr{H}_{>u_0}}(\mathcal{S}).
	\end{equation}
	The associated Type~II$_{\infty}$ von Neumann entropy is
	\begin{equation}
		S_\omega\big(u;\widehat{\mathcal{A}}_{\mathscr{H}_{>u_0}}(\mathcal{S})\big)
		:= -\mathrm{tr}_\mathcal{S}\!\Big(\hat{\rho}_\omega(u;\mathcal{S})\,\log\hat{\rho}_{\omega}(u;\mathcal{S})\Big).
	\end{equation}
	
	The analysis of \cref{sec:genentropyderiv} carries over to the angle-dependent case in a patchwise fashion. In particular, under the same assumptions (perturbative regime, local KMS property, nesting of algebras, and sharply peaked edge mode wavefunctionals), the modular Hamiltonian of \(\hat{\rho}_\omega(u;\mathcal{S})\) takes the form
	\begin{equation}
		\log\hat{\rho}_\omega(u;\mathcal{S})
		\approx
		-\beta\,\hat{\mathscr{A}}_\mathcal{S}(\infty)
		+ \hat{h}_{\Omega|\Psi}(u;\mathcal{S})
		- \hat{h}_\Omega(\infty;\mathcal{S})
		- \hat{h}_\Psi(u;\mathcal{S}),
	\end{equation}
	where \(\hat{\mathscr{A}}_\mathcal{S}(u)\) is the area operator localized to \(\mathcal{S}\),
	\begin{equation}
		\hat{\mathscr{A}}_\mathcal{S}(u)
		:= \int_\mathcal{S} d^{d-2}x\,\hat{\mu}(u,x^A),
	\end{equation}
	and the \(\hat{h}\)'s are the appropriate Connes cocycles restricted to \(\mathcal{S}\). From this, one finds
	\begin{equation}
		S_\omega\big(u;\widehat{\mathcal{A}}_{\mathscr{H}_{>u_0}}(\mathcal{S})\big)
		\;\approx\;
		\int_\mathcal{S} d^{d-2}x
		\left[
		\frac{\langle\hat{\mu}(u,x^A)\rangle_\omega}{4 G_N}
		+ s_{\text{bulk}}(u,x^A;\omega)
		\right],
	\end{equation}
	up to the same state-independent constants and mild smearing in \(u\) discussed in the main text. Here \(s_{\text{bulk}}(u,x^A;\omega)\) is the local bulk entropy density at angle \(x^A\).
	
	In other words, the Type~II\(_\infty\) von Neumann entropy of the angle-dependent crossed product algebra \(\widehat{\mathcal{A}}_{\mathscr{H}_{>u_0}}(\mathcal{S})\) coincides with the generalized entropy of the horizon localized to the patch \(\mathcal{S}\) and averaged over shifts $\Delta u$ in the position $u_0$ of the cut:
	\begin{equation}
		S_\omega\big(u;\widehat{\mathcal{A}}_{\mathscr{H}_{>u_0}}(\mathcal{S})\big)
		\;\approx\;
		\bar{S}_{\text{gen}}(u;\mathcal{S},\omega),
	\end{equation}
	where \(S_{\text{bulk}}(u;\mathcal{S},\omega)\) is the ordinary bulk von Neumann entropy associated with the restriction of \(\omega\) to the bulk algebra in the causal development of \(\mathcal{S}\). The expression inside the angular integral can be interpreted as a generalized entropy density, built from the local area density operator \(\hat{\mu}(u,x^A)\) and the bulk entropy density.
	
	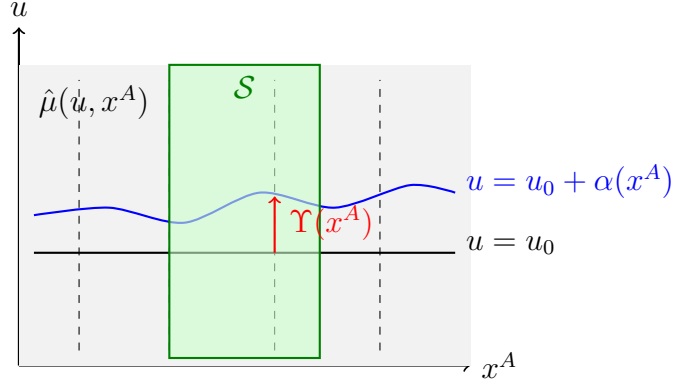
\begin{figure}[t]
		\centering
		\begin{tikzpicture}[scale=1.0]
			\draw[->,thick] (0,0) -- (0,4.5) node[above] {$u$};
			\draw[->,thick] (0,0) -- (6,0) node[right] {$x^A$};
			
			\fill[gray!10] (0,0) rectangle (6,4);
			
			\draw[thick] (0.2,1.5) -- (5.8,1.5);
			\node[anchor=west] at (5.8,1.6) {$u = u_0$};
			
			\draw[thick,blue]
			plot[smooth] coordinates {
				(0.2,2.0)
				(1.2,2.1)
				(2.2,1.9)
				(3.2,2.3)
				(4.2,2.1)
				(5.2,2.4)
				(5.8,2.3)
			};
			\node[anchor=west,blue] at (5.8,2.45) {$u = u_0 + \alpha(x^A)$};
			
			\foreach \x in {0.8,2.0,3.4,4.8} {
				\draw[dashed] (\x,0.2) -- (\x,3.8);
			}
			
			\fill[green!20,opacity=0.6] (2.0,0.1) -- (4.0,0.1) -- (4.0,4.0) -- (2.0,4.0) -- cycle;
			\draw[green!50!black,thick] (2.0,0.1) -- (4.0,0.1) -- (4.0,4.0) -- (2.0,4.0) -- cycle;
			\node[green!50!black] at (3.0,3.7) {$\mathcal{S}$};
			
			\draw[red,->,thick] (3.4,1.5) -- (3.4,2.25);
			\node[red,anchor=west] at (3.45,1.9) {$\Upsilon(x^A)$};
			
			\node[black] at (1.0,3.5) {$\hat{\mu}(u,x^A)$};
			
		\end{tikzpicture}
		\caption{Schematic depiction of angle-dependent supertranslations and local patches on the horizon. The horizontal axis labels the null generators by angle \(x^A\); the vertical axis is the affine parameter \(u\). A reference cut \(u=u_0\) (black) is shifted by an angle-dependent amount \(\alpha(x^A)\) (blue), generated by the edge mode field \(\Upsilon(x^A)\) along each generator. A finite angular patch \(\mathcal{S}\) is shown in green; the crossed product algebra \(\widehat{\mathcal{A}}_{\mathscr{H}_{>u_0}}(\Sigma)\) associated with this patch is a Type~II\(_\infty\) factor with a canonical trace \(\mathrm{tr}_\mathcal{S}\), whose von Neumann entropy reproduces the generalized entropy localized to \(\mathcal{S}\). The local area density operator \(\hat{\mu}(u,x^A)\) integrates over \(\mathcal{S}\) to give the area operator \(\hat{\mathscr{A}}_\mathcal{S}(u)\).}
		\label{fig:angle-dependent-ST}
	\end{figure}
	
	\section{Monotonicity of algebraic entropy under nesting\label{sec:montonocityentropy}}
	
	We now justify that the algebraic entropy
	\begin{equation}
		S(\hat{\psi};\widehat{\mathcal{M}}_{\mathscr{H}_{>u}})
		= - \operatorname{tr}\!\big[\rho_{\hat{\psi}}(u)\,\log\rho_{\hat{\psi}}(u)\big]
		\label{eq:typeII-entropy-def}
	\end{equation}
	is monotone under nesting of the Type~II$_\infty$ horizon algebra.
	
	Let $(\widehat{\mathcal{M}},\operatorname{tr})$ be a Type~II$_\infty$ factor equipped with a faithful
	normal semifinite trace $\operatorname{tr}$, and let
	\(\widehat{\mathcal{N}}\subset\widehat{\mathcal{M}}\) be a von Neumann subalgebra.  
	Given a normal state $\omega$ on $\widehat{\mathcal{M}}$ with finite entropy, there exists a unique
	density operator $\rho\in \widehat{\mathcal{M}}$ such that
	\begin{equation}
		\omega(\hat{\mathscr{O}}) = \operatorname{tr}(\rho\,\hat{\mathscr{O})}, \
		\hat{\mathscr{O}}\in\widehat{\mathcal{M}},
	\end{equation}
	with $\rho\ge 0$ and $\operatorname{tr}(\rho)=1$.  
	The restriction $\omega|_{\mathcal{N}}$ is again normal, and can be written in the same
	form using a density operator $\rho_{\mathcal{N}}\in\widehat{\mathcal{N}}$:
	\begin{equation}
		\omega|_{\mathcal{N}}(\hat{\mathscr{O}}) = \operatorname{tr}(\rho_{\mathcal{N}}\,\hat{\mathscr{O}}), \ \hat{\mathscr{O}}\in\widehat{\mathcal{N}}.
	\end{equation}
	
	Because $\operatorname{tr}$ is tracial, there exists a unique normal
	$\operatorname{tr}$–preserving conditional expectation
	\begin{equation}
		E_{\mathcal{N}}\colon\widehat{\mathcal{M}}\mapsto\widehat{\mathcal{N}},
	\end{equation}
	characterized by
	\begin{equation}
		\operatorname{tr}\!\big(E_{\mathcal{N}}(\hat{X})\,\hat{\mathscr{O}}\big)
		= \operatorname{tr}(\hat{X}\,\hat{\mathscr{O}}), \ \hat{X}\in\widehat{\mathcal{M}},~\hat{\mathscr{O}}\in\widehat{\mathcal{N}}.
	\end{equation}
	In particular,
	\begin{equation}
		\rho_{\mathcal{N}} = E_{\mathcal{N}}(\rho),
	\end{equation}
	since both sides implement the same restricted state.
	
	Define the convex function
	\begin{equation}
		\varphi(t) = t\log t, \ t>0.
	\end{equation}
	It is a standard fact that $\varphi$ is operator convex on $(0,\infty)$.
	Jensen's operator inequality for the unital completely positive map
	$E_{\mathcal{N}}$ then gives
	\begin{equation}
		\varphi\!\big(E_{\mathcal{N}}(\rho)\big)
		\;\leq\; E_{\mathcal{N}}\!\big(\varphi(\rho)\big).
	\end{equation}
	Applying the trace and using $\operatorname{tr}\circ E_{\mathcal{N}}=\operatorname{tr}$,
	we obtain
	\begin{equation}
		\operatorname{tr}\!\big(\rho_{\mathcal{N}}\log\rho_{\mathcal{N}}\big)
		= \operatorname{tr}\!\big[\varphi(E_{\mathcal{N}}(\rho))\big]
		\;\leq\; \operatorname{tr}\!\big[E_{\mathcal{N}}(\varphi(\rho))\big]
		= \operatorname{tr}\!\big(\rho\log\rho\big).
	\end{equation}
	Thus the entropies
	\begin{equation}
		S(\omega;\widehat{\mathcal{M}}) := -\operatorname{tr}(\rho\log\rho),\ 
		S(\omega;\widehat{\mathcal{N}}) := -\operatorname{tr}(\rho_{\mathcal{N}}\log\rho_{\mathcal{N}})
	\end{equation}
	satisfy
	\begin{equation}
		S(\omega;\widehat{\mathcal{N}}) \;\geq\; S(\omega;\widehat{\mathcal{M}}).
		\label{eq:entropy-monotone-inclusion}
	\end{equation}
	In words: restricting a state from $\widehat{\mathcal{M}}$ to a subalgebra
	$\widehat{\mathcal{N}}\subset\widehat{\mathcal{M}}$ can only increase the algebraic von Neumann
	entropy defined with respect to the trace.
	
	Specializing to the one-parameter family of Type~II$_\infty$ horizon
	algebras~$\widehat{\mathcal{M}}_{\mathscr{H}_{>u}}$ constructed above, the isotony property
	\begin{equation}
		U(\delta u)\,\widehat{\mathcal{M}}_{\mathscr{H}_{>u_0}}\,U(-\delta u)\subset
		\widehat{\mathcal{M}}_{\mathscr{H}_{>u_0}}, \ \delta u\ge 0,
	\end{equation}
	together with \Eqref{eq:entropy-monotone-inclusion} implies that
	\begin{equation}
		S\big(\widehat{\psi};\widehat{\mathcal{M}}_{\mathscr{H}_{>u}}\big)
		\;\geq\; S\big(\widehat{\psi};\widehat{\mathcal{M}}_{\mathscr{H}_{>u_0}}\big),
	\end{equation}
	so the horizon entropy \eqref{eq:typeII-entropy-def} is monotone
	non-decreasing along a one-parameter family of nested subalgebras.

	\bibliographystyle{JHEP}
	
	\bibliography{hamiltonians-null-surfaces,asycps,softhair}
	
\end{document}